\documentclass{iopart}
\usepackage{iopams}
\usepackage{braket,amsmath,color,graphicx,cite}
\usepackage[margin=2cm]{geometry}
\usepackage[colorlinks,linkcolor={blue},citecolor={blue}]{hyperref}
\allowdisplaybreaks
\begin{document}

\title{Kondo frustration via charge fluctuations: a route to Mott localisation}

\author{Abhirup Mukherjee$^1$, N. S. Vidhyadhiraja$^2$, A. Taraphder$^3$ and Siddhartha Lal$^1$}
\eads{\mailto{am18ip014@iiserkol.ac.in}, \mailto{raja@jncasr.ac.in}, \mailto{arghya@phy.iitkgp.ernet.in}, \mailto{slal@iiserkol.ac.in}}

\address{$^1$Department of Physical Sciences, Indian Institute of Science Education and Research-Kolkata, W.B. 741246, India}

\address{$^2$Theoretical Sciences Unit, Jawaharlal Nehru Center for Advanced Scientific Research, Jakkur, Bengaluru 560064, India}

\address{$^3$Department of Physics, Indian Institute of Technology Kharagpur, Kharagpur 721302, India}

\date{\today}

\begin{abstract}
We propose a minimal effective impurity model that captures the phenomenology of the Mott-Hubbard metal-insulator transition (MIT) of the half-filled Hubbard model on the Bethe lattice in infinite dimensions as observed by dynamical mean field theory (DMFT). This involves extending the standard Anderson impurity model Hamiltonian to include an explicit Kondo coupling $J$, as well as a local on-site correlation \(U_b\) on the conduction bath site connected directly to the impurity. For the case of attractive local bath correlations ($U_{b}<0$), the extended Anderson impurity model (e-SIAM) 
sheds new light on several aspects of the DMFT phase diagram. For example, the $T=0$ metal to insulator quantum phase transition (QPT) is preceded by 
an excited state quantum phase transition (ESQPT) where the local moment eigenstates are emergent in the low-lying spectrum.
Long-ranged fluctuations are observed near both the QPT and ESQPT, suggesting that they are the origin of the quantum critical scaling observed recently at high temperatures in DMFT simulations.
The \(T=0\) gapless excitations at the QCP display particle-hole interconversion processes, and exhibit power-law behaviour in self-energies and two-particle correlations. These are signatures of non-Fermi liquid behaviour that emerge from the partial breakdown of the Kondo screening.
\end{abstract}

\maketitle

\section{Introduction}

The rich physics of metal-insulator transitions in strongly-correlated systems has been an active subject of study for quite some time~\cite{Mott_RMP_1968,milligan_1985,imada1998metal}, yet much still remains to be understood. It involves diverse aspects such as spin and charge fluctuations, quasiparticle renormalisation effects, anomalous metallic phases and unconventional superconductivity, and has been studied using an equally diverse array of methods like mean-field theory, renormalisation group approaches, numerical techniques like exact diagonalisation, quantum Monte Carlo and dynamical mean-field theory, and many others. In particular, dynamical mean-field theory (DMFT)~\cite{kuramoto1987,Cox1988,metzner_volhardt_1989,zhang_1993,georges1996,parcollet_2004,maier_2005,kotliar_rmp_2006,ohashi_2008} obtains an exact solution of the Mott metal-insulator transition (MIT) 1/2-filled Hubbard model~\cite{Mott_1949,gutzwiller_1963,kanamori_1963,hubbard1963electron,brinkman_rice_1970} on the Bethe lattice with infinite coordination number, in terms of an Anderson impurity model with a self-consistently determined bath 
obtained by requiring, in an iterative manner, that its local Greens function be equal to that of the impurity site. The above-mentioned transition can be captured by the local spectral function through (i) the continuous appearance of a Mott gap, followed by (ii) the sharpening and vanishing of the central Kondo resonance. These two features are often referred to, respectively, as the Mott-Hubbard~\cite{hubbard1963electron} and Brinkman-Rice~\cite{brinkman_rice_1970} scenarios of the Mott MIT. The exact nature of the solution arises from the fact that all non-local contributions to the lattice self-energy are observed to vanish upon taking the limit of an infinite coordination number for the lattice model. 
Thus, the simplification is that the dynamics of any local site on the lattice is determined completely by a quantum impurity problem~\cite{georges1996,Logan_2016}.
It must also be noted that this exact solution precludes any long-range order in the system, and corresponds to the case of a maximally frustrated Hubbard model involving long-range and frustrating inter-site hopping such that both the metallic and insulating phases remain paramagnetic~\cite{vucicevic_2013}. Due to the exact and non-perturbative nature of the DMFT solution for the MIT in \(d=\infty\), the method has been extended to models of strongly correlated electrons in finite spatial dimensions~\cite{park2008,rohringer_2018,maier_2005}, as well as the study of the electronic properties of various correlated materials ~\cite{lichenstein_1998,kotliar_2006,held_2007_ldadmft}.

Despite this progress, a key aspect of the DMFT solution for the Hubbard model in $d=\infty$ remains to be understood. During the search for a self-consistent impurity model, the conduction bath is modified drastically in order to become correlated~\cite{held2008}. 
The numerical implementation of self-consistency, however, precludes a deeper understanding of the precise nature of the correlations present in the conduction bath of the impurity model, and its implications for the electron dynamics of the associated bulk lattice (Hubbard) model. 
Below, we lay out the specific questions addressed by us, and summarise our results at the end of this section.\\[5pt]
{\bf i.}~Is there a minimal but effective quantum impurity model Hamiltonian that describes the Mott MIT of the $1/2$-filled Hubbard model on the Bethe lattice in \(d=\infty\)?\\[5pt]
{\bf ii.}~What are the fluctuations that destroy the metal and lead to the insulating phase? Can we obtain a universal theory for these competing tendencies?\\[5pt]
{\bf iii.}~The coexistence of metallic and insulating phases at $T=0$ within DMFT shows that the insulating solution is present within the many-body spectrum of the metallic phase. Can an associated impurity model Hamiltonian display the emergence of the insulating state prior to the transition?\\[5pt]
{\bf iv.}~Does this explain the spinodals and the first-order line obtained at $T>0$ in DMFT? What is the origin of the quantum critical fluctuations observed recently above the finite temperature second-order critical point~\cite{terletska_mott_2011,vucicevic_2013}?\\[5pt]
{\bf v.}~Is it possible to obtain a low-energy theory for the local gapless excitations precisely at the MIT, where the metal is on the brink of destruction? How do these excitations compare with those of the local Fermi liquid, e.g., in terms of self-energies and two-particle correlation functions?\\
\par
The essence of our approach is to model phenomenologically the lattice self-energy obtained from DMFT in the form of additional bath correlations within an extended Anderson impurity model.
In addition to the usual on-site repulsion ($U$) and single-particle hybridisation ($V$) between the impurity and the conduction bath of the Anderson impurity model (eq.~\eqref{basic-siam}), we introduce (i) an additional on-site correlation ($U_{b}$) on the bath site with which the impurity couples, and (ii) an antiferromagnetic Kondo coupling ($J$) between the impurity and the conduction bath (eq.~\eqref{GIAM-ham}). 
We note that a similar impurity model-based approach was taken towards understanding the physics of the heavy fermions several years ago by Si and Kotliar~\cite{si_kotliar_1993,kotliarsi_1993}. 
We postpone a comparison of our work with theirs to the discussions section. The rest of the work is structured as follows. Sec.~\eqref{def-ham} describes the extended model that we will study, and the unitary renormalisation group (URG) method that we employ to study it is presented in Sec.~\eqref{section-urg}. In Sec.~\eqref{urg-theory} and \eqref{desc-mit}, we describe the phase diagram and various characteristics of the impurity phase transition. In Sec.~\eqref{dmft}, we use our extended model to explain various features of the coexistence region observed in DMFT. In Sec.~\eqref{excitations}, we describe the effect of the impurity on the low-lying excitations of the bath, near and at the transition. We conclude in Sec.~\eqref{concl} with some discussions and possible future directions. For the convenience of readers, we first present below a brief summary of our main results.

\subsection*{\bf Summary of our main results}

\begin{itemize}
\item {\it Presence of a local metal-insulator transition}: At a critical value of the parameter $r=-U_{b}/J$, the effective impurity model shows a transition from a Kondo screened phase into an unscreened local moment phase. The quantum critical point (QCP) involves a degeneration of the Kondo singlet and the local moment states.
\item {\it The physics of Kondo screening and local pairing drives the transition:}
The transition involves the frustration of the Kondo screening of the impurity by enhanced local pairing fluctuations in the bath, and can be described by a universal theory written in terms of $J$ and $U_b$.
\item {\it Emergence of insulating solutions in the metallic phase:}
Our analysis reveals that at a certain value of the parameter $r$ prior to the transition, the single-particle hybridisation parameter ($V$) turns irrelevant (in the RG sense), and this leads to the emergence of the local moment solutions within the many-particle spectrum through an excited state quantum phase transition (ESQPT).
\item {\it Critical fluctuations and the coexistence region}: 
We observe the appearance of long-ranged quantum fluctuations extending into the conduction bath in the vicinity of both the ESQPT at $r=r_{c1}$ and QPT at $r=r_{c2}$. We believe that these are the likely origin of the critical fluctuations observed above the finite temperature second-order critical point in DMFT~\cite{terletska_mott_2011,vucicevic_2013}. The two-step process at $T=0$ also provides a natural explanation for the coexistence of metallic and insulating features in the phase diagram, in the regime $r_{c1}<r<r_{c2}$.
\item {\it Emergence of non-Fermi liquid excitations at the QCP}:
Precisely at the QCP, the local  Fermi liquid is replaced by a quasi-local non-Fermi liquid (NFL) that spans the impurity, zeroth and first sites of the conduction bath. The NFL results from a degeneracy between the local moment and singlet states, and leads to (i) ``Andreev scattering" of incoming states into orthogonal outgoing states, (ii) anomalous power-law behaviour in the self-energies and two-particle correlations with universal exponents, and (iii) a fractional entanglement entropy of the impurity.
\item {\it Correlated Fermi liquid excitations in the Hubbard sidebands}:
A many-body perturbation theoretic treatment of the Hubbard sidebands reveals that they are comprised of the holon-doublon excitations created by the hybridisation of the impurity site with the conduction bath. These excitations consist of decoupled local Fermi liquids for the holons and doublons at the lowest order, which, at higher orders, become coupled via correlated holon-doublon scattering between impurity and bath.
\end{itemize}

\section{Model Hamiltonian and method}

\subsection{The extended Anderson impurity model}
\label{def-ham}

The single-impurity Anderson model (SIAM)~\cite{anderson_1961,anderson_1978} consists of a single impurity site with local repulsive correlation \(U\) hybridising with a non-interacting fermionic conduction bath through a (momentum-independent) single-particle transfer whose coupling is \(V\). For the case of a half-filled impurity site, the Hamiltonian of the SIAM is given by
\begin{eqnarray} 
	\mathcal{H}_\text{A} = -\frac{U}{2} \left(\hat n_{d \uparrow} - \hat n_{d \downarrow}\right)^2 + \sum_{\vec k,\sigma} \epsilon_{\vec k} \tau_{\vec k,\sigma} + V\sum_\sigma \left(c^\dagger_{d\sigma}c_{0\sigma} + \text{h.c.}\right),
	\label{basic-siam}
\end{eqnarray}
where \(\tau_{\vec k,\sigma} \equiv c^\dagger_{\vec k,\sigma}c_{\vec k,\sigma} - 1/2\) indicate the occupancy of the single-particle momentum state \(\ket{\vec k}\). Also, \(c_{d\sigma}\) and \(c_{0\sigma} = \sum_k c_{k\sigma}\) are the fermionic annihilation operators of spin \(\sigma\) for the impurity and conduction bath site to which it couples (henceforth referred to as the {\it zeroth site}) respectively.  The conduction bath is typically considered to possess a constant (i.e., energy-independent) density of states. 

The SIAM (along with its \(U \to \infty\) limit, the Kondo model) has been studied using several analytical and numerical techniques~\cite{anderson1969exact,anderson1970exact,anderson1970,haldane1978scaling,jefferson_1977,wilson1975,hrk_wilson_1980,andrei_1980,andreiKondoreview,Wiegmann_1981,tsvelickKondoreview,kotliar_1996,Duki_2011,borda_2008,streib_2013,anirban_kondo}. 
The general conclusion for the positive \(U\) case at \(T=0\) is that on the particle-hole symmetric (that is, half-filled) line, the impurity local moment is always screened by the conduction electrons (referred to as the Kondo cloud~\cite{sorensen_erik_affleck_1996,affleck_ian_2001,simon_pascal_2003,martin2010,martin2019,Goldhaber-Gordon1998,Cronenwett1998,Schmid_Weis1998,pustilnik_glazman_2004,Borzenets2020,neel_berndt_2008,Zhao2005}). Enhanced spin-flip scattering at low-energies leads to the formation of a macroscopic singlet ground state and local Fermi liquid gapless excitations~\cite{nozaki2012,mora_2015}.
In order to enhance the SIAM, we introduce two extra two-particle interaction terms into the Hamiltonian:
\begin{itemize}
	\item a spin-exchange term \(J \vec{S}_d\cdot\vec{S}_0\) between the impurity spin \(\vec S_d\) and the spin \(\vec S_0\) of the zeroth site, and 
	\item a local particle-hole symmetric correlation term \(-U_b \left(\hat n_{0 \uparrow} - \hat n_{0 \downarrow}\right)^2\) on the bath zeroth site.
\end{itemize}

With these additional terms, the Hamiltonian of the {\it extended single-impurity Anderson model} (henceforth referred to as the e-SIAM) is, at particle-hole symmetry, given by
\begin{eqnarray}
	\label{GIAM-ham}
	\mathcal{H}_\text{E-A} = \mathcal{H}_\text{A} + J \vec{S}_d\cdot\vec{S}_0 - \frac{1}{2}U_b \left(\hat n_{0 \uparrow} - \hat n_{0 \downarrow}\right)^2~.
\end{eqnarray}
All the terms in the Hamiltonian have been depicted schematically in the left panel of Fig.~\eqref{zeromode-bare}. The additional interaction terms \(J\) and \(U_b\) enjoy particle-hole, SU(2)-spin and U(1)-charge symmetries. The e-SIAM Hamiltonian (eq.~\eqref{GIAM-ham}) therefore preserves all the local symmetries of the half-filled Hubbard model on a lattice, and can potentially serve as an effective auxiliary quantum impurity model (with a correlated bath) describing the local physics of the latter.

\subsection{The unitary renormalisation group method}\label{section-urg}
\begin{figure}[!htb]
	$\vcenter{\hbox{\includegraphics[width=0.45\textwidth]{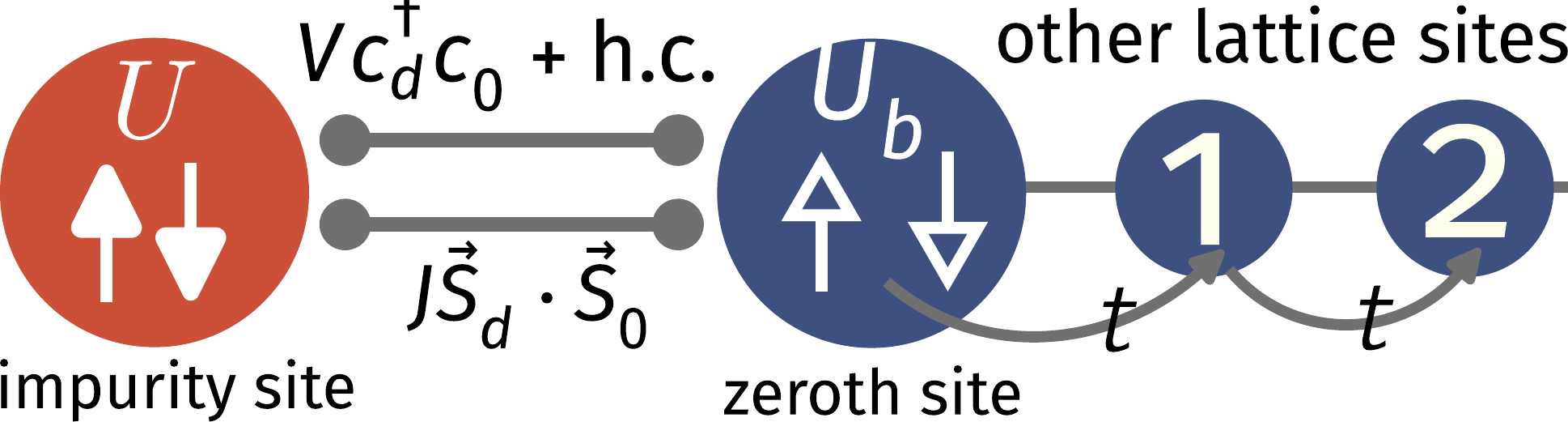}}}$
	\hspace*{\fill}
	$\vcenter{\hbox{\includegraphics[width=0.3\textwidth]{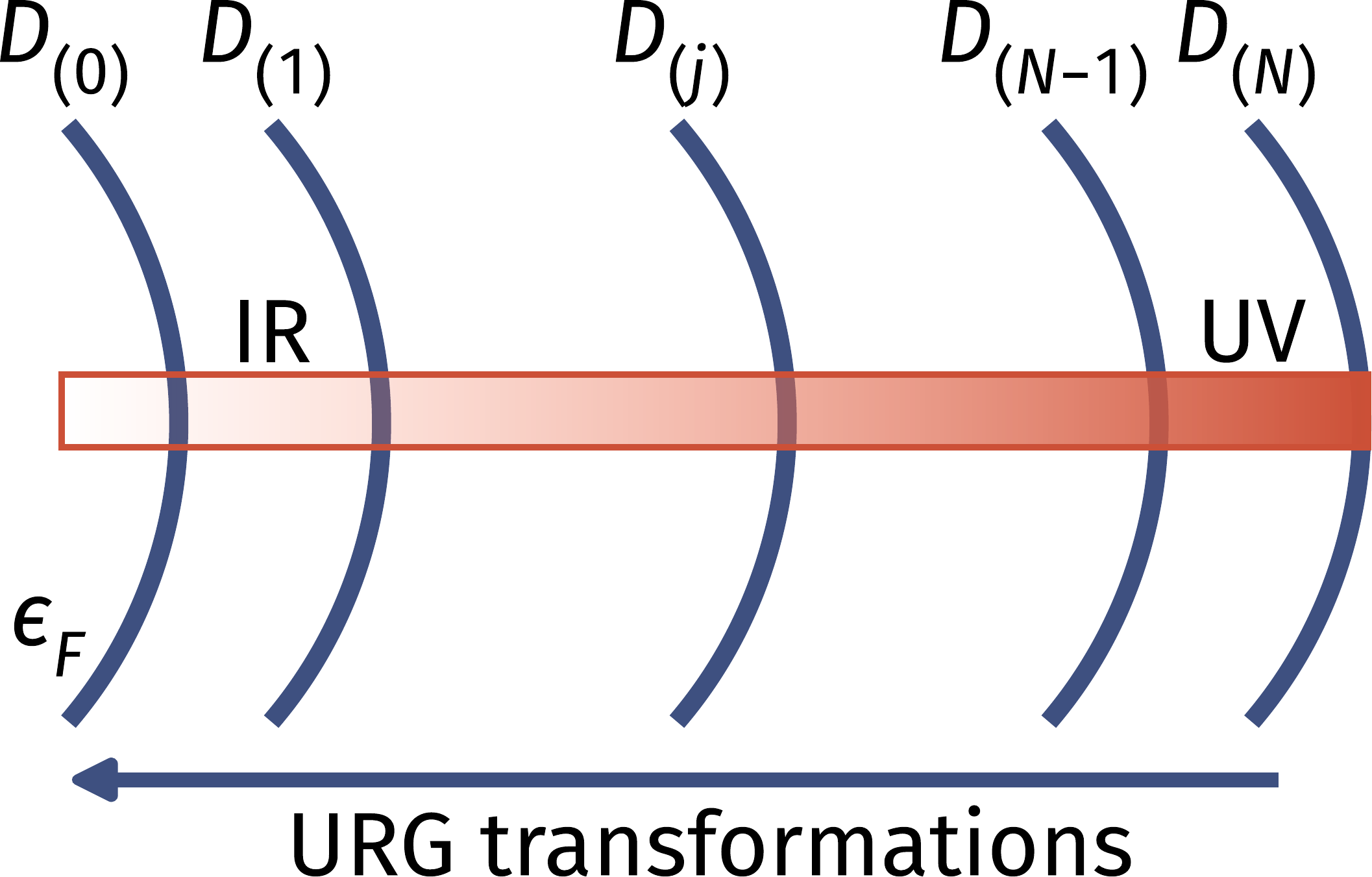}}}$
	\caption{{\it Left:} Schematic 1D representation of the extended SIAM Hamiltonian. The red sphere is the impurity site with on-site Hubbard term \(U\). It is connected via two couplings \(V\) and \(J\) to the bath zeroth site (large blue sphere) that has its own on-site Hubbard term \(U_b\). The small blue spheres make up the rest of the bath, connected through the single-particle hopping \(t\). {\it Right:} High energy - low energy scheme defined and used in the URG method. The states away from the Fermi surface form the UV subspace and are decoupled first, leading to a Hamiltonian which is more block-diagonal and comprised of only the IR states near the Fermi surface.}
	\label{zeromode-bare}
\end{figure}

In order to obtain the various low-energy phases of the e-SIAM, we perform a scaling analysis of the associated Hamiltonian (eq.~\eqref{GIAM-ham}) using the recently developed unitary renormalisation group (URG) method \cite{anirbanurg1,anirbanurg2}. The method has been applied successfully on a wide variety of problems of correlated fermions~\cite{santanukagome,1dhubjhep,anirbanmott2,anirbanmott1,siddharthacpi,anirban_kondo,patra_mck,anirbanurg1,anirbanurg2}. The method proceeds by resolving quantum fluctuations in high-energy degrees of freedom, leading to a low-energy Hamiltonian with renormalised couplings and new emergent degrees of freedom. Typically, for a system with Fermi energy \(\epsilon_F\) and bandwidth \(D_0\), the sequence of isoenergetic shells \(\left\{D_{(j)}\right\}, D_{(j)}\in \left[\epsilon_F, D_0\right] \) define the states whose quantum fluctuations we sequentially resolve. The momentum states lying on shells \(D_{(j)}\) that are far away from the Fermi surface comprise the UV states, while those on shells near the Fermi surface comprise the IR states. This scheme is shown in the right panel of Fig.~\eqref{zeromode-bare}.

As a result of the URG transformations, the Hamiltonian \(H_{(j)}\) at a given RG step \(j\) involves scattering processes between the \(k-\)states that have energies lower than \(D_{(j+1)}\). The unitary transformation \(U_{(j)}\) is then defined so as to remove the number fluctuations of the currently most energetic set of states \(D_{(j)}\)~\cite{anirbanurg1,anirbanurg2}:
\begin{eqnarray}
	H_{(j-1)} = U_{(j)} H_{(j)} U^\dagger_{(j)}~, \text{such that} ~\left[H_{(j-1)}, \hat n_{j}\right] =0~.
\end{eqnarray}
The eigenvalue of $\hat{n}_{j}$ has, thus, been rendered an integral of motion (IOM) under the RG transformation.

The unitary transformations can be expressed in terms of a generator \(\eta_{(j)}\) that has fermionic algebra~\cite{anirbanurg1,anirbanurg2}:
\begin{eqnarray}
	\label{unitary}
	U_{(j)} = \frac{1}{\sqrt 2}\left(1 + \eta_{(j)} - \eta_{(j)}^\dagger\right)~,~ \quad\left\{ \eta_{(j)},\eta_{(j)}^\dagger \right\} = 1~,
\end{eqnarray}
where \(\left\{\cdot\right\}\) is the anticommutator. The unitary operator \(U_{(j)}\) that appears in Eq.~\eqref{unitary} can be cast into the well-known general form \(U = e^\mathcal{S}, \mathcal{S} = \frac{\pi}{4}\left( \eta^\dagger_{(j)} - \eta_{(j)} \right)\) that a unitary operator can take, defined by an anti-Hermitian operator \(\mathcal{S}\). The generator \(\eta_{(j)}\) is given by the expression~\cite{anirbanurg1,anirbanurg2}
\begin{eqnarray}
	\eta^\dagger_{(j)} = \frac{1}{\hat \omega_{(j)} - \text{Tr}\left(H_{(j)} \hat n_{j}\right) } c^\dagger_{j} \text{Tr}\left(H_{(j)}c_{j}\right)~.
\end{eqnarray}
The operators \(\eta_{(j)},\eta^\dagger_{(j)}\) behave as the many-particle analogues of the single-particle field operators \(c_j,c^\dagger_j\) - they change the occupation number of the single-particle Fock space \(\ket{n_j}\).  The important operator \(\hat \omega_{(j)}\) originates from the quantum fluctuations that exist in the problem because of the non-commutation of the kinetic energy terms and the interaction terms in the Hamiltonian:
\begin{eqnarray}
	\hat \omega_{(j)} = H_{(j-1)} - H^i_{(j)}~.
	\label{omega}
\end{eqnarray}
\(H^i_{(j)}\) is the part of \(H_{(j)}\) that commutes with \(\hat n_j\) but does {\it not} commute with at least one \(\hat n_l\) for \(l < j\). The RG flow continues up to energy \(D^*\), where a fixed point is reached from the vanishing of the RG function. 
Detailed comparisons of the URG with other methods (e.g., the functional RG, spectrum bifurcation RG etc.) can be found in Refs.~\cite{anirbanmott1,anirbanurg1}. More information on the unitary RG method is provided in Section 1 of the Supplementary Materials~\cite{supp_mat}.

\section{\(T=0\) scaling theory for the e-SIAM}
\label{urg-theory}

\subsection{URG equations for the e-SIAM}

The derivation of the RG equations for the extended Anderson impurity model (e-SIAM) Hamiltonian is shown in Section 2 of the Supplementary Materials~\cite{supp_mat}. The bath coupling \(U_b\) is marginal. We provide below the RG equations of the remaining couplings for a given quantum fluctuation scale \(\omega\):
\begin{eqnarray}
	\Delta U &= 4V^2 n_j\left(\frac{1}{d_1} - \frac{1}{d_0}\right) - n_j\frac{J^2}{d_2},\\
	\Delta V &= -\frac{3n_j V}{8}\left[J\left(\frac{1}{d_2} + \frac{1}{d_1}\right) +  \frac{4U_b}{3}\sum_{i=1}^4 \frac{1}{d_i}\right],\nonumber \\
	\Delta J &= -\frac{n_j J\left(J + 4U_b\right)}{d_2}~,\label{rg-eqn}
\end{eqnarray}
where the denominators \(d_i\) are given by
\begin{eqnarray}\label{rg-eqtn1}
	d_0 = \omega - \frac{D}{2} + \frac{U_b}{2} - \frac{U}{2},~d_1 = \omega - \frac{D}{2} + \frac{U_b}{2} + \frac{U}{2} + \frac{J}{4}~,\\
	d_2 = \omega - \frac{D}{2} + \frac{U_b}{2} + \frac{J}{4}~,d_3 = \omega - \frac{D}{2} + \frac{U_b}{2}~.
\end{eqnarray}
The symbols used in the RG equations have the following meanings: \(\Delta U\) represents the renormalisation of the coupling \(U\) in going from the \(j^\text{th}\) Hamiltonian to the \(\left( j-1 \right) ^\text{th}\) Hamiltonian by decoupling the isoenergetic shell at energy \(D_{(j)}\) (see right panel of Fig.\eqref{zeromode-bare}). \(n_j\) is the number of electronic states on the shell \(D_{(j)}\). We note that the labels \(U_0,J_0,V_0\) that appear in various figures (and elsewhere in the text) represent the bare values of the associated couplings \(U,J\) and \(V\). The RG equations reduce, in the perturbative regime of the couplings $U, V$ and $J$, to the well-known ``poor man's" scaling forms obtained for the  SIAM~\cite{haldane1978scaling} and the single-channel Kondo model~\cite{anderson1970} respectively.

 The RG fixed point Hamiltonian describes the low-energy phase of the system. In general, if the RG fixed point is reached at an energy scale \(D^*\), the fixed point Hamiltonian \(\mathcal{H}^*\) is obtained simply from the fixed point values of the couplings (and by noting that the states above \(D^*\) are now part of the IOMs):
\begin{eqnarray}
	\label{fixed-point-ham}
	\mathcal{H}^* = \sum_{\sigma,\vec k}^{|\epsilon_{\vec k}| < D^*} \epsilon_{\vec k} \tau_{\vec k,\sigma} + V^*\sum_\sigma \left( c^\dagger_{d\sigma}c_{0\sigma} + \text{h.c.}\right) + J^* \vec{S}_d\cdot\vec{S}_0\nonumber - \frac{1}{2}U^*\left(\hat n_{d \uparrow} - \hat n_{d \downarrow}\right)^2 - \frac{1}{2}U_b \left(\hat n_{0 \uparrow} - \hat n_{0 \downarrow}\right)^2~.
\end{eqnarray}
The fixed point values of the couplings are obtained by solving the RG equations numerically.

\begin{figure}[!htbp]
	$\vcenter{\hbox{\includegraphics[width=0.48\textwidth]{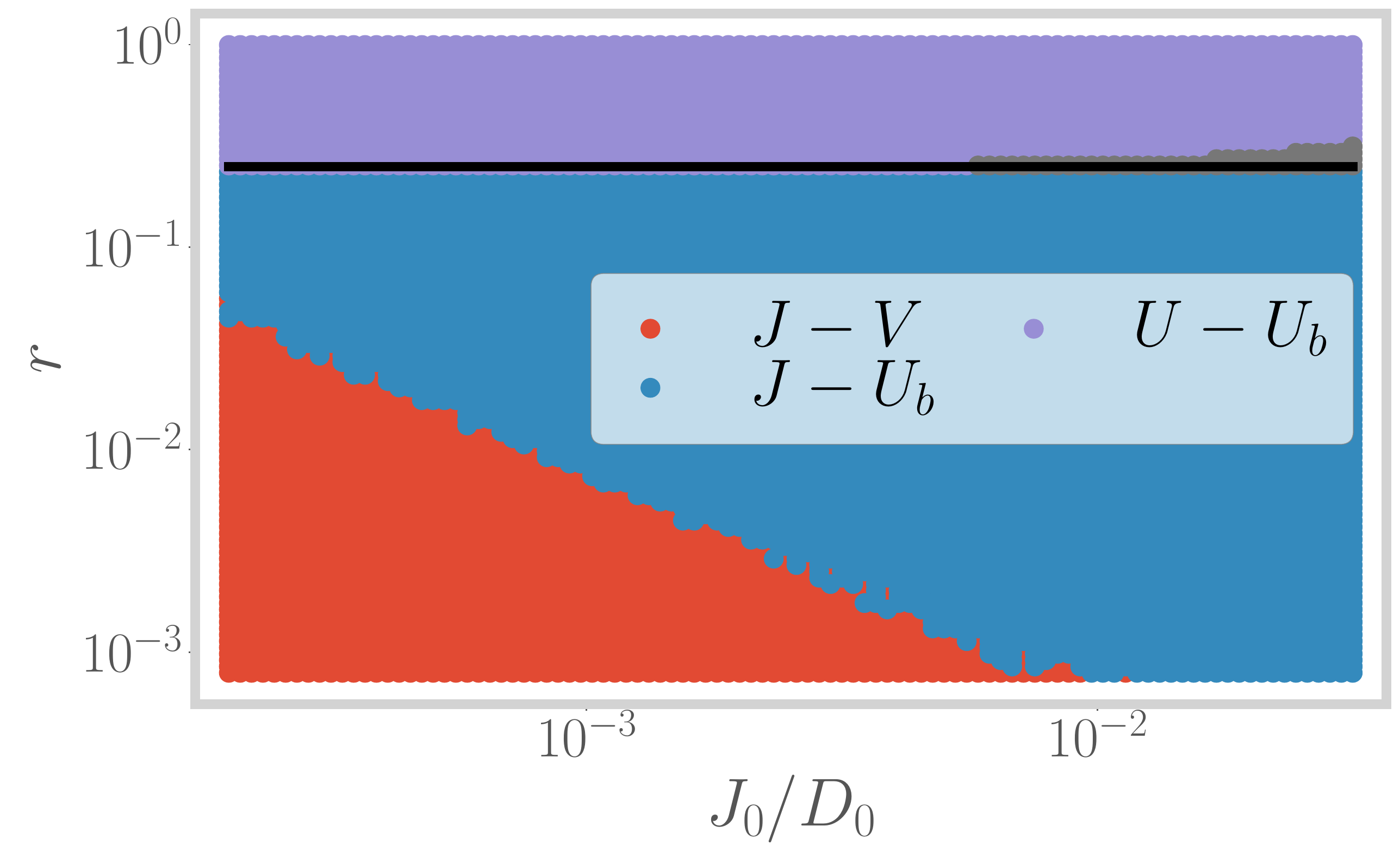}}}$
	\hspace*{\fill}
	$\vcenter{\hbox{\includegraphics[width=0.48\textwidth]{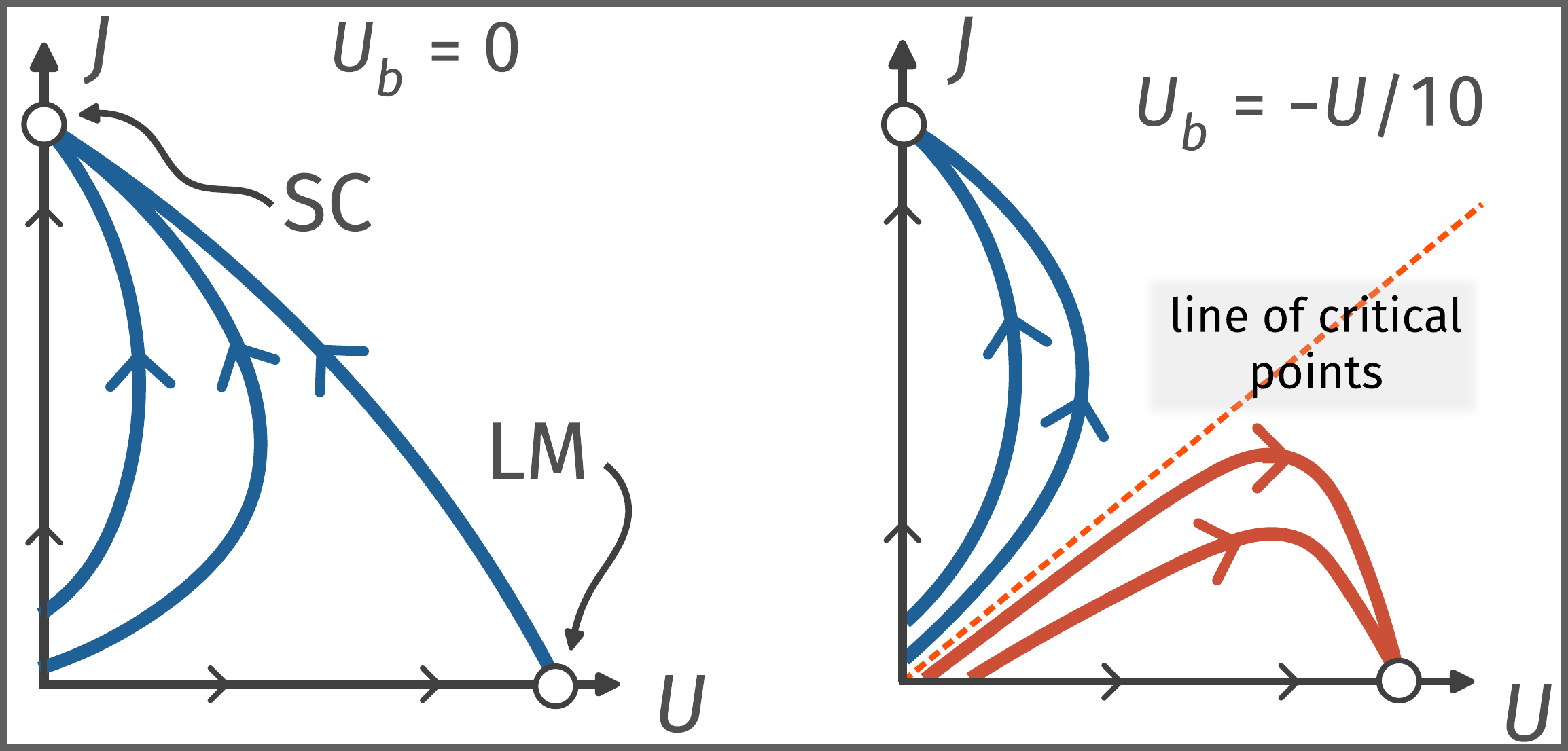}}}$
	\caption{{\it Left:} Phase diagram of the e-SIAM, in the space of \(r\) and \(J_0/D_0\). \(U_b\) is set to \(-U_0/10\). The various phases are described in the text. {\it Right:} Schematic nature of RG flows for \(U_b=0\) and \(U_b = -U/10\). Blue curves represent RG flows towards the strong-coupling (SC) fixed point, while red curves represent RG flows towards the local moment (LM) fixed point.}
	\label{phase-diag}
\end{figure}

\subsection{Phase diagram}

We will work in the low-energy regime where the quantum fluctuation scale \(\omega\) is such that all the denominators are negative: \(d_i < 0 ~\forall~i\). Moreover, we constrain the impurity and local bath correlations (\(U\) and \(U_b\) respectively) through the relation \(U_{b} = -U/10\). While the precise value of the factor of \(1/10\) is unimportant, we have 
chosen a factor considerably smaller than 1 in order to demonstrate that an interesting body of results can be obtained with a value of $|U_{b}|$ that is much smaller than that of $U$. Further, the negative sign in the above relation is significant, as we shall see below that the MIT is obtained for the case of \(U_b\) being negative (i.e., attractive on-site correlations on the zeroth site of the bath). The relation between $U_{b}$ and $U$ is motivated on phenomenological grounds such that the MIT occurring in the bulk lattice model (obtained upon increasing the on-site Hubbard repulsion to large values) corresponds, in the auxiliary model mapping within DMFT, to the local MIT observed upon tuning the impurity correlation $U$ (and the related bath correlation \(U_b\)) within the proposed effective impurity model Hamiltonian. The physical significance of attractive on-site correlations in the bath lies in providing a mechanism for the frustration of Kondo screening (in the RG equation for an antiferromagnetic Kondo coupling $J(>0)$, eq.~\eqref{rg-eqn}) within an Anderson impurity coupled to a single channel of conduction electrons. We provide a detailed discussion of this point in the concluding section of our work.

\begin{table*}[!htbp]
\centering
\begin{tabular}{|cl|c|c|}
\hline
& Regime & Low-energy effective Hamiltonian & impurity ground-state\\[5pt]
\hline
1. & \(0 < r < r_{c1}\) &  \(\text{K.E}^* + V^*\sum_\sigma \left( c^\dagger_{d\sigma}c_{0\sigma} + \text{h.c.}\right) + J^* \vec{S}_d\cdot\vec{S}_0\) & \(\frac{1}{\sqrt 2}\left(\ket{\uparrow_d}\ket{\downarrow_0} - \ket{\downarrow_d} \ket{\uparrow_0} + \ket{2_d}\ket{0_0} + \ket{0_d} \ket{2_0}\right)\)\\[10pt]
2. & \(r_{c1} < r < r_{c2}\) &  \(\text{K.E}^* + J^* \vec{S}_d\cdot\vec{S}_0 - \frac{1}{2}U_b \left(\hat n_{0 \uparrow} - \hat n_{0 \downarrow}\right)^2\) & \(\frac{1}{\sqrt 2}\left(\ket{\uparrow_d}\ket{\downarrow_0} - \ket{\downarrow_d} \ket{\uparrow_0}\right)\)\\[10pt]
3. & \(r_{c2} < r\) &  \(\text{K.E}^* - \frac{U^*}{2}\left(\hat n_{d \uparrow} - \hat n_{d \downarrow}\right)^2 - \frac{U_b}{2} \left(\hat n_{0 \uparrow} - \hat n_{0 \downarrow}\right)^2\) & \(\left\{ \ket{\uparrow_d},\ket{\downarrow_d}\right\}\)\\[5pt]
\hline
\end{tabular}
\caption{Effective Hamiltonians and ground-states of the three important parts of the phase diagram in Fig.~\eqref{phase-diag}. The ground-state is simplified to capture only the configuration of the impurity site and at most the bath zeroth site. \(\text{K.E.}^*\) represents the kinetic energy of the conduction electron states residing within the fixed point window \(D^*\).}
\label{phases-table}
\end{table*}

\begin{figure*}[!htbp]
\centering
\includegraphics[width=\textwidth]{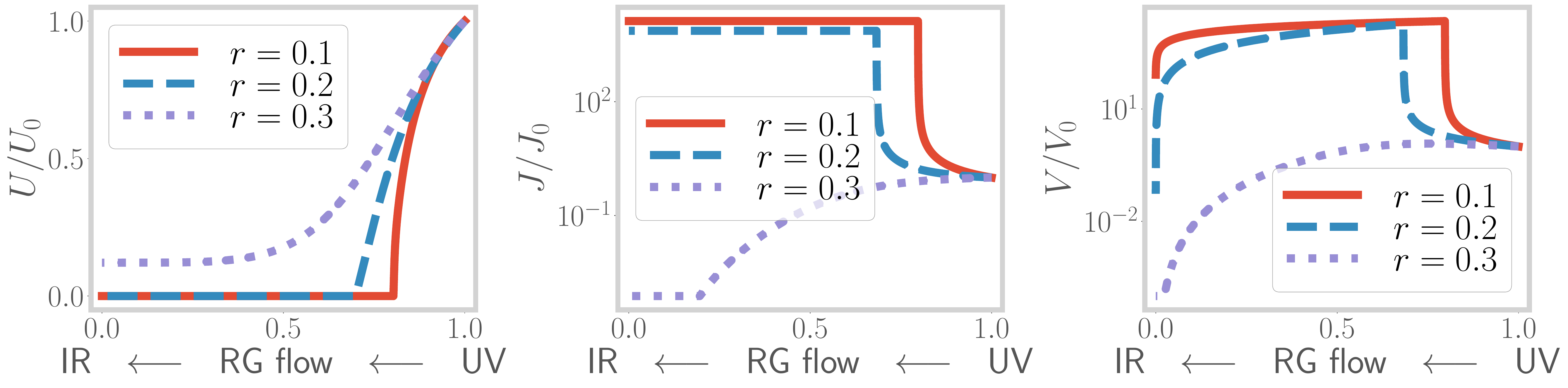}
\caption{Variation of couplings \(U,V\) and \(J\) along the RG transformations, for three values of the transition-tuning ratio \(r = -U_b/J_0\). The x-axis represents the distance of the running cutoff from the Fermi surface; the rightmost point is the first RG step (UV) and the leftmost point is the final RG step (IR). The red curves represent the flows for \(r < r_{c1}\), where both \(V\) and \(J\) are relevant. The blue curves represent the RG flows for \(r_{c1} < r < r_{c2}\), where \(J\) is relevant but \(V\) is irrelevant. The violet curves represent RG flows for \(r > r_{c2}\), where both \(V\) and \(J\) are irrelevant but \(U\) flows to a finite value.}
\label{rg-flow}
\end{figure*}

The phase diagram is shown in the left panel of Fig.~\eqref{phase-diag} in terms of the parameter $r=|U_{b}|/J$ (y-axis) and the ratio of the bare Kondo coupling ($J_{0}$) to the bare conduction bath bandwidth ($D_{0}$) (x-axis), we first define two important points in the space of couplings. These are values of the parameter \(r\) where there is a qualitative change in the nature of RG flows, and hence in the low-energy physics of the model
\begin{itemize}
	\item \(r = r_{c1} \left(= -\left(\frac{U_b}{J}\right)_{c1} = \frac{3}{20} = \left(\frac{U}{10 J}\right)_{c1}>0\right)\): At this point, the coupling \(V\) becomes irrelevant,
	\item \(r = r_{c2} \left(= -\left(\frac{U_b}{J}\right)_{c2} = \frac{1}{4} = \left(\frac{U}{10 J}\right)_{c2}\right)\): At this point, the coupling \(J\) also turns irrelevant. Note that $r_{c2} > r_{c1}$.
\end{itemize}
We will use these two values of the parameter $r$ as checkpoints around which we can describe the low-energy physics. 
There are three important parts in Fig.~\eqref{phase-diag}:
\begin{itemize}
	\item red region, \(0 < r < r_{c1}\): the \(J-V\) model; {\it \(V\) and \(J\) are both relevant, but $U$ is irrelevant}; spin and charge delocalisation on the impurity; spin-charge mixing in ground-state
	\item blue region, \(r_{c1} < r < r_{c2}\): the \(J-U_b\) model; {\it $J$ is relevant, but \(V\) and $U$ are both irrelevant}; charge localisation and spin delocalisation on the impurity; singlet ground-state
	\item violet region, \(r_{c2} < r\): the \(U-U_b\) model; {\it $U$ is relevant, but \(V\) and \(J\) are both irrelevant}; spin and charge localisation on the impurity; local moment ground-state.
\end{itemize}
Typical RG flows that lead to these phases are shown in Fig.~\eqref{rg-flow}. We also note that the grey region shown in the top right corner of Fig.\eqref{phase-diag} corresponds to a model in which all three couplings ($J$, $U$ and $V$) are RG irrelevant. We find, however, that this phase is an artefact of solving the RG equations for an impurity coupled to a finite-sized conduction bath, and it gradually disappears upon increasing the bath size.

\subsection{Phase transition at \(r_{c2}\)}
The effective Hamiltonians and corresponding impurity ground-states for various phases have been listed in table ~\eqref{phases-table}. The variation in the ground-state has been checked by numerically solving the fixed point Hamiltonian for various bare values of the couplings, and is shown in the left panel of Fig.~\eqref{spec_func}.
A sharp change in the ground-state from spin-singlet to local moment shows that the blue and violet phases are separated by an {\it impurity delocalisation-localisation transition} (black line in Fig.~\eqref{phase-diag}). The transition occurs at finite values of the correlations: \(r_{c2} = -\left(\frac{U_b}{J}\right)_{c2} = \frac{1}{4} = \left(\frac{U}{10 J}\right)_{c2}\). This is a stark contrast from the standard SIAM, where the transition can happen only at on-site correlation \(U \to\infty\). Therefore, the presence of the critical point at finite values of the various couplings transforms the landscape of RG phase diagram shown schematically in the right panel of Fig.~\eqref{phase-diag} (right panel): the RG flows split into two classes - those that flow towards the strong-coupling Kondo screened fixed point and those that flow towards the local moment fixed point.

By a simple rewriting of the RG equation for \(J\) (eq.~\eqref{rg-eqn}), the impurity transition can be seen to arise from a competition between the Kondo screening physics of \(J\) and the local pairing physics of \(U_b\):
\begin{eqnarray}
\Delta J = \overbrace{\frac{(J + 2U_b)^2 n_j}{|d_{2}|}}^\text{usual Kondo physics} - \overbrace{\frac{(2U_b)^2 n_j}{|d_{2}|}}^\text{competing pairing physics}~,
\end{eqnarray}
where $d_{2}=\omega - D/2 + U_b/2 + J/4$. The competition between the effective Kondo term \((J + 2U_b)^2\) and the competing pairing term \(-4U_b^2\) leads to the presence of two stable phases - one that is Kondo screened and one that remains unscreened.

In the DMFT treatment of the $1/2$-filled Hubbard model on the Bethe lattice in infinte dimensions, the vanishing of non-local contributions to the lattice self-energy means that the lattice Greens function can be computed self-consistently by solving a local quantum impurity problem~\cite{georges1996}. In the rest of the work, we provide extensive evidence that the local transition observed in the e-SIAM is very similar to the Mott MIT observed in DMFT. We conclude thereby that the e-SIAM models faithfully the round-trip excursions of an electron on the Bethe lattice, and that the impurity phase transition observed in the e-SIAM offers a local description of the Mott MIT in the bulk Hubbard model. With this evidence in mind, we will henceforth refer to the transition at \(r_{c2}\) as a {\it local metal-insulator transition}.

\section{Descriptors of the local MIT}
\label{desc-mit}

The present section gives more clarity on the nature of the impurity phase transition in the form of additional descriptors of the transition such as impurity spectral function and measures of entanglement, computed from the e-SIAM Hamiltonian.
\subsection{Evolution of the impurity spectral function}

\begin{figure}[!htb]
	\includegraphics[width=0.45\textwidth]{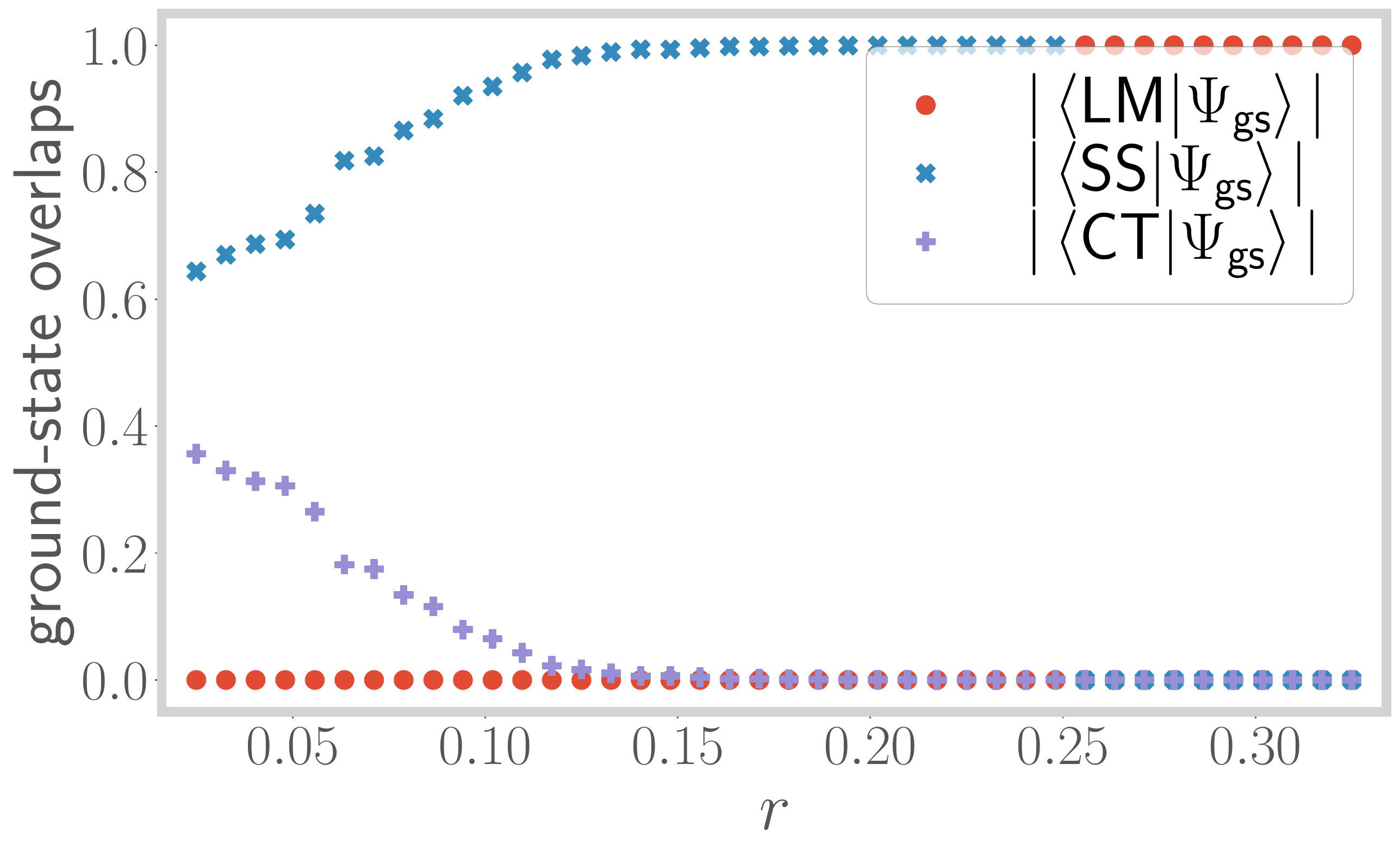}
	\hspace*{\fill}
	\includegraphics[width=0.45\textwidth]{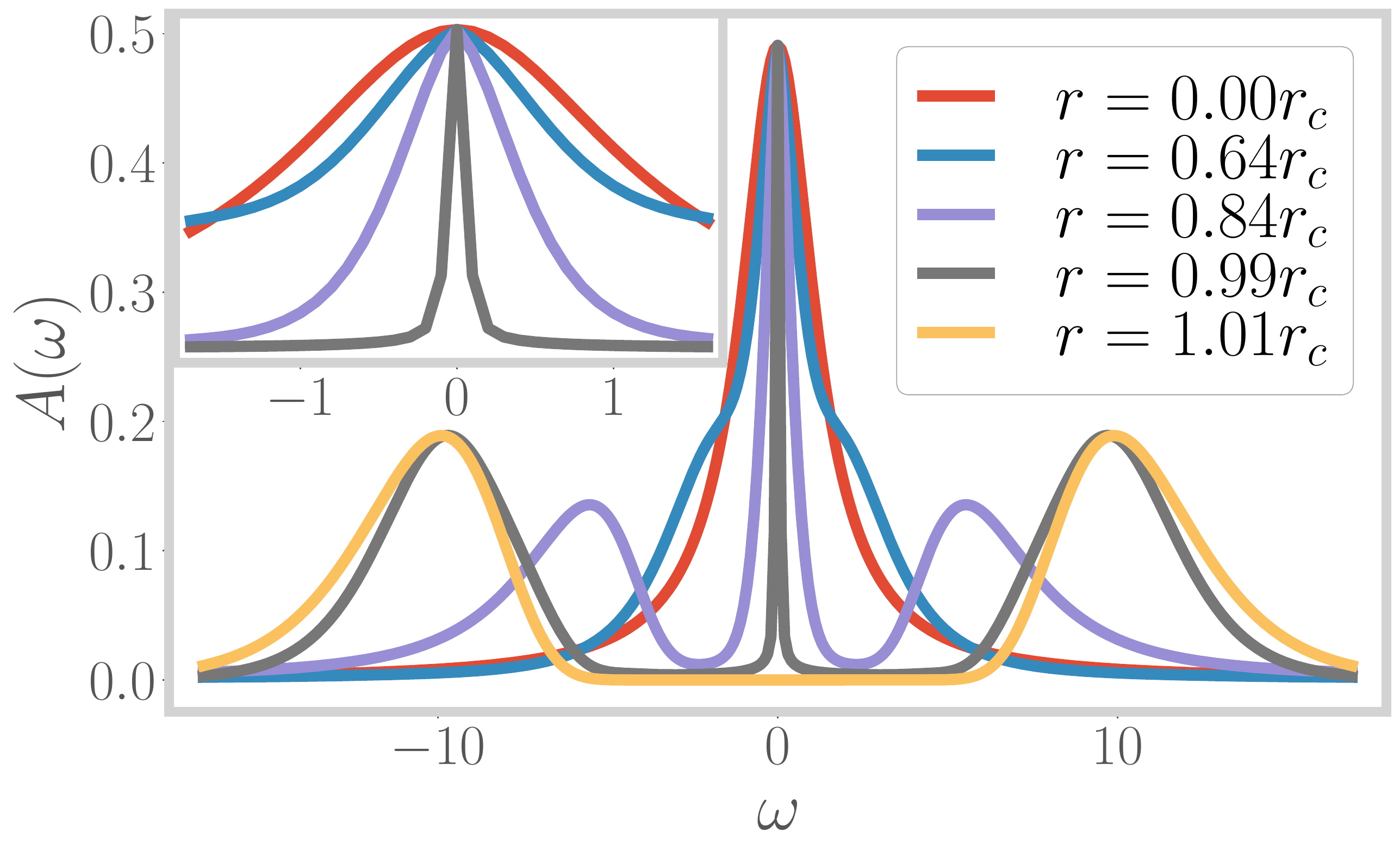}
	\caption{{\it Left:} Overlap of the RG fixed point ground state \(\ket{\Psi}_\text{gs}\) with the spin-singlet state \(\ket{\text{SS}}\), the zero charge member of the charge triplet states \(\ket{\text{CT}}\) and the local moment states \(\ket{\text{LM}}\), for all three regimes of the model. Beyond the critical point \(r_{c2} = 0.25\), the overlap with the entangled states vanish, indicating the transition to a decoupled local moment. {\it Right:} Variation of the impurity spectral function from \(r = 0\) to \(r > r_{c2}\). At small \(r\), the central peak is broad, but at larger \(r\), it sharpens, and the difference in spectral weight is used in creating the Hubbard sidebands. For \(r > r_{c2}\), the central peak vanishes.}
	\label{spec_func}
\end{figure}

The impurity spectral function of the standard SIAM (eq.~\eqref{basic-siam}) always displays a central peak at finite values of \(U\) (along with Hubbard sidebands at sufficiently large \(U\)), indicating the presence of gapless local Fermi liquid excitations on the impurity site~\cite{wilson1975,nozieres1974fermi,costi_hewson_1990}. To demonstrate the impurity localisation transition, we compute the impurity local spectral function of the e-SIAM (eq.~\eqref{GIAM-ham}). This involves numerically diagonalising the effective Hamiltonian at various energy scales along the RG flow and computing the spectral weight at a range of frequencies from UV to IR. We find that the spectral function (shown in the right panel of Fig.~\eqref{spec_func}) displays three notable features, in agreement with DMFT results~\cite{georges1996}:
\begin{itemize}
	\item The appearance of a {\it preformed gap} (flattening of spectral function between the central peak and sidebands) as \(r\) crosses \(r_{c1} \left(= -\left(\frac{U_b}{J}\right)_{c1} = \frac{3}{20} = \left(\frac{U}{10 J}\right)_{c1}\right)\): The preformed gap simply indicates the separation of the spin (central peak) and charge (sidebands) degrees of freedom beyond \(r_{c1}\). This separation is brought about by the irrelevance of \(V\).
	\item The sharpening of the central peak, and concomitant increase in the preformed gap, as $r$ is increased in the range $r_{c1} < r < r_{c2}$.
	\item The vanishing of the central peak and appearance of a {\it hard gap} as \(r\) crosses \(r_{c2}\): The irrelevance of \(J\) beyond \(r_{c2}\) destroys the Kondo screening and localises the impurity moment. This involves the destabilisation of the singlet ground state, and the stabilisation of the local moment states in its place.
\end{itemize}

\begin{figure}[!htb]
	\includegraphics[width=0.45\textwidth]{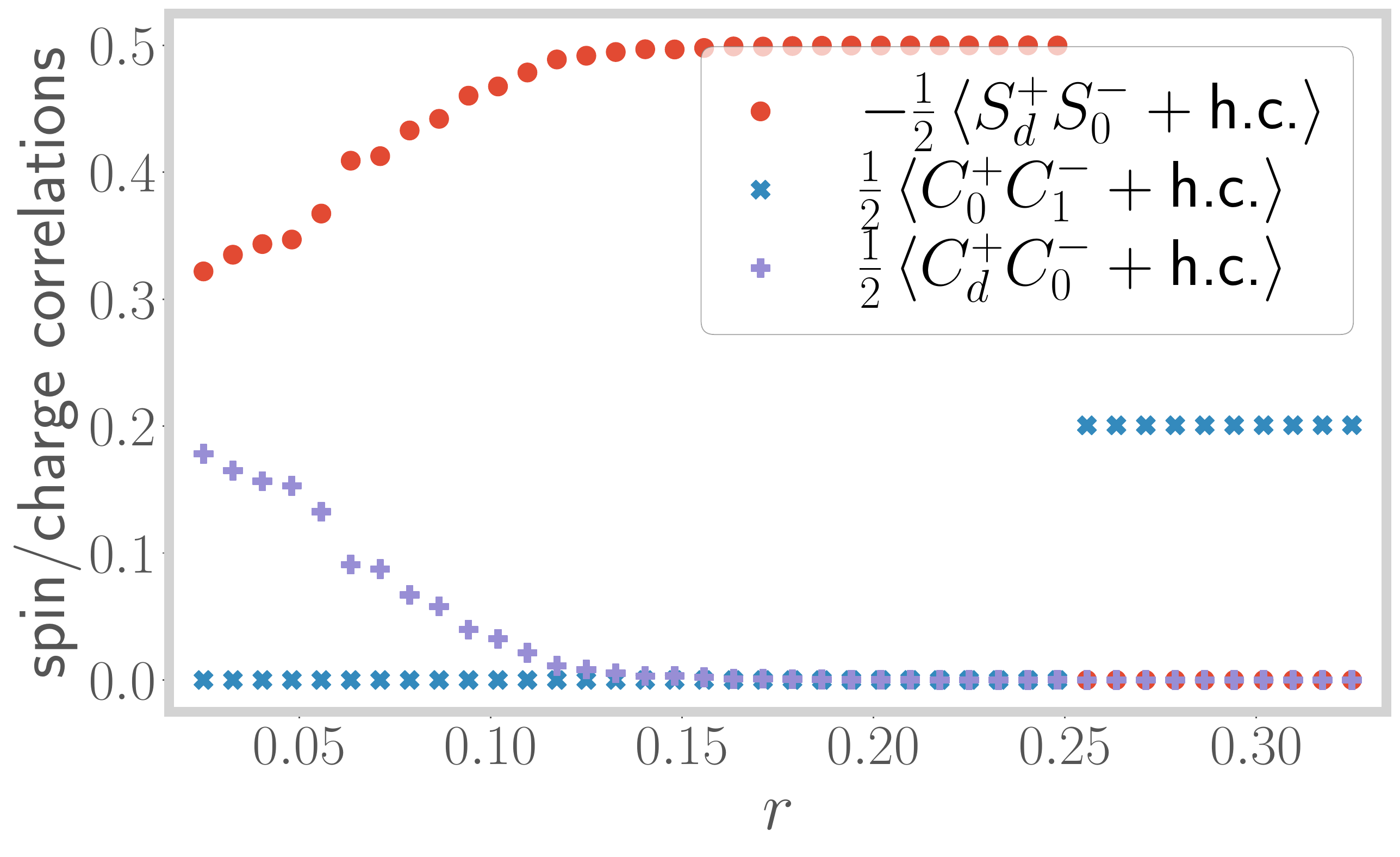}
	\hspace*{\fill}
	\includegraphics[width=0.45\textwidth]{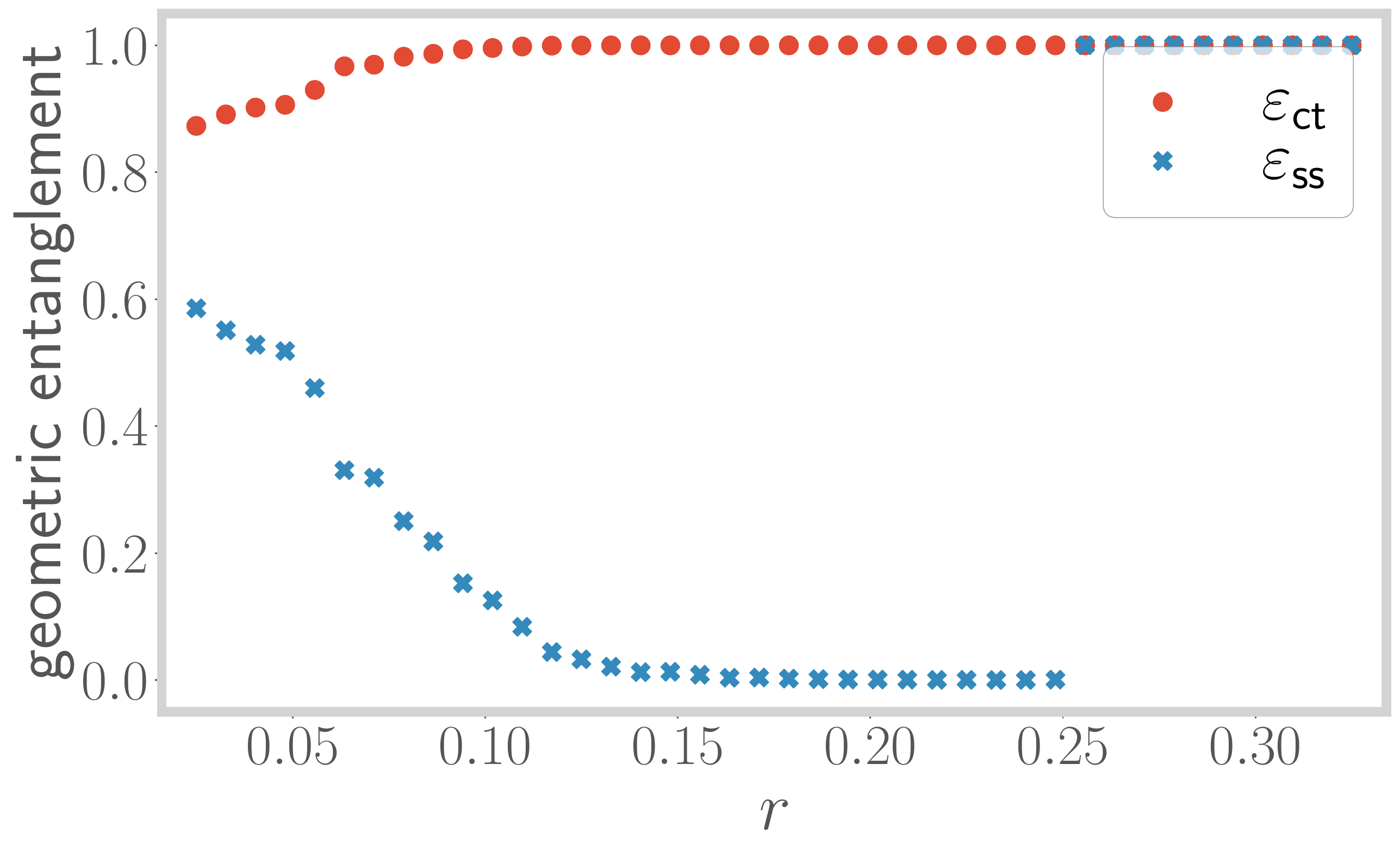}
	\caption{{\it Left:} Variation of impurity-bath spin-flip correlation (red) and charge isospin-flip correlation (violet), as well as intra-bath charge isospin-flip correlation (blue), from \(r \sim 0\) to \(r > r_{c2}\). The imp.-bath isospin correlation vanishes at \(r_{c1}\), indicating the change in RG relevance of \(V\). The imp.-bath spin correlation vanishes at \(r_{c2}\), indicating the marginality of \(J\) at that point. The intra-bath correlation picks up after the transition, showing the decoupling of the impurity from the bath. {\it Right:} Variation of the geometric entanglement \(\varepsilon_\text{ss}\) (blue) w.r.t. the singlet state and that w.r.t the charge triplet zero state, \(\varepsilon_\text{ct}\) (red), with \(r\). The former becomes maximum (unity) at \(r_{c1}\), showing that the true ground-state has no charge content. The latter discontinuously jumps to unity at \(r_{c2}\), therefore acting as an order parameter for the transition.}
	\label{gstate-correlations}
\end{figure}

These features are also reflected in ground-state correlation measures like spin-flip and charge isospin-flip correlations (see left panel of Fig.~\eqref{gstate-correlations}), which are defined as follows:
\begin{eqnarray}
\hspace*{-1cm}
\frac{1}{2}\left(\braket{S_{i}^{+}S_{j}^{-}} + \text{h.c.}\right) = \frac{1}{2}\left(\braket{c^\dagger_{i \uparrow}c_{i \downarrow}c^\dagger_{j \downarrow}c_{j \uparrow}} + \text{h.c.}\right)~,~
\frac{1}{2}\left(\braket{C_{i}^{+}C_{j}^{-}} + \text{h.c.}\right) = \frac{1}{2}\left(\braket{c^\dagger_{i \uparrow}c^\dagger_{i \downarrow}c_{j \downarrow}c_{j \uparrow}} + \text{h.c.}\right)~.
\end{eqnarray}
At small values of \(r\), both the correlations are large as the ground-state has both spin and charge content (row 1 of table~\eqref{phases-table}). The subsequent decrease in the impurity-bath charge flip correlation (violet points of Fig.~\eqref{gstate-correlations}), and the simultaneous increase in the impurity-bath spin-flip correlation (red points of Fig.~\eqref{gstate-correlations}), can be attributed to the irrelevance of \(V\) at \(r\simeq r_{c1}\) and the appearance of the preformed gap in the spectral function. Beyond \(r_{c2}\), the spin-flip correlation sharply drops to zero due to the irrelevance of \(J\). They are replaced by intra-bath correlations like the charge isospin-flip correlation between the zeroth site and the first site (blue points of Fig.~\eqref{gstate-correlations}). Such intra-bath correlations are promoted by the bath on-site term \(U_b\). The sudden change in the nature of the ground state, and resulting correlations, at $r_{c2}$ indicates the presence of a quantum critical point whose nature will be analysed further in subsequent sections. 

\subsection{Using entanglement to track correlations across the transition}

We will now show that a certain measure of entanglement behaves as an order parameter for the transition. We can define a geometric measure of entanglement in terms of wavefunctions \(\ket{\psi_1}\) and \(\ket{\psi_2}\)~\cite{shimony1995degree,wei2003geometric,horodecki2009quantum}:
\begin{eqnarray}
	\varepsilon\left(\psi_1,\psi_2\right) = 1 - |\braket{\psi_1 | \psi_2}|^2~.
\end{eqnarray}
From this definition, if \(\ket{\psi_1}\) corresponds to a separable state, the entanglement content of the state \(\ket{\psi_2}\) is small if it's overlap with \(\ket{\psi_1}\) is large.
For brevity, we will use the notation \(\varepsilon_\text{ss} \equiv \varepsilon\left(\psi_\text{ss},\psi^{(2)}_\text{gs}\right), \varepsilon_\text{ct} \equiv \varepsilon\left(\psi_\text{ct},\psi^{(2)}_\text{gs}\right)\) to represent the geometric entanglement between the e-SIAM ground-state \(\ket{\psi_\text{gs}}\) and the singlet state \(\ket{\psi_\text{ss}}\) or the charge triplet zero state \(\ket{\psi_\text{ct}}\).
The latter two states are shown in Table \eqref{phases-table}.
The entanglement measures $\varepsilon_{\text{ss}}$ and $\varepsilon_{\text{ct}}$ can be related to the impurity Greens function through the following equation (details can be found in Sec. 3 of the Supplementary Materials~\cite{supp_mat}):
\begin{eqnarray}
	G_d(\omega) = \sum_n &\left[\left(1 - \varepsilon_\text{ss} \right) G_{\Phi_\text{ss},\Phi_\text{ss}}(\omega,n) + \left(1 - \varepsilon_\text{ct} \right) G_{\Phi_\text{ct},\Phi_\text{ct}}(\omega,n) \right.\nonumber \\
			     &\left.+ 2\sqrt{\left(1 - \varepsilon_\text{ss} \right)}\sqrt{\left(1 - \varepsilon_\text{ct} \right)} G_{\Phi_\text{ss},\Phi_\text{ct}}(\omega,n)\right]~,
	\label{greens-func}
\end{eqnarray}
where
\begin{eqnarray}
	G_{\psi_1,\psi_2}(\omega,n) = \frac{1}{2}\frac{\braket{\psi_1 | c_{d\sigma} | \Psi_n}\braket{\Psi_n | c^\dagger_{d\sigma} | \psi_2} + \text{h.c.}}{\omega + E_\text{gs} - E_n} + \frac{1}{2}\frac{\braket{\psi_1 | c^\dagger_{d\sigma} | \Psi_n}\braket{\Psi_n | c_{d\sigma} | \psi_2} + \text{h.c.}}{\omega - E_\text{gs} + E_n}~.
\end{eqnarray}

This expression displays that the evolution of the impurity Greens function $G_{d}$ 
with $r$ (and related correlation functions shown in the left panel of Fig.\eqref{gstate-correlations}) is dependent on that of the entanglement measures $\varepsilon_{\text{ss}}$ and $\varepsilon_{\text{ct}}$ (shown in the right panel of Fig.~\eqref{gstate-correlations}). 
Indeed, we find that the geometric entanglement \(\varepsilon_\text{ct}\) with the charge sector increases towards unity as the transition at \(r_{c2}\) is approached, and remains unity after the transition. A more sensitive measure is the singlet entanglement \(\varepsilon_\text{ss}\): it initially decreases to zero at \(r_{c1}\) owing to the irrelevance of \(V\), but rises discontinuously to unity at the transition and becomes equal to the charge entanglement in the local moment phase.
Therefore, \(\varepsilon_\text{ss}\) {\it acts as an order parameter} for the local MIT at \(r_{c2}\). We find that the cross-term \(\sqrt{\left(1 - \varepsilon_\text{ss} \right)}\sqrt{\left(1 - \varepsilon_\text{ct} \right)}\) decreases monotonically to zero with increasing \(r\) (not shown), displaying a continual decrease in the mixing of the spin and charge sectors and the {\it immobilisation of the doublons and holons on the impurity site}. 

In general, any one-particle or two-particle fluctuation \(\left<O_1 O_2^\dagger \right>\) that acts on the combined Hilbert space of the impurity and the zeroth site can be expressed in terms of these entanglement measures \(\varepsilon_\text{ss},\varepsilon_\text{ct}\). The detailed derivations and expressions are given in Section 3 of the Supplementary Materials~\cite{supp_mat}. As a demonstration, consider the spin-spin correlation $\langle S_{d}^{+}S_{0}^{-}\rangle$ between the impurity and the zeroth site shown in the left panel of Fig.~\eqref{gstate-correlations}. The general expression given in Section 3 of the Supplementary Materials~\cite{supp_mat} is of the form
\begin{eqnarray}
	\langle S_{d}^{+}S_{0}^{-}\rangle =& \left(1 - \varepsilon_\text{ss}\right) \braket{\Phi_\text{ss} | O_2 O_1^\dagger | \Phi_\text{ss}} + \left(1 - \varepsilon_\text{ct}\right) \braket{\Phi_\text{ct} | O_2 O_1^\dagger | \Phi_\text{ct}} \nonumber \\
				      &+ \sqrt{1 - \varepsilon_\text{ss}}\sqrt{1 - \varepsilon_\text{ct}}\left(\braket{\Phi_\text{ss} | O_2 O_1^\dagger | \Phi_\text{ct}} + \braket{\Phi_\text{ct} | O_2 O_1^\dagger | \Phi_\text{s}}\right)~.
\end{eqnarray}
From the expression given, we find that only the singlet overlap is non-zero, such that $\langle S_{d}^{+}S_{0}^{-}\rangle$ is directly proportional to the quantity \(1-\varepsilon_\text{ss}\). As the entanglement measure \(\varepsilon_\text{ss}\) increases towards the transition (right panel of Fig.~\eqref{gstate-correlations}), the quantity \(1 - \varepsilon_\text{ss}\) decreases, in turn leading to the decrease in the magnitude of the correlation observed in the left panel of Fig.~\eqref{gstate-correlations}. For another explicit connection between correlations and entanglement measures, we present relations between the quantum Fisher information (QFI)~\cite{Hauke2016} and many-particle Greens functions in Section 3 of the Supplementary Materials~\cite{supp_mat}. There, we also plot the QFI for a number of two-particle operators as a function of \(r\), notably the ones corresponding to the degree of compensation for the impurity (\(\braket{\vec{S}_d\cdot\vec{S}_0}\)) and the impurity magnetisation (\(\braket{S_d^z}\)).
These two quantities (and hence the corresponding QFI) are important because they track the local MIT and act as order parameters for the transition, and the QFI corresponding to these two operators quantify the quantum fluctuations present in the system corresponding to these order parameters.
We show in Sec. 3 of the Supplementary Materials that the two phases on either side of the transition are characterised by distinct values of this pair of QFI: while the QFI corresponding to the degree of compensation is zero in the Kondo screened phase, it becomes non-zero in the local moment phase, and the opposite is true for the QFI arising from the impurity magnetisation.
The phase precisely at the transition is distinct from those on either side, because it displays a non-zero value for both of the QFI.
While it is expected that a critical point would show enhanced fluctuations of multiple kinds (giving rise to universality), it is enlightening to find that this is also reflected in a measure of many-particle entanglement.

\section{Universal theory for the local metal-insulator transition}\label{insights}

\begin{figure}[htpb]
	\hspace*{\fill}
	\includegraphics[height=0.07\textheight]{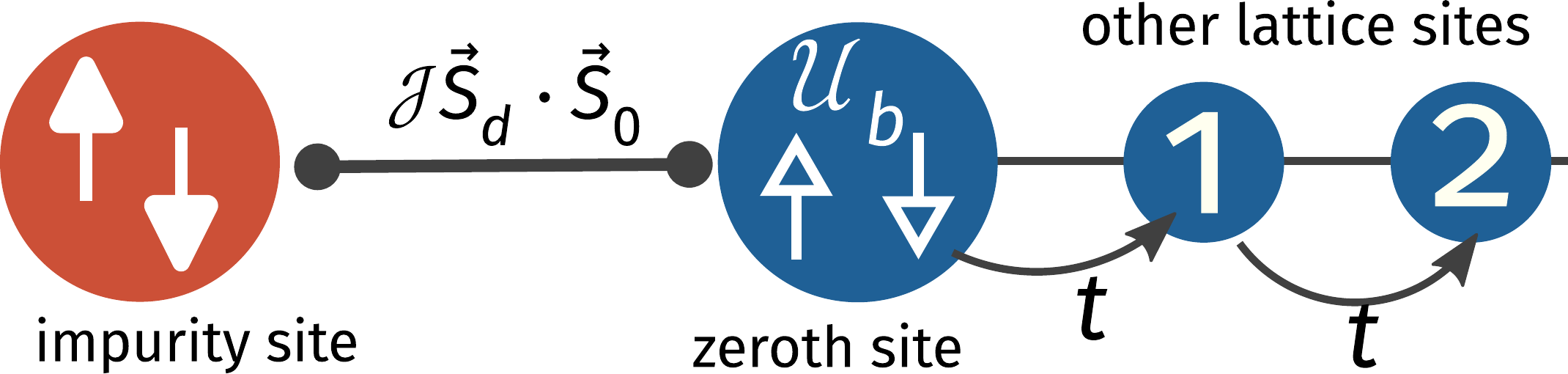}
	\hspace*{\fill}
	\includegraphics[height=0.07\textheight]{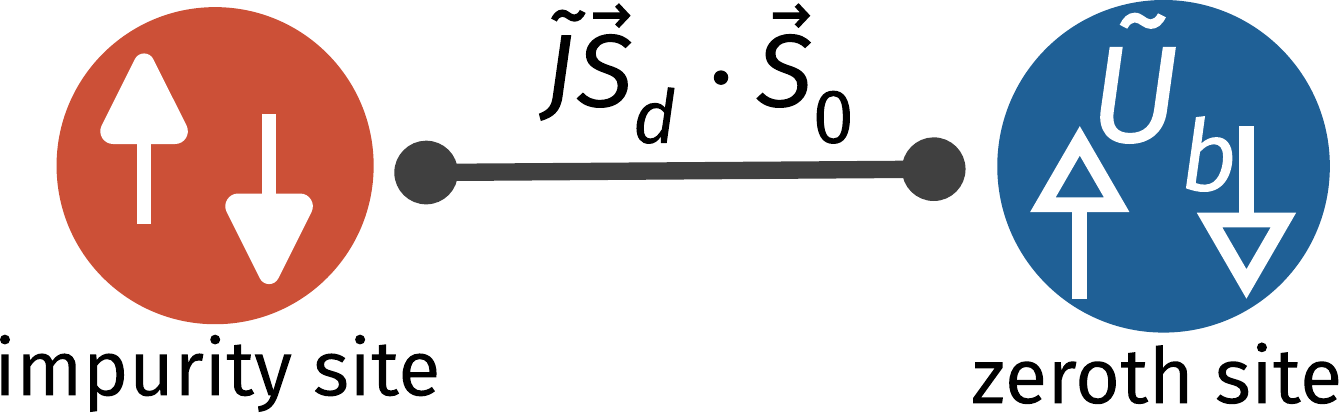}
	\hspace*{\fill}
	\caption{{\it Left:} Schematic 1D construction of universal theory for the metal-insulator transition, obtained by integrating out the charge states of the impurity near the transition. {\it Right:} Zero bandwidth limit of the low-energy effective Hamiltonian for \(r\) between \(r_{c1}\) and \(r_{c2}\) (second row in table~\eqref{phases-table}). The bath is reduced to just a single degree of freedom - the zeroth site that is directly coupled to the impurity site.}
	\label{universal-theory}
\end{figure}

In order to identify the competing tendencies near the transition at \(r_{c2}\), we will now obtain the minimal effective Hamiltonian that displays the same transition. This involves integrating out the degrees of freedom that do not affect the low-energy physics near \(r_{c2}\).
We note that, due to the irrelevance of \(V\), there is no scattering between the spin and charge states on the impurity at low energies.
Further, the impurity charge states \(\ket{0}\) and \(\ket{\uparrow_d \downarrow_d}\) have been pushed to the high energy Hubbard sidebands because of the large value of \(U\) close to \(r_{c2}\).
As a result, the impurity charge states can be safely decoupled from the spin states through a Schrieffer-Wolff transformation. Up to second order in $V^{2}/U$, this transformation accounts for the effects of \(U\) and \(V\) by generating an additional (Kondo) spin-exchange term \(\delta J \left(\sim \frac{V^2}{U+U_b}\right)\), as well as an additional on-site correlation \(\delta U_b\left(\sim \frac{2V^2}{U+U_b+J/2} - \frac{8V^{2}}{U-U_{b}}\right)\) on the zeroth site of the conduction bath. In this way, we obtain the following renormalised effective Hamiltonian:
\begin{eqnarray}
	\label{universal}
	H_\text{MIT} = \mathcal{J} \vec{S}_d\cdot\vec{S}_0 - \frac{1}{2}\mathcal{U}_b\left(\hat n_{0 \uparrow} - \hat n_{0 \downarrow}\right)^2 + H_\text{K.E.}~,
\end{eqnarray}
where \(\mathcal{J} = J + \delta J\) is the renormalised s-d interaction and \(\mathcal{U}_b = U_b + \delta U_b\) is the renormalised local correlation on the bath zeroth site.
A schematic 1D construction of the effective Hamiltonian is shown in the left panel of Fig.~\eqref{universal-theory}.
We note that a similar approach towards extracting an effective theory for the Mott MIT has been employed in the past (~\cite{moeller_1995,held_2013}). However, those works essentially led to renormalised Kondo models that do not possess any frustration of the Kondo screening. Thus, they cannot display an impurity transition in the absence of the requirement of self-consistency, and can describe only the physics of the metallic regime. Importantly, the effective \(\mathcal{J}-\mathcal{U}_b\) model (eq.~\eqref{universal}) is consistent with the IR fixed point Hamiltonian obtained in the appropriate regime \(r_{c1} < r < r_{c2}\) (row 2 of table~\eqref{phases-table}).

The RG equations for \(\mathcal{J}\) and \(\mathcal{U}_b\) of the simplified effective impurity model eq.~\eqref{universal} can be obtained by setting \(U = V = 0\) in the RG equations \eqref{rg-eqn}:
\begin{eqnarray}
\label{JUb-rg}
\Delta \mathcal{J} = -\frac{n_j \mathcal{J}\left(\mathcal{J} + 4\mathcal{U}_b\right)}{d_2}~~,~~ \Delta\mathcal{U}_b = 0~.
\end{eqnarray}
For \(\mathcal{J} + 4\mathcal{U}_b > 0\), the Kondo coupling is relevant and the low-energy phase is a paramagnetic local Fermi liquid with gapless excitations. But for \(\mathcal{J} + 4\mathcal{U}_b < 0\), the Kondo coupling becomes irrelevant and the ground state is a decoupled local moment that is isolated from the bath. Such an effective picture of the transition can be understood if one notes that a straightforward way to destroy the Kondo screening is to inhibit the coordinated spin-fluctuations between the impurity and the bath. Such frustration of Kondo screening is precisely the effect of the \(\mathcal{U}_b\) term: it promotes charge fluctuations on the bath site coupled directly to the impurity, reducing thereby the spectral weight for spin-flip scatterings between the impurity and bath.

\begin{figure}[htpb]
	\includegraphics[width=0.41\textwidth]{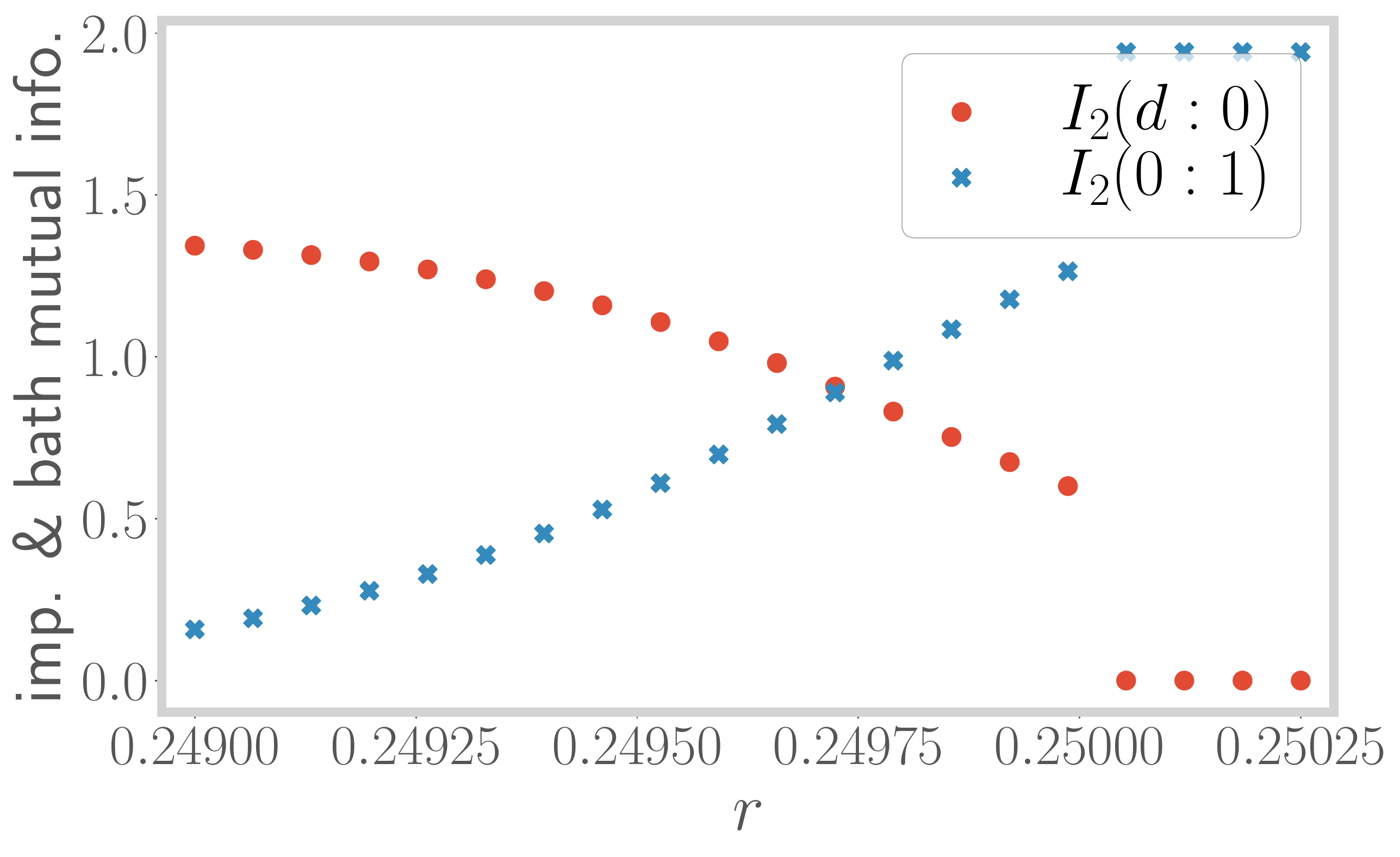}
	\hspace*{\fill}
	\includegraphics[width=0.48\textwidth]{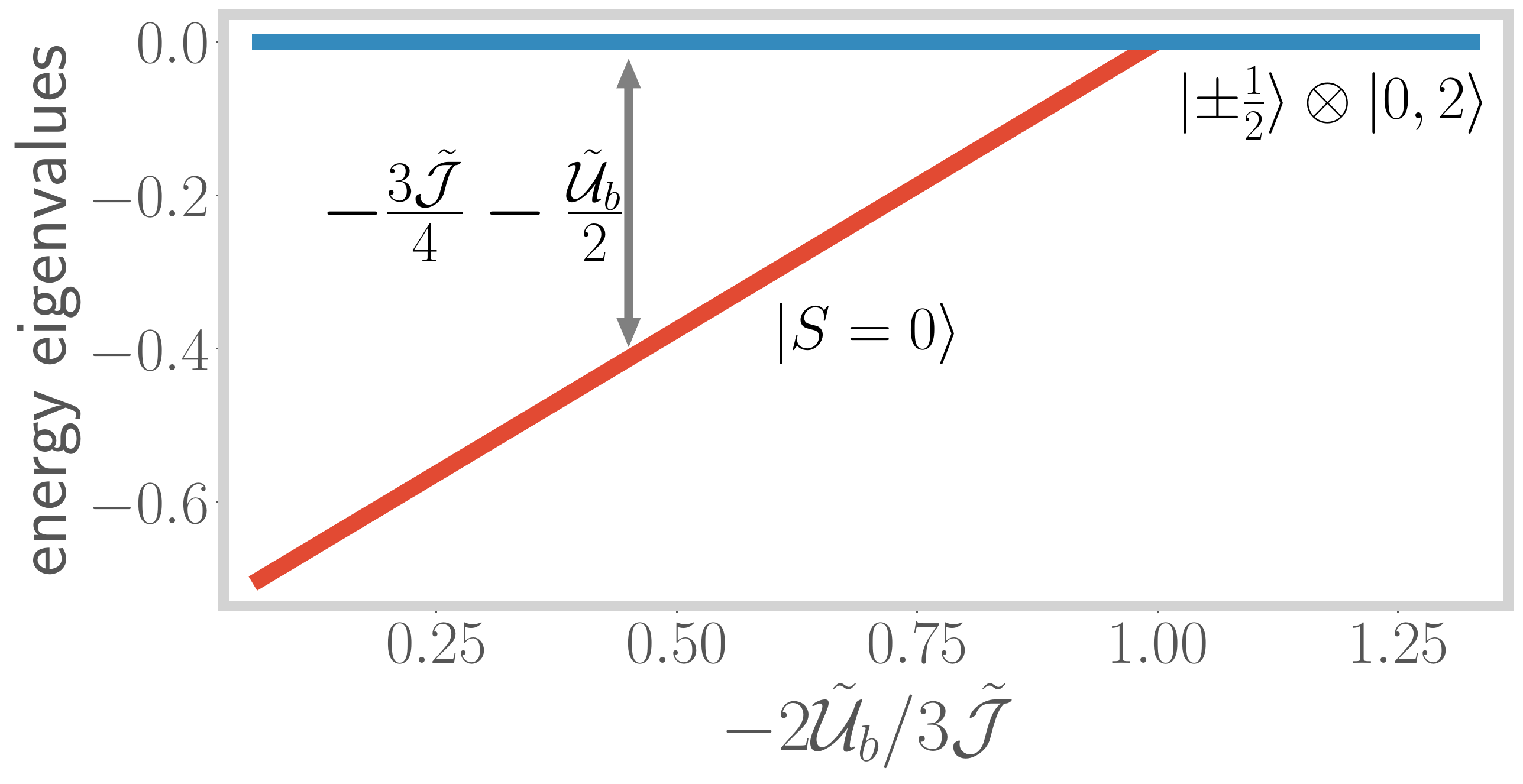}
	\caption{{\it Left:} Very close to the transition, the mutual information \(I_2(d:0)\) (red curve) between impurity and the zeroth site reduces, while that between the zeroth and the first site (\(I_2(0:1)\), blue curve) increases, showing the redistribution of entanglement. {\it Right:} Spectrum of the zero bandwidth Hamiltonian of the right panel of Fig.~\eqref{universal-theory}. The blue line represents the local moment states at zero energy, while the red line represents the singlet state at an energy that continuously increases upon increasing \(r\).}
	\label{twosite-spectrum}
\end{figure}

Indeed, the local correlation $\mathcal{U}_b$ encourages entanglement between the sites of the bath, and makes the formation of the impurity-bath singlet difficult. In other words, the \(\mathcal{J}-\mathcal{U}_b\) model displays the destabilisation of the singlet by redistributing the entanglement from the impurity+bath system to purely within the bath (see left panel of Fig.~\eqref{twosite-spectrum}). Given the simplicity of these arguments, 
our analysis makes the case that local pairing fluctuations of the bath in eq.~\eqref{universal} offer a universal mechanism by which to frustrate the Kondo effect, and lead thereby to an impurity transition that is the local counterpart of the Mott MIT obtained by auxiliary model approaches such as DMFT. We note that similar conclusions were reached in Refs.\cite{si_kotliar_1993,kotliarsi_1993,kotliar_si_toulouse_1996} for the emergence of non-Fermi liquid phases at critical points in the mixed valence regime of the periodic Anderson model.

Further insight into the destabilisation of the singlet can be obtained from a zero-bandwidth approximation of the bath. Under such an approximation, the IR effective Hamiltonian obtained 
from the RG flow (row 2 of Fig.~\eqref{phases-table} takes the form of a two-site model with modified couplings (shown in the right panel of Fig.~\eqref{universal-theory}):
\begin{eqnarray}
\mathcal{\tilde J} \vec{S}_d\cdot\vec{S}_0 - \frac{1}{2}\mathcal{\tilde U}_b\left(\hat n_{0 \uparrow} - \hat n_{0 \downarrow}\right)^2 \label{zero-bw-pic}~.
\end{eqnarray}
The spectrum of this model is shown in the right panel of Fig.~\eqref{twosite-spectrum}. For \(|\mathcal{\tilde U}_b|/\mathcal{\tilde J} < 3/2\), the singlet ground-state (red) is separated from the excited local moment states (blue) by a gap of \(\frac{-3\mathcal{\tilde J}}{4} - \frac{\mathcal{\tilde U}_b}{2}\). As we tune \(r\) towards \(r_{c2}\), \(|\mathcal{\tilde U}_b|\) increases and the fixed point value \(\mathcal{\tilde J}\) decreases, leading to an overall reduction in the gap. At the critical point, the gap closes and the states become degenerate at zero energy; this is the equivalent of the quantum critical point at $r_{c2}$ of the full impurity model that was discussed earlier. We note that the vanishing of the energy of the metallic state was also observed by Brinkman and Rice from a Gutzwiller-type variational calculation of the Hubbard model at half-filling~\cite{brinkman_rice_1970}.

\subsection{Local pairing and the destruction of Kondo screening} 
\begin{figure}[!htpb]
	\centering
	$\vcenter{\hbox{\includegraphics[width=0.43\textwidth]{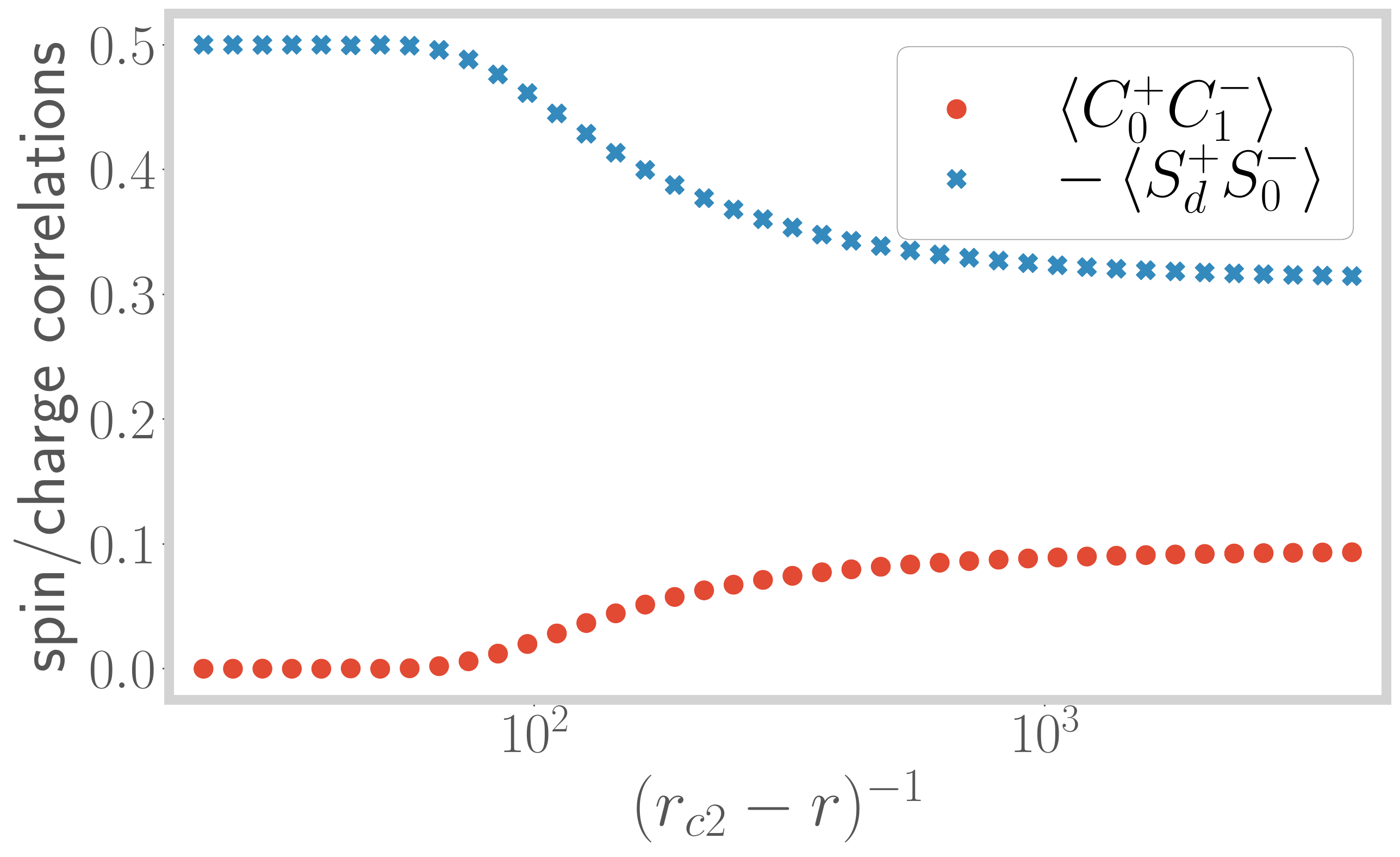}}}$
	\hspace*{\fill}
	$\vcenter{\hbox{\includegraphics[width=0.45\textwidth]{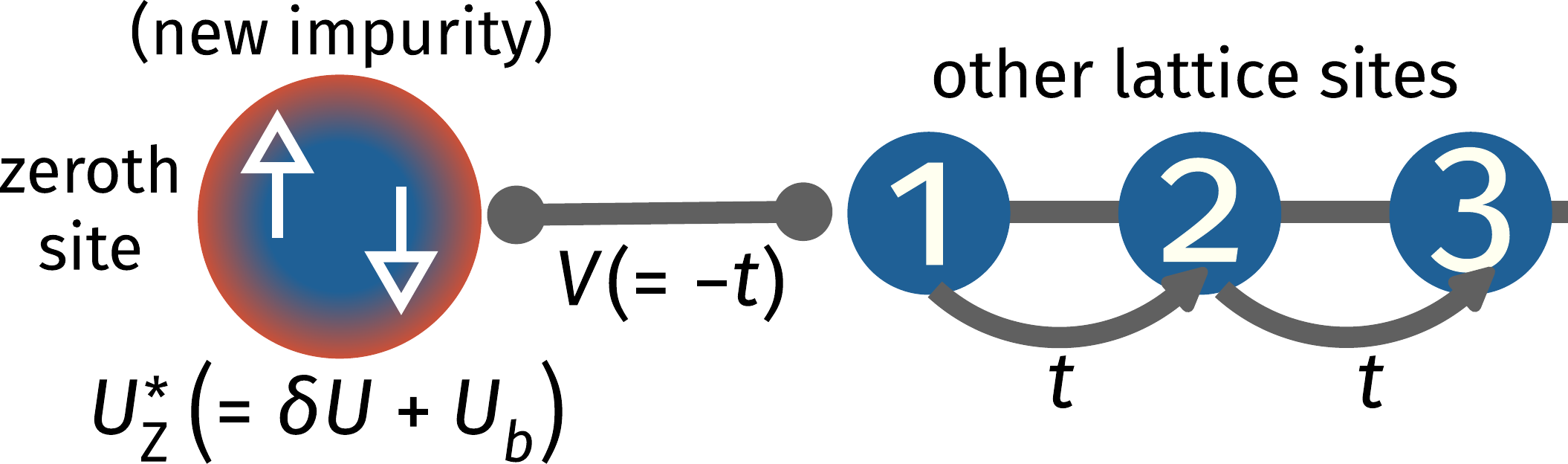}}}$
	\caption{{\it Left:} Spin-flip correlation between the impurity and the bath (blue), and charge isospin-flip correlation between the bath zeroth and first sites (red), both as a function of the tuning parameter \(r\), very close to the transition. The operators are defined as \(S_i^+ = c^\dagger_{i \uparrow}c_{i \downarrow}\) and \(C_i^+ = c^\dagger_{i \uparrow}c^\dagger_{i \downarrow}\). {\it Right:} New effective SIAM Hamiltonian generated upon integrating out the coupling between the impurity and the bath zeroth site. In the new SIAM, the zeroth site acts as the new impurity site with a renormalised on-site correlation \(U_Z^* = U_b + \delta U_Z^*\).}
	\label{spin-charge-corr}
\end{figure}

Very close to the transition at \(r_{c2}\), we find signatures of the breakdown of the Kondo cloud in terms of decaying impurity-zeroth site correlations (blue curve in the left panel of Fig.~\eqref{spin-charge-corr}) and enhanced intra-bath correlations. Remarkably, we find an {\it increase in pairing correlations} between the bath zeroth site and first site (red curve in the left panel of Fig.~\eqref{spin-charge-corr}). As discussed above, these are observed to be responsible for the destruction of the Kondo screening: spin and charge degrees of freedom are mutually exclusive, and the increased charge fluctuations on the bath zeroth site suppress the spin-flip scattering processes between the impurity and the zeroth site. We will show later that these fluctuations lead to the destruction of the local Fermi liquid excitations in our model, and replace them with non-Fermi local excitations in the neighbourhood of \(r_{c2}\).
We believe that the observed growth in non-local pairing correlations near the transition at \(r_{c2}\) is likely tied to a putative low-energy divergence of the local pairing susceptibility. A similar observation was made in Ref.\cite{si_kotliar_1993} from an auxiliary model-based analysis of a hole-doped extended Hubbard model in infinite dimensions that is pertinent to the physics of the heavy fermions. Similar to the conclusions of Ref.\cite{si_kotliar_1993}, we expect the pairing fluctuations of the bath to become dominant upon tuning the e-SIAM away from half-filling, signalling an instability towards a superconducting state.

\subsection{Emergent self-consistency in the e-SIAM}

As is well-established, the DMFT self-consistency equation is equivalent to requiring that the impurity Greens function become equal to the local Greens function in the bath~\cite{georges1996}.
Such a condition is also used in the projective self-consistent technique of Moeller et al.~\cite{moeller_1995} for the states within the central Kondo resonance.
We will now show that our model displays a qualitatively similar emergent feature.
To proceed, by employing a one-step URG transformation, we integrate out the impurity site from the rest of the fixed-point Hamiltonian of eq.~\eqref{fixed-point-ham}.
The details are shown in Sec. 4 of the Supplementary Materials~\cite{supp_mat}. The essential idea is similar to that of the Schrieffer-Wolff transformation: removing the impurity-bath couplings \(J\) and \(V\) generates an additional repulsive correlation \(\delta U_z^* \) on the zeroth site. The new low-energy model \(H_\text{Z}\) is therefore an Anderson impurity model with a net local correlation \(U_\text{Z}^* = \delta U_Z^* + U_b\) on the zeroth site (which is now the ``new impurity site"), and a single-particle hybridisation \(V_\text{Z}^* = -t\) coupling with a conduction bath formed by the remaining sites. This new impurity model is depicted schematically in the right panel of Fig.~\eqref{spin-charge-corr}.
\begin{eqnarray}
	H_\text{Z}^*  = \underbrace{-\frac{1}{2}U_\text{Z}^*\left(\hat n_{0\uparrow} - \hat n_{0\downarrow}\right)^2}_\text{new impurity = \(0^\text{th}\) site} + \underbrace{V_\text{Z}^*\sum_{\left<j,0 \right>}\left(c^\dagger_{0\sigma}c_{j\sigma} + \text{h.c.}\right)}_\text{hopping between \(0^\text{th}\) site \& new bath} + \underbrace{(-t) \sum_{\left<i,j \right>}\left(c^\dagger_{i\sigma}c_{j\sigma} + \text{h.c.}\right)}_\text{K.E. of new bath}~.
\end{eqnarray}

Since, for \(r < r_{c2}\), the model always flows to strong-coupling at low-energies, the largest energy scales are \(J^*\) and \(V^*\). Using this, the effective correlation \(U_Z^*\) on the zeroth site can be expressed to leading order as
\begin{eqnarray}
	\label{new-imp-correlation}
	U_\text{Z}^* \simeq \frac{J^*}{4}\frac{1}{1 + 2\gamma} - V^* \frac{\gamma}{\gamma^2 - \frac{1}{4}},~ \text{ where }\gamma\equiv V^*/J^*~.
\end{eqnarray}
From \(r = 0\) to \(r = r_{c1}\), the factor \(\gamma\) decreases due to the gradual removal of single-particle hopping from the impurity site (irrelevance of \(V\)), leading to an increase in the correlation \(U_Z^*\). This is simply a restatement of the fact that the scattering processes that create the central impurity resonance induce a repulsive correlation on the bath zeroth site. The fact that we end up with a standard SIAM on the zeroth site once the impurity site has been integrated out means that the spectral function of this new impurity will again go undergo a sharpening of the central peak (and the appearance of the Hubbard sidebands) upon increasing the parameter $r$ of the original e-SIAM. This ensures that the impurity and zeroth site spectral functions look qualitatively similar up to \(r_{c1}\). One can now repeat iteratively this process - decoupling the zeroth site generates a standard SIAM with a repulsive correlation on the first site, and so on. The fact that excursions starting from any point along the bath can be described by a positive \(U\) SIAM is, therefore, the emergent self-consistency in our model. Similar indications of a correlated spectral function on lattice sites far away from the impurity were also observed in Ref.\cite{martin2019} from finite-U slave boson calculations of the SIAM. Beyond \(r_{c1}\), the irrelevance of \(V\) means that \(U_Z^*\) reduces to just \(J^*/4\). As \(r\) is now increased towards \(r_{c2}\), the correlation decreases because \(J\) is moving towards its critical point, indicating that the bath zeroth site is moving away from its local moment regime. This is another reflection of the increase in pairing fluctuations of the bath, as well as the lowering of spin-flip fluctuations between the impurity and the bath.

For \(r > r_{c2}\), the impurity site decouples from the bath in the impurity model. This impurity model can be promoted to a bulk model as follows. Recall that in the auxiliary model mapping, any lattice site \(\vec r_i\) of the bulk lattice can act as the impurity, and one can think of the impurity model with the impurity at \(\vec r_i\) as a representation of the excursion of an electron from any such site \(\vec r_i\) into the rest of the bath. When all such impurity models undergo the transition at \(r = r_{c2}\) simultaneously, the result is the decoupling of all sites from their respective baths and a paramagnetic bulk insulator is obtained.

\section{Coexistence of local metallic and insulating phases in the e-SIAM}
\label{dmft}

The DMFT solution of the Hubbard model on the Bethe lattice exhibits a coexistence region of metallic and insulating solutions between two spinodal lines \(U_{c1}(T)\) and \(U_{c2}(T)\)~\cite{georges1996}, with the metallic solution having lower internal energy and the insulating solution being a metastable state at a higher energy~\cite{moeller_1995,georges1996,bulla_1999,georges_2004_dmft,georges_krauth_1993}.
The \(T=0\) transition is observed to be second order in nature and occurs through a merging of the metallic and insulating solutions at \(U_{c2}\)~\cite{rozenberg_1994_zerotemp,moeller_1995}. On the other hand, at $T>0$, the MIT happens along the first-order line \(U_c(T) \left(U_{c1} < U_c <U_{c2} \right)\) where the free energies of the two solutions become equal, and the two spinodal lines merge into a second order critical point at a sufficiently high temperature \(T_c\). 
We will now show that our Hamiltonian-based approach gives a clear picture of various aspects of the  physics that lead to the first-order transition at $T>0$.

\subsection{Excited state quantum phase transition at \(r_{c1}\): the Mott-Hubbard scenario}

We have already discussed in detail the nature of the phase transition at \(r_{c2}\): the zero-bandwidth picture provided (around eq.~\eqref{zero-bw-pic}) shows the merging of the metallic and insulating solutions at that point, enabling identification of \(r_{c2}\) with the $T=0$ continuous phase transition observed at \(U_{c2}\) in the DMFT phase diagram. Moreover, the finite temperature line \(r_{c2}(T) = U_{c2}(T)\) marks the boundary beyond which the spin-singlet is no longer an eigenstate of the quantum-mechanical spectrum.

The focus of this subsection is the physics of the other important point \(r_{c1}\). As has been mentioned before, this point marks the RG irrelevance of the single-particle hybridisation amplitude \(V\), and leads to the exclusion of the charge states from the ground-state (see left panel of Fig.~\eqref{spec_func}). This exclusion means that there is now one fewer scattering channel by which the impurity electron can hybridise with the bath. In turn, this leads to a {\it partial localisation of the impurity}, and can be thought of as the first step towards the more complete localisation that occurs at \(r_{c2}\).

Apart from the change in the ground-state, the physics at \(r_{c1}\) also involves a phase transition in certain excited states of the spectrum. To expose this, we consider the following states in the zero-bandwidth spectrum of the impurity model given in eq.~\eqref{GIAM-ham}:
\begin{eqnarray}\label{esqpt-1}
	\ket{1,\sigma,\pm} &=& \alpha_\pm \ket{\sigma_d}\ket{0_0} + \sqrt{1 - \alpha_\pm^2}\ket{0_d}\ket{\sigma_0}, ~\ket{3,\sigma,\pm} = \alpha_\pm \ket{\sigma_d}\ket{2_0} + \sqrt{1 - \alpha_\pm^2}\ket{2_d}\ket{\sigma_0} ~ ~\left(\sigma=\uparrow,\downarrow\right) ~,\nonumber\\
	E_+ &=& -\frac{U_0}{4} + \sqrt{V_0^2 + \frac{U_0^2}{16}}, ~~E_- = -\frac{U_\text{im}}{4} - \sqrt{V_{\text{im}}^2 + \frac{U_{\text{im}}^2}{16}},\nonumber \\
	\alpha_+ &=& \frac{V_0}{\sqrt{V_0^2 + \left(E_+ + \frac{U_0}{2}\right)^2 }}, ~ ~\alpha_- = \frac{-V_\text{im}}{\sqrt{V_\text{im}^2 + \left(E_- + \frac{U_{\text{im}}}{2}\right)^2}}~.\label{esqpt-2}
\end{eqnarray}
\(E_\pm\) is the energy of the states \(\ket{1(3),\sigma,\pm}\) and the subscripts $d$ and $0$ refer to the impurity and bath zeroth site respectively. The states \(\ket{1(3),\sigma,+}\) are high-energy states, so their energy \(E_+\) and coefficient \(\alpha_+\) involve the bare single-particle hybridisation \(V_0\) and impurity on-site repulsion $U_{0}$. On the other hand, the other states \(\ket{1(3),\sigma,-}\) are closer to the IR energy scale, and involve renormalised intermediate-scale couplings \(V_\text{im}\) and $U_{\text{im}}$. 
Both \(E_+\) and \(E_-\) are four-fold degenerate because of the SU(2) spin \((\uparrow \leftrightarrow \downarrow)\) and particle-hole \((\ket{0} \leftrightarrow \ket{2})\) symmetries. For example, the degenerate subspace corresponding to \(E_+\) is the set of states \(\left\{\ket{1,\uparrow,+}, \ket{1,\downarrow,+}, \ket{3,\uparrow,+}, \ket{3,\downarrow,+}\right\}\). 

\begin{figure}[htpb]
	\includegraphics[width=0.43\textwidth]{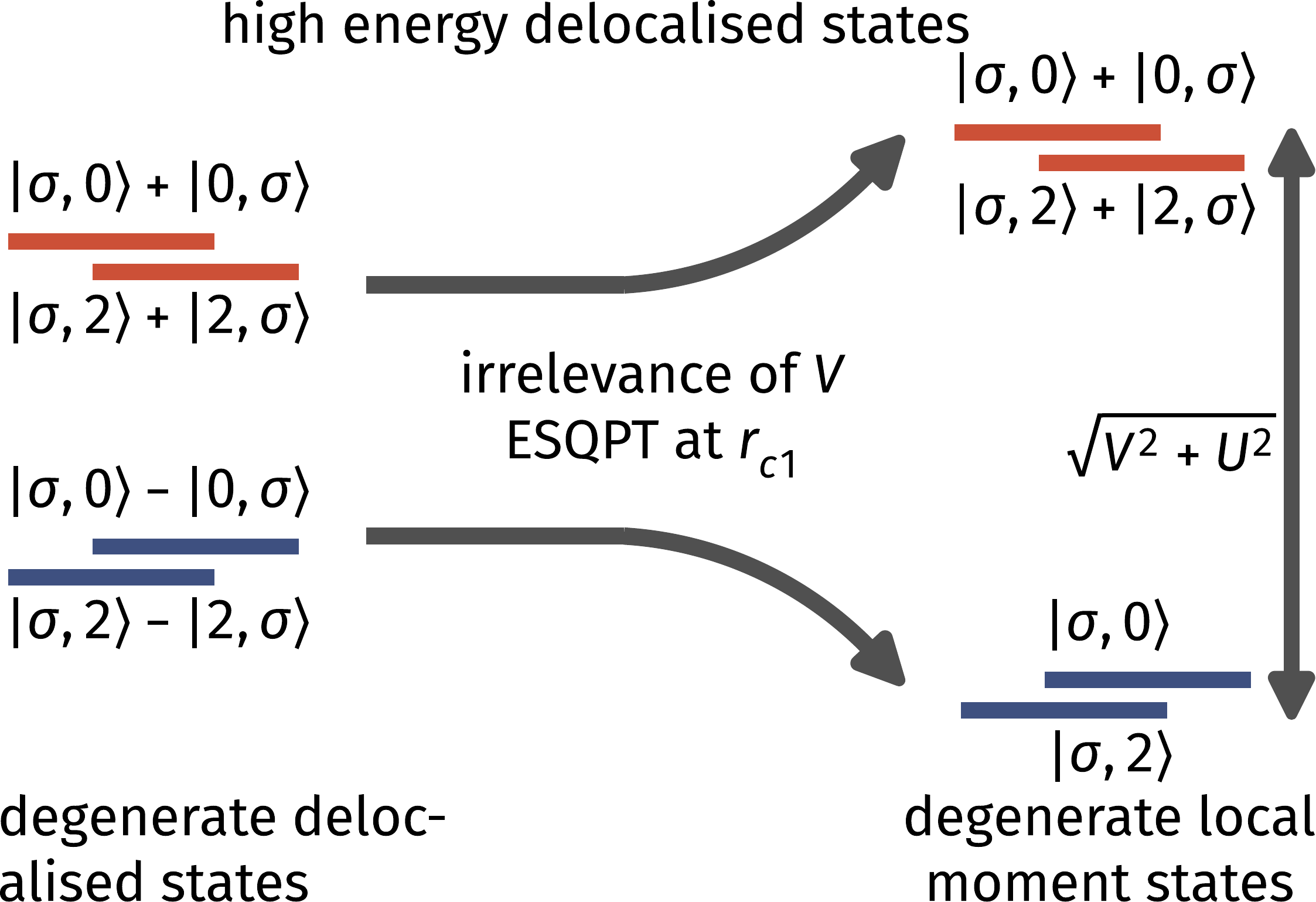}
	\hspace*{\fill}
	\includegraphics[width=0.43\textwidth]{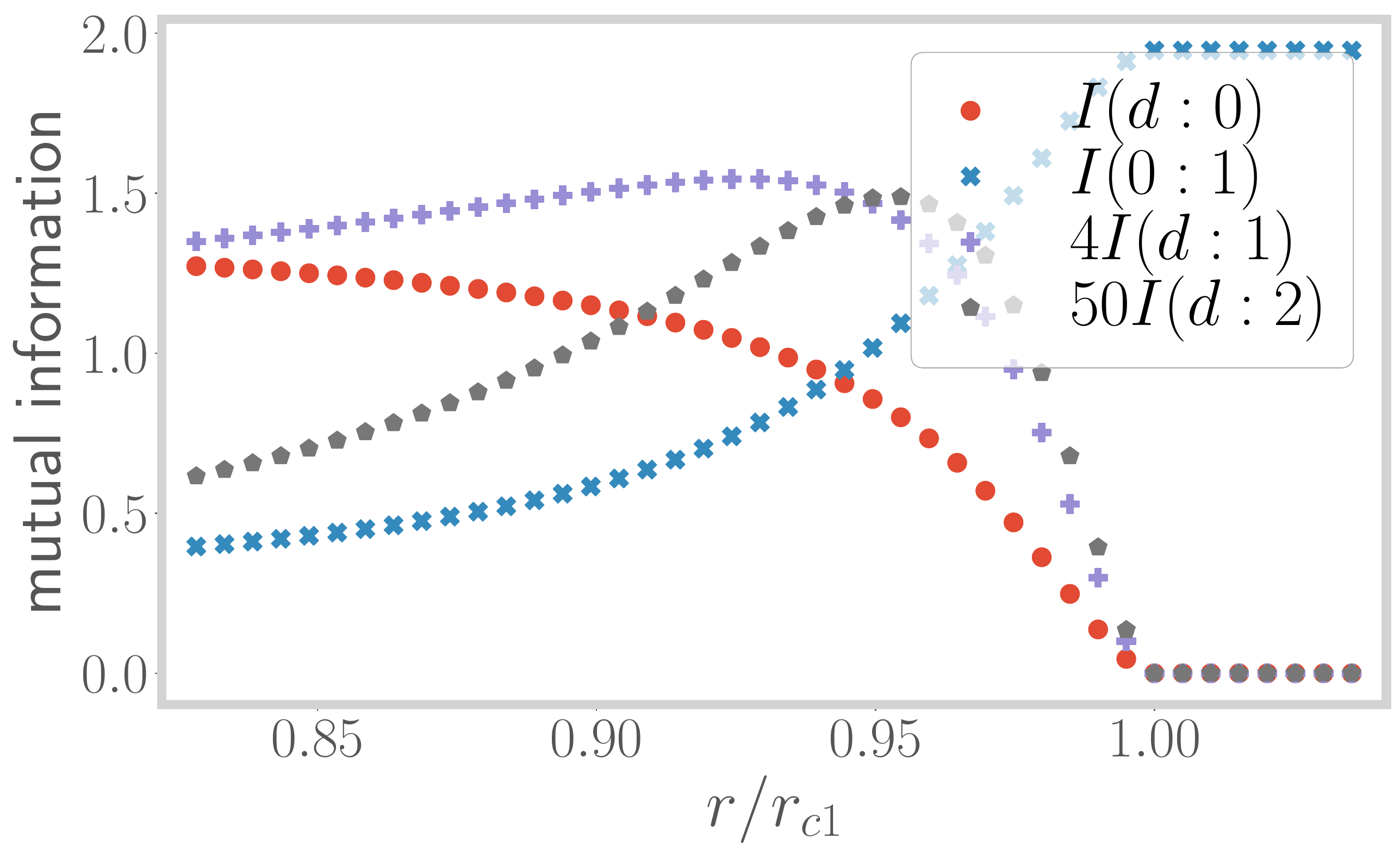}
	\caption{{\it Left:} Evolution of the excited states mentioned in Eqs.~\eqref{esqpt-1} through \eqref{esqpt-2} as \(r\) is tuned through \(r_{c1}\). The high-energy states (red) retain their impurity-bath coupling, while their low-energy counterparts (blue) become disentangled and form a degenerate set of local moment states. The high-energy states allow delocalisation into the bath and lead to the broad Hubbard sidebands. {\it Right:} Variation of mutual information (MI) between various parties, across the ESQPT at \(r_{c1}\). The MI (red) between the impurity and the zeroth site vanishes at \(r_{c1}\), showing the localisation of the impurity. The same between the bath zeroth site and the next site becomes maximum at \(r_{c1}\). The MI between the impurity and the other sites (purple and gray) of the bath show an initial rise near \(r_{c1}\), indicating that entanglement is becoming long-ranged near the ESQPT.}
	\label{esqpt-schematic}
\end{figure}

All these states are in general delocalised - they involve the impurity hybridising with the bath via \(V\). At \(r = r_{c1}\), however, the coupling \(V\) becomes irrelevant, so that the coefficient \(\alpha_-\) of the low-energy state vanishes beyond that point. The low-energy states \(\ket{1(3),\sigma,-}\) thus become localised at \(r_{c1}\) and give rise to a set of excited and {\it degenerate local moment states}, i.e., $\alpha_- \to 0$ as $r \to r_{c1}$ leads to
\begin{eqnarray}
\ket{1(3),\sigma,-} \to \begin{cases}
	\ket{\uparrow_d}\ket{0_0}, \ket{\uparrow_d}\ket{0_0}\\
	\ket{\downarrow_d}\ket{2_0}, \ket{\downarrow_d}\ket{2_0}\\
\end{cases}~.
\end{eqnarray}
The other coefficient \(\alpha_+\) remains non-zero because, as mentioned earlier, \(V\) is non-zero in the UV scales of the RG flow. As a result, the high-energy states \(\ket{1(3),\sigma,+}\) remain delocalised, and are separated from the localised states by an energy-scale
\begin{eqnarray}
\lim_{r \to r_{c1}^-} \left(E_+ - E_-\right) &\simeq & \sqrt{V_0^2 + \left(\frac{U_0}{4}\right)^2} + \sqrt{V_{\text{im}}^2 + \left(\frac{U_{\text{im}}}{4}\right)^2} + \frac{U_{\text{im}}-U_0}{4}~.
\end{eqnarray}
This is shown schematically in the left panel of Fig.~\eqref{esqpt-schematic}, and corresponds to the preformed gap in the zero mode spectrum. The point \(r_{c1}\) therefore represents {\it a delocalisation-localisation excited state quantum phase transition} (ESQPT) where degenerate local moment states are emergent as excited states in the many-body spectrum. This localisation of the impurity at $r_{c1}$ is shown in the form of excited state mutual information in the right panel of Fig.~\eqref{esqpt-schematic}. This ESQPT acts as a precursor to the QPT at $r_{c2}$, where the local moment states become degenerate with the spin-singlet ground state. The local moment states are stabilised as ground states in the insulating phase for $r>r_{c2}$. 

\subsection{Theory for the charge excitations in the Hubbard sidebands}
Beyond \(r_{c1}\), excitations into {\it the local moment states \(\ket{1,\sigma,-}\) and \(\ket{3,\sigma,-}\) reside at the edge of the central peak} in the impurity spectral function (purple curve in the right panel of Fig.\eqref{spec_func}) but provide no spectral weight due to the lack of electron mobility. This explains the development of a preformed gap in the impurity spectral function. That these local moment states have to reside at the edge of the central peak becomes clear when we note that as the width of the central peak shrinks continuously, the local moment states must also recede towards zero frequency and finally replace the zero frequency peak at \(r = r_{c2}\) in order to give rise to the insulating local moment phase for \(r > r_{c2}\). On the other hand, the {\it still-delocalised high-energy states \(\ket{1,\sigma,+}\) and \(\ket{3,\sigma,+}\) are pushed into the Hubbard sidebands}, and their hybridisation with the bath through \(V\) and \(t\) is responsible for the broadening of the sidebands. 

This isolation of the delocalised states into the sidebands means that charge delocalisation processes are now excluded from the physics at low-energies. As accessing the sidebands involves excitations at exorbitantly high energy-scales, such processes can only happen virtually and involve very short time-scales. The central Kondo resonance observed at low energies, therefore, does not support any charge delocalisation, and metallic excitations propagate only through spin-flip scattering processes of the impurity. This reveals that the Mott insulator comes about through a local binding of doublons and holons~\cite{Mott_1949,kohn1964theory,castellani_1979}. Closely related to this is the Mott-Hubbard scenario of the MIT, which is equivalent to our observation of the appearance of an optical gap in the spectrum after \(r_{c1}\).

The fact that the sidebands are broad indicates that there are gapless excitations propagating from the impurity site into the conduction bath, but whose energy lies outside the Mott gap. These can also be viewed as metallic excitations of the bath that are induced by the coupled impurity. In order to expose the nature of these excitations, we perform a calculation similar to the local Fermi liquid calculation of Nozières~\cite{nozieres1974fermi}. This involves considering the states at \(\omega \sim \pm U/2\) as the ground-states of the Hubbard sidebands, and then computing how excitations into the bath renormalise the ground-state subspace. Up to second order in the hopping strength \(t\), the effective Hamiltonian for the excitations of the bath can be expressed as
\begin{eqnarray}
H_\text{eff}^{(2)} = \frac{4t^2(1 - \alpha_+^2)}{E_+ - E_\text{gs}}C_\text{tot}^z C_1^z - t\sum_{i >0,\sigma}\left(c^\dagger_{i\sigma}c_{i+1\sigma} + \text{h.c.}\right) ~,\label{hodoLFL}
\end{eqnarray}
where \(E_+\) and \(\alpha_+\) have been defined in eq.~\eqref{esqpt-2}, \(C_i^z\ket{2(0)} = +(-)\frac{1}{2}\ket{2(0)}\) is the z-component of the charge isospin operator at a particular site $i$, and \(\vec C_\text{tot} = \vec C_d + \vec C_0\) is the total isospin for the impurity and zeroth sites.
\(E_\text{gs} = U/2\) is the two-site energy of the ground-state subspace within the sideband.
These excitations are of the local Fermi liquid kind - the absence of any isospin-flip scattering term promotes the independent delocalisation of the doublon and holon states into the bath. The second order effective Hamiltonian can therefore be written as the sum of two decoupled parts corresponding to holon and doublon propagation respectively, and suggesting adiabatic continuity with the \(U=0\) Hubbard model~\cite{hrk_1990_prl}. 

However, fourth order corrections lead to the appearance of scattering processes that convert holon and doublons into one another, resulting in the local Fermi liquid becoming correlated
\begin{eqnarray}
	H_\text{eff}^{(4)} = \frac{\gamma^4 \alpha_+^2 \beta^2}{\left( 1 - \alpha_+^2 \right) \left( E_+^\prime - E_\text{gs} \right) } \left[- C_\text{tot}^z C_\text{tot}^2 C_1^z + \sqrt{2}\mathcal{P}_\text{tot}^{4}\left(C_0^+ - C_d^+\right)C_1^- + \text{h.c.}\right]~,\label{holonLFL}
\end{eqnarray}
where \(E_+^\prime = \frac{U}{4} - \frac{3J}{8} + \sqrt{4V^2 + \left(\frac{U}{4} + \frac{3J}{8}\right)^2}\) is the energy of the state containing the charge isospin triplet zero, and \(\beta = \frac{2V}{\sqrt{4V^2 + \left(U + 3J/4\right)^2}}\). \(\mathcal{P}^4_\text{tot}\) projects on to the \(n_d + n_0 = 4\) subspace.
The scattering processes in eq.~\eqref{holonLFL} are clearly non-Fermi liquid in nature as they allow the inter-conversion of the charge isospin eigenstates, and therefore reduce the lifetime of the quasiparticle excitations of the local Fermi liquid obtained in eq.~\eqref{holonLFL}. Additional details pertaining to the calculation of this effective Hamiltonian are present in Sec. 5 of the Supplementary Materials~\cite{supp_mat}.

\subsection{Hysteresis and the first-order line}

\begin{figure}[!htpb]
	\centering
	\includegraphics[width=0.35\textwidth]{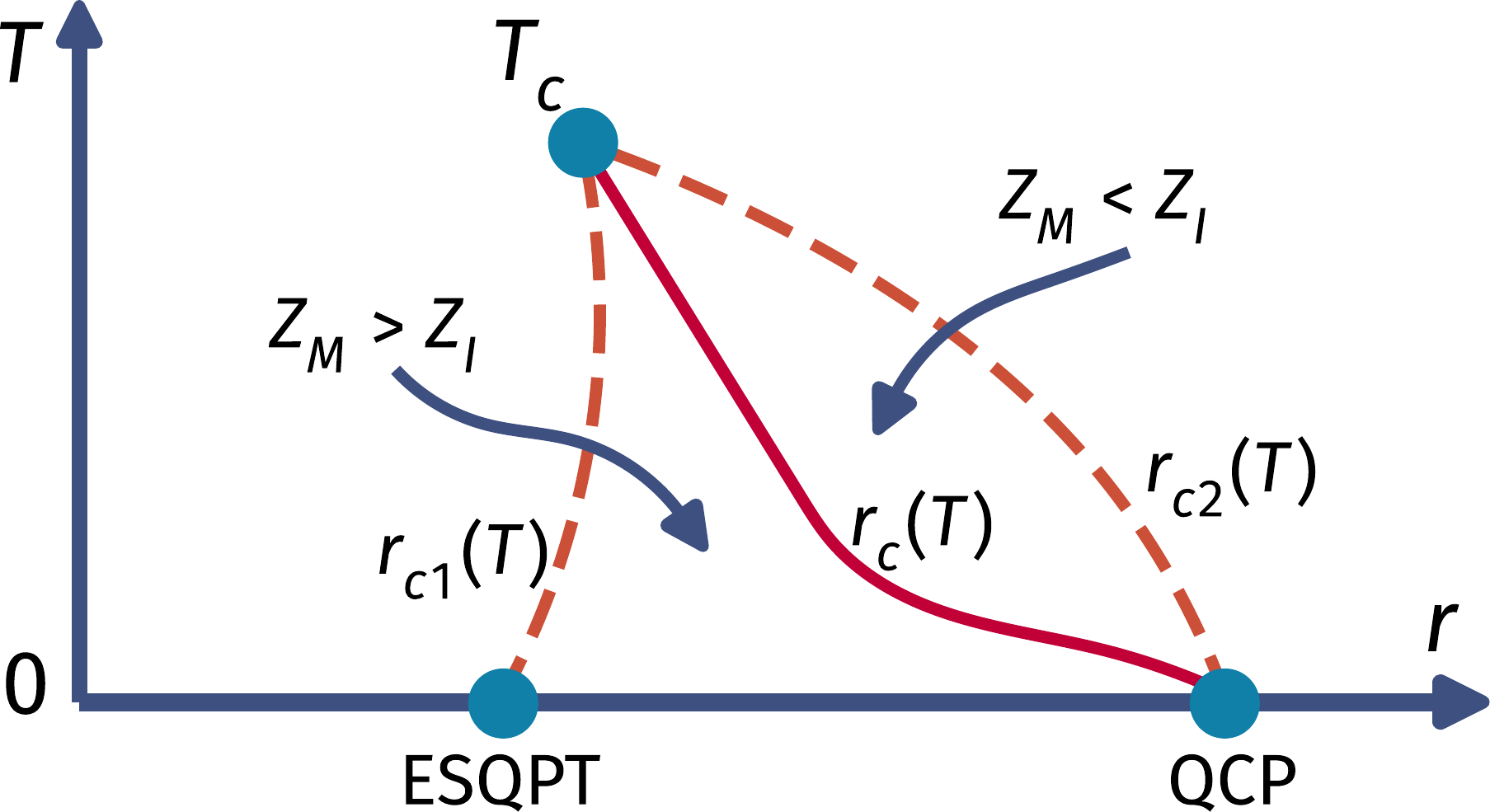}
	\caption{Qualitative structure of the finite temperature coexistence region of the \(J-U_b\) model. The dotted lines on the left and the right represent the spinodals \(r_{c1}\) and \(r_{c2}\) where the insulating and metallic solutions become unstable, respectively. The solid red line represents the first-order line where the free energies and the partition functions of the two solutions become equal, \(Z_M = Z_I\).}
	\label{coexistence-schematic}
\end{figure}

As discussed above, the localisation of charge in the emergent local moment states \(\ket{1,\sigma,-}\) and \(\ket{3,\sigma,-}\) for \(r > r_{c1}\) means they do not contribute any spectral weight to the impurity spectral function. Instead, they lead to an emergent preformed/optical gap between the central Kondo resonance and the Hubbard sidebands on each side. This {\it optical gap} finally becomes the true Mott gap at \(r = r_{c2}\) when the central peak disappears.
On the other hand, upon starting from \(r > r_{c2}\) and reducing \(r\), the system is initially in the (doubly degenerate) local moment ground states, and the impurity DOS is of course gapped. In the rest of the subsection, we will show that the presence of these two transitions leads to finite temperature behaviour for the e-SIAM that is qualitatively similar to that of the Hubbard model as seen from DMFT (that was described at the beginning of this section, and can also be seen in Fig.~\ref{coexistence-schematic}).

At non-zero temperatures, the system is described by not only the ground-state but also the excitations. The metallic and insulating solutions become temperature dependent, with energies \(E_M(r,T)\) and \(E_I(r,T)\) respectively. The curve \(r_{c1}(T)\) at a given temperature again represents the values of the parameter \(r\) at which the local moment states enter the spectrum as excited eigenstates, while \(r_{c2}(T)\) marks the point where the metallic ground-state leaves the spectrum: \(E_M(r_{c2}(T), T) = E_I(r_{c2}(T), T)\). As a result, within the parameter range \(r_{c1}(T) < r(T) < r_{c2}(T)\) at a finite temperature \(T\), the low-energy part of the spectrum (states within the central Kondo resonance) involves both the metallic and insulating solutions. These two low-lying solutions then govern the partition function and hence the free energy:
\begin{eqnarray}
	Z(r,T) \simeq Z_M(r,T) + Z_I(r,T),\quad F(r,T) \simeq -k_B T \ln \left(Z_M(r,T) + Z_I(r,T)\right),\label{free energy eq}
\end{eqnarray}
where \(Z_M(r,T) = \exp{\left(-\beta \mathcal{V}\left(E_M(r,T) - TS_M(r,T)\right)\right)}\) is the partition function corresponding to the metallic state with energy \(E_M(r,T)\) and entropy \(S_M(r,T)\) per unit volume \(\mathcal{V}\), and \(Z_I = \exp{\left(-\beta \mathcal{V}\left(E_I(r,T) - TS_I(r,T)\right)\right)}\) is similarly the partition function corresponding to the local moment states.

In the thermodynamic limit, only the larger of \(Z_M\) and \(Z_I\) contributes to the partition function $Z(r,T)$. Given a temperature \(T\), there exists a value \(r_c(T)\) lying in the range \(r_{c1}(T) < r_{c}(T) < r_{c2}(T)\) at which the partition functions \(Z_M(T)\) and \(Z_I(T)\) become equal: \(Z_M(r_c(T),T) = Z_I(r_c(T),T)\). This condition defines a curve \(\left\{ (T, r_c(T)) \right\} \) at which the system undergoes a first-order transition between the metallic and insulating states:
\begin{eqnarray}
	\lim_{\mathcal{V} \to \infty} F(r,T) = 
	\begin{cases}
	-k_B T \ln Z_M(r,T) \text{ if } r < r_c ~ (Z_M > Z_I)\\
	-k_B T \ln Z_I(r,T) \text{ if } r > r_c ~ (Z_M < Z_I)
	\end{cases}~.
\end{eqnarray}
The first-order curve \(r_c(T)\) is determined from the partition function equality (or equivalently, the free energy equality) condition between the metal and the insulator:
\begin{eqnarray}
	T = \frac{E_M(r_c,T) - E_I(r_c,T)}{S_M(r_c,T) - S_I(r_c,T)}~.
\end{eqnarray}
At \(T = 0\), the first-order line \(r_c(T)\) becomes identical to the critical point \(r_{c2}\), as \(E_M(r_c) = E_I(r_c)\) at $T=0$. This simply means that the thermal entropy plays no role at zero temperature, and the transition is purely quantum-mechanical in nature. As temperature is now increased, the large entropy \(S_I\) of the doubly degenerate local moment states comes into play and the insulating state is able to take over at a smaller value of \(r\). This suggests that a first-order transition can take place in the thermodynamic limit at $r_c(T) < r_{c2} (T)$. 
This has been shown schematically in Fig.~\eqref{coexistence-schematic}.

Further, away from the thermodynamic limit, the free energy in eq.~\eqref{free energy eq} admits contributions from the stable metallic state as well as the metastable insulating state, and results in the coexistence of the two phases. This is because, depending on whether we are approaching from \(r_{c2}^+\) or \(r_{c1}^-\), the system will remain mostly in the minimum of either \(Z_I\) or \(Z_M\) respectively. There exist finite activation barriers in the free energy for passage into the other minimum, leading to hysteresis. The transition is observed only at \(r_{c1}(T)\) or \(r_{c2}(T)\) where the insulating and metallic solutions, respectively, become unstable. These two lines, therefore, mark the spinodals of the coexistence region. At a sufficiently high temperature \(T_c\), the metallic and insulating solutions become indistinguishable at the first spinodal \(r_{c1}\): \(E_M(r_{c1},T_c) = E_I(r_{c1},T_c), S_M(r_{c1},T_c) = S_I(r_{c1},T_c)\). We recall that of these two equalities, the first (\(E_M = E_I\)) also corresponds to the locus of the second spinodal (corresponding to $r_{c2}$). Moreover, combining the two equalities (energy and entropy) leads to the free energy equality condition, marking the locus of the first-order line \(r_c(T)\). This has the consequence that the two spinodals and the first-order line merge into a second order critical point \(r_c(T_c)\) at temperature \(T_c\). This concludes our discussion of the finite temperature behaviour of the extended SIAM model and its qualitative similarities with the DMFT phase diagram for the half-filled Hubbard model on the Bethe lattice.

\subsection{Zero temperature origin of critical fluctuations above the second order point}
\label{rc2-corrs}

\begin{figure}[htpb]
	\centering
	\includegraphics[width=0.45\textwidth]{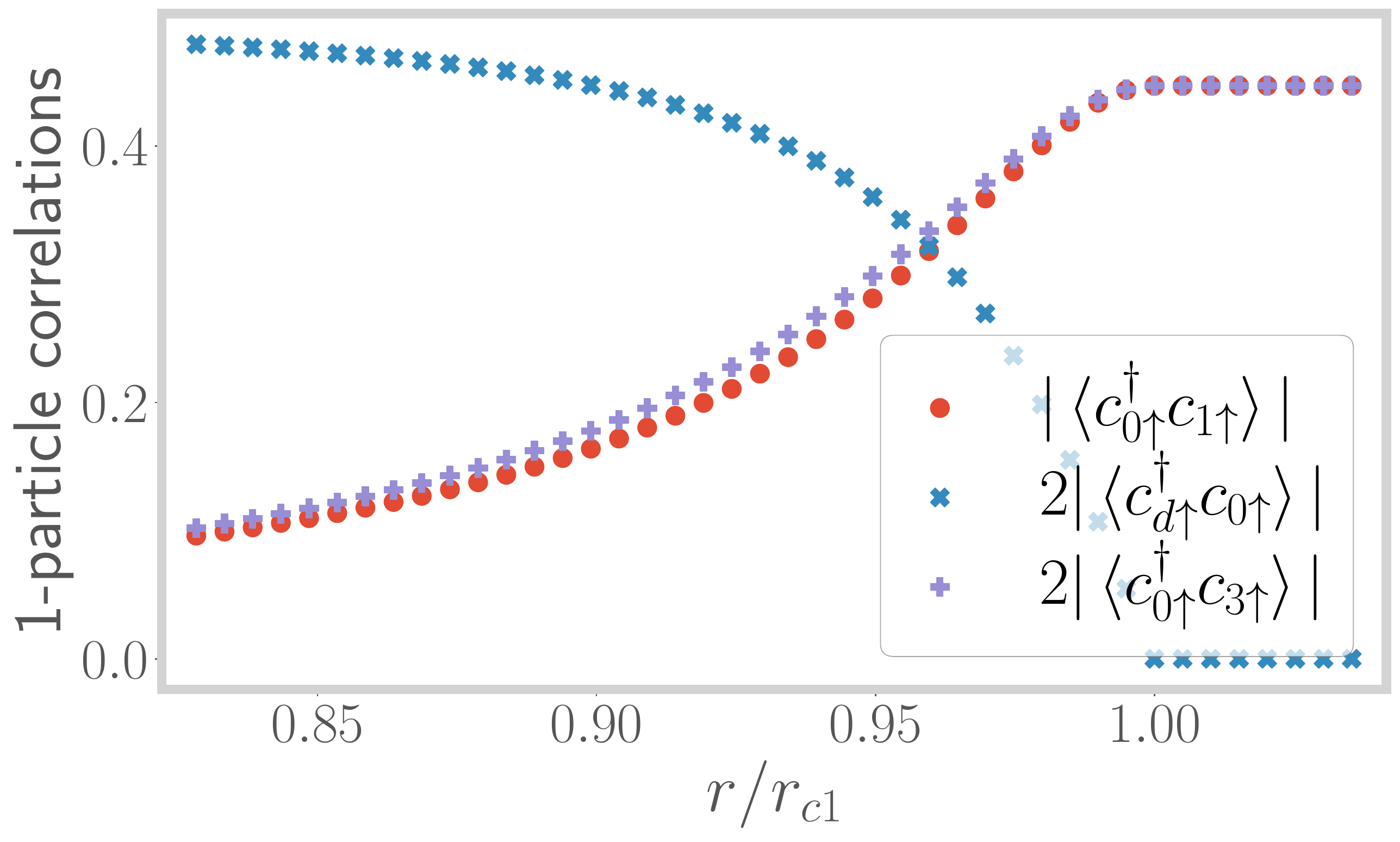}\\
	\includegraphics[width=0.45\textwidth]{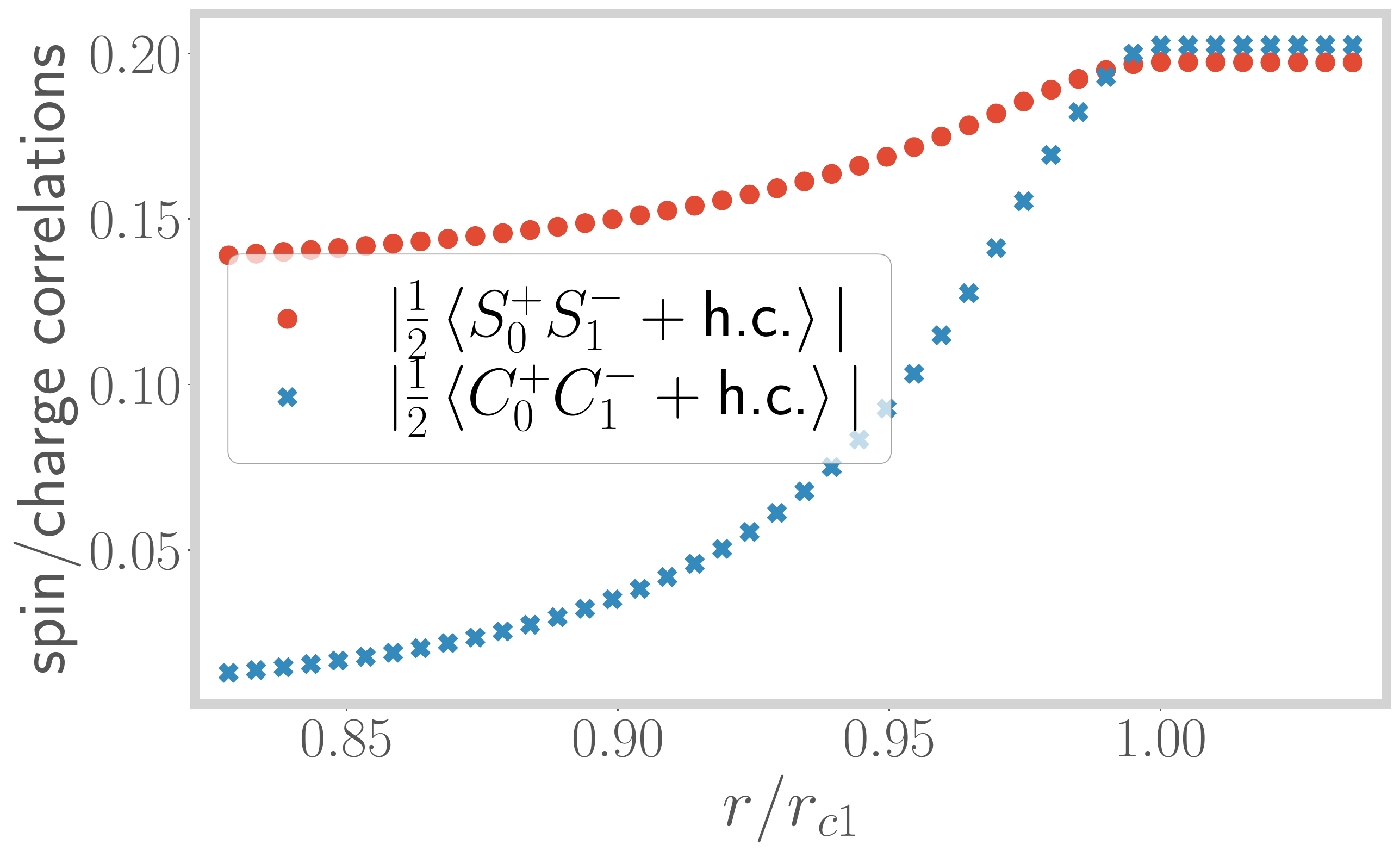}
	\hspace*{\fill}
	\includegraphics[width=0.45\textwidth]{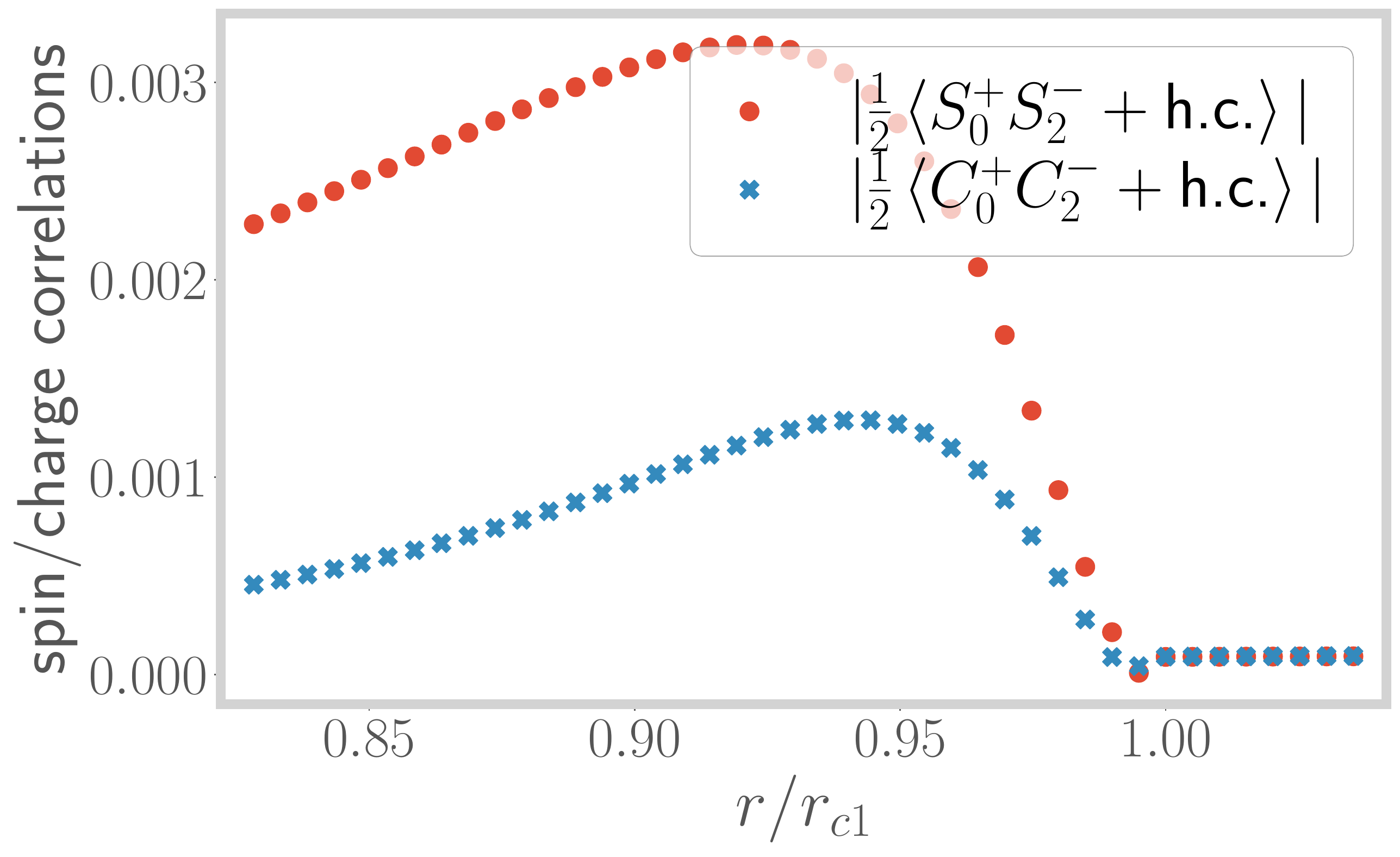}
	\caption{{\it Top:} Variation of the one particle correlations in the state \(\ket{1,\sigma,-}\) between impurity and bath zeroth site(blue), as well as within the bath (red and violet). The former vanishes, depicting the excited state localisation transition at \(r_{c1}\). {\it Left:} Variation of the spin-flip (red) and charge isospin-flip (blue) correlations between impurity and bath, close to \(r_{c1}\). They both increase, indicating that the impurity is now less strongly coupled with the bath. {\it Right:} Variation of the spin-flip (red) and charge isospin-flip (blue) correlations between the zeroth and second sites of the bath, close to \(r_{c1}\). Both show an initial increase, leading to the propagation of relatively long-range correlations into the bath.}
	\label{Uc1-sc-02}
\end{figure}

Above the second order point \(r_c(T_c)\), the DMFT phase diagram shows a rapid crossover from the paramagnetic metallic solution to the paramagnetic insulating solution~\cite{georges1996,limelette_2003}. Remarkably, signatures of quantum critical scaling in this crossover region have been recently predicted from theoretical analyses~\cite{terletska_mott_2011,vucicevic_2013}, as well as detected experimentally in transport measurements of several organic compounds~\cite{Kagawa2005,Furukawa2015}. In Refs.\cite{terletska_mott_2011,vucicevic_2013}, this has been ascribed to the existence of a \textit{hidden quantum criticality} in the maximally frustrated $1/2$-filled Hubbard model on the Bethe lattice with infinite coordination number. In order to locate the zero temperature origin of these signatures of quantum criticality, we inspect carefully the quantum-mechanical fluctuations near the two important points \(r_{c1}\) and \(r_{c2}\) in our model.

Close to \(r_{c1}\), we compute correlations in the state \(\ket{1,\sigma,-}\).
As shown in the top panel of Fig.~\eqref{Uc1-sc-02}, we find that while the one-particle correlation between the impurity and the bath vanishes, the same between the zeroth and first sites of the bath picks up. Importantly, these correlations extend beyond the immediate neighbourhood of the impurity. This is observed in, for example, the one-particle correlation between the zeroth site and the second site (purple curve in top panel of Fig.~\eqref{Uc1-sc-02}). Some other signatures of correlations in the bath are shown in Fig.~\eqref{Uc1-sc-02}: while the left panel displays increased spin-flip and charge isospin-flip correlations between the zeroth and first sites of the bath, a similar phenomenon is observed between the zeroth and second sites of the bath in the right panel. 
Our findings are consistent with the presence of scale-invariant solutions obtained from recent NRG-DMFT calculations ~\cite{Eisenlohr_2019} that correspond to the metastable insulating solutions in the coexistence region.

\begin{figure}[htpb]
	\includegraphics[width=0.45\textwidth]{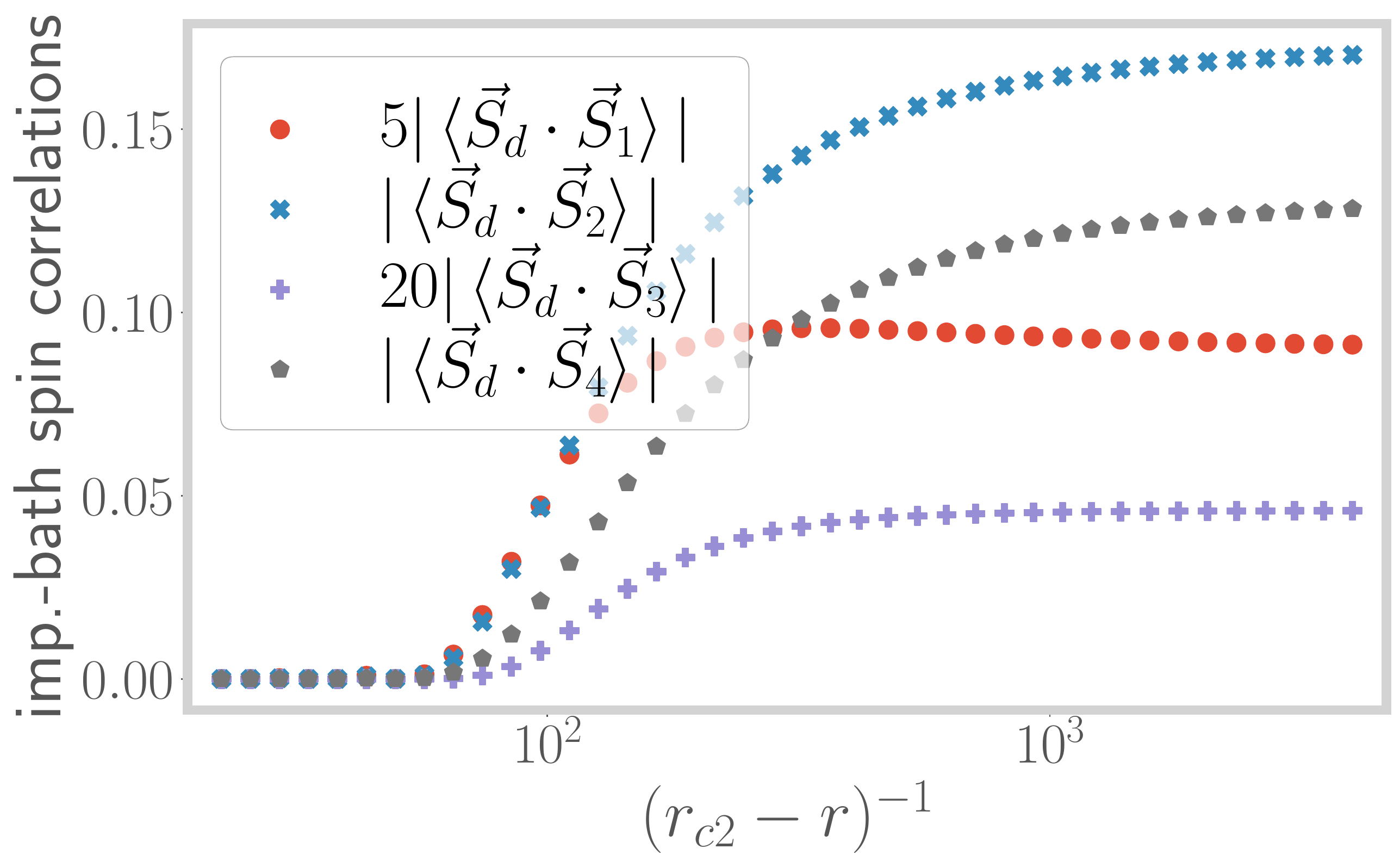}
	\hspace*{\fill}
	\includegraphics[width=0.45\textwidth]{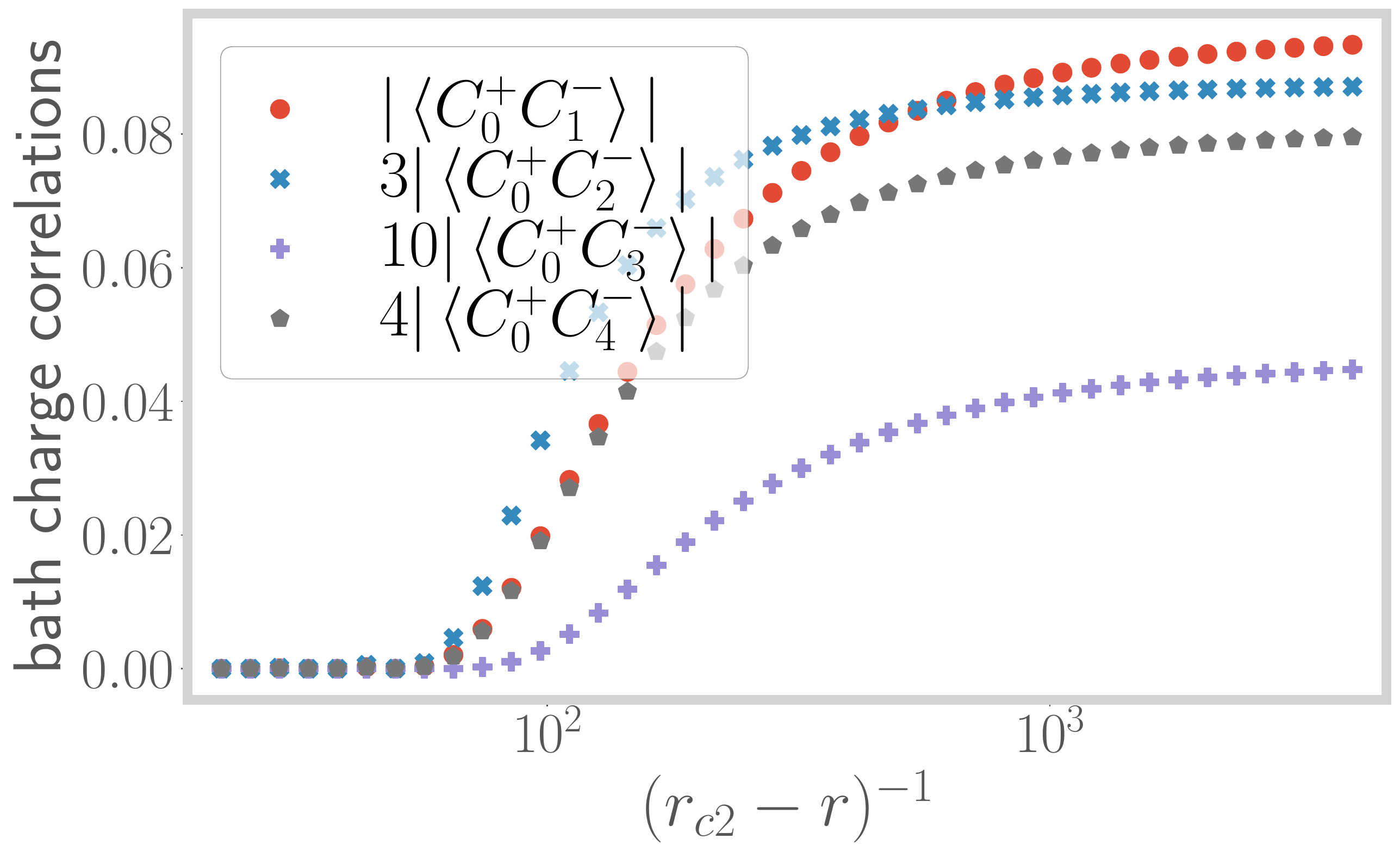}
	\caption{{\it Left:} Variation of the spin-spin correlations between the impurity and bath sites 1 through 4, close to \(r_{c2}\). They show an increase, revealing the distribution of entanglement into the bath. {\it Right:} Variation of pairing correlations within the bath. These correlations show an increase, because of the weakening of the Kondo singlet.}
	\label{Uc2-ss}
\end{figure}

The growth of similar long-ranged correlations within the bath (and leading away from the immediate neighbourhood of the impurity) 
near \(r_{c2}\) as well.
We recall that correlations between the impurity and the zeroth site are observed to decrease (see left panel of Fig.\eqref{gstate-correlations}).
Instead, as shown in Fig.~\eqref{Uc2-ss}, non-trivial two-particle correlations arise between the impurity and bath zeroth sites with bath sites that are farther away.
The spreading of the spin-spin correlations shown in the left panel of Fig.~\eqref{Uc2-ss} indicates a ``stretching" of the Kondo singlet state prior to its destruction. We will show later that these enhanced correlations also result in a diverging quasiparticle mass at \(r_{c2}\), indicating a breakdown of the local Fermi liquid metal. Further, the right panel of Fig.~\eqref{Uc2-ss} indicates that longer-ranged pairing correlations develop between the bath zeroth site and bath sites farther away upon approaching the transition.

In general, within the metallic phase and away from \(r_{c1}\) or \(r_{c2}\), the impurity is strongly coupled to the bath zeroth site and the impurity-bath entanglement follows an area law characteristic of the local nature of the impurity problem. However, near the excited state and ground state transitions at \(r_{c1}\) or \(r_{c2}\) respectively, these results indicate a spreading of entanglement into the bath: more and more bath sites beyond the zeroth site get correlated with the impurity site as the ESQPT or the QCP is approached. This is corroborated in Fig. 3 of the Supplementary Materials~\cite{supp_mat}.
This suggests a significant enhancement of the entanglement beyond the area-law, and is consistent with the critical quantum fluctuations observed in the various two-particle correlations. 

We have argued above that the physics of zero temperature points \(r_{c1}\) and \(r_{c2}\) survive at finite temperatures in the form of the two spinodal lines in the phase diagram. The results of this subsection then indicate that it might be possible to experimentally detect the signatures of such critical quantum fluctuations around the spinodals even at \(T > 0\). We note that signatures of second-order criticality have been indeed observed recently in vanadium sesquioxide, in the form of critical slowing down and critical opalescence~\cite{satyaki_2020_PRL}. It is, therefore, tempting to speculate that these signatures arise from the presence of critical quantum fluctuations.

\section{Non-Fermi liquid signatures at the MIT}
\label{excitations}

\subsection{Death of the local Fermi liquid: the Brinkman-Rice scenario}

\begin{figure}[htpb]
	\includegraphics[width=0.45\textwidth]{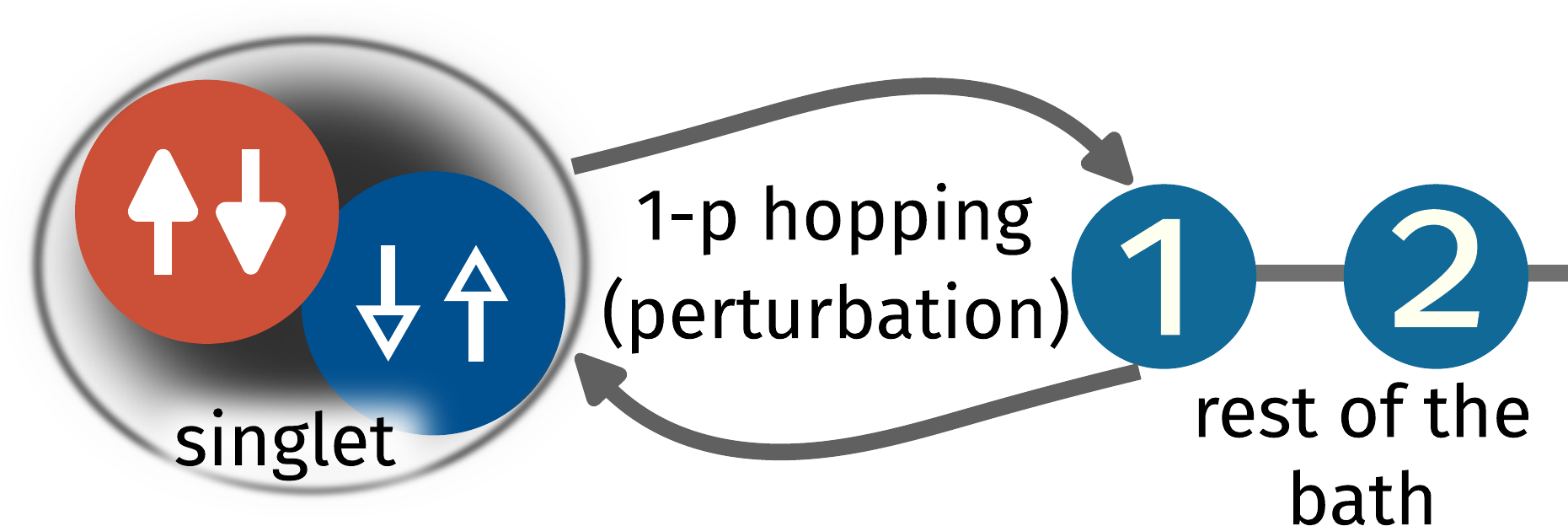}
	\hspace*{\fill}
	\includegraphics[width=0.4\textwidth]{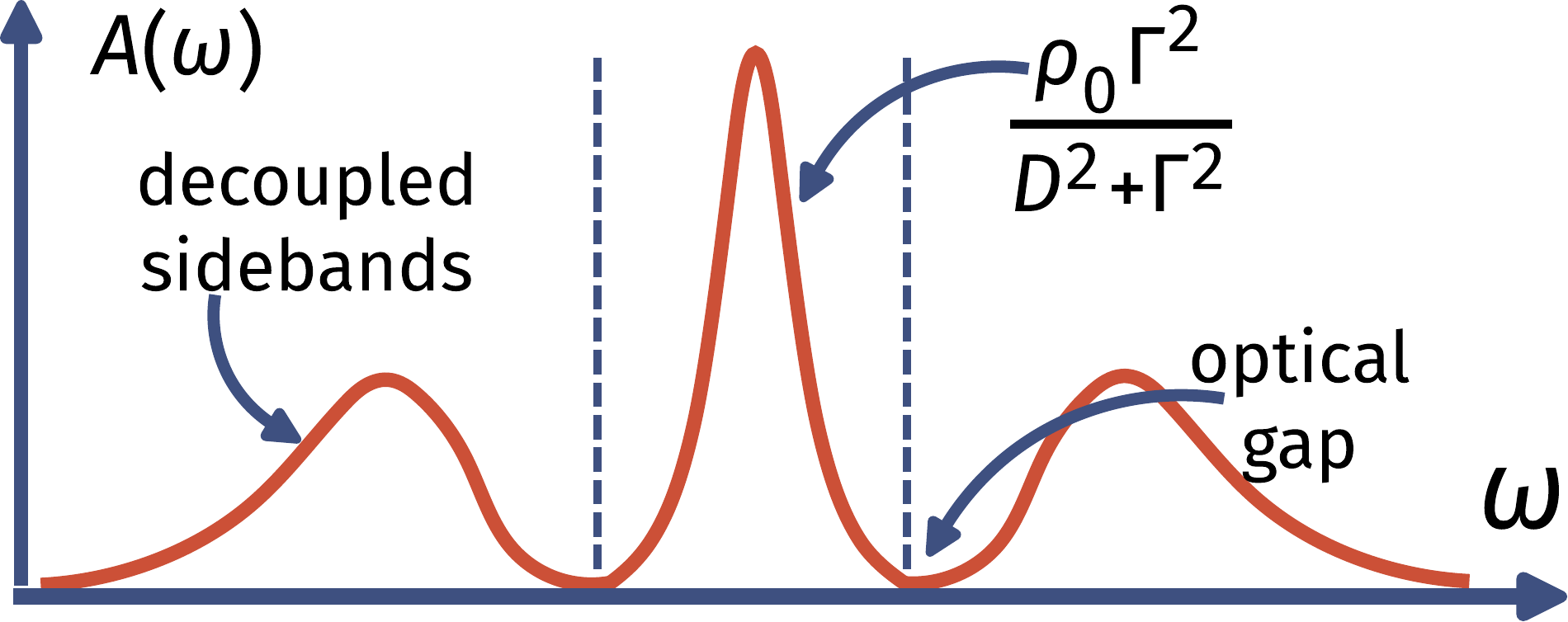}
	\caption{{\it Left:} Setup for obtaining the low-energy excitations above the strong-coupling ground-state. The effects of the rest of the bath is included by treating the hopping between the zeroth site (blue sphere on the left) and the first site (sphere labelled 1) as a perturbation on top of the singlet ground-state formed by the impurity (red sphere) and the zeroth site. {\it Right:} At \(r_{c1}\), the Hubbard sidebands have become isolated from the central peak in the local spectral function because of the finite optical gap. Removing the UV excitations that couple the low-energy and high-energy bands then leads to a theory of the Kondo model with a Lorentzian DOS.}
	\label{lorentzian}
\end{figure}

The low-lying metallic excitations of the bath can be obtained by considering the singlet ground-state of the model (of energy \(\sim -3\mathcal{\tilde J}/4\)) in the metallic phase, and then studying the effect of an electron hopping term (with coupling \(t\)) between the singlet and the rest of the bath as a perturbation (see left panel of Fig.~\eqref{lorentzian}). Such a strong-coupling expansion in powers of \(t^2/\mathcal{\tilde J}\) leads to the usual local Fermi liquid effective Hamiltonian in the Kondo model~\cite{nozieres1974fermi,wilson1975,Noz_blandin_1980,anirban_kondo}:
\begin{eqnarray}
	\label{lfl-ham}
	H_\text{LFL} = \mathcal{F} \hat n_{1 \uparrow} \hat n_{1 \downarrow} + H_\text{KE}~, ~ ~ \mathcal{F} \sim t^4 / \mathcal{\tilde J}^3~,
\end{eqnarray}
where \(\hat n_{1 \sigma}\) are the number operators for the first site (site adjacent to the zeroth site of the conduction bath), \(H_\text{KE}\) is the kinetic energy arising from the nearest-neighbour hopping among all sites in the bath apart from the zeroth site, and \(\mathcal{F}\) is the local Fermi liquid correlation strength. The Kondo singlet (formed between the impurity spin and the bath zeroth site) has decoupled from the rest of the lattice, and eq.~\eqref{lfl-ham} describes the effective Hamiltonian for the rest of the bath beyond the zeroth site. 

Following an identical approach, we find that the effective Hamiltonian for the low-lying excitations in the metallic phase is given by
\begin{eqnarray}
	H_\text{LFL} = \mathcal{F}\left[\hat n_{1 \uparrow} \hat n_{1 \downarrow} + \left(1 - \hat n_{1 \uparrow}\right) \left(1 - \hat n_{1 \downarrow}\right) \right] + H_\text{KE},\quad \text{ where }\mathcal{F} = \frac{2t^4}{\mathcal{\tilde J}\left(3\mathcal{\tilde J}/4 + \mathcal{\tilde U}_b\right)^2}~.
\end{eqnarray}
We note that for our extended SIAM, the local Fermi liquid correlation \(\mathcal{F}\) diverges as the transition is approached at \(3\mathcal{\tilde J}/4 + \mathcal{\tilde U}_b \to 0\). The divergence of \(\mathcal{F}\) leads to the divergence of the renormalised mass of the quasiparticles, and can also be obtained from a Gutzwiller variational calculation as shown by Brinkman and Rice~\cite{brinkman_rice_1970}. This shows the breakdown of perturbation theory and is indicative of the fact that the ground-state is about to change at the metal-insulator transition.

The death of the local Fermi liquid is also seen from the vanishing of the dynamically-generated low-energy Kondo screening scale \(T_K\). Towards obtaining $T_K$ close to the transition, we follow the approach used by Moeller et al. 1995~\cite{moeller_1995} and Held et al. 2013~\cite{held_2013}. Near the transition, they obtained a Kondo model from the SIAM by applying a Schrieffer-Wolff transition that removes the charge fluctuations of the impurity site and retains the physics of only the Kondo coupling \(J\). This amounts to removing the side-peaks from the impurity spectral function and focusing on the low-energy central peak (right panel of Fig.~\eqref{lorentzian}). Held et al. then integrated the RG equation for this Kondo model by using a Lorentzian DOS \(\rho(D)\) in the bath: \(\rho(D) = \frac{\rho_0 \Gamma^2}{D^2 + \Gamma^2}\).
This is motivated by the fact that the central peak of the impurity spectral function is a Lorentzian, and the bath becomes equivalent to the impurity site under self-consistency.

We implement the same approach on the e-SIAM, such that the Schrieffer-Wolff transformation leads to a \(J-U_b\) model with a Lorentzian electronic DOS in the bath. The expression of the Kondo temperature for such a system has been derived in Sec. 6 of the Supplementary Materials~\cite{supp_mat}. Very close to the transition, we have \(r \to 0.25^{-}\), and the Kondo temperature is given by
\begin{eqnarray}
	T_K = \frac{D_0}{k_B} \exp\left[-\frac{\ln\left( r_{c2} - r \right)}{4 U_b \rho_0}\right]~,
\end{eqnarray}
where \(r_{c2} = 0.25\) and \(D_0\) is the bare bandwidth. Note that the pre-factor of the logarithm is positive as \(U_b\) is negative: \(-4U_b\rho_0 = |4U_b \rho_0|\). As we approach the transition, the parameter \(r\) takes the limit \(r \to r_{c2}^-\), and the Kondo temperature scale vanishes:
\begin{eqnarray}
	\lim_{r \to r_{c2}^-} T_K = \frac{D_0}{k_B} \lim_{r \to r_{c2}^-}\left(r_{c2} - r\right)^{4|U_b\rho_0|} \to 0~.
\end{eqnarray}
Following the renormalised perturbation theory approach of Hewson~\cite{hewson1993,coleman2015}, the imaginary part of the self-energy of the local Fermi liquid quasiparticles is given by
\begin{eqnarray}
	\text{Im}\left[\Sigma(\omega)\right] \sim \frac{\mathcal{F}^2}{D_{0}} \frac{\omega^2}{T_{K}^{2}}~, \label{sigma_rc2}
\end{eqnarray}
while the quasiparticle residue is given by \(Z \sim T_K\). The divergence of \(\text{Im}\left[\Sigma(\omega)\right]\) at the transition and the vanishing of \(T_K\) and the quasiparticle residue are important indicators of the loss of the local Fermi liquid excitations and the breakdown of Kondo screening.

The vanishing of the Kondo temperature scale also leads to the divergence of thermodynamic quantities such as the impurity contribution to local spin susceptibility \(\chi_\text{imp}\) and the specific heat coefficient \(\gamma_\text{imp}\). As the low-energy theory is a local Fermi liquid, \(\chi_\text{imp}\) and \(\gamma_\text{imp}\) retain their Fermi liquid forms but involve the highly renormalised Kondo temperature scale~\cite{wilson1975,hewson_1993_prl}:
\begin{eqnarray}
	\chi_\text{imp} = \frac{w}{4k_B T_K}, \quad\lim_{r \to r_{c2}^-}\chi_\text{imp} = \frac{w}{4k_B}\left(r_{c2} - r\right)^{-4|U_b\rho_0|}, \nonumber\\
	\gamma_\text{imp} = \frac{\pi^2 k_B^2 w}{6 T_K}, \quad \lim_{r \to r_{c2}^-}\gamma_\text{imp} = \frac{\pi^2 k_B^2 w}{6}\left(r_{c2} - r\right)^{-4|U_b\rho_0|}~,
\end{eqnarray}
where \(w \sim 0.4128\) is the Wilson number for the Kondo model~\cite{wilson1975,anirban_kondo}. These results also show that the ratio of \(\chi_\text{imp}\) and \(\gamma_\text{imp}\), referred to as the Wilson ratio \(R \equiv \frac{4\pi^2 k_B^2 \chi_\text{imp}}{3 \gamma_\text{imp}}\), remains pinned at the single-channel Kondo value of \(R_\text{LFL} = 2\) for \(r \to r_{c2}-\). This shows that the low-frequency Landau quasiparticles are able to survive until very close to the QCP. As we will show in the next subsection, the metallic excitations precisely at the QCP acquire non-Fermi liquid character and involve additional correlations between holons and doublons. These additional correlations are expected to lead to an enhancement of the Wilson ratio. Evidence for this can be found in the DMFT calculation of a local Wilson ratio very close to the Brinkman-Rice transition, \(R_\text{loc} = 2.9 \pm 0.2\)~\cite{georges1996}.

\subsection{Emergence of non-Fermi liquid excitations at the transition}

In order to obtain an effective Hamiltonian for the excitations precisely at the critical point \(r=1/4\), we first note that the ground-state subspace of the zero bandwidth $J-U_{b}$ effective model (eq.~\eqref{zero-bw-pic}) becomes degenerate at this point:
\begin{eqnarray}
\frac{1}{\sqrt 2}\left(\ket{\uparrow}_d\ket{\downarrow}_0 - \ket{\downarrow}_d\ket{\uparrow}_0\right)~,~ \ket{\sigma}_d\ket{0}_0~,~\ket{\sigma}\ket{2}_0~,~~~ (\sigma=\uparrow,\downarrow)~,
\end{eqnarray}
with all the states lying at zero energy. Here, the first ket (with subscript \(d\)) represents the impurity spin configuration while the second ket (with subscript \(0\)) represents the configuration of the bath zeroth site). We now diagonalise these states in the presence of the electron hopping \(t\) into the rest of the conduction bath, and obtain the effective Hamiltonian for the low-lying excitations mediated by \(t\). This is described in detail in Sec. 7 of the Supplementary Materials~\cite{supp_mat}. We note that the ground-state is four-fold degenerate at energy \(-t\) in the presence of the hopping, and comprises the following states:
\begin{eqnarray}
	\label{degenerate-nfl}
	\ket{N_\text{tot}=2,S^z_\text{tot}=0}~,~\ket{N_\text{tot}=4,S^z_\text{tot}=0}~, ~\ket{N_\text{tot}=3,S^z_\text{tot}=\frac{1}{2}}~,~\ket{N_\text{tot}=3,S^z_\text{tot}=-\frac{1}{2}}~,
\end{eqnarray}
where \(N_\text{tot} = \hat n_d + \hat n_0 + \hat n_1\) and \(S^z_\text{tot} = S^z_d + S^z_0 + S^z_1\) are the total number operator and total magnetisation operator respectively for the impurity, bath zeroth and first sites taken together. It is worth observing that
\begin{itemize}
	\item the presence of the unscreened states \(\ket{N_\text{tot}=3,S^z_\text{tot}=\pm\frac{1}{2}}\) in the ground-state subspace is a result of the inexact screening at the critical point, and
	\item the presence of a degenerate ground-state manifold indicates that these excitations of the bath will likely be of the non-Fermi liquid (NFL) kind~\cite{varma2002singular,si_kotliar_1993,Kotliar_1993}.
\end{itemize}
The precise form of the effective Hamiltonian for this non-Fermi liquid, as well as the behaviour of various correlation functions, is obtained by considering the ground-state manifold in conjunction with the excited states at energy \(-t\). 

Since the total Hamiltonian conserves the total spin \(S_\text{tot}^z\), the effective Hamiltonian separates into two sectors, \(S_\text{tot}^z = 0\) and \(|S_\text{tot}^z| = \frac{1}{2}\). Detailed calculations on obtaining this effective Hamiltonian from the RG fixed point theory are shown in Sec. 7 of the Supplementary Materials~\cite{supp_mat}. In order to highlight certain distinct features, we present simplified effective Hamiltonians only for the \(S_\text{tot}^z = 0\) and \(S_\text{tot}^z = \frac{1}{2}\) sectors. We first take a look at the \(S^z_\text{tot}=0\) sector:
\begin{eqnarray}
	H_\text{eff}^{S^z_\text{tot}=0} = t\vec{S}_d\cdot\left(\vec{\mathcal{S}}_{1,-} + \vec{\mathcal{S}}_{3,-}\right)
	\label{unscreened_sector1}~.
\end{eqnarray}
The effective spin-1/2 ladder operators \(\vec{\mathcal{S}}_{1,\pm}\) act on the positive and negative parity sectors (labelled by $\pm$) of the \(\hat n_0 + \hat n_1 = 1\) subspace spanned by the states
\begin{eqnarray}
	\ket{\Uparrow}_{1,\pm} = \ket{\uparrow}_0\ket{0}_1 \pm \ket{0}_0\ket{\uparrow}_1, \quad \ket{\Downarrow}_{1,\pm} = \ket{\downarrow}_0\ket{0}_1 \pm \ket{0}_0\ket{\downarrow}_1~,
\end{eqnarray}
leading to a Pauli matrix representation of $\vec{\mathcal{S}}_{1,\pm}$ in that basis.
The Pauli matrix operators \(\vec{\mathcal{S}}_{3,\pm}\) similarly act on the \(\hat n_0 + \hat n_1 = 3\) subspace spanned by the states \(\ket{\Uparrow}_{3,\pm} = \ket{\uparrow}_0\ket{2}_1 \pm \ket{2}_0\ket{\uparrow}_1\) and \(\ket{\Downarrow}_{3,\pm} = \ket{\downarrow}_0\ket{2}_1 \pm \ket{2}_0\ket{\downarrow}_1\), obtained by applying the transformation \(\ket{0} \to {2}\) on the \(\hat n_0 + \hat n_1 = 1\) counterparts. 
We now point out two interesting features of this simplified effective Hamiltonian in eq.~\eqref{unscreened_sector1}.
\begin{itemize}
	\item The fact that the impurity is now interacting with emergent spins \(\vec{\mathcal{S}}_1\) and \(\vec{\mathcal{S}}_3\) that span over two lattice sites (0 and 1) instead of just the zeroth site indicates that the entanglement between the impurity and bath is now extended beyond the bath zeroth site, and that {\it the singlet is being stretched}. 
	\item The presence of two spins that are trying to simultaneously screen the impurity spin introduces frustration in the dynamics of the impurity. This is reminiscent of the 2-channel Kondo effect, and the connection is made more precise in the concluding section of our work. 
\end{itemize}
Both these features are precursors to the ultimate and complete destruction of screening in the local moment regime, and reflect the fact that the impurity-bath system {\it is now in an over-screened state}.

We now focus on the effective Hamiltonian of the \(S^z_\text{tot} \neq 0\) subspace:
\begin{eqnarray}
	 H_\text{eff}^{S^z_\text{tot}\neq 0} = -\frac{1}{2}t\left(S_d^- \mathcal{B}^+_\uparrow + \text{h.c.}\right) \label{unscreened_sector2}~.
\end{eqnarray}
The Pauli matrix operators \(\mathcal{B}^+_\uparrow\) and \(\mathcal{B}^-_\uparrow\) flip between the doublet of states \(\ket{\uparrow}_0\ket{\uparrow}_1\) and \(\frac{1}{\sqrt 2}\left( \ket{2}_0\ket{0}_1 + \ket{0}_0\ket{2}_1 \right) \):
\begin{eqnarray}\label{andreev}
	\mathcal{B}^-_\uparrow\ket{\uparrow}_0\ket{\uparrow}_1 = \frac{1}{\sqrt 2}\left( \ket{2}_0\ket{0}_1 + \ket{0}_0\ket{2}_1 \right),\nonumber \\
	\mathcal{B}^+_\uparrow\frac{1}{\sqrt 2}\left( \ket{2}_0\ket{0}_1 + \ket{0}_0\ket{2}_1 \right) = \ket{\uparrow}_0\ket{\uparrow}_1~.
\end{eqnarray}
The structure of the effective Hamiltonian in eq.~\eqref{unscreened_sector2} leads to another drastic difference from the behaviour of the local Fermi liquid: it allows for the electrons to be ``Andreev scattered" into an orthogonal state by hopping from the first site into the zeroth site. This is easily seen by considering the scattering processes in eq.~\eqref{andreev}: a state \(\ket{\uparrow}_0\ket{\uparrow}_1\) (i.e., with site 1 in $\uparrow$ configuration) has a finite probability of being flipped into the \(\ket{\downarrow}_0\ket{2}_1\) state (i.e., with site 1 in the doublon configuration). {\it An incident electron \(c^\dagger_{1 \uparrow}\) can therefore emerge as a doublon \(c^\dagger_{1 \uparrow}c^\dagger_{1 \downarrow}\) or a hole \(c_{1 \uparrow}\) upon scattering from the zeroth site}. This is a direct consequence of the entanglement between the singlet and the first site at the critical point that was absent in the metallic regime of the e-SIAM. We note that similar orthogonal scattering processes also occur in a two-channel Kondo problem between effective pseudo-particles and pseudo-holes~\cite{von_delft_1998}.

At this point, it is worth noting that the polarised ground-state \(\ket{S_\text{tot}^z=\frac{1}{2}}\) can be written as an equal superposition of the singlet state \(\ket{\text{SS}}_{d0}\otimes\ket{\uparrow}_1\) and the local moment states \(\frac{1}{\sqrt 2}\left(\ket{\uparrow, 0, 2} - \ket{\uparrow, 2, 0}\right)\). This symmetry-broken ground-state, along with its counterpart $\ket{S_\text{tot}^z=-\frac{1}{2}}$, act as a bridge between the ground-states of the metallic and insulating phases. They are thus important in displaying the breakdown of the Kondo cloud at the QCP, and the consequences arising from it. We will hence consider these states in the next subsection and demonstrate some exotic non-Fermi liquid properties of these states, e.g., inexact screening of the impurity, fractional impurity magnetisation and fractional impurity entanglement entropy.

Finally, by mapping the quantum impurity problem of the e-SIAM onto that of a classical Coulomb gas~\cite{anderson1969exact,si_kotliar_1993}, it can be shown that the local self-energies \(\Sigma_{dd}(\omega)\) and \(\Sigma_{00}(\omega)\) of the impurity and zeroth sites respectively and certain two-particle correlation functions have power-law behaviours in the frequency domain:
\begin{eqnarray}\label{exponents}
	\text{Re}\left[\Sigma_{dd}\right](\omega) \sim |\omega|^{\gamma_{dd}}, ~ ~\text{Re}\left[\Sigma_{00}\right](\omega) \sim |\omega|^{\gamma_{00}}~,\\
\left<S_d^+\right> \sim |\omega|^{(\alpha_1 - 1)/2}, ~ ~\left<c^\dagger_{0 \uparrow} c^\dagger_{ 0\downarrow}\right> \sim |\omega|^{(\alpha_3 - 1)/2}~.
\end{eqnarray}
The precise forms of the exponents in terms of the conduction electron scattering phase shifts, as well as several other technical details of the calculation, are provided in Sec. 8 of the Supplementary Materials~\cite{supp_mat}. While the various exponents are found to be non-universal functions of the fixed point value of $J$ and the coupling $U_{b}$ for $r\to r_{c2}-$, we argue in subsection \eqref{excess} that they assume universal values precisely at the QCP ($r=r_{c2}$). The algebraic behaviour of these local correlations
is reminiscent of critical behaviour ascribed to the class of local quantum criticality~\cite{Coleman_Si_2001,Si2001}.

\subsection{Impurity magnetisation and entanglement entropy}
More indications of non-Fermi liquid behaviour at the QCP is obtained from a calculation of the impurity magnetisation and the impurity entanglement entropy (\(S_\text{EE}(d)\)) (shown in Sec. 9 of the Supplementary Materials~\cite{supp_mat}) for the symmetry-broken states \(\ket{S_\text{tot}^z = \sigma/2}\). This 
leads to a fractional entanglement entropy (in units of \(\log 2\)) for each of the two states:
\begin{eqnarray}
	\rho_\text{imp} &=& \begin{pmatrix} \frac{1}{2} + m_\text{imp}^z & 0 \\ 0 & \frac{1}{2} - m_\text{imp}^z \end{pmatrix} = \begin{pmatrix} \frac{3}{4} & 0 \\ 0 & \frac{1}{4} \end{pmatrix}~,\nonumber\\ 
	S_\text{EE}(d) &=& -\text{Tr}\left(\rho_\text{imp} \ln \rho_\text{imp}\right) \simeq 0.81 \log 2~,
\end{eqnarray}
where the impurity magnetisation \(m_\text{imp}^z\) takes the value \(m_\text{imp}^z = 1/4\) at the QCP (i.e., half the value for a local moment).
Further, \(S_\text{EE}(d)\) can be written in terms of an effective impurity degeneracy \(g_\text{imp}\), which we define using the impurity magnetisation: \(g_\text{imp} \equiv 1 + 2|m_\text{imp}^z|\), such that it takes the  expected values of 1 and 2 in the local Fermi liquid and local moment phases (with \(m_\text{imp}^z\) having values 0 and \(\frac{1}{2}\) respectively). The effective degeneracy in the polarised subspace \(S_\text{tot}^z = \pm 1/2\) at the QCP then turns out to be \(3/2\), which is half-way between a unique state and a doublet. This corresponds to partial screening of the impurity degrees of freedom at the QCP, in contrast to complete screening in the metallic phase (\(g_\text{imp} = 1\)) and the absence of screening in the insulating phase (\(g_\text{imp} = 2\)).

In Sec. 9 of the Supplementary Materials~\cite{supp_mat}, we show that the incomplete magnetisation (and hence the fractional value of \(S_\text{EE}\)) arises from the mixing of the local Fermi liquid ground-state \(\ket{\phi} = \otimes_{k <= k_F} \ket{k \uparrow}\ket{k \downarrow}\) and the gapless excitations above it, \(\ket{e}_\sigma = e^\dagger_\sigma \ket{\phi}, e \equiv \sum_{k \in FS} c^\dagger_{k\sigma}\), leading to the decay of the local Fermi liquid quasiparticles~\cite{varma2002singular}. This is captured by the modified ground-state \(\ket{-}\) and lowest-lying excited states \(\ket{+}\) at the QCP:
\begin{eqnarray}\label{new-ground-state}
	\ket{\pm} = \frac{1}{\sqrt 2}\left(\ket{SS}\otimes\ket{e}_\sigma \pm \ket{LM}_\sigma\otimes\ket{\phi}\right)~.
\end{eqnarray}
The spin-singlet state \(\ket{SS} \sim \sum_\sigma \sigma\ket{\sigma}_d\ket{\bar\sigma}_0\) and the local moment state \(\ket{LM}_\sigma = \ket{\sigma}_d \ket{2}_0\) represent the configurations of the impurity and the bath zeroth sites. Such a ground-state \(\ket{-}\) should be contrasted with the local Fermi liquid ground-state \(\ket{SS}\otimes\ket{\phi}\) that is stable in the metallic phase for \(r_{c1} < r < r_{c2}\). This change in the ground-state manifests in the vanishing of the quasiparticle residue of the local Fermi liquid excitations: \(Z = |\braket{+ | e^\dagger_{\sigma} | -}|^2 = |\bra{+} \left(\ket{LM}_\sigma\otimes \ket{e}_\sigma\right)|^2 = 0\), consistent with the orthogonality catastrophe~\cite{anderson1967infrared} described below eq.~\eqref{sigma_rc2}.
As the local Fermi liquid quasiparticles are rendered unstable, they are replaced by stable composite three-site excitations of the form, \(S_d^- c_{0 \bar\sigma} e^\dagger_\sigma\). This is further reflected in the fact that the corresponding residue for such three-site composite excitation is non-zero at the QCP: \(|\braket{+|S_d^- c_{0 \bar\sigma} e^\dagger_\sigma|-}|^2 > 0\).

\subsection{Friedel scattering phase shift and fractional excess charge}\label{excess}
The orthogonality catastrophe faced by the infrared excitations of the conduction bath at the QCP points towards the presence of an ``unrenormalised" phase shift~\cite{anderson1990} of the scattering \(k-\)states. Indeed, using the Friedel sum rule~\cite{friedel_1956,langer1961friedel,langreth1966}, we obtain a \(\pi/2\) phase shift for the extended states at the Fermi surface for the QCP, which is only half the unitary limit. This can be seen by recalling that the ground-state in eq.~\eqref{new-ground-state} is an equal admixture of a decoupled local moment state and a singlet state. Through the singlet state, the impurity contributes an excess charge of unity to the Fermi surface of the conduction bath (because of the presence of gapless spin-flip fluctuations). The local moment state does not contribute any excess charge, because the impurity is decoupled from the bath. The net excess charge then comes out to be \(n_\text{exc} = \frac{1}{2}\left( 1 + 0 \right) = \frac{1}{2}\). From the Friedel sum rule, the phase shift at the fixed point is \(\delta^* = \pi n_\text{exc} = \pi /2\), as mentioned above. Expectedly, \(n_\text{exc}\) is zero in the local moment phase. Also, as the exponents of the algebraic form for the local quantities shown in eq.~\eqref{exponents} are functions of the scattering phase shift, we find that these exponents take universal values precisely at the QCP.

The values of the excess charge at the QCP and in the local moment phase lead to important corollaries. Firstly, the jump in the excess charge by unity in going from \(r < r_{c2}\) to \(r > r_{c2}\) is in fact a reduction in the Luttinger volume \(V_L\) of the conduction bath~\cite{luttinger1960fermi,martin1982fermi}. \(V_L\) counts the number of extended states present at and below the Fermi volume, and corresponds to a  topological quantum number~\cite{oshikawa2000topological,seki2017topological,Heath2020,anirbanurg1}. Since the excess charge \(n_\text{exc}\) calculated above is simply the impurity contribution to \(V_L\) at the Fermi surface, the difference in the excess charge across the transition implies that the two phases acquire different values for the invariant \(V_L\). Thus, $n_\text{exc}$ tracks the change ($\Delta V_{L}$) in the Luttinger volume, and the transition is topological in nature. Secondly, in a recent work, Sen et al.~\cite{sen_mitchell_2020} have shown that the Mott MIT of the infinite-dimensional Hubbard model proceeds through the dissociation of domain walls in a fictitious Su-Schrieffer-Heeger chain connected to the physical lattice sites. If one couples their observation with the fact that the two dissociated domain walls at the ends of the SSH chain are together known to host a single charge~\cite{SSHRMP1988,stone_1985}, it becomes evident that the fractional excess charge we obtain at the QCP corresponds to the state that is localised in the SSH chain near the physical lattice site as obtained by Sen et al. The well-known bulk-boundary correspondence of the SSH model (see, e.g., \cite{Asboth_2016}) suggests that the excess charge $n_\text{exc}$ and the change $\Delta V_{L}$ in the Luttinger volume are the respective topological invariants of the boundary and bulk in the e-SIAM that are tied to one another. Further, the appearance of a half-quantised $n_\text{exc}$ at the MIT of the e-SIAM corresponds, in the SSH model, to the dissociation of a domain wall-anti domain wall pair with each carrying a half charge~\cite{SSHRMP1988,stone_1985}.

We show in Sec. 9 of the Supplementary Materials~\cite{supp_mat} that it is in fact  possible to connect the scattering phase shift ($\delta^{*}$), impurity magnetisation ($m_{\text{imp}}^{z}$), effective degeneracy ($\tilde{g}$) and excess charge ($n_{\text{exc}}$) in a single relation:
\begin{eqnarray}\label{phase-mag}
	\delta^* = \pi n_\text{exc} = \left[1 + \left(\frac{2-\tilde g}{\tilde g-1}\right)^2\right]^{-1} = \frac{4{m_{\text{imp}}^z}^2}{1 - 4|m_{\text{imp}}^z| + 8{m_{\text{imp}}^z}^2}~.
\end{eqnarray}
As we now discuss, this provides a unified picture of the effect of frustration on the impurity spin. As we have seen earlier, the presence of \(U_b\) in the Hamiltonian of the e-SIAM introduces local moment states into the spectrum and leads to states with non-vanishing impurity magnetisation at the QCP. This non-zero magnetisation can be interpreted as a partial screening of the impurity spin, in turn resulting in an effective impurity degeneracy \((1 < \tilde g < 2\) between that of a unique screened state and an unscreened local moment. The partial screening also manifests in only half the excess charge being contributed to the conduction bath. This reduction in the excess charge acts as a change in the boundary conditions felt by the conduction electrons coupled to the impurity, and manifests as a phase shift that is less than the unitarity limit of \(\pi\).

\section{Discussions and Outlook}
\label{concl}
In summary, we have shown that an attractive correlation on the bath zeroth site is enough to frustrate the Kondo effect and stabilise the local moment phase. 
The destruction of the Kondo cloud, and the associated local Fermi liquid, occurs through {\it pairing fluctuations in the bath and proximate to the impurity} (left panel of Fig.~\eqref{spin-charge-corr}).
This is reminiscent of a subdominant superconducting tendency that was observed in the half-filled Hubbard model at $T=0$ from a unitary RG treatment in Refs.\cite{anirbanmott1,Mukherjee_mott_merg}.
We find that the critical point displays non-Fermi liquid behaviour with a vanishing quasiparticle residue \(Z\). 
The strong agreement of our results with several aspects of DMFT suggests that the local self-energy obtained self-consistently in that method can be represented quite faithfully through the e-SIAM. It will, thus, be important to test the predictions offered by our approach directly within the DMFT method. We note that the non-Fermi liquid at the critical point displays a {\it partial correlation of doublons and holons} that is distinct from not only the correlated Fermi liquid metal (unbound holons and doublons), but also the paramagnetic insulator (comprised of bound holons and doublons). The excitations that propagate through the two unscreened states in the ground-state subspace (Eq.~\eqref{degenerate-nfl}) involve the simultaneous creation of a doublon and a holon: in these channels, the doublons and holons cannot propagate in the absence of each other. Such an incomplete correlation is an intermediate step towards complete confinement in the local moment phase. It appears interesting to experimentally test some of these ideas on a mesoscopic quantum dot~\cite{Iftikhar2015,Iftikhar2018} which is, apart from the usual electron tunnel coupling to an electronic reservoir, additionally coupled through the proximity effect to a superconducting lead.

We stress here that the e-SIAM analysed in this work represents an impurity model that (i) has a single correlated channel of conduction electrons, (ii) is consistent with the symmetries of a half-filled Hubbard model (SU(2)-spin, U(1)-charge and particle-hole), and (iii) shows a local metal-insulator transition. The model, therefore, provides a minimal route towards obtaining a Hamiltonian-based understanding of DMFT. Further, at the level of the renormalisation group flows that involve the frustration of Kondo screening, the effect of the attractive on-site interaction \(U_b\) introduced by us is equivalent to that of additional bath correlations that have been introduced in related impurity models. These include a single impurity Anderson model with multiple anisotropic conduction channels studied by Giamarchi et al.~\cite{giamarchi_varma_1993}, and a periodic Anderson model enhanced with explicit s-d coupling (as well as a density-density interaction) between the conduction and impurity bands, studied by Si and Kotliar~\cite{si_kotliar_1993,kotliar_si_toulouse_1996,Si_kotliar_NFL_1993} as well as by Ruckenstein et al.,\cite{ruckenstein1991}. In the former \(K-\)channel model of Ref.\cite{giamarchi_varma_1993}, the \(z-\)component (\(V_l, ~ l\in\left[1, K-1\right] \)) of the Kondo interaction term in the \(K-1\) additional conduction channels acts a source of frustration for the spin-flip term \(\gamma_0\) of the Kondo interaction in the original \(l=0\) channel, leading to a breakdown of Kondo screening. The \(U_b\) coupling in the e-SIAM acts as a similar source of frustration: the equivalence lies at the level of the renormalisation group equations, such that both \(V_l\) and \(U_b\) oppose the RG relevance of \(\gamma_0\) and \(J\) respectively. This can be made more precise by comparing the RG equations of our work (eqs.\eqref{rg-eqn}) and that of Giamarchi et al.~\cite{giamarchi_varma_1993}: the mapping between the two frustration terms \(U_b \) and \(V_l\) is found to be of the form \(U_b \sim \sum_{l=1}^{K-1}V_l^2 \rho\), where \(\rho\) is the conduction bath DOS. A similar relationship can be found between the RG equations obtained by us and those for the extended periodic Anderson model studied by Si and Kotliar~\cite{si_kotliar_1993}. In this sense, our analysis applies to a wide variety of models where the Kondo effect is stable in a certain regime of  parameters, but is destroyed in other parameter regimes by some form of quantum-mechanical frustration of the impurity spin degree of freedom. 

Extensions of the present work involve analysing the mixed valence regime of the impurity site (i.e., away from half-filling) within the e-SIAM auxiliary model. This allows the possibility that the finite-temperature critical end-point of the DMFT first-order transition could turn into a quantum critical point from the merging of the $T=0$ ESQPT and QPT observed in the present work. Recent DMFT calculations~\cite{vucicevic2015} show a shrinking of the coexistence region with hole-doping, but fall short of revealing a QCP. If such a QCP does exist, can it harbour pair fluctuations between the impurity and zeroth bath sites that become dominant upon doping, signalling thereby a putative superconducting instability of a related bulk model? We recall that a superconducting state of matter was indeed observed to be emergent from a QCP in the hole-doped Hubbard model at $T=0$ via a unitary RG treatment in Refs.\cite{anirbanmott2,Mukherjee_mott_merg}.
The phenomenon of electronic differentiation in \(k-\)space can perhaps be captured by considering cluster variants of the e-SIAM, i.e., multiple impurities connected to one another through single-particle hopping and/or RKKY-like interactions~\cite{Ferrero2007,sakai_2009}.
In general, the presence of multiple and varied classes of correlations in the auxiliary model - localisation from Mott physics, delocalisation from spin and charge fluctuations, and pairing from local attractive correlations - makes this a strong candidate for an auxiliary model that carries the potential for describing the emergence of a variety of novel phases of correlated quantum matter.

\ack
AM thanks IISER Kolkata for funding through a junior and a senior research fellowship. SL thanks the SERB, Govt. of India for funding through MATRICS grant MTR/2021/000141 and Core Research Grant CRG/2021/000852. The authors gratefully acknowledge discussions with G. Martins, E. Vernek, E. Anda, B. Bansal, S. Kundu, S. R. Hassan, M. S. Laad, R. K. Singh, S. Sen, D. Jaiswal-Nagar, T. Banerjee and N. Mohanta.

\section*{References}

\bibliography{esiam-manuscript}

\begin{thebibliography}{100}

\bibitem{Mott_RMP_1968}
N.~F. Mott.
\newblock Metal-insulator transition.
\newblock {\em Rev. Mod. Phys.}, 40:677--683, Oct 1968.

\bibitem{milligan_1985}
R~F Milligan and G~A Thomas.
\newblock The metal-insulator transition.
\newblock {\em Annual Review of Physical Chemistry}, 36(1):139--158, 1985.

\bibitem{imada1998metal}
Masatoshi Imada, Atsushi Fujimori, and Yoshinori Tokura.
\newblock Metal-insulator transitions.
\newblock {\em Reviews of modern physics}, 70(4):1039, 1998.

\bibitem{kuramoto1987}
Y~Kuramoto and T~Watanabe.
\newblock Theory of momentum-dependent magnetic response in heavy-fermion
  systems.
\newblock In {\em Proceedings of the Yamada Conference XVIII on
  Superconductivity in Highly Correlated Fermion Systems}, pages 80--83.
  Elsevier, 1987.

\bibitem{Cox1988}
D.~L. Cox and N.~Grewe.
\newblock Transport properties of the anderson lattice.
\newblock {\em Zeitschrift f{\"u}r Physik B Condensed Matter}, 71(3):321--340,
  Sep 1988.

\bibitem{metzner_volhardt_1989}
Walter Metzner and Dieter Vollhardt.
\newblock Correlated lattice fermions in $d=\ensuremath{\infty}$ dimensions.
\newblock {\em Phys. Rev. Lett.}, 62:324--327, Jan 1989.

\bibitem{zhang_1993}
X.~Y. Zhang, M.~J. Rozenberg, and G.~Kotliar.
\newblock Mott transition in the d=\ensuremath{\infty} hubbard model at zero
  temperature.
\newblock {\em Phys. Rev. Lett.}, 70:1666--1669, Mar 1993.

\bibitem{georges1996}
Antoine Georges, Gabriel Kotliar, Werner Krauth, and Marcelo~J Rozenberg.
\newblock Dynamical mean-field theory of strongly correlated fermion systems
  and the limit of infinite dimensions.
\newblock {\em Reviews of Modern Physics}, 68(1):13, 1996.

\bibitem{parcollet_2004}
O.~Parcollet, G.~Biroli, and G.~Kotliar.
\newblock Cluster dynamical mean field analysis of the mott transition.
\newblock {\em Phys. Rev. Lett.}, 92:226402, Jun 2004.

\bibitem{maier_2005}
Thomas Maier, Mark Jarrell, Thomas Pruschke, and Matthias~H. Hettler.
\newblock Quantum cluster theories.
\newblock {\em Rev. Mod. Phys.}, 77:1027--1080, Oct 2005.

\bibitem{kotliar_rmp_2006}
G.~Kotliar, S.~Y. Savrasov, K.~Haule, V.~S. Oudovenko, O.~Parcollet, and C.~A.
  Marianetti.
\newblock Electronic structure calculations with dynamical mean-field theory.
\newblock {\em Rev. Mod. Phys.}, 78:865--951, Aug 2006.

\bibitem{ohashi_2008}
Takuma Ohashi, Tsutomu Momoi, Hirokazu Tsunetsugu, and Norio Kawakami.
\newblock Finite temperature mott transition in hubbard model on anisotropic
  triangular lattice.
\newblock {\em Phys. Rev. Lett.}, 100:076402, Feb 2008.

\bibitem{Mott_1949}
N~F Mott.
\newblock The basis of the electron theory of metals, with special reference to
  the transition metals.
\newblock {\em Proceedings of the Physical Society. Section A}, 62(7):416--422,
  jul 1949.

\bibitem{gutzwiller_1963}
Martin~C. Gutzwiller.
\newblock Effect of correlation on the ferromagnetism of transition metals.
\newblock {\em Phys. Rev. Lett.}, 10:159--162, Mar 1963.

\bibitem{kanamori_1963}
Junjiro Kanamori.
\newblock {Electron Correlation and Ferromagnetism of Transition Metals}.
\newblock {\em Progress of Theoretical Physics}, 30(3):275--289, 09 1963.

\bibitem{hubbard1963electron}
John Hubbard.
\newblock Electron correlations in narrow energy bands.
\newblock In {\em Proceedings of the royal society of london a: mathematical,
  physical and engineering sciences}, volume 276, pages 238--257. The Royal
  Society, 1963.

\bibitem{brinkman_rice_1970}
W.~F. Brinkman and T.~M. Rice.
\newblock Application of gutzwiller's variational method to the metal-insulator
  transition.
\newblock {\em Phys. Rev. B}, 2:4302--4304, Nov 1970.

\bibitem{Logan_2016}
David~E Logan and Martin~R Galpin.
\newblock Mott insulators and the doping-induced mott transition within dmft:
  exact results for the one-band hubbard model.
\newblock {\em Journal of Physics: Condensed Matter}, 28(2):025601, dec 2015.

\bibitem{vucicevic_2013}
J.~Vučičević, H.~Terletska, D.~Tanasković, and V.~Dobrosavljević.
\newblock Finite-temperature crossover and the quantum widom line near the mott
  transition.
\newblock {\em Phys. Rev. B}, 88:075143, Aug 2013.

\bibitem{park2008}
Hyowon Park, Kristjan Haule, and Gabriel Kotliar.
\newblock Cluster dynamical mean field theory of the mott transition.
\newblock {\em Physical review letters}, 101(18):186403, 2008.

\bibitem{rohringer_2018}
G.~Rohringer, H.~Hafermann, A.~Toschi, A.~A. Katanin, A.~E. Antipov, M.~I.
  Katsnelson, A.~I. Lichtenstein, A.~N. Rubtsov, and K.~Held.
\newblock Diagrammatic routes to nonlocal correlations beyond dynamical mean
  field theory.
\newblock {\em Rev. Mod. Phys.}, 90:025003, May 2018.

\bibitem{lichenstein_1998}
A.~I. Lichtenstein and M.~I. Katsnelson.
\newblock Ab initio calculations of quasiparticle band structure in correlated
  systems: Lda++ approach.
\newblock {\em Phys. Rev. B}, 57:6884--6895, Mar 1998.

\bibitem{kotliar_2006}
G.~Kotliar, S.~Y. Savrasov, K.~Haule, V.~S. Oudovenko, O.~Parcollet, and C.~A.
  Marianetti.
\newblock Electronic structure calculations with dynamical mean-field theory.
\newblock {\em Rev. Mod. Phys.}, 78:865--951, Aug 2006.

\bibitem{held_2007_ldadmft}
K.~Held.
\newblock Electronic structure calculations using dynamical mean field theory.
\newblock {\em Advances in Physics}, 56(6):829--926, 2007.

\bibitem{held2008}
Karsten Held, Andrey~A Katanin, and Alessandro Toschi.
\newblock Dynamical vertex approximationan introduction.
\newblock {\em Progress of Theoretical Physics Supplement}, 176:117--133, 2008.

\bibitem{terletska_mott_2011}
H.~Terletska, J.~Vučičević, D.~Tanasković, and V.~Dobrosavljević.
\newblock Quantum critical transport near the mott transition.
\newblock {\em Phys. Rev. Lett.}, 107:026401, Jul 2011.

\bibitem{si_kotliar_1993}
Qimiao Si and Gabriel Kotliar.
\newblock Metallic non-fermi-liquid phases of an extended hubbard model in
  infinite dimensions.
\newblock {\em Phys. Rev. B}, 48:13881--13903, Nov 1993.

\bibitem{kotliarsi_1993}
Gabriel Kotliar and Qimiao Si.
\newblock Quantum chemistry, anomalous dimensions, and the breakdown of fermi
  liquid theory in strongly correlated systems.
\newblock {\em Physica Scripta}, T49:165, 1993.

\bibitem{anderson_1961}
P.~W. Anderson.
\newblock Localized magnetic states in metals.
\newblock {\em Phys. Rev.}, 124:41--53, Oct 1961.

\bibitem{anderson_1978}
P.~W. Anderson.
\newblock Local moments and localized states.
\newblock {\em Rev. Mod. Phys.}, 50:191--201, Apr 1978.

\bibitem{anderson1969exact}
Philip~W Anderson and Gideon Yuval.
\newblock Exact results in the kondo problem: equivalence to a classical
  one-dimensional coulomb gas.
\newblock {\em Physical Review Letters}, 23(2):89, 1969.

\bibitem{anderson1970exact}
Philip~W Anderson, G~Yuval, and DR~Hamann.
\newblock Exact results in the kondo problem. ii. scaling theory, qualitatively
  correct solution, and some new results on one-dimensional classical
  statistical models.
\newblock {\em Physical Review B}, 1(11):4464, 1970.

\bibitem{anderson1970}
PW~Anderson.
\newblock A poor man's derivation of scaling laws for the kondo problem.
\newblock {\em Journal of Physics C: Solid State Physics}, 3(12):2436, 1970.

\bibitem{haldane1978scaling}
FDM Haldane.
\newblock Scaling theory of the asymmetric anderson model.
\newblock {\em Physical Review Letters}, 40(6):416, 1978.

\bibitem{jefferson_1977}
J~H Jefferson.
\newblock a renormalisation group approach to the mixed valence problem.
\newblock {\em Journal of Physics C: Solid State Physics}, 10(18):3589--3599,
  sep 1977.

\bibitem{wilson1975}
Kenneth~G. Wilson.
\newblock The renormalization group: Critical phenomena and the kondo problem.
\newblock {\em Rev. Mod. Phys.}, 47:773--840, Oct 1975.

\bibitem{hrk_wilson_1980}
H.~R. Krishna-murthy, J.~W. Wilkins, and K.~G. Wilson.
\newblock Renormalization-group approach to the anderson model of dilute
  magnetic alloys. i. static properties for the symmetric case.
\newblock {\em Phys. Rev. B}, 21:1003--1043, Feb 1980.

\bibitem{andrei_1980}
N.~Andrei.
\newblock Diagonalization of the kondo hamiltonian.
\newblock {\em Phys. Rev. Lett.}, 45:379--382, Aug 1980.

\bibitem{andreiKondoreview}
N~Andrei, K~Furuya, and J~H Lowenstein.
\newblock Solution of the kondo problem.
\newblock {\em Rev. Mod. Phys.}, 55:331, 1983.

\bibitem{Wiegmann_1981}
P~B Wiegmann.
\newblock Exact solution of the s-d exchange model (kondo problem).
\newblock {\em Journal of Physics C: Solid State Physics}, 14(10):1463--1478,
  apr 1981.

\bibitem{tsvelickKondoreview}
A~M Tsvelick and P~B Wiegmann.
\newblock Exact results in the theory of magnetic alloys.
\newblock {\em Adv. in Phys.}, 32:453, 1983.

\bibitem{kotliar_1996}
Gabriel Kotliar and Qimiao Si.
\newblock Toulouse points and non-fermi-liquid states in the mixed-valence
  regime of the generalized anderson model.
\newblock {\em Phys. Rev. B}, 53:12373--12388, May 1996.

\bibitem{Duki_2011}
Solomon~F. Duki.
\newblock Solvable limit for the su($n$) kondo model.
\newblock {\em Phys. Rev. B}, 83:134423, Apr 2011.

\bibitem{borda_2008}
L.~Borda, A.~Schiller, and A.~Zawadowski.
\newblock Applicability of bosonization and the anderson-yuval methods at the
  strong-coupling limit of quantum impurity problems.
\newblock {\em Phys. Rev. B}, 78:201301, Nov 2008.

\bibitem{streib_2013}
Simon Streib, Aldo Isidori, and Peter Kopietz.
\newblock Solution of the anderson impurity model via the functional
  renormalization group.
\newblock {\em Phys. Rev. B}, 87:201107, May 2013.

\bibitem{anirban_kondo}
Anirban Mukherjee, Abhirup Mukherjee, N.~S. Vidhyadhiraja, A.~Taraphder, and
  Siddhartha Lal.
\newblock Unveiling the kondo cloud: Unitary renormalization-group study of the
  kondo model.
\newblock {\em Phys. Rev. B}, 105:085119, Feb 2022.

\bibitem{sorensen_erik_affleck_1996}
Erik~S. S\o{}rensen and Ian Affleck.
\newblock Scaling theory of the kondo screening cloud.
\newblock {\em Phys. Rev. B}, 53:9153--9167, Apr 1996.

\bibitem{affleck_ian_2001}
Ian Affleck and Pascal Simon.
\newblock Detecting the kondo screening cloud around a quantum dot.
\newblock {\em Phys. Rev. Lett.}, 86:2854--2857, Mar 2001.

\bibitem{simon_pascal_2003}
Pascal Simon and Ian Affleck.
\newblock Kondo screening cloud effects in mesoscopic devices.
\newblock {\em Phys. Rev. B}, 68:115304, Sep 2003.

\bibitem{martin2010}
C.~A. Busser, G.~B. Martins, L.~C. Ribeiro, E.~Vernek, E.~V. Anda, and
  E.~Dagotto.
\newblock {\em Phys. Rev. 9}, 81:045111, 2010.

\bibitem{martin2019}
L.~C. Ribeiro, G.~B. Martins, G.~Gomez-Silva, and E.~V. Anda.
\newblock {\em Phys. Rev. B}, 99:085139, 2019.

\bibitem{Goldhaber-Gordon1998}
D.~Goldhaber-Gordon, Hadas Shtrikman, D.~Mahalu, David Abusch-Magder,
  U.~Meirav, and M.~A. Kastner.
\newblock Kondo effect in a single-electron transistor.
\newblock {\em Nature}, 391(6663):156--159, Jan 1998.

\bibitem{Cronenwett1998}
Sara~M. Cronenwett, Tjerk~H. Oosterkamp, and Leo~P. Kouwenhoven.
\newblock A tunable kondo effect in quantum dots.
\newblock {\em Science}, 281(5376):540--544, 1998.

\bibitem{Schmid_Weis1998}
Jörg Schmid, Jürgen Weis, Karl Eberl, and Klaus {v. Klitzing}.
\newblock A quantum dot in the limit of strong coupling to reservoirs.
\newblock {\em Physica B: Condensed Matter}, 256-258:182--185, 1998.

\bibitem{pustilnik_glazman_2004}
Michael Pustilnik and Leonid Glazman.
\newblock Kondo effect in quantum dots.
\newblock {\em Journal of Physics: Condensed Matter}, 16(16):R513, apr 2004.

\bibitem{Borzenets2020}
Ivan V.~Borzenets, Jeongmin Shim, Jason C.~H. Chen, Arne Ludwig, Andreas~D.
  Wieck, Seigo Tarucha, H.-S. Sim, and Michihisa Yamamoto.
\newblock Observation of the kondo screening cloud.
\newblock {\em Nature}, 579(7798):210--213, Mar 2020.

\bibitem{neel_berndt_2008}
N.~N\'eel, J.~Kr\"oger, R.~Berndt, T.~O. Wehling, A.~I. Lichtenstein, and M.~I.
  Katsnelson.
\newblock Controlling the kondo effect in ${\mathrm{cocu}}_{n}$ clusters atom
  by atom.
\newblock {\em Phys. Rev. Lett.}, 101:266803, Dec 2008.

\bibitem{Zhao2005}
Aidi Zhao, Qunxiang Li, Lan Chen, Hongjun Xiang, Weihua Wang, Shuan Pan, Bing
  Wang, Xudong Xiao, Jinlong Yang, J.~G. Hou, and Qingshi Zhu.
\newblock Controlling the kondo effect of an adsorbed magnetic ion through its
  chemical bonding.
\newblock {\em Science}, 309(5740):1542--1544, September 2005.

\bibitem{nozaki2012}
Masahiro Nozaki, Shinsei Ryu, and Tadashi Takayanagi.
\newblock Holographic geometry of entanglement renormalization in quantum field
  theories.
\newblock {\em Journal of High Energy Physics}, 2012(10):193, 2012.

\bibitem{mora_2015}
Christophe Mora, Cătălin~Paşcu Moca, Jan von Delft, and Gergely Zaránd.
\newblock Fermi-liquid theory for the single-impurity anderson model.
\newblock {\em Phys. Rev. B}, 92:075120, Aug 2015.

\bibitem{anirbanurg1}
Anirban Mukherjee and Siddhartha Lal.
\newblock Holographic unitary renormalisation group for correlated electrons-i:
  a tensor network approach.
\newblock {\em Nuclear Physics B}, 960:115170, 2020.

\bibitem{anirbanurg2}
Anirban Mukherjee and Siddhartha Lal.
\newblock Holographic unitary renormalisation group for correlated
  electrons-ii: insights on fermionic criticality.
\newblock {\em Nuclear Physics B}, 960:115163, 2020.

\bibitem{santanukagome}
Santanu Pal, Anirban Mukherjee, and Siddhartha Lal.
\newblock Correlated spin liquids in the quantum kagome antiferromagnet at
  finite field: a renormalization group analysis.
\newblock {\em New Journal of Physics}, 21(2):023019, feb 2019.

\bibitem{1dhubjhep}
Anirban Mukherjee, Siddhartha Patra, and Siddhartha Lal.
\newblock Fermionic criticality is shaped by fermi surface topology: a case
  study of the tomonaga-luttinger liquid.
\newblock {\em Journal of High Energy Physics}, 04:148, 2021.

\bibitem{anirbanmott2}
Anirban Mukherjee and Siddhartha Lal.
\newblock Scaling theory for mott{\textendash}hubbard transitions-{II}: quantum
  criticality of the doped mott insulator.
\newblock {\em New Journal of Physics}, 22(6):063008, jun 2020.

\bibitem{anirbanmott1}
Anirban Mukherjee and Siddhartha Lal.
\newblock Scaling theory for mott{\textendash}hubbard transitions: I. t = 0
  phase diagram of the 1/2-filled hubbard model.
\newblock {\em New Journal of Physics}, 22(6):063007, jun 2020.

\bibitem{siddharthacpi}
Siddhartha Patra and Siddhartha Lal.
\newblock Origin of topological order in a cooper-pair insulator.
\newblock {\em Phys. Rev. B}, 104:144514, Oct 2021.

\bibitem{patra_mck}
Siddhartha Patra, Abhirup Mukherjee, Anirban Mukherjee, N.~S. Vidhyadhiraja,
  A.~Taraphder, and Siddhartha Lal.
\newblock Frustration shapes multi-channel kondo physics: a star graph
  perspective.
\newblock {\em pre-print}, 2022.

\bibitem{supp_mat}
Supplementary Materials are available online.

\bibitem{nozieres1974fermi}
P~Nozieres.
\newblock A ``fermi-liquid" description of the kondo problem at low
  temperatures.
\newblock {\em Journal of Low Temperature Physics}, 17:31, 1974.

\bibitem{costi_hewson_1990}
T.A. Costi and A.C. Hewson.
\newblock A new approach to the calculation of spectra for strongly correlated
  systems.
\newblock {\em Physica B: Condensed Matter}, 163(1):179--181, 1990.

\bibitem{shimony1995degree}
Abner Shimony.
\newblock Degree of entanglement.
\newblock {\em Annals of the New York Academy of Sciences}, 755(1):675--679,
  1995.

\bibitem{wei2003geometric}
Tzu-Chieh Wei and Paul~M Goldbart.
\newblock Geometric measure of entanglement and applications to bipartite and
  multipartite quantum states.
\newblock {\em Physical Review A}, 68(4):042307, 2003.

\bibitem{horodecki2009quantum}
Ryszard Horodecki, Pawe{\l} Horodecki, Micha{\l} Horodecki, and Karol
  Horodecki.
\newblock Quantum entanglement.
\newblock {\em Reviews of modern physics}, 81(2):865, 2009.

\bibitem{Hauke2016}
Philipp Hauke, Markus Heyl, Luca Tagliacozzo, and Peter Zoller.
\newblock Measuring multipartite entanglement through dynamic susceptibilities.
\newblock {\em Nature Physics}, 12(8):778--782, Aug 2016.

\bibitem{moeller_1995}
Goetz Moeller, Qimiao Si, Gabriel Kotliar, Marcelo Rozenberg, and Daniel~S.
  Fisher.
\newblock Critical behavior near the mott transition in the hubbard model.
\newblock {\em Phys. Rev. Lett.}, 74:2082--2085, Mar 1995.

\bibitem{held_2013}
K.~Held, R.~Peters, and A.~Toschi.
\newblock Poor man's understanding of kinks originating from strong electronic
  correlations.
\newblock {\em Phys. Rev. Lett.}, 110:246402, Jun 2013.

\bibitem{kotliar_si_toulouse_1996}
Gabriel Kotliar and Qimiao Si.
\newblock Toulouse points and non-fermi-liquid states in the mixed-valence
  regime of the generalized anderson model.
\newblock {\em Phys. Rev. B}, 53:12373--12388, May 1996.

\bibitem{bulla_1999}
R.~Bulla.
\newblock Zero temperature metal-insulator transition in the
  infinite-dimensional hubbard model.
\newblock {\em Phys. Rev. Lett.}, 83:136--139, Jul 1999.

\bibitem{georges_2004_dmft}
Antoine Georges.
\newblock Strongly correlated electron materials: Dynamical mean‐field theory
  and electronic structure.
\newblock {\em AIP Conference Proceedings}, 715(1):3--74, 2004.

\bibitem{georges_krauth_1993}
Antoine Georges and Werner Krauth.
\newblock Physical properties of the half-filled hubbard model in infinite
  dimensions.
\newblock {\em Phys. Rev. B}, 48:7167--7182, Sep 1993.

\bibitem{rozenberg_1994_zerotemp}
Marcelo~J. Rozenberg, Goetz Moeller, and Gabriel Kotliar.
\newblock The metal–insulator transition in the hubbard model at zero
  temperature ii.
\newblock {\em Modern Physics Letters B}, 08(08n09):535--543, 1994.

\bibitem{kohn1964theory}
Walter Kohn.
\newblock Theory of the insulating state.
\newblock {\em Physical review}, 133(1A):A171, 1964.

\bibitem{castellani_1979}
C.~Castellani, C.~Di Castro, D.~Feinberg, and J.~Ranninger.
\newblock New model hamiltonian for the metal-insulator transition.
\newblock {\em Phys. Rev. Lett.}, 43:1957--1960, Dec 1979.

\bibitem{hrk_1990_prl}
H.~R. Krishnamurthy, C.~Jayaprakash, Sanjoy Sarker, and Wolfgang Wenzel.
\newblock Mott-hubbard metal-insulator transition in nonbipartite lattices.
\newblock {\em Phys. Rev. Lett.}, 64:950--953, Feb 1990.

\bibitem{limelette_2003}
P.~Limelette, A.~Georges, D.~Jérome, P.~Wzietek, P.~Metcalf, and J.~M. Honig.
\newblock Universality and critical behavior at the mott transition.
\newblock {\em Science}, 302(5642):89--92, 2003.

\bibitem{Kagawa2005}
F.~Kagawa, K.~Miyagawa, and K.~Kanoda.
\newblock Unconventional critical behaviour in a quasi-two-dimensional organic
  conductor.
\newblock {\em Nature}, 436(7050):534--537, Jul 2005.

\bibitem{Furukawa2015}
Tetsuya Furukawa, Kazuya Miyagawa, Hiromi Taniguchi, Reizo Kato, and Kazushi
  Kanoda.
\newblock Quantum criticality of mott transition in organic materials.
\newblock {\em Nature Physics}, 11(3):221--224, Mar 2015.

\bibitem{Eisenlohr_2019}
Heike Eisenlohr, Seung-Sup~B. Lee, and Matthias Vojta.
\newblock Mott quantum criticality in the one-band hubbard model: Dynamical
  mean-field theory, power-law spectra, and scaling.
\newblock {\em Phys. Rev. B}, 100:155152, Oct 2019.

\bibitem{satyaki_2020_PRL}
Satyaki Kundu, Tapas Bar, Rajesh~Kumble Nayak, and Bhavtosh Bansal.
\newblock Critical slowing down at the abrupt mott transition: When the
  first-order phase transition becomes zeroth order and looks like second
  order.
\newblock {\em Phys. Rev. Lett.}, 124:095703, Mar 2020.

\bibitem{Noz_blandin_1980}
{Nozi\`eres, Ph.} and {Blandin, A.}
\newblock Kondo effect in real metals.
\newblock {\em J. Phys. France}, 41(3):193--211, 1980.

\bibitem{hewson1993}
A.~C. Hewson.
\newblock {\em The Kondo Problem to Heavy Fermions}.
\newblock Cambridge University Press, 1993.

\bibitem{coleman2015}
Piers Coleman.
\newblock {\em Introduction to many-body physics}.
\newblock Cambridge University Press, 2015.
\newblock Chapter:18.

\bibitem{hewson_1993_prl}
A.~C. Hewson.
\newblock Renormalized perturbation expansions and fermi liquid theory.
\newblock {\em Phys. Rev. Lett.}, 70:4007--4010, Jun 1993.

\bibitem{varma2002singular}
CM~Varma, Z~Nussinov, and Wim Van~Saarloos.
\newblock Singular or non-fermi liquids.
\newblock {\em Physics Reports}, 361(5-6):267--417, 2002.

\bibitem{Kotliar_1993}
Gabriel Kotliar and Qimiao Si.
\newblock Quantum chemistry, anomalous dimensions, and the breakdown of fermi
  liquid theory in strongly correlated systems.
\newblock {\em Physica Scripta}, T49A:165--171, jan 1993.

\bibitem{von_delft_1998}
Jan von Delft, Gergely Zar\'and, and Michele Fabrizio.
\newblock Finite-size bosonization of 2-channel kondo model: A bridge between
  numerical renormalization group and conformal field theory.
\newblock {\em Phys. Rev. Lett.}, 81:196--199, Jul 1998.

\bibitem{Coleman_Si_2001}
P~Coleman, C~Pépin, Qimiao Si, and R~Ramazashvili.
\newblock How do fermi liquids get heavy and die?
\newblock {\em Journal of Physics: Condensed Matter}, 13(35):R723, aug 2001.

\bibitem{Si2001}
Qimiao Si, Silvio Rabello, Kevin Ingersent, and J.~Lleweilun Smith.
\newblock Locally critical quantum phase transitions in strongly correlated
  metals.
\newblock {\em Nature}, 413(6858):804--808, Oct 2001.

\bibitem{anderson1967infrared}
Philip~W Anderson.
\newblock Infrared catastrophe in fermi gases with local scattering potentials.
\newblock {\em Physical Review Letters}, 18(24):1049, 1967.

\bibitem{anderson1990}
P.~W. Anderson.
\newblock ``luttinger-liquid'' behavior of the normal metallic state of the 2d
  hubbard model.
\newblock {\em Phys. Rev. Lett.}, 64:1839--1841, Apr 1990.

\bibitem{friedel_1956}
J.~Friedel.
\newblock On some electrical and magnetic properties of metallic solid
  solutions.
\newblock {\em Canadian Journal of Physics}, 34(12A):1190--1211, 1956.

\bibitem{langer1961friedel}
JS~Langer and V~Ambegaokar.
\newblock Friedel sum rule for a system of interacting electrons.
\newblock {\em Physical Review}, 121(4):1090, 1961.

\bibitem{langreth1966}
David~C. Langreth.
\newblock Friedel sum rule for anderson's model of localized impurity states.
\newblock {\em Phys. Rev.}, 150:516--518, Oct 1966.

\bibitem{luttinger1960fermi}
JM~Luttinger.
\newblock Fermi surface and some simple equilibrium properties of a system of
  interacting fermions.
\newblock {\em Physical Review}, 119(4):1153, 1960.

\bibitem{martin1982fermi}
Richard~M Martin.
\newblock Fermi-surfae sum rule and its consequences for periodic kondo and
  mixed-valence systems.
\newblock {\em Physical Review Letters}, 48(5):362, 1982.

\bibitem{oshikawa2000topological}
Masaki Oshikawa.
\newblock Topological approach to luttinger's theorem and the fermi surface of
  a kondo lattice.
\newblock {\em Physical Review Letters}, 84(15):3370, 2000.

\bibitem{seki2017topological}
Kazuhiro Seki and Seiji Yunoki.
\newblock Topological interpretation of the luttinger theorem.
\newblock {\em Physical Review B}, 96(8):085124, 2017.

\bibitem{Heath2020}
Joshuah~T Heath and Kevin~S Bedell.
\newblock Necessary and sufficient conditions for the validity of luttinger’s
  theorem.
\newblock {\em New Journal of Physics}, 22(6):063011, jun 2020.

\bibitem{sen_mitchell_2020}
Sudeshna Sen, Patrick~J. Wong, and Andrew~K. Mitchell.
\newblock The mott transition as a topological phase transition.
\newblock {\em Phys. Rev. B}, 102:081110, Aug 2020.

\bibitem{SSHRMP1988}
A.~J. Heeger, S.~Kivelson, J.~R. Schrieffer, and W.~P. Su.
\newblock Solitons in conducting polymers.
\newblock {\em Rev. Mod. Phys.}, 60:781--850, Jul 1988.

\bibitem{stone_1985}
M.~Stone.
\newblock Elementary derivation of one-dimensional fermion-number
  fractionalization.
\newblock {\em Phys. Rev. B}, 31:6112--6115, May 1985.

\bibitem{Asboth_2016}
János~K. Asbóth, László Oroszlány, and András Pályi.
\newblock {\em A Short Course on Topological Insulators}.
\newblock Springer International Publishing, 2016.

\bibitem{Mukherjee_mott_merg}
Anirban Mukherjee and Siddhartha Lal.
\newblock Superconductivity from repulsion in the doped 2d electronic hubbard
  model: an entanglement perspective.
\newblock {\em Journal of Physics: Condensed Matter}, 34(27):275601, apr 2022.

\bibitem{Iftikhar2015}
Z.~Iftikhar, S.~Jezouin, A.~Anthore, U.~Gennser, F.~D. Parmentier, A.~Cavanna,
  and F.~Pierre.
\newblock Two-channel kondo effect and renormalization flow with macroscopic
  quantum charge states.
\newblock {\em Nature}, 526(7572):233--236, Oct 2015.

\bibitem{Iftikhar2018}
Z.~Iftikhar, A.~Anthore, A.~K. Mitchell, F.~D. Parmentier, U.~Gennser,
  A.~Ouerghi, A.~Cavanna, C.~Mora, P.~Simon, and F.~Pierre.
\newblock Tunable quantum criticality and super-ballistic transport in a
  \&\#x201c;charge\&\#x201d; kondo circuit.
\newblock {\em Science}, 360(6395):1315--1320, 2018.

\bibitem{giamarchi_varma_1993}
T.~Giamarchi, C.~M. Varma, A.~E. Ruckenstein, and P.~Nozi\`eres.
\newblock Singular low energy properties of an impurity model with finite range
  interactions.
\newblock {\em Phys. Rev. Lett.}, 70:3967--3970, Jun 1993.

\bibitem{Si_kotliar_NFL_1993}
Qimiao Si and G.~Kotliar.
\newblock Fermi-liquid and non-fermi-liquid phases of an extended hubbard model
  in infinite dimensions.
\newblock {\em Phys. Rev. Lett.}, 70:3143--3146, May 1993.

\bibitem{ruckenstein1991}
AE~Ruckenstein and CM~Varma.
\newblock A theory of marginal fermi-liquids.
\newblock {\em Physica C: Superconductivity}, 185:134--140, 1991.

\bibitem{vucicevic2015}
J.~Vu\ifmmode \check{c}\else \v{c}\fi{}i\ifmmode \check{c}\else
  \v{c}\fi{}evi\ifmmode~\acute{c}\else \'{c}\fi{},
  D.~Tanaskovi\ifmmode~\acute{c}\else \'{c}\fi{}, M.~J. Rozenberg, and
  V.~Dobrosavljevi\ifmmode~\acute{c}\else \'{c}\fi{}.
\newblock Bad-metal behavior reveals mott quantum criticality in doped hubbard
  models.
\newblock {\em Phys. Rev. Lett.}, 114:246402, Jun 2015.

\bibitem{Ferrero2007}
Michel Ferrero, Lorenzo~De Leo, Philippe Lecheminant, and Michele Fabrizio.
\newblock Strong correlations in a nutshell.
\newblock {\em Journal of Physics: Condensed Matter}, 19(43):433201, oct 2007.

\bibitem{sakai_2009}
Shiro Sakai, Yukitoshi Motome, and Masatoshi Imada.
\newblock Evolution of electronic structure of doped mott insulators:
  Reconstruction of poles and zeros of green's function.
\newblock {\em Phys. Rev. Lett.}, 102:056404, Feb 2009.

\end{thebibliography}


\begin{thebibliography}{10}

\bibitem{santanukagome}
Santanu Pal, Anirban Mukherjee, and Siddhartha Lal.
\newblock Correlated spin liquids in the quantum kagome antiferromagnet at
  finite field: a renormalization group analysis.
\newblock {\em New Journal of Physics}, 21(2):023019, feb 2019.

\bibitem{anirbanmott1}
Anirban Mukherjee and Siddhartha Lal.
\newblock Scaling theory for mott{\textendash}hubbard transitions: I. t = 0
  phase diagram of the 1/2-filled hubbard model.
\newblock {\em New Journal of Physics}, 22(6):063007, jun 2020.

\bibitem{anirbanmott2}
Anirban Mukherjee and Siddhartha Lal.
\newblock Scaling theory for mott{\textendash}hubbard transitions-{II}: quantum
  criticality of the doped mott insulator.
\newblock {\em New Journal of Physics}, 22(6):063008, jun 2020.

\bibitem{1dhubjhep}
Anirban Mukherjee, Siddhartha Patra, and Siddhartha Lal.
\newblock Fermionic criticality is shaped by fermi surface topology: a case
  study of the tomonaga-luttinger liquid.
\newblock {\em Journal of High Energy Physics}, 04:148, 2021.

\bibitem{siddharthacpi}
Siddhartha Patra and Siddhartha Lal.
\newblock Origin of topological order in a cooper-pair insulator.
\newblock {\em Phys. Rev. B}, 104:144514, Oct 2021.

\bibitem{anirban_mott_ent}
Anirban Mukherjee and Siddhartha Lal.
\newblock Superconductivity from repulsion in the doped 2d electronic hubbard
  model: an entanglement perspective.
\newblock {\em J. Phys.: Condens. Matter}, 34:275601, 2022.

\bibitem{anirban_kondo}
Anirban Mukherjee, Abhirup Mukherjee, N.~S. Vidhyadhiraja, A.~Taraphder, and
  Siddhartha Lal.
\newblock Unveiling the kondo cloud: Unitary renormalization-group study of the
  kondo model.
\newblock {\em Phys. Rev. B}, 105:085119, Feb 2022.

\bibitem{anderson1970}
PW~Anderson.
\newblock A poor man's derivation of scaling laws for the kondo problem.
\newblock {\em Journal of Physics C: Solid State Physics}, 3(12):2436, 1970.

\bibitem{anirbanurg1}
Anirban Mukherjee and Siddhartha Lal.
\newblock Holographic unitary renormalisation group for correlated electrons-i:
  a tensor network approach.
\newblock {\em Nuclear Physics B}, 960:115170, 2020.

\bibitem{shimony1995degree}
Abner Shimony.
\newblock Degree of entanglement.
\newblock {\em Annals of the New York Academy of Sciences}, 755(1):675--679,
  1995.

\bibitem{wei2003geometric}
Tzu-Chieh Wei and Paul~M Goldbart.
\newblock Geometric measure of entanglement and applications to bipartite and
  multipartite quantum states.
\newblock {\em Physical Review A}, 68(4):042307, 2003.

\bibitem{horodecki2009quantum}
Ryszard Horodecki, Pawe{\l} Horodecki, Micha{\l} Horodecki, and Karol
  Horodecki.
\newblock Quantum entanglement.
\newblock {\em Reviews of modern physics}, 81(2):865, 2009.

\bibitem{braunstein1994statistical}
Samuel~L Braunstein and Carlton~M Caves.
\newblock Statistical distance and the geometry of quantum states.
\newblock {\em Physical Review Letters}, 72(22):3439, 1994.

\bibitem{Hauke2016}
Philipp Hauke, Markus Heyl, Luca Tagliacozzo, and Peter Zoller.
\newblock Measuring multipartite entanglement through dynamic susceptibilities.
\newblock {\em Nature Physics}, 12(8):778--782, Aug 2016.

\bibitem{moeller_1995}
Goetz Moeller, Qimiao Si, Gabriel Kotliar, Marcelo Rozenberg, and Daniel~S.
  Fisher.
\newblock Critical behavior near the mott transition in the hubbard model.
\newblock {\em Phys. Rev. Lett.}, 74:2082--2085, Mar 1995.

\bibitem{held_2013}
K.~Held, R.~Peters, and A.~Toschi.
\newblock Poor man's understanding of kinks originating from strong electronic
  correlations.
\newblock {\em Phys. Rev. Lett.}, 110:246402, Jun 2013.

\bibitem{anderson1969exact}
Philip~W Anderson and Gideon Yuval.
\newblock Exact results in the kondo problem: equivalence to a classical
  one-dimensional coulomb gas.
\newblock {\em Physical Review Letters}, 23(2):89, 1969.

\bibitem{si_kotliar_1993}
Qimiao Si and Gabriel Kotliar.
\newblock Metallic non-fermi-liquid phases of an extended hubbard model in
  infinite dimensions.
\newblock {\em Phys. Rev. B}, 48:13881--13903, Nov 1993.

\bibitem{giamarchi2004}
Thierry Giamarchi.
\newblock {\em Quantum physics in one dimension}.
\newblock Clarendon Oxford, 2004.

\bibitem{Logan_2000}
Matthew~T Glossop and David~E Logan.
\newblock Single-particle dynamics of the anderson model: a local moment
  approach.
\newblock {\em Journal of Physics: Condensed Matter}, 14(26):6737, jun 2002.

\bibitem{logan_2014}
David~E. Logan, Adam~P. Tucker, and Martin~R. Galpin.
\newblock Common non-fermi liquid phases in quantum impurity physics.
\newblock {\em Phys. Rev. B}, 90:075150, Aug 2014.

\bibitem{Logan_2015}
David~E Logan and Martin~R Galpin.
\newblock Mott insulators and the doping-induced mott transition within {DMFT}:
  exact results for the one-band hubbard model.
\newblock {\em Journal of Physics: Condensed Matter}, 28(2):025601, dec 2015.

\bibitem{kopp_chakravarty_2007}
Angela Kopp, Xun Jia, and Sudip Chakravarty.
\newblock Replacing energy by von neumann entropy in quantum phase transitions.
\newblock {\em Annals of Physics}, 322(6):1466--1476, 2007.

\bibitem{friedel_1956}
J.~Friedel.
\newblock On some electrical and magnetic properties of metallic solid
  solutions.
\newblock {\em Canadian Journal of Physics}, 34(12A):1190--1211, 1956.

\bibitem{langer1961friedel}
JS~Langer and V~Ambegaokar.
\newblock Friedel sum rule for a system of interacting electrons.
\newblock {\em Physical Review}, 121(4):1090, 1961.

\bibitem{langreth1966}
David~C. Langreth.
\newblock Friedel sum rule for anderson's model of localized impurity states.
\newblock {\em Phys. Rev.}, 150:516--518, Oct 1966.

\end{thebibliography}

\end{document}


\title{Supplementary Materials for ``Kondo frustration via charge fluctuations: a route to Mott localisation"}

\author{Abhirup Mukherjee$^1$, N. S. Vidhyadhiraja$^2$, A. Taraphder$^3$ and Siddhartha Lal$^1$}
\eads{\mailto{am18ip014@iiserkol.ac.in}, \mailto{raja@jncasr.ac.in}, \mailto{arghya@phy.iitkgp.ernet.in}, \mailto{slal@iiserkol.ac.in}}
\address{$^1$Department of Physical Sciences, Indian Institute of Science Education and Research-Kolkata, W.B. 741246, India}
\address{$^2$Theoretical Sciences Unit, Jawaharlal Nehru Center for Advanced Scientific Research, Jakkur, Bengaluru 560064, India}
\address{$^3$Department of Physics, Indian Institute of Technology Kharagpur, Kharagpur 721302, India}

\section{The unitary renormalisation group method}

The non-trivial nature of the impurity problem arises from the fact that the many-particle correlation between the impurity spin and the conduction electrons induces a many-particle interaction among the conduction electron states.
This means that within the conduction band, the states at high and low energies get entangled with each other, leading to what is called UV-IR mixing.
One of the methods of obtaining the low-energy physics while taking into account the effect of the high energy degrees of freedom is the renormalisation group (RG) approach.
The specific variant of RG used in this work is the unitary renormalisation group (URG) method developed by some of us in Refs.~\cite{santanukagome,anirbanmott1,anirbanmott2,1dhubjhep,siddharthacpi,anirban_mott_ent,anirban_kondo}.
The URG proceeds by applying unitary transformations on the Hamiltonian in order to decouple the high-energy \(k-\)states, leading to a reduction in the effective bandwidth and changes in the couplings for the low-energy degrees of freedom in the process.
This leads to a sequence of renormalised Hamiltonians, and hence the coupling RG equations.
While this is similar in spirit to the Poor Man's scaling approach of Anderson as applied to the single channel Kondo model~\cite{anderson1970}, there are several differences between the methods that will be apparent when we describe the method in more detail below.

We start by defining a UV-IR scheme for the single-particle electronic states \(\vec k = (k_F + |\vec k|) \hat k\).
We label the single-particle \(\vec k-\)states in terms of their normal distance $\Lambda = |\vec k|$ from the Fermi surface (FS) and the orientation of the unit vectors $\hat{s} = \hat k$, where $\hat{s}=\frac{\nabla\epsilon_{\mathbf{k}}}{|\nabla\epsilon_{\mathbf{k}}|}|_{\epsilon_{\mathbf{k}}=E_{F}}$.
Each \(\hat s\) represents a direction that is normal to the FS.
Combining these two labels with the spin index \(\sigma=\uparrow,\downarrow\), each single-particle state can be uniquely labelled as $\ket{j,l} = \ket{(k_F + \Lambda_j) \hat s,\sigma}, l\equiv (\hat{s},\sigma)$.
The $\Lambda$'s are arranged as follows: $\Lambda_{N}>\Lambda_{N-1}>\ldots>0$, such that \(\Lambda_N\) is farthest from the Fermi surface and is hence the most energetic (UV) while \(\Lambda_0\) is closest to the Fermi surface and is least energetic (IR).
The URG proceeds by disentangling these states \(\Lambda_j\), starting from those near the UV and gradually scaling towards the IR.
This leads to the Hamiltonian flow equation~\cite{anirbanurg1}
\begin{eqnarray}
\centering
\label{urg_map}
H_{(j-1)}=U_{(j)}H_{(j)}U^{\dagger}_{(j)}~,
\end{eqnarray}
where the unitary operation $U_{(j)}$ is the unitary map at RG step $j$. 
$U_{(j)}$ disentangles all the electronic states 
$|\mathbf{k}_{\Lambda_{j}\hat{s}_{m}},\sigma\rangle$
on the isogeometric curve and has the form~\cite{anirbanmott1,anirbanurg1}
\begin{eqnarray}
\centering U_{(j)}=\prod_{l}U_{j,l}, U_{j,l}=\frac{1}{\sqrt{2}}\sum_l [1+\eta_{j,l}-\eta^{\dagger}_{j,l}]~,
\end{eqnarray}
where \(l\) sums over the states on the isoenergetic shell at distance \(\Lambda_j\), and $\eta_{j,l}$ are electron-hole transition operators following the algebra
\begin{eqnarray}
	\left\{\eta_{j,l},\eta_{j,l}^{\dagger}\right\} = 1~,~\left[\eta_{j,l},\eta_{j,l}^{\dagger}\right] = 2\hat n_{j,l} - 1~.
\end{eqnarray}
The transition operator can be expressed in terms of the off-diagonal part of the Hamiltonian, \(H^X_{j,l} = Tr_{j,l}(c^{\dagger}_{j,l}H_{j})c_{j,l} + \text{h.c.}\), and the diagonal part \(H^D_{j,l}\) (kinetic energy and self-energies):
\begin{eqnarray}
	\eta_{j,l}&=&Tr_{j,l}(c^{\dagger}_{j,l}H_{j,l})c_{j,l}\frac{1}{\hat{\omega}_{j,l}-Tr_{j,l}(H^{D}_{j,l}\hat{n}_{j,l})\hat{n}_{j,l}}~.~~\label{e-TransOp}
\end{eqnarray}
The off-diagonal operator \(Tr_{j,l}(c^{\dagger}_{j,l}H_{j,l})c_{j,l}\) in the numerator of \(\eta_{j,l}\) contains all possible scattering vertices that  change the configuration of the Fock state \(\ket{j,l}\)~\cite{anirbanurg1}. The generic forms of $H^{D}_{j,l}$ and $H^{X}_{j,l}$ are as follows
\begin{eqnarray}
H^{D}_{j,l}=&\sum_{\Lambda\hat{s},\sigma}\epsilon^{j,l}\hat{n}_{\mathbf{k}_{\Lambda\hat{s}},\sigma}+\sum_{\alpha}\Gamma_{\alpha}^{4,(j,l)}\hat{n}_{\mathbf{k}\sigma}\hat{n}_{\mathbf{k}'\sigma'} +\sum_{\beta}\Gamma_{\beta}^{6,(j,l)}\hat{n}_{\mathbf{k}\sigma}\hat{n}_{\mathbf{k}'\sigma'}\hat{n}_{\mathbf{k}''\sigma''}+\ldots~,\nonumber \\
H^{X}_{j,l}=&\sum_{\alpha}\Gamma_{\alpha}^{2}c^{\dagger}_{\mathbf{k}\sigma}c_{\mathbf{k}'\sigma'}+\sum_{\beta}\Gamma_{\beta}^{4}c^{\dagger}_{\mathbf{k}\sigma}c^{\dagger}_{\mathbf{k}'\sigma'}c_{\mathbf{k}_{1}'\sigma_{1}'}c_{\mathbf{k}_{1}\sigma_{1}}+\ldots~.
\end{eqnarray}
The indices \(\alpha\) and \(\beta\) are strings that denote the quantum numbers of the incoming and outgoing electronic states at a particular interaction vertex \(\Gamma^n_{\alpha}\) or  \(\Gamma^m_{\beta}\). The operator $\hat{\omega}_{j,l}$ accounts for the quantum fluctuations arising from the non-commutation between different parts of the renormalised Hamiltonian and has the following form~\cite{anirbanurg1}
\begin{eqnarray}
\hat{\omega}_{j,l}&=&H^{D}_{j,l}+H^{X}_{j,l}-H^{X}_{j,l-1}~.\label{qfOp}
\end{eqnarray}
Upon disentangling electronic states $\hat{s},\sigma$ along a isogeometric curve at distance $\Lambda_{j}$, the following effective Hamiltonian $H_{j,l}$ is generated 
\begin{eqnarray}
H_{j,l}=\prod_{m=1}^{l}U_{j,m}H_{(j)}[\prod_{m=1}^{l}U_{j,m}]^{\dagger}~.
\end{eqnarray}

Accounting for only the leading tangential scattering processes, as well as other momentum transfer processes along the normal direction $\hat{s}$, the renormalised Hamiltonian \(H_{(j-1)}\) has the form~\cite{anirbanurg1}
\begin{eqnarray}
Tr_{j,(1,\ldots,2n_{j})}(H_{(j)})+\sum_{l=1}^{2n_{j}}\lbrace c^{\dagger}_{j,l}Tr_{j,l}(H_{(j)}c_{j,l}),\eta_{j,l}\rbrace\tau_{j,l}~.\label{HRG}
\end{eqnarray}
The RG fixed point is reached when the denominator in eq.~\eqref{e-TransOp} vanishes, at a certain energy scale \(\Lambda^*\).
The vanishing of the denominator can be shown to be concomitant with the vanishing of the off-diagonal component \(H^X\)~\cite{anirbanurg1}, and the fixed point value of the quantum fluctuation operator is equal to one of the eigenvalues of the Hamiltonian.
The fixed point Hamiltonian \(H^*\) consists of the renormalised, still-entangled degrees of freedom \(\Lambda_j\) that lie inside the window \(\Lambda^*\): \(\Lambda_j < \Lambda^*\), as well as the integrals of motion (IOMs) that were decoupled by the URG transformations along the way.
The IOMs have been stripped of any number fluctuations and are therefore diagonal in the basis of the number operators for each of the decoupled degrees of freedom.
\begin{eqnarray}
	H^* = \sum_{\Lambda_j < \Lambda^*} H^*(\Lambda_j) + \sum_{\Lambda_j > \Lambda^*}H_\text{IOMs}(\Lambda_j)~.
\end{eqnarray}
The effective Hamiltonian can be used to construct the corresponding thermal density matrix and hence the partition function at a temperature \(T\):
\begin{eqnarray}
Z^* = \mathrm{Tr}\left[ \hat{\rho}^*\right] = \mathrm{Tr}\left[ e^{-\beta \hat{H}^{*}}\right] = \mathrm{Tr}\left[ U^{\dagger} e^{-\beta \hat{H}^{*}} U\right] = \mathrm{Tr}\left[ e^{-\beta \hat{H}}\right] = Z
\label{partfunc}~,
\end{eqnarray}
where \(\beta = 1/k_B T\), \(U = \prod_{1}^{j^{*}}U_{(j)}\), $H$ is the bare Hamiltonian and $j^{*}$ is the RG step at which the IR stable fixed point is reached. The unitary transformations preserve the partition function along the RG flow.

\section{Derivation of renormalisation group equations for the extended-SIAM}

The Hamiltonian we are working with is the extended Anderson impurity model (described in Section III of the main manuscript):
\begin{eqnarray}
	\mathcal{H} = -\frac{1}{2}U \left(\hat n_{d \uparrow} - \hat n_{d \downarrow}\right)^2 + \sum_{\vec k,\sigma} \epsilon_{\vec k} \tau_{\vec k,\sigma} + J \vec{S}_d\cdot\vec{S}_0 + V\sum_\sigma \left( c^\dagger_{d\sigma}c_{0\sigma} + \text{h.c.}\right) - \frac{1}{2}U_b \left(\hat n_{0 \uparrow} - \hat n_{0 \downarrow}\right)^2~.\quad
\end{eqnarray}
The renormalisation in the Hamiltonian \(H_{(j)}\) at the \(j^\text{th}\) RG step, upon decoupling an electronic state \(q\beta\), is given by the expression:
\begin{eqnarray}
	\left(\Delta H_{(j)}\right)_{\vec q,\beta} = H_{(j-1)} - H_{(j)} = c^\dagger_{\vec q,\beta} T_{\vec q,\beta} \frac{1}{\omega_e - H_D} T^\dagger_{\vec q,\beta}c_{\vec q,\beta} + T^\dagger_{\vec q,\beta} c_{\vec q,\beta} \frac{1}{\omega_h - H_D} c^\dagger_{\vec q,\beta}T_{\vec q,\beta}~,
\end{eqnarray}
where \(\omega_{e,h}\) are the quantum fluctuation scales for the electron and hole scattering channels, \(H_D\) is the part of the Hamiltonian that is diagonal in \(k-\)space, and \(T_{\vec q,\beta} + T_{\vec q,\beta}^\dagger\) is the part of the Hamiltonian that does not commute with \(\hat n_{\vec q,\beta}\) (in short, it is the part that is off-diagonal with respect to a particular state \(\vec q,\beta\)). This off-diagonal part is made up of contributions from \(V,J\) and \(U_b\): \(T_{\vec q \uparrow} = \left[V c^\dagger_{d \uparrow}  + J \sum_{\vec k}S_d^+ c^\dagger_{\vec k \downarrow}  + U_b \sum_{\vec k_1,\vec k_2,\vec k_3}c^\dagger_{\vec k_1 \uparrow} c^\dagger_{\vec k_3 \downarrow} c_{\vec k_4 \downarrow}\right] c_{\vec q \uparrow}\),
while the diagonal part is given by \(H_D = \epsilon_{\vec q} \tau_{q\uparrow} + \frac{1}{4}J S_d^z\left(\hat n_{q \uparrow} - \hat n_{q \downarrow}\right) - \frac{1}{2}U_b \left(\tau_{q \uparrow} - \tau_{q \downarrow}\right)^2\). The contribution from \(U_b\) is obtained by taking the Hartree contributions corresponding to the states \(\ket{q\sigma}\) from the full interacting term in the Hamiltonian.
Note that we have ignored potential scattering terms in the off-diagonal part \(T_{\vec q\beta}\). In order to allow both spin-flip and charge transfer scattering processes, we start from initial states with \(\tau_{q \uparrow} = -\tau_{q \downarrow}\). As a result, the contribution to \(H_D\) from the kinetic energy is \(\epsilon_q \tau_{q \uparrow} = \left(\pm D\right)\times\left(\pm\frac{1}{2}\right) = D/2\), while the contribution from \(U_b\) is \(-\frac{1}{2}U_b\left(\pm\frac{1}{2} - \left( \mp\frac{1}{2} \right) \right)^2 = -U_b/2 \).

\subsection{Renormalisation of \(U_b\)}
\(U_b\) can renormalise only via itself. The relevant renormalisation term in the particle sector is
\begin{eqnarray}
	U_b^2 \sum_{q\beta}\sum_{k_1,k_2,k_3,\atop{k_1^\prime,k_2^\prime,k_3^\prime}} c^\dagger_{q\beta}c_{k_1\beta}c^\dagger_{k_3\overline\beta}c_{k_1^\prime\overline\beta}\frac{1}{\omega - H_D}c^\dagger_{k_2^\prime\overline\beta}c_{k_3^\prime\overline\beta}c^\dagger_{k_2\beta}c_{q\beta}~.
\end{eqnarray}
In order to renormalise \(U_b\), we need to contract one more pair of momenta. There are two choices. The first is by setting \(k_3 = k_3^\prime = q\). The two internal states, then, are \(q\beta\) and \(q\overline\beta\). As discussed above, the intermediate state energy is \(-U_b/4\). We therefore have
\begin{eqnarray}
	\frac{U_b^2 n_j}{\omega - D/2 + \frac{U_b}{2}}\sum_{\beta}\sum_{k_1,k_2,k_1^\prime,k_2^\prime} c_{k_1\beta}c_{k_1^\prime\overline\beta}c^\dagger_{k_2^\prime\overline\beta}c^\dagger_{k_2\beta} = \frac{U_b^2 n_j}{\omega - D/2 + \frac{U_b}{2}}\sum_{\beta}\sum_{k_1,k_2,k_1^\prime,k_2^\prime} c^\dagger_{k_2^\prime\overline\beta}c_{k_1^\prime\overline\beta}c^\dagger_{k_2\beta}c_{k_1\beta}~.
\end{eqnarray}
Another way to contract the momenta is by setting \(k_1^\prime = k_2^\prime = q\), which gives a renormalisation of
\begin{eqnarray}
	\frac{U_b^2 n_j}{\omega - D/2 + \frac{U_b}{2}}\sum_{\beta}\sum_{k_1,k_2,k_3,k_3^\prime} c_{k_1\beta}c^\dagger_{k_3 \overline\beta}c_{k_3\prime\overline\beta}c^\dagger_{k_2\beta} = -\frac{U_b^2 n_j}{\omega - D/2}\sum_{\beta}\sum_{k_1,k_2,k_3,k_3^\prime} c^\dagger_{k_3 \overline\beta}c_{k_3\prime\overline\beta}c^\dagger_{k_2\beta}c_{k_1\beta}~.
\end{eqnarray}
The two contributions cancel each other. The same cancellation happens in the hole sector as well.

\subsection{Renormalisation of the impurity correlation \(U\)}
The coupling \(U\) is renormalised by two kinds of vertices: \(V^2\) and \(J^2\). We will consider these processes one at a time. For convenience, we define \(\epsilon_d = -U/2\).

The renormalisation arising from the first kind of terms, in the particle sector, is
\begin{eqnarray}
	\sum_{q\beta}c^\dagger_{q\beta}c_{d\beta}\frac{V^2}{\omega - H_D}c^\dagger_{d\beta}c_{q\beta} &= \sum_{q\beta}V^2 \hat n_{q\beta} \left( 1 - \hat n_{d\beta} \right)\left( \frac{1-\hat n_{d \overline\beta }}{\omega_0 - E_0} + \frac{\hat n_{d \overline\beta}}{\omega_1 - E_1}\right) \nonumber\\
												      &= V^2 n_j\sum_{\beta}\left( 1 - \hat n_{d\beta} \right)\left( \frac{1-\hat n_{d \overline\beta }}{\omega_0 - E_0} + \frac{\hat n_{d \overline\beta}}{\omega_1 - E_1}\right)~.
\end{eqnarray}
\(q\) runs over the momentum states that are being decoupled at this RG step: \(|q| = \Lambda_j\). \(E_{1,0}\) are the diagonal parts of the Hamiltonian at \(\hat n_{d\overline \beta}=1,0\) respectively: \(E_1 = \frac{D}{2}\) and \(E_0 = \frac{D}{2} + \epsilon_d - \frac{J}{4}\). In order to relate \(\omega_0\) and \(\omega_1\) with the common fluctuation scale \(\omega\) for the conduction electrons, we will replace these quantum fluctuation scales by the current renormalised values of the single-particle self-energy for the initial state from which we started scattering. For \(\hat n_{d\overline\beta}=0\), there is no additional self-energy because the impurity does not have any spin: \(\omega_0 = \omega\). For \(\hat n_{d\overline\beta} = 1\), we have an additional self-energy of \(\epsilon_d\) arising from the correlation on the impurity: \(\omega_1 = \omega + \epsilon_d\).
Substituting the values of \(E_{0,1}\) and \(\omega_{0,1}\), we get
\begin{eqnarray}
	\label{ren_ed_Vp}
	V^2 n_j\sum_{\beta}\left( 1 - \hat n_{d\beta} \right)\left( \frac{1-\hat n_{d \overline\beta }}{\omega - \frac{D}{2} - \epsilon_d + \frac{J}{4}} + \frac{\hat n_{d \overline\beta}}{\omega - \frac{D}{2} + \epsilon_d}\right)~.
\end{eqnarray}
Performing a similar calculation for the hole sector gives the contribution:
\begin{eqnarray}
	\label{ren_ed_Vh}
	V^2 n_j\sum_{\beta}\hat n_{d\beta}\left( \frac{1-\hat n_{d \overline\beta }}{\omega - \frac{D}{2} + \epsilon_d} + \frac{\hat n_{d \overline\beta}}{\omega - \frac{D}{2} - \epsilon_d + \frac{J}{4}}\right)~.
\end{eqnarray}
We now come to the second class of terms: spin-spin. We first look at the particle sector:
\begin{eqnarray}
	\label{ren_ed_Jpp}
	\frac{J^2}{4}\sum_{q\beta}c^\dagger_{d\overline\beta}c_{d\beta}c^\dagger_{q\beta}c_{-q\overline\beta} \frac{1}{\omega - H_D}c^\dagger_{d\beta}c_{d\overline\beta}c^\dagger_{q\overline\beta}c_{q\beta} = \frac{J^2}{4} n_j\frac{1}{\omega - \frac{D}{2} + \frac{J}{4}} \sum_{\beta}\hat n_{d\overline\beta}\left( 1 - \hat n_{d\beta} \right)~.
\end{eqnarray}
The diagonal part in the denominator was simple to deduce in this case, because the nature of the scattering requires the spins \(S_d^z\) and \(\frac{\beta}{2}\left(\hat n_{q\beta} - \hat n_{q \overline\beta}\right)\) to be anti-parallel. This ensures that the intermediate state has an energy of \(E = \frac{D}{2} + \epsilon_d - \frac{J}{4}\), and the quantum fluctuation scale is \(\omega^\prime = \omega + \epsilon_d\), such that \(\omega^\prime - E = \omega - \frac{D}{2} + \frac{J}{4}\). In the hole sector, we have
\begin{eqnarray}
	\label{ren_ed_Jph}
	\frac{J^2}{4} n_j\frac{1}{\omega - \frac{D}{2} + \frac{J}{4}} \sum_{\beta}\hat n_{d\beta}\left( 1 - \hat n_{d\overline\beta} \right)~.
\end{eqnarray}
From eqs.~\eqref{ren_ed_Vp}, \eqref{ren_ed_Vh}, \eqref{ren_ed_Jpp} and \eqref{ren_ed_Jph}, we write
\begin{eqnarray}
	\Delta U &= \Delta \epsilon_2 + \Delta \epsilon_0 - 2\Delta \epsilon_1 \nonumber\\
		 &= -\frac{4V^2 n_j}{\omega + \frac{U_b}{2} - \frac{D}{2} - U/2} + \frac{4V^2 n_j}{\omega + \frac{U_b}{2} - \frac{D}{2} + U/2 + \frac{J}{4}} - \frac{J^2n_j}{\omega + \frac{U_b}{2} - \frac{D}{2} + \frac{J}{4}}~,
\end{eqnarray}
where we have restored the contribution from \(U_b\) in the denominator.

\subsection{Renormalisation of the hybridisation \(V\)}
Renormalisation of \(V\) happens through two kinds of processes: \(VJ\) and \(VU_b\). Within the first kind, the scattering can be either via \(S_d^z\) or through \(S_d^\pm\). For the first kind, we have the following contribution in the particle sector:
\begin{eqnarray}
	\sum_{q\beta} Vc^\dagger_{q\beta} c_{d\beta} \frac{1}{\omega - H_D}\frac{1}{4}J \sum_{k} \left(\hat n_{d\beta} - \hat n_{d\overline\beta}\right) c^\dagger_{k\beta}c_{q\beta} \nonumber\\
	= \frac{1}{4}V J n_j \frac{1}{2}\left(\frac{1}{\omega^\prime_1 - E} + \frac{1}{\omega^\prime_2 - E}\right)\sum_{k\beta} \left(1 - \hat n_{d\overline\beta}\right) c_{d\beta}c^\dagger_{k\beta}~.
\end{eqnarray}
The transformation from \(\frac{1}{\omega - H_D}\) to \(\frac{1}{2}\left(\frac{1}{\omega^\prime_1 - E} + \frac{1}{\omega^\prime_2 - E}\right)\) is made so that we can account for both the initial state and the final state energies through the two fluctuation scales \(\omega^\prime_1\) and \(\omega_2^\prime\) respectively; we calculate the denominators for both the initial and final states and then take the mean of the two (hence the factor of half in front). This was not required previously because, in the earlier scattering processes, the impurity returned to its initial state at the end, at least in terms of \(\epsilon_d \left( \hat n_{d \uparrow} - \hat n_{d \downarrow} \right)^2 \), and so we had \(\omega_1^\prime = \omega_2^\prime = \omega^\prime\).
Substituting the energies and the \(\omega-\)scales, we get
\begin{eqnarray}
	-\left(\frac{\frac{n_j}{4}V J \frac{1}{2}}{\omega - \frac{D}{2} + \frac{J}{4}} + \frac{\frac{n_j}{4}V J \frac{1}{2}}{\omega - \frac{D}{2} - \epsilon_d + \frac{J}{4}}\right)\sum_{k\beta}\left(1 - \hat n_{d\overline\beta}\right) c^\dagger_{k\beta} c_{d\beta}~.
\end{eqnarray}
One can generate another such process by exchanging the single-particle process and the spin-exchange process:
\begin{eqnarray}
	\sum_{q\beta} \frac{1}{4}J \sum_{k} \left(\hat n_{d\beta} - \hat n_{d\overline\beta}\right) c^\dagger_{q\beta}c_{k\beta} \frac{1}{\omega - H_D} V c^\dagger_{d\beta} c_{q\beta}~.
\end{eqnarray}
This is simply the Hermitian conjugate of the previous contribution. Combining this with the previous then gives
\begin{eqnarray}
	-\frac{n_j}{8}V J \left(\frac{1}{\omega - \frac{D}{2} + \frac{J}{4}} + \frac{1}{\omega - \frac{D}{2} - \epsilon_d + \frac{J}{4}}\right) \sum_{k\beta}\left(1 - \hat n_{d\overline\beta}\right) \times\left(c^\dagger_{d\beta} c_{k\beta} + \text{h.c.}\right)~.
\end{eqnarray}
If we similarly calculate the contributions from the spin-exchange processes involving \(S_d^\pm\), we get
\begin{eqnarray}
	-\frac{1}{4}V J n_j \left(\frac{1}{\omega - \frac{D}{2} + \frac{J}{4}} + \frac{1}{\omega - \frac{D}{2} - \epsilon_d + \frac{J}{4}}\right) \sum_{k\beta} \left(1 - \hat n_{d\beta}\right)  \left(c^\dagger_{k\overline\beta} c_{d\overline\beta} + \text{h.c.}\right)~.
\end{eqnarray}
The contributions from the hole sector are obtained by making the transformation \(\hat n_{d\overline\beta} \to 1 - \hat n_{d\overline\beta}\) on the particle sector contributions. The total renormalisation to \(V\) from \(VJ\) processes are
\begin{eqnarray}
	-\frac{3n_j}{8}V J \left(\frac{1}{\omega +U_b/4 - \frac{D}{2} + \frac{J}{4}} + \frac{1}{\omega +U_b/4 - \frac{D}{2} + U/2 + \frac{J}{4}}\right)\sum_{k\beta}\left(c^\dagger_{d\beta} c_{k\beta} + \text{h.c.}\right)~.
\end{eqnarray}
The renormalisation in \(V\) from \(U_b\) arises through terms of \(V U_b\) and \(U_b V\) kind. The first term gives
\begin{eqnarray}
	-\sum_{k\beta} c^\dagger_{d\beta} c_{k\beta} &\left[\frac{\hat n_{d\overline\beta}}{2}\left(\frac{n_jU_b V}{\omega - \frac{D}{2} - \frac{U}{2} + \frac{U_b}{4}} + \frac{n_jU_b V}{\omega - \frac{D}{2} + \frac{U_b}{4}}\right) \right.\nonumber\\
						     &\left.+ \frac{1-\hat n_{d\overline\beta}}{2}\left(\frac{n_jU_b V}{\omega - \frac{D}{2} + \frac{U_b}{4} + \frac{U}{2} + \frac{J}{4}} + \frac{n_jU_b V}{\omega - \frac{D}{2} + \frac{U_b}{4} + \frac{J}{4}}\right)\right]~.
\end{eqnarray}
The second term is of the form \(\sum_{q\beta}\sum_{k}U_b V c^\dagger_{q\beta}c_{d\beta} \frac{1}{\omega - H_D} \hat n_{q\overline\beta} c^\dagger_{k\beta}c_{q\beta}\),
and this is just the Hermitian conjugate of the previous term, so these two terms together lead to
\begin{eqnarray}
	-n_jU_b V\sum_{k\beta} \left(c^\dagger_{d\beta} c_{k\beta} + \text{h.c.}\right)&\left[\frac{\hat n_{d\overline\beta}}{2}\left(\frac{1}{\omega - \frac{D}{2} - \frac{U}{2} + \frac{U_b}{4}} + \frac{1}{\omega - \frac{D}{2} + \frac{U_b}{4}}\right) \right.\nonumber\\
											     &\left. + \frac{1-\hat n_{d\overline\beta}}{2}\left(\frac{1}{\omega - \frac{D}{2} + \frac{U_b}{4} + \frac{U}{2} + \frac{J}{4}} +\frac{1}{\omega - \frac{D}{2} + \frac{U_b}{4} + \frac{J}{4}}\right)\right]~.
\end{eqnarray}
In the hole sector, we have
\begin{eqnarray}
	-n_jU_b V\sum_{k\beta} c^\dagger_{k\beta} \left[\frac{\hat n_{d\overline\beta}}{2}\left(\frac{1}{\omega_1 - E_1} + \frac{1}{\omega^\prime_1 - E_1}\right) + \frac{1-\hat n_{d\overline\beta}}{2}\left(\frac{1}{\omega_0 - E_0} + \frac{1}{\omega_0^\prime - E_0}\right)\right] c_{d\beta}~.
\end{eqnarray}
\(E_1 = D/2 - U_b/4 - U/2 - J/4,~ ~ ~ E_0 = D/2 - U_b/4 - K/4\). The fluctuation scales are \(\omega_1 = \omega = \omega_0^\prime,~ ~ ~ \omega_1^\prime = \omega - U/2 = \omega_0\). Substituting these gives
\begin{eqnarray}
	-\sum_{k\beta} c^\dagger_{d\beta} c_{k\beta} &\left[\frac{1 - \hat n_{d\overline\beta}}{2}\left(\frac{n_jU_b V}{\omega - \frac{D}{2} - \frac{U}{2} + \frac{U_b}{4}} + \frac{n_jU_b V}{\omega - \frac{D}{2} + \frac{U_b}{4}}\right) + \frac{\hat n_{d\overline\beta}}{2}\left(\frac{n_jU_b V}{\omega - \frac{D}{2} + \frac{U_b}{4} + \frac{U}{2} + \frac{J}{4}} \right. \right. \nonumber\\
						     &\left.\left.+ \frac{n_jU_b V}{\omega - \frac{D}{2} + \frac{U_b}{4} + \frac{J}{4}}\right)\right]~.
\end{eqnarray}
The other term, obtained by exchanging \(V\) and \(U_b\), gives the Hermitian conjugate, so the overall contribution from the hole sector is the same as the total contribution from the particle sector, but with \(\hat n_{d\overline\beta} \to 1 - \hat n_{d\overline\beta}\). Combining both the sectors, we get
\begin{eqnarray}
	-\sum_{k\beta} \left(c^\dagger_{d\beta} c_{k\beta} + \text{h.c.}\right) &\left[\left(\frac{n_jU_b V/2}{\omega - \frac{D}{2} - \frac{U}{2} + \frac{U_b}{4}} + \frac{n_jU_b V/2}{\omega - \frac{D}{2} + \frac{U_b}{4}}\right) + \left(\frac{n_jU_b V/2}{\omega - \frac{D}{2} + \frac{U_b}{4} + \frac{U}{2} + \frac{J}{4}} \right.\right.\nonumber\\
										&\left.\left.+ \frac{n_jU_b V/2}{\omega - \frac{D}{2} + \frac{U_b}{4} + \frac{J}{4}}\right)\right]~.
\end{eqnarray}

\subsection{Renormalisation of the spin-exchange coupling \(J\)}
The term \(J \vec{S_d}\cdot\vec{S}_0\) can be split into three parts: \(J^z S_d^z, \frac{1}{2}J^+ S_d^+ S_0^-\) and \(\frac{1}{2}J^- S_d^- S_0^+\). We will only calculate the renormalisation in \(J^+\), which will be equal to that of \(J^-,J^z\) due to spin-rotation symmetry. The terms that renormalise \(J^+\) are of the form \(S_d^+ S_d^z\) and \(S_d^z S_d^+\). In the particle sector, we have
\begin{eqnarray}
	-\sum_{q} \sum_{kk^\prime}\frac{1}{4}J^2 S_d^+ c^\dagger_{q\downarrow}c_{k^\prime \uparrow} \frac{1}{\omega - H_D}S_d^z c^\dagger_{k \downarrow}c_{q \downarrow} = n_j \frac{1}{4}J^2 \left(-\frac{1}{2}S_d^+\right) \sum_{kk^\prime}c^\dagger_{k \downarrow}c_{k^\prime \uparrow} \frac{1}{\omega + \frac{U_b}{2} - \frac{D}{2} + \frac{J}{4}}~,\nonumber\\
	\sum_{q} \sum_{kk^\prime} \frac{1}{4}J^2 S_d^z c^\dagger_{q \uparrow}c_{k^\prime \uparrow} \frac{1}{\omega - H_D} S_d^+ c^\dagger_{k\downarrow}c_{q \uparrow} = -n_j \frac{1}{4}J^2 \left(\frac{1}{2}S_d^+\right) \sum_{kk^\prime}c^\dagger_{k \downarrow}c_{k^\prime \uparrow} \frac{1}{\omega + \frac{U_b}{2} - \frac{D}{2} + \frac{J}{4}}~.\qquad
\end{eqnarray}
The denominator is determined using \(E = \frac{D}{2} + \epsilon_d - \frac{J}{4}\) and \(\omega^\prime = \omega + \epsilon_d\).
In the hole sector, we similarly have
\begin{eqnarray}
	\sum_{q} \sum_{kk^\prime}\frac{1}{4}J^2 S_d^+ c^\dagger_{k\downarrow}c_{q \uparrow} \frac{1}{\omega - H_D}S_d^z c^\dagger_{q \uparrow}c_{k^\prime \uparrow} = n_j \frac{1}{4}J^2 \left(-\frac{1}{2}S_d^+\right) \sum_{kk^\prime}c^\dagger_{k \downarrow}c_{k^\prime \uparrow} \frac{1}{\omega + \frac{U_b}{2} - \frac{D}{2} + \frac{J}{4}}~,\nonumber\\
	-\sum_{q} \sum_{kk^\prime} \frac{1}{4}J^2 S_d^z c^\dagger_{k \downarrow}c_{q \downarrow} \frac{1}{\omega - H_D} S_d^+ c^\dagger_{q\downarrow}c_{k^\prime \uparrow} = -n_j \frac{1}{4}J^2 \left(\frac{1}{2}S_d^+\right) \sum_{kk^\prime}c^\dagger_{k \downarrow}c_{k^\prime \uparrow} \frac{1}{\omega + \frac{U_b}{2} - \frac{D}{2} + \frac{J}{4}}~.
\end{eqnarray}
The renormalisation due to \(U_b\) also happens through multiple terms. One of the terms is
\begin{eqnarray}
	\frac{1}{2} J U_b \sum_{q} \sum_{k,k^\prime} S_d^+ c^\dagger_{q \downarrow} c_{k \uparrow} \frac{1}{\omega - H_D} \hat n_{q \uparrow} c^\dagger_{k^\prime \downarrow}c_{q \downarrow} = -\frac{1}{2}\frac{J U_b n_j}{\omega - \frac{D}{2} + \frac{U_b}{2} + \frac{J}{4}} \sum_{k,k^\prime} S_d^+ c^\dagger_{k^\prime \downarrow} c_{k \uparrow}~.
\end{eqnarray}
The factor of half in front is the same half factor that appears in front of the \(S_1^+ S_2^-, S_1^-S_2^+\) terms when we rewrite \(\vec{S}_1\cdot\vec{S}_2\) in terms of \(S^z, S^\pm\). Another term is obtained by switching \(J\) and \(U_b\):
\begin{eqnarray}
	\frac{1}{2} J U_b \sum_{q} \sum_{k,k^\prime} \hat n_{q \downarrow} c^\dagger_{q \uparrow} c_{k \uparrow} \frac{1}{\omega - H_D}S_d^+ c^\dagger_{k^\prime \downarrow} c_{q \uparrow} = -\frac{1}{2}\frac{J U_b n_j}{\omega - \frac{D}{2} + \frac{U_b}{2} + \frac{J}{4}} \sum_{k,k^\prime} S_d^+ c^\dagger_{k^\prime \downarrow} c_{k \uparrow}~.
\end{eqnarray}

The corresponding terms in the hole sector are
\begin{eqnarray}
	-\frac{1}{2}\frac{J U_b n_j}{\omega - \frac{D}{2} + \frac{U_b}{2} + \frac{J}{4}} \sum_{k,k^\prime} S_d^+ c^\dagger_{k^\prime \downarrow} c_{k \uparrow}~, -\frac{1}{2}\frac{J U_b n_j}{\omega - \frac{D}{2} + \frac{U_b}{2} + \frac{J}{4}} \sum_{k,k^\prime} S_d^+ c^\dagger_{k^\prime \downarrow} c_{k \uparrow}~.
\end{eqnarray}

\subsection{Final URG equations}
In summary, the renormalisation in the couplings takes the form:
\begin{eqnarray}
	\Delta U &= 4V^2 n_j\left(\frac{1}{d_1} - \frac{1}{d_0}\right) - n_j\frac{J^2}{d_2},\quad &\Delta V = -\frac{3n_j V}{8}\left[J\left(\frac{1}{d_2} + \frac{1}{d_1}\right) +  \frac{4U_b}{3}\sum_{i=1}^4 \frac{1}{d_i}\right],\nonumber \\
	\Delta J &= -\frac{n_j J\left(J + 4U_b\right)}{d_2}~,\quad &\Delta U_b = 0\label{rg-eqn}~,
\end{eqnarray}
where the denominators \(d_i\) are given by
\begin{eqnarray}\label{rg-eqtn1}
	d_0 = \omega - \frac{D}{2} + \frac{U_b}{2} - \frac{U}{2},~d_1 = \omega - \frac{D}{2} + \frac{U_b}{2} + \frac{U}{2} + \frac{J}{4}~,\\
	d_2 = \omega - \frac{D}{2} + \frac{U_b}{2} + \frac{J}{4}~,d_3 = \omega - \frac{D}{2} + \frac{U_b}{2}~.
\end{eqnarray}
For the sake of completeness, we present the RG equation for a charge isospin Kondo coupling $K$ between the impurity and the bath: 
\begin{eqnarray}
\Delta K = -\frac{n_j K\left(K + 4U_b\right)}{\omega - \frac{D}{2} + \frac{U_b}{2} + \frac{K}{4}}~.
\end{eqnarray}
Indeed, the RG equation for $K$ is observed to be identical in form to that for the spin Kondo coupling $J$, obtained through the transformation \(J \to K\). This indicates that charge fluctuations between the bath zeroth and first sites (that are incited by an attractive interaction $U_{b}$) lead to the RG irrelevance of $K (>0)$, and thereby a destabilisation of the charge Kondo effect similar to that presented in the main manuscript for $J$.

\section{Expressing correlation functions in terms of entanglement}
\label{general-ent}

We will first relate the impurity Greens function to the geometric entanglement. Given a ground state \(\ket{\Psi}_\text{gs}\) and a spectrum of energies \(\left\{ E_n \right\} \), the retarded impurity Greens function (in time domain) is defined as \(G_{d\sigma}(t) = -i\theta(t)\braket{\left\{c_{d\sigma}(t),c^\dagger_{d\sigma}\right\}}\). It can be given a spectral representation in terms of the eigenstates \(\left\{\ket{\Psi}_n\right\}\) of the e-SIAM:
\begin{eqnarray}
	\label{green_spectral}
	G_{d\sigma}(\omega) = \frac{1}{Z}\sum_n \left[ \frac{|\braket{\Psi_\text{gs} | c_{d\sigma} | \Psi_n}|^2}{\omega + E_\text{gs} - E_n} + \frac{|\braket{\Psi_n | c_{d\sigma} | \Psi_\text{gs}}|^2}{\omega - E_\text{gs} + E_n}\right] ~,
\end{eqnarray}
where \(Z \equiv \sum_n e^{-\beta E_n}\) is the partition function.
We will now insert a complete basis into the expression. The basis will be the set of eigenstates of the Hamiltonian  \(H = H_1 + H_2\), where \(H_1\) is the two-site \(J-U_b\) Hamiltonian formed by the impurity and zeroth sites, and \(H_2\) is a tight-binding Hamiltonian formed by the remaining sites. Since the Hamiltonians are decoupled, the eigenstates \(\ket{\Psi}_{m,n}\) will be direct product states formed by combining the eigenstates \(\ket{\phi}_m,\ket{\psi}_n\) of the individual Hamiltonians \(H_1\) and \(H_2\) respectively: \(\ket{\Phi}_{m,n} = \ket{\phi}_m\otimes\ket{\psi}_n\). Inserting this basis leads to the expression:
\begin{eqnarray}
	\label{summation}
	\ket{\Psi}_\text{gs} = \sum_{m,n} \ket{\Phi}_{m,n} \left(\bra{\phi_{m}}\otimes\bra{\psi_n}\right)\ket{\Psi_\text{gs}}~.
\end{eqnarray}
The ground state of \(H_1\) is the spin-singlet: \(\ket{\phi}_0 = \ket{\text{ss}}\). We denote the ground state of the tight-binding Hamiltonian \(H_2\) by \(\ket{\psi}_0\). Because of the highly renormalised couplings \(V\) and \(J\), the impurity site is almost maximally entangled with the zeroth site, such that the ground state \(\ket{\Psi}_\text{gs}\) in the \(r < 0.25\) regime can be thought of as a direct product of the two-site ground state, \(c\ket{\text{ss}} + \sqrt{1-c^2}\ket{\text{ct}}\), in direct product with the tight-binding ground state:
\begin{eqnarray}
	\ket{\Psi}_\text{gs} \simeq \left(c\ket{\text{ss}} + \sqrt{1-c^2}\ket{\text{ct}}\right)  \otimes \ket{\psi}_0 = \ket{\Phi}_\text{ss} + \ket{\Phi}_\text{ct}~,
\end{eqnarray}
where \(\ket{\Phi}_\text{ss(ct)} = \ket{\text{ss}\left( \text{ct} \right)}\otimes\ket{\psi}_0 \).
This suggests that not all terms in the summation of eq.~\eqref{summation} contribute; out of all \(\left\{ \ket{\psi}_n \right\} \), only the ground state \(n=0\) contributes, while only \(\ket{ss}\) and \(\ket{ct}\) contribute from the set \(\left\{ \ket{\phi}_m \right\} \). The summation then simplifies to
\begin{eqnarray}
	\label{gs-ss}
	\ket{\Psi}_\text{gs} = \ket{\Phi}_{ss}\braket{\text{ss}|\Psi^{(2)}_\text{gs}} + \ket{\Phi}_{ct}\braket{\text{ct}|\Psi^{(2)}_\text{gs}}~,
\end{eqnarray}
where \(\ket{\Psi^{(2)}_\text{gs}}\) is the two-site part of the ground state and can be obtained by starting with the full ground state \(\ket{\Psi}_\text{gs}\) and tracing over the other lattice sites of the system.

We assume that the global phases of the wavefunctions are real, and since the internal weights of the wavefunctions are real as well, the overlaps \(\braket{\text{ss}|\Psi^{(2)}_\text{gs}},\braket{\text{ct}|\Psi^{(2)}_\text{gs}}\) are also real. We can use these overlaps to define a geometric measure of entanglement~\cite{shimony1995degree,wei2003geometric,horodecki2009quantum}:
\begin{eqnarray}
	\varepsilon\left(\psi_1,\psi_2\right) = 1 - |\braket{\psi_1 | \psi_2}|^2~.
\end{eqnarray}
If \(\ket{\psi_1}\) is thought of as a separable state, then \(\ket{\psi_2}\) should be less entangled if its overlap with \(\ket{\psi_1}\) is large, which is indeed borne out by the definition.
The overlaps then become \(\braket{\text{ss}|\Psi^{(2)}_\text{gs}} = \sqrt{1 - \varepsilon\left(ss,\Psi^{(2)}_\text{gs}\right)}\), and similarly for the state \(\ket{\text{ct}}\). For brevity, we will use the notation \(\varepsilon_\text{ss} \equiv \varepsilon\left(ss,\Psi^{(2)}_\text{gs}\right), \varepsilon_\text{ct} \equiv \varepsilon\left(ct,\Psi^{(2)}_\text{gs}\right)\). The retarded impurity Greens function for spin \(\sigma\) can now be written in terms of these measures:
\begin{eqnarray}\label{greens_func_entng}
	G_{d\sigma}(\omega) = \frac{1}{Z}\sum_n &\left[\left(1 - \varepsilon_\text{ss} \right) \left(\frac{|\braket{\Phi_\text{ss} | c_{d\sigma} | \Psi_n}|^2}{\omega + E_\text{gs} - E_n} + \frac{|\braket{\Psi_n | c_{d\sigma} | \Phi_\text{ss}}|^2}{\omega - E_\text{gs} + E_n}\right) + \left(1 - \varepsilon_\text{ct} \right) \frac{|\braket{\Phi_\text{ct} | c_{d\sigma} | \Psi_n}|^2}{\omega + E_\text{gs} - E_n}\right.\nonumber\\
		&\left. + \left(1 - \varepsilon_\text{ct} \right)\frac{|\braket{\Psi_n | c_{d\sigma} | \Phi_\text{ct}}|^2}{\omega - E_\text{gs} + E_n} + \sqrt{\left(1 - \varepsilon_\text{ss} \right)}\sqrt{\left(1 - \varepsilon_\text{ct} \right)} \frac{\braket{\Phi_\text{ss} | c_{d\sigma} | \Psi_n}\braket{\Psi_n | c^\dagger_{d\sigma} | \Phi_\text{ct}} + \text{h.c.}}{\omega + E_\text{gs} - E_n} \right. \nonumber\\
		&\left.+ \sqrt{\left(1 - \varepsilon_\text{ss} \right)}\sqrt{\left(1 - \varepsilon_\text{ct} \right)}\frac{\braket{\Phi_\text{ct} | c_{d\sigma} | \Psi_n}\braket{\Psi_n | c^\dagger_{d\sigma} | \Phi_\text{ss}} + \text{h.c.}}{\omega - E_\text{gs} + E_n}\right]~.
\end{eqnarray}

We now generalise this to any real-space two-particle correlation involving operators \(O_1, O_2\) that are at most two-particle operators and act on the combined Hilbert space of the impurity site and the zeroth site. The so-called lesser Greens function corresponding to these operators is defined as 
\begin{eqnarray}\label{lesser-gf}
	G^<_{O_1^\dagger, O_2}(t) \equiv i\braket{O_1^\dagger O_2(t)}, \quad G^<_{O_1^\dagger, O_2}(t\to \infty) = i\braket{\Psi_\text{gs}|O_1^\dagger O_2 | \Psi_\text{gs}} ~.
\end{eqnarray}
We focus on the static case (\(t \to \infty\)) here. Following eq.~\eqref{gs-ss}, we can rewrite the ground states in terms of the entanglement measures mentioned above.
\begin{eqnarray}
	\ket{\Psi}_\text{gs} = \ket{\Phi}_{ss}\sqrt{1 - \varepsilon_\text{ss}} + \ket{\Phi}_{ct}\sqrt{1 - \varepsilon_\text{ss}}~,
\end{eqnarray}
where \(\ket{\Phi}_\text{ss(ct)}\) are the zero spin and zero charge isospin eigenstates of the Hamiltonian \(H = H_1 + H_2\) defined below eq.~\eqref{green_spectral}.
Substituting this in the static lesser Greens function gives
\begin{eqnarray}
	\label{gen-entng}
	\frac{1}{i}G^<_{O_1^\dagger, O_2}(t\to \infty) =& \left(1 - \varepsilon_\text{ss}\right) \braket{\Phi_\text{ss} | O_2 O_1^\dagger | \Phi_\text{ss}} + \left(1 - \varepsilon_\text{ct}\right) \braket{\Phi_\text{ct} | O_2 O_1^\dagger | \Phi_\text{ct}} \nonumber \\
				      &+ \sqrt{1 - \varepsilon_\text{ss}}\sqrt{1 - \varepsilon_\text{ct}}\left(\braket{\Phi_\text{ss} | O_2 O_1^\dagger | \Phi_\text{ct}} + \braket{\Phi_\text{ct} | O_2 O_1^\dagger | \Phi_\text{s}}\right)~.
\end{eqnarray}
This can be extended to generalised retarded Greens functions \(G^R_{O_1^\dagger,O_2}(t) \equiv -i\theta(t) \braket{\left\{O_1^\dagger(t), O_2\right\}}\), whose spectral representation is of the form
\begin{eqnarray}
	\label{retarded}
	G^R_{O_1^\dagger,O_2}(\omega) = \frac{1}{Z}\sum_n \left[ \frac{\braket{\Psi_\text{gs} | O_2 | \Psi_n}\braket{\Psi_n | O_1^\dagger | \Psi_\text{gs}}}{\omega + E_\text{gs} - E_n} + \frac{\braket{\Psi_n | O_2 | \Psi_\text{gs}}\braket{\Psi_\text{gs} | O_1^\dagger | \Psi_n}}{\omega - E_\text{gs} + E_n}\right] ~.
\end{eqnarray}
Following an approach very similar to the one that led to eq.~\eqref{greens_func_entng}, we can cast the generalised Greens function in terms of the geometric entanglement measures:
\begin{eqnarray}
	G^R_{O_1^\dagger,O_2}(\omega) = \frac{1}{Z}\sum_n &\left[\left(1 - \varepsilon_\text{ss} \right) \left(\frac{\left(O_1\right)_{\text{ss},n}^* \left( O_2 \right)_{\text{ss}, n}}{\omega + E_\text{gs} - E_n} + \frac{\left(O_1\right)_{n,\text{ss}}^* \left( O_2 \right)_{n,\text{ss}}}{\omega - E_\text{gs} + E_n}\right) + \left(1 - \varepsilon_\text{ct} \right) \frac{\left(O_1\right)_{\text{ct},n}^* \left( O_2 \right)_{\text{ct}, n}}{\omega + E_\text{gs} - E_n}\right.\nonumber\\
						&\left. + \left(1 - \varepsilon_\text{ct} \right)\frac{\left(O_1\right)_{n,\text{ct}}^* \left( O_2 \right)_{n,\text{ct}}}{\omega - E_\text{gs} + E_n} + \sqrt{\left(1 - \varepsilon_\text{ss} \right)}\sqrt{\left(1 - \varepsilon_\text{ct} \right)} \frac{\left(O_2\right)_{\text{ss},n} \left( O_1 \right)_{\text{ct},n}^* + \text{h.c.}}{\omega + E_\text{gs} - E_n} \right. \nonumber\\
		&\left.+ \sqrt{\left(1 - \varepsilon_\text{ss} \right)}\sqrt{\left(1 - \varepsilon_\text{ct} \right)}\frac{\left(O_2\right)_{\text{ct},n} \left( O_1 \right)_{\text{ss},n}^* + \text{h.c.}}{\omega - E_\text{gs} + E_n}\right]~,
\end{eqnarray}
where \(\left( O_2 \right)_{\text{ss},n} \) is the matrix element \(\braket{\Phi_\text{ss} | O_2 | \Psi_n}\), and similar definitions exist for \(\left( O_1 \right)_{\text{ss},n} \) and its \(\ket{\Phi_\text{ct}}\) counterparts.

Closely related to geometric entanglement is the quantum Fisher information (QFI) \(F_Q\)~\cite{braunstein1994statistical,Hauke2016}, which quantifies how sensitive a state \(\rho\) is to unitary transformations generated by an observable \(\hat O\).
For a pure state \(\rho = \ket{\psi}\bra{\psi}\), \(F_Q (\psi,\hat O)\) is defined in terms of the variance of the operator $\hat O$ (i.e., the connected correlation function):
\begin{eqnarray}
	F_Q(\psi,\hat O) = 4 \Delta (\hat O)^{2}= 4\left(\braket{\psi|\hat O^2|\psi} - \braket{\psi|\hat O|\psi}^2\right)~,
\end{eqnarray}
and it provides an upper bound on the precision that can be achieved in measuring the parameter \(\theta\) that is dual to the observable \(\hat O\): \(\mathcal{N}\left( \Delta \theta \right)^2 \geq F_Q^{-1} \), \(\mathcal{N}\) being the number of independent measurements~\cite{braunstein1994statistical,Hauke2016}. \(F_Q (\psi,\hat O)\) is thus a measure of the entanglement arising from quantum fluctuation content in $\ket{\psi}$ (considered with respect to an eigenstate of $\hat O$). This can be made more manifest by considering a traceless operator \(\hat M\) with eigenstates \(\left\{\ket{m}\right\}\). Without loss of generality, we can rescale the operator such that \(\sum m^2 = 1/2\). There are several examples of such operators in the context of the present work, such as the impurity magnetisation operator (\(S_d^z\)), the Kondo spin-flip operator (\(S_d^+ S_0^- + \text{h.c.}\)) and the Kondo isospin-flip operator (\(C_d^+ C_0^- + \text{h.c.}\)). If \(\varepsilon_m(\psi) \equiv 1 - |\braket{m | \psi}|^2\) is the geometric entanglement between a given state \(\ket{\psi}\) and the eigenstates of \(\hat M\), the QFI corresponding to \(\hat M\) in the state \(\psi\) can be written as
\begin{eqnarray}\label{QFI-geo-ent}
	F_Q(\psi,\hat M) = 4\left[\frac{1}{2} - \sum_m m^2\varepsilon_m - \left(\sum_m m\varepsilon_m\right)^2 \right] ~.
\end{eqnarray}
The total geometric entanglement \(\sum_m \varepsilon_m\) is constrained to be \(d_M - 1\), where \(d_M\) is the dimension of the operator \(\hat M\).
Because each eigenvalue \(m\) has a magnitude of at most \(1/\sqrt 2\) (all \(m^2\) must add up to \(1/2\)), the maximum magnitude of the last two terms in eq.~\eqref{QFI-geo-ent} (and hence the minimum value of the QFI) is attained for the case when some of the \(\varepsilon_m\) are zero.
However, because the eigenstates are orthogonal, only one of them can have perfect overlap with the state \(\ket{\psi}\) and only one \(\varepsilon_m\) can be zero at a time.
The case of minimum QFI therefore corresponds to \(\varepsilon_{m^*} = 0, \varepsilon_{m\neq m^*}=1\), leading to \(F_Q = 0\).
The opposite situation arises when all the geometric entanglement measures are non-zero and equal, \(\varepsilon_m = 1/d_M\), leading to a maximum QFI value of \(F_Q = 2/d_M\).
A uniformly spread geometric entanglement distribution, therefore, leads to a larger QFI, while a more focused distribution leads to a smaller QFI.
This sums up the relation between the quantum Fisher information and geometric entanglement.

If the operator \(\hat M\) is such that its eigenstates are separable states from the perspective of a certain party, the vanishing QFI arises when \(\ket{\psi}\) has zero geometric entanglement with respect to one of the separable states, and the QFI saturates when \(\ket{\psi}\) has a finite geometric entanglement with all of the states.
A large QFI can thus be associated with more entanglement. 
Moreover, if the eigenstates of \(\hat M\) are symmetry-broken states, a vanishing QFI indicates that \(\ket{\psi}\) is very close to such a symmetry-broken state, while a large QFI indicates that \(\ket{\psi}\) is a uniform superposition of such symmetry-broken states and therefore preserves the symmetry as a whole. A larger QFI points towards the presence of more quantum fluctuations and the lack of susceptibility towards symmetry-breaking.

\begin{figure}[htpb]
	\centering
	\includegraphics[width=0.48\textwidth]{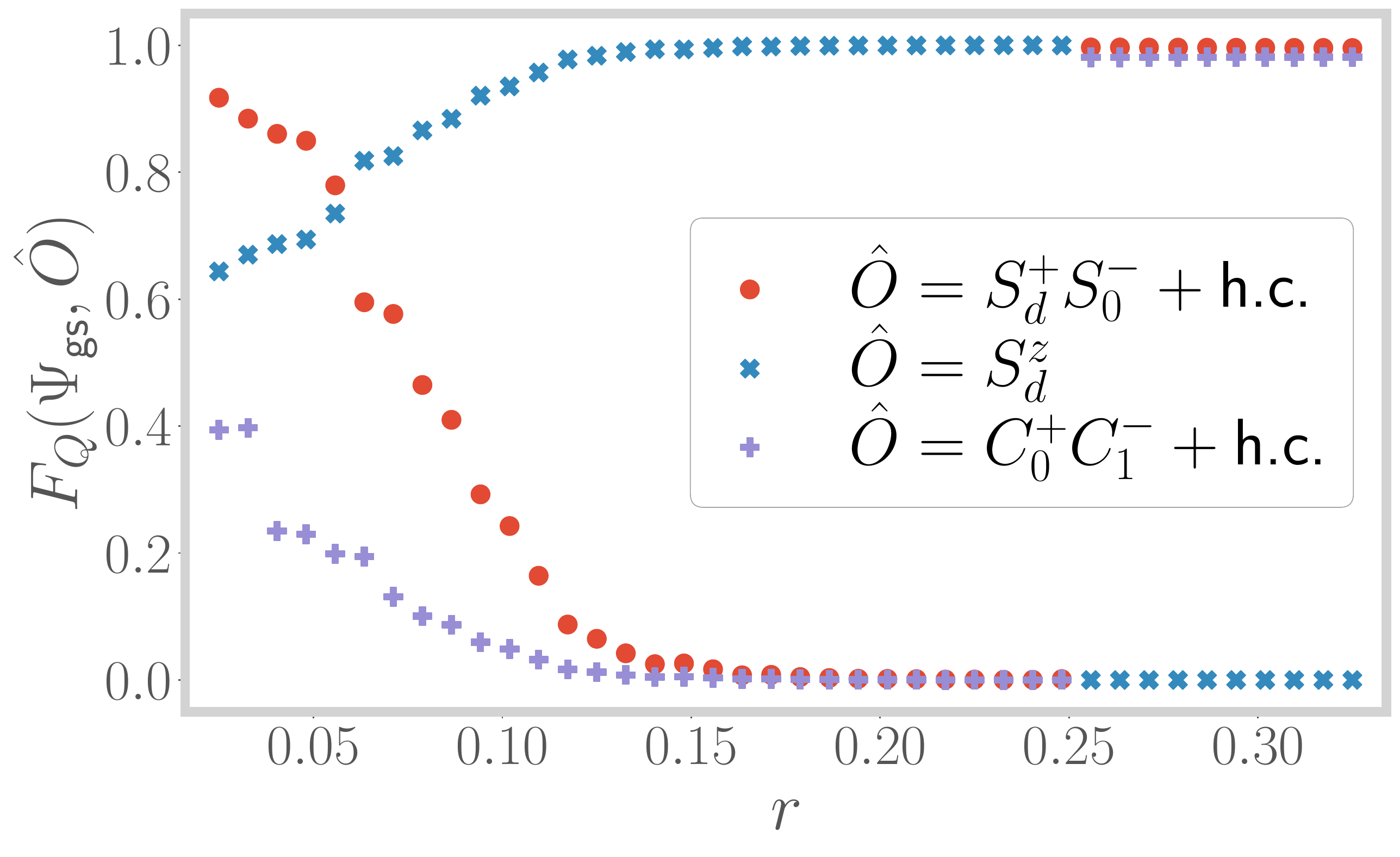}
	\includegraphics[width=0.48\textwidth]{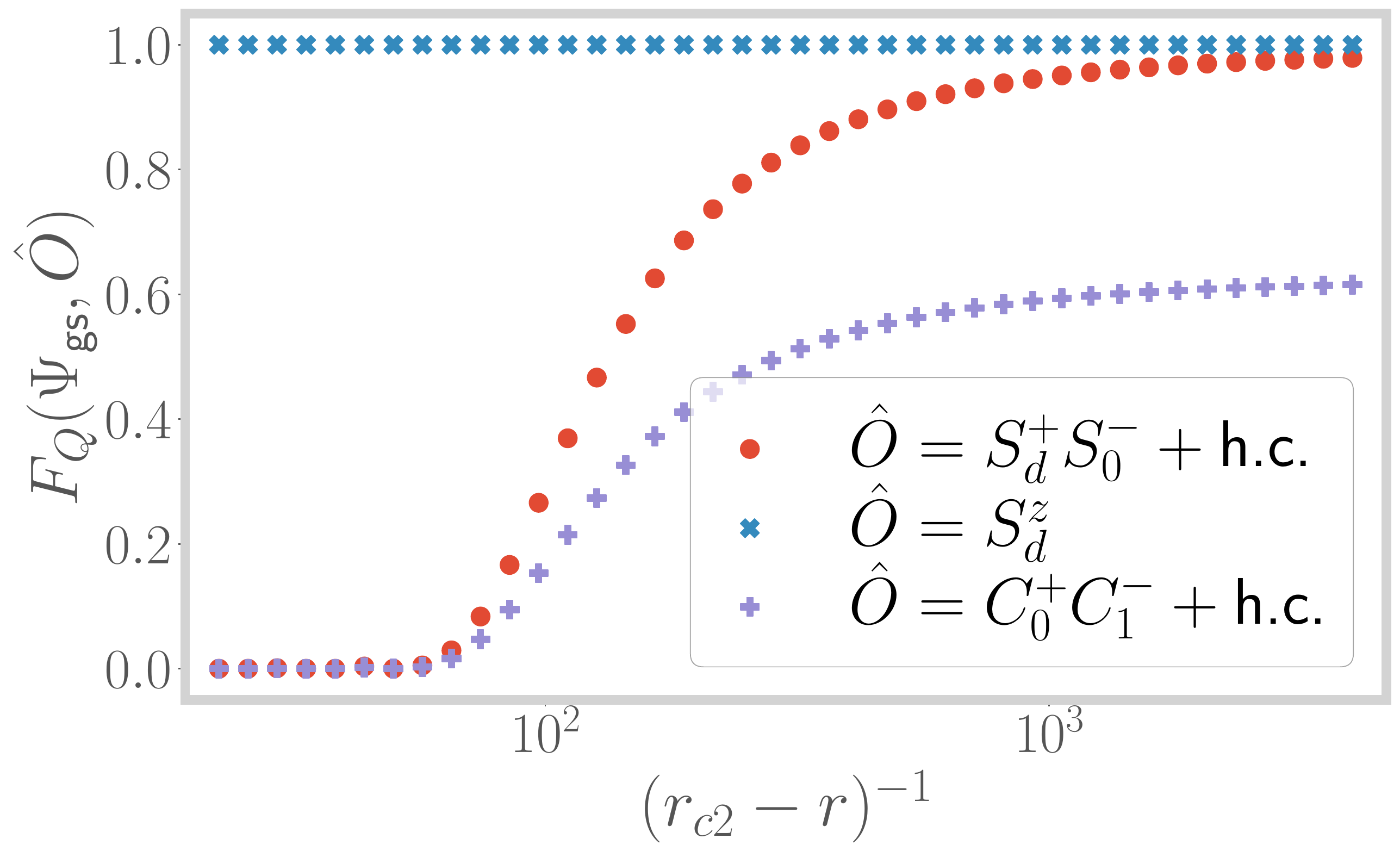}
	\caption{{\it Left:} Variation of the quantum Fisher information for three separate fluctuations, across the entire range of \(r\). {\it Right:} The QFI corresponding to the same three operators, but very close to the QCP at \(r_{c2}\).}
	\label{QFI-esiam}
\end{figure}

In order to demonstrate these ideas in the present problem, we computed the QFI in the ground-state of the e-SIAM for a number of Hermitian operators, as a function of the parameter \(r=-U_b/J_0\).
These are shown in the left panel of Fig.~\eqref{QFI-esiam} (left panel). 
In order to explain the behaviour depicted in the figure, we point out again that the ground-states for \(r \ll r_{c2}\), \(r \lesssim r_{c2}\) and \(r > r_{c2}\) are \(\ket{\text{SS}} + \ket{\text{CT}}\), \(\ket{\text{SS}}\) and \(\ket{\text{LM}}\) respectively, where \(\ket{\text{SS}}\), \(\ket{\text{CT}}\) and $\ket{\text{LM}}$ are the spin-singlet state, charge triplet and local moment states respectively.
We focus on the spin-flip QFI, corresponding to the operator \(\hat O = S_d^+ S_0^- + \text{h.c.}\).
\begin{itemize}
	\item For \(r < r_{c1}\), the ground-state involves both \(\ket{SS}\) and \(\ket{CT}\), meaning that neither of the two entanglement measures \(\varepsilon_\text{SS}\) or \(\varepsilon_\text{CT}\) will be zero.
	As mentioned in the preceding paragraph, this leads to the maximal QFI, as can be seen in the figure as well.
	\item At \(r = r_{c1}\), the charge triplet content vanishes and the ground-state is purely a singlet beyond that point.
	As a result, in the range \(r_{c1} < r < r_{c2}\), the geometric entanglement corresponding to the singlet state is zero, leading to a minimal and vanishing QFI.
	This is again shown in the figure.
	\item Finally, for \(r > r_{c2}\), the local moment states become the ground-states, where the bath zeroth site displays all four configurations as a superposition, because of the hopping into the rest of the bath.
	This state is again not an eigenstate of the spin-flip operator, and the geometric entanglement will be uniformly distributed across the states, none being zero.
	This explains the large QFI in the local moment phase.
\end{itemize}
Similar explanations hold for the single-particle hopping FQI between the impurity and the zeroth site, and the charge isospin FQI between the zeroth site and the first site.
In Fig.~\eqref{QFI-esiam} (right panel), we show the QFI for the same three operators very close to the MIT (i.e., for $r\lesssim r_{c2}$). 
Two important quantities that track the local MIT and act as order parameters for the transition are the degree of compensation for the impurity (\(\braket{\vec{S}_d\cdot\vec{S}_0}\)) and the impurity magnetisation (\(\braket{S_d^z}\)), and the first two QFI (red and blue curves) in Fig.~\ref{QFI-esiam} are therefore particularly important because that they quantify the quantum fluctuations present in the system corresponding to these order parameters.
Our analysis shows that the two phases on either side of the transition are characterised by distinct values of this pair of QFI: while the QFI corresponding to the degree of compensation zero in the Kondo screened phase, it becomes non-zero in the local moment phase, and the opposite is true for the QFI arising from the impurity magnetisation.
The phase precisely at the transition is distinct from those on either side, because it displays a non-zero value for both the QFI.
While it is expected that a critical point would show enhanced fluctuations of multiple kinds (giving rise to universality), it is enlightening to find that this is also reflected in a measure of many-particle entanglement.

All three QFIs are observed to saturate to finite, non-zero values at the MIT, indicating that the non-Fermi liquid state therein possesses quantum fluctuations of all three varieties. 

We will now relate the QFI to some other measures of correlation. The QFI corresponding to an observable can be directly related to the static lesser Greens functions associated with that observable:
\begin{eqnarray}
	\label{lesser-QFI}
	F_Q(\psi,\hat O_2) = 4\left(\braket{\hat O_2 \hat O_1^\dagger}\bigg|_{\hat O_1^\dagger=\hat O_2} - \braket{\hat O_2 \hat O_1^\dagger}^2\bigg|_{\hat O_1=1}\right) = 4\left[G^<_{O_2, O_2}(t\to \infty) - \left(G^<_{1, O_2}(t\to \infty)\right)^2\right]
\end{eqnarray}
In the context of the present work, eq.~\eqref{gen-entng} allows relating the QFI to the geometric entanglement measured with respect to the spin-singlet and charge triplet states. By combining eqs.~\eqref{gen-entng} and \eqref{lesser-QFI}, we get
\begin{eqnarray}
	F_Q(\psi,\hat O_2) &= 4\left[ \left(1 - \varepsilon_\text{ss}\right) \braket{\Phi_\text{ss} | \hat O_2^2 | \Phi_\text{ss}} + \left(1 - \varepsilon_\text{ct}\right) \braket{\Phi_\text{ct} | \hat O_2^2 | \Phi_\text{ct}} \right.\nonumber \\
			   &\left.+ \sqrt{1 - \varepsilon_\text{ss}}\sqrt{1 - \varepsilon_\text{ct}}\left(\braket{\Phi_\text{ss} | \hat O_2^2 | \Phi_\text{ct}} + \braket{\Phi_\text{ct} | \hat O_2^2 | \Phi_\text{s}}\right) - \left\{\left(1 - \varepsilon_\text{ss}\right) \braket{\Phi_\text{ss} | \hat O_2 | \Phi_\text{ss}} \right.\right.\nonumber\\
			   &\left.\left. + \left(1 - \varepsilon_\text{ct}\right) \braket{\Phi_\text{ct} | \hat O_2 | \Phi_\text{ct}} + \sqrt{1 - \varepsilon_\text{ss}}\sqrt{1 - \varepsilon_\text{ct}}\left(\braket{\Phi_\text{ss} | \hat O_2 | \Phi_\text{ct}} + \braket{\Phi_\text{ct} | \hat O_2 | \Phi_\text{s}}\right)\right\}^2 \right]~.
\end{eqnarray}
Further, the retarded Greens function defined above in eq.~\eqref{retarded} can very generally be related to the QFI.
By using the simplified forms \(\frac{i}{2}G^R_{O_2,O_2}(t \to \infty) = \braket{\left(O_2\right)^2}\) and \(\frac{i}{2}G^R_{1,O_2}(t \to \infty) = \braket{O_2}\), we get
\begin{eqnarray}
	F_Q(\psi,\hat O_2) = 2iG^R_{O_2, O_2}(t\to \infty) + \left(G^R_{1, O_2}(t\to \infty)\right)^2~.
\end{eqnarray}

For the sake of completeness, we generalise the above relations to finite temperatures by adopting the approach of Hauke et al.~\cite{Hauke2016}. For a mixed state at inverse temperature \(\beta\) characterised by the density matrix \(\rho = \sum_n f_n \ket{n}\bra{n}\) where \(\left(E_n,\ket{n}\right)\) are the eigenvalue-eigenstate pairs and \(f_n = e^{-\beta E_n}/\sum_n e^{-\beta E_n}\) are the Boltzmann weights, the FQI can be assigned a spectral representation of the form~\cite{Hauke2016}
\begin{eqnarray}
	F_Q(\rho, \hat O) = 2\sum_{m,n} \frac{\left(f_m - f_n\right)^2}{f_m + f_n}|\braket{m|\hat O|n}|^2~.
\end{eqnarray}
By using the identity \(\frac{f_m - f_n}{f_m + f_n} = \int_{-\infty}^\infty d\omega \delta(\omega + E_m - E_n)\tanh \left( \beta \omega/2 \right) \), the finite temperature FQI can be written in the form
\begin{eqnarray}\label{fqi-spectral}
	F_Q(\rho, \hat O) = 2\int_{-\infty}^\infty d\omega \tanh \left( \beta \omega/2 \right) \sum_{m,n} \left(f_m - f_n\right) \delta(\omega + E_m - E_n)|\braket{m|\hat O|n}|^2~.
\end{eqnarray}
In order to relate this with a correlation function, we now point out that the lesser Greens function \(G^<\) defined in eq.~\eqref{lesser-gf}, as well as the greater Greens function \(G^>\) that we will now define, also admit spectral representations in terms of the eigenstates \(\ket{n}\) of the full problem:
\begin{eqnarray}\label{lesser-spectral}
	G^<_{O^\dagger, O}(t) \equiv i\braket{\hat O^\dagger \hat O(t)} \implies & G^<_{O^\dagger, O}(\omega) = 2\pi i \sum_{m,n}f_n \delta(\omega + E_m - E_n) |\braket{m|\hat O^\dagger|n}|^2~,\\
	G^>_{O^\dagger, O}(t) \equiv -i\braket{\hat O(t) \hat O^\dagger} \implies & G^>_{O^\dagger, O}(\omega) = -2\pi i \sum_{m,n}f_m \delta(\omega + E_m - E_n) |\braket{m|\hat O|n}|^2~.
\end{eqnarray}
If \(\hat O\) is a Hermitian operator, eqs.~\eqref{lesser-spectral} can be used to rewrite the QFI in terms of \(G^\lessgtr_{O^\dagger, O}(\omega)\). Indeed, by combining eqs.~\eqref{lesser-spectral} and eq.~\eqref{fqi-spectral} with the assumption \(\hat O = \hat O^\dagger\), we get
\begin{eqnarray}
	F_Q(\rho, \hat O) = -\frac{1}{\pi i}\int_{-\infty}^\infty d\omega \tanh \left( \beta \omega/2 \right)\left(G^>_{O, O}(\omega) + G^<_{O, O}(\omega)\right)~. 
\end{eqnarray}
This constitutes the general relation between the finite temperature QFI for a general Hermitian many-particle operator \(\hat O\) and the corresponding many-particle correlations at zero and finite frequencies.

The Kubo linear response function associated with the operator \(\hat O(t)\) due to a perturbation from the same operator is associated with a correlation function/susceptibility \(\mathcal{C}_{\hat O}(t,t^\prime) = i\theta(t - t^\prime)\left< \left[\hat O(t), \hat O(t^\prime)\right] \right>\). By going to the frequency domain, the imaginary part of such a function can be related to the sum of the lesser and greater Greens functions:
\begin{eqnarray}
	\hspace*{2.2cm} \mathcal{C}_{\hat O}(\omega) &= i\int_0^\infty d(t-t^\prime) e^{i\omega(t-t^\prime)}\left( \left<\hat O(t) \hat O(t^\prime) \right> - \left<\hat O(t^\prime) \hat O(t) \right>\right)\nonumber \\
	\implies \mathcal{C}_{\hat O}(\omega) - \mathcal{C}_{\hat O}(\omega)^* &= i\int_0^\infty d(t-t^\prime)\left( e^{i\omega(t-t^\prime)}-e^{-i\omega(t-t^\prime)} \right)\left( \left<\hat O(t) \hat O(t^\prime) \right> - \left<\hat O(t^\prime) \hat O(t) \right>\right)\nonumber\\
						&= i\int_{-\infty}^{\infty}d(t-t^\prime) e^{i\omega(t-t^\prime)} \left( \left<\hat O(t) \hat O(t^\prime) \right> - \left<\hat O(t^\prime) \hat O(t) \right>\right)\nonumber \\
						&= -\left( G^>_{O,O}(\omega) + G^<_{O,O}(\omega) \right) 
\end{eqnarray}
This allows us to connect the QFI with the imaginary part of the Kubo correlation function/susceptibility obtained in Ref.\cite{Hauke2016}:
\begin{eqnarray}
F_Q(\rho, \hat O) = \frac{1}{\pi i}\int_{-\infty}^\infty d\omega \tanh \left( \beta \omega/2 \right)\left(\mathcal{C}_{\hat O}(\omega) - \mathcal{C}_{\hat O}(\omega)^*\right) ~.
\end{eqnarray}

\section{One-shot decoupling of impurity from zeroth site}
\label{app-imp-remove}
We will decouple the impurity states from the fixed point Hamiltonian using a single URG transformation. This will of course generate correlations on the zeroth site. The resulting Hamiltonian will be an AIM with the hopping between the zeroth site and the rest of the chain acting as the effective hybridisation. Schematically, we will have
\begin{eqnarray}
	H^* = H_\text{imp} + V_\text{imp-0} + H_\text{0} + H_\text{0-1} + H_\text{rest} ~ {\xrightarrow{\text{decouple imp.}}} & E_\text{imp} + \tilde H_\text{0} + H_\text{0-1} + H_\text{rest} \nonumber\\
															       &= H_\text{new imp} + V_\text{new imp - rest} + H_\text{rest}~.
\end{eqnarray}

Since the impurity is not coupled with any site beyond the zeroth site, the parts of the Hamiltonian that involve "rest" will not change in the process. This also means that the decoupling can be performed by looking at the smaller Hamiltonian
\begin{eqnarray}
	H_\text{imp+0} = H_\text{imp} + H_\text{0} + V_\text{imp-0} =& -\frac{U^*}{2}\left( \hat n_{d \uparrow} - \hat n_{d \downarrow} \right)^2 - U_b \left( \hat n_{0 \uparrow} - \hat n_{0 \downarrow} \right)^2 + J^*S_d^z S_0^z\nonumber\\
								    & + {V^*}\sum_\sigma \left(c^\dagger_{d\sigma}c_{0\sigma} + \text{h.c.}\right) + \frac{J^*}{2}\left( c^\dagger_{d \uparrow}c_{d \downarrow} c^\dagger_{0 \downarrow} c_{0 \uparrow} + \text{h.c.} \right)~.
\end{eqnarray}

\subsection{Renormalisation from \({V^*}\)}
From the off-diagonal term involving \({V^*}\), we generate the following term, in the particle sector for \(0\):
\begin{eqnarray}
\sum_\sigma c^\dagger_{0\sigma}c_{d\sigma} \frac{{V^*}^2}{\tilde \omega - H_d} c^\dagger_{d\sigma}c_{0\sigma} = \sum_\sigma c^\dagger_{0\sigma}c_{d\sigma} \frac{{V^*}^2}{\tilde \omega + \frac{U^*}{2}\left( 1 - \hat n_{d\bar\sigma} \right)^2 + U_b \hat n_{0\bar\sigma} + \frac{J^*}{4}\left( 1 - \hat n_{d\bar\sigma} \right) \hat n_{0\bar\sigma} } c^\dagger_{d\sigma}c_{0\sigma}~.
\end{eqnarray}
where \(\tilde \omega\) is the quantum fluctuation operator for the impurity site and \(H_d = -\frac{U^*}{2}\left( \hat n_{d \uparrow} - \hat n_{d \downarrow} \right)^2 - U_b \left( \hat n_{0 \uparrow} - \hat n_{0 \downarrow} \right)^2 + J^*S_d^z S_0^z \), is the diagonal part of the Hamiltonian for the imp+0 system. In order to resolve the operators in the denominator, we expand the unity in the numerator using the identity \(1 = \hat n_{d\bar\sigma}\hat n_{0\bar\sigma} + \hat n_{d\bar\sigma}\hat h_{0\bar\sigma} + \hat h_{d\bar\sigma}\hat n_{0\bar\sigma} + \hat h_{d\bar\sigma}\hat h_{0\bar\sigma}\), where \(\hat h = 1 - \hat n\) is hole operator. On Substituting this, we get
\begin{eqnarray}
	\sum_\sigma  c^\dagger_{0\sigma}c_{d\sigma} \frac{{V^*}^2}{\tilde \omega - H_d} c^\dagger_{d\sigma}c_{0\sigma} &= {V^*}^2 \sum_\sigma c^\dagger_{0\sigma}c_{d\sigma} \frac{\hat h_{d\bar\sigma}\hat h_{0\bar\sigma} + \hat h_{d\bar\sigma}\hat n_{0\bar\sigma} + \hat n_{d\bar\sigma}\hat h_{0\bar\sigma} + \hat n_{d\bar\sigma}\hat n_{0\bar\sigma}}{\tilde \omega + \frac{U^*}{2}\hat h_{d\bar\sigma} + U_b \hat n_{0\bar\sigma} + \frac{J^*}{4}\hat h_{d\bar\sigma} \hat n_{0\bar\sigma} } c^\dagger_{d\sigma}c_{0\sigma}\nonumber\\
														       &= {V^*}^2 \sum_\sigma \hat h_{d\sigma} \hat n_{0\sigma}\left[\frac{\hat h_{d\bar\sigma}\hat h_{0\bar\sigma}}{\tilde\omega_{00} + \frac{U^*}{2}} + \frac{\hat h_{d\bar\sigma}\hat n_{0\bar\sigma}}{\tilde\omega_{01} + \frac{U^*}{2} + U_b + \frac{J^*}{4}} + \frac{\hat n_{d\bar\sigma}\hat h_{0\bar\sigma}}{\tilde\omega_{10}} + \frac{\hat n_{d\bar\sigma}\hat n_{0\bar\sigma}}{\tilde\omega_{11} + U_b}\right]~.\qquad
\end{eqnarray}
\(\tilde\omega_{(0,1),(0,1)}\) represents the quantum fluctuation scale corresponding to the configuration in the numerator.

In order to "freeze" the impurity dynamics, we will substitute \(\hat n_{d\sigma} = \hat n_{d\bar\sigma} = \frac{1}{2}\), because of the \({Z}_2\) symmetry and the particle-hole symmetry of the impurity levels. This gives
\begin{eqnarray}
	\frac{1}{4}{V^*}^2 \sum_\sigma \hat n_{0\sigma} \left[\frac{\hat h_{0\bar\sigma}}{\tilde\omega_{00} + \frac{U^*}{2}} + \frac{\hat n_{0\bar\sigma}}{\tilde\omega_{01} + \frac{U^*}{2} + U_b + \frac{J^*}{4}} + \frac{\hat h_{0\bar\sigma}}{\tilde\omega_{10}} + \frac{\hat n_{0\bar\sigma}}{\tilde\omega_{11} + U_b}\right]~.
\end{eqnarray}

The state that most closely represents the metallic ground state is \(\bar \omega_{10}\), we take that as the reference quantum fluctuation scale \(\bar\omega\). It is of the order of \(\bar\omega \sim -\frac{J^*}{4} - \frac{U^*}{2} - U_b\). We will now relate the other scales to \(\bar\omega\) by expressing them in terms of the energy of the initial state:
\begin{eqnarray}
	\tilde\omega_{00} \sim -U_b = \bar\omega + \frac{J^*}{4} + \frac{U^*}{2}, ~\tilde\omega_{01} \sim 0 = \bar\omega + \frac{J^*}{4} + \frac{U^*}{2} + U_b, ~\tilde\omega_{11} \sim -\frac{U^*}{2} = \bar\omega + \frac{J^*}{4} + U_b~.
\end{eqnarray}
Substituting these gives
\begin{eqnarray}
	\frac{1}{4}{V^*}^2 \sum_\sigma \hat n_{0\sigma} \left[\hat h_{0\bar\sigma}\alpha_1 + \hat n_{0\bar\sigma}\alpha_2\right]~,
\end{eqnarray}
where \(\alpha_1 = \left(\bar\omega + U^* + \frac{J^*}{4}\right)^{-1} + \left(\bar\omega\right)^{-1}\) and \(\alpha_2 = \left(\bar\omega + U^* + 2U_b + \frac{J^*}{2}\right)^{-1} + \left(\bar\omega + 2U_b + \frac{J^*}{4}\right)^{-1}\).

Because of the particle-hole symmetry on the impurity as well as in the bath, the renormalisation from the hole sector is obtained simply by transforming \(\hat n \leftrightarrow \hat h\):
\begin{eqnarray}
	\frac{1}{4}{V^*}^2 \sum_\sigma \hat h_{0\sigma} \left[\hat n_{0\bar\sigma}\alpha_1 + \hat h_{0\bar\sigma}\alpha_2\right]~.
\end{eqnarray}
The total renormalisation arising from \(V\) is therefore 
\begin{eqnarray}
	\frac{1}{4}{V^*}^2 \sum_\sigma \left[\alpha_1 \left( \hat n_{0\sigma}\hat h_{0\bar\sigma} + \hat h_{0\sigma}\hat n_{0\bar\sigma}\right) + \alpha_2 \left( \hat n_{0\sigma}\hat n_{0\bar\sigma} + \hat h_{0\sigma}\hat h_{0\bar\sigma}\right)\right] = \frac{1}{2}{V^*}^2 \left(\alpha_1 - \alpha_2\right) \left(\hat n_{0 \uparrow} - \hat n_{0 \downarrow}\right)^2 + \text{constant}~.\qquad
\end{eqnarray}

\subsection{Renormalisation from \(J^*\)}
The renormalisation arising from decoupling the Kondo coupling has two terms. The first term arises when the spin of the zeroth site is initially up:
\begin{eqnarray}
	\frac{{J^*}^2}{4}c^\dagger_{d \downarrow}c_{d \uparrow} c^\dagger_{0 \uparrow}c_{0 \downarrow} \frac{1}{\tilde \omega - H_d} c^\dagger_{0 \downarrow}c_{0 \uparrow} c^\dagger_{d \uparrow}c_{d \downarrow} &= c^\dagger_{d \downarrow}c_{d \uparrow} c^\dagger_{0 \uparrow}c_{0 \downarrow} \frac{{J^*}^2/4}{\tilde \omega + \frac{U^*}{2} + U_b + \frac{J^*}{4}} c^\dagger_{0 \downarrow}c_{0 \uparrow} c^\dagger_{d \uparrow}c_{d \downarrow} \nonumber\\
																										   &= \frac{{J^*}^2}{4}\frac{\hat n_{d \downarrow} \hat h_{d \uparrow} \hat n_{ 0 \uparrow} \hat h_{0 \downarrow}}{\tilde \omega + \frac{U^*}{2} + U_b + \frac{J^*}{4}} ~.
\end{eqnarray}
The quantum fluctuation scale \(\tilde \omega\) for this process can be similarly related to \(\bar\omega\): \(\tilde\omega \sim -\frac{U^*}{2} - U_b - \frac{J^*}{4} = \bar\omega\). This gives
\begin{eqnarray}
	\frac{{J^*}^2}{4}\frac{\hat n_{d \downarrow} \hat h_{d \uparrow} \hat n_{ 0 \uparrow} \hat h_{0 \downarrow}}{\bar\omega + \frac{U^*}{2} + U_b + \frac{3J^*}{4}} = \frac{{J^*}^2}{16}\frac{\hat n_{ 0 \uparrow} \hat h_{0 \downarrow}}{\bar\omega + \frac{U^*}{2} + U_b + \frac{J^*}{4}}~,
\end{eqnarray}
as the renormalisation for this configuration. At the final step, we substituted \(\hat n_{d\sigma} = \hat h_{d\sigma} = \frac{1}{2}\). For the other configuration where the spin of the zeroth site is down, we get
\begin{eqnarray}
	\frac{{J^*}^2}{16}\frac{\hat n_{ 0 \downarrow} \hat h_{0 \uparrow}}{\bar\omega + \frac{U^*}{2} + U_b + \frac{J^*}{4}}~.
\end{eqnarray}
Adding both sectors, we get
\begin{eqnarray}
	\frac{{J^*}^2}{16}\frac{1}{\bar\omega + \frac{U^*}{2} + U_b + \frac{J^*}{4}} \left(\hat n_{ 0\uparrow} - \hat n_{0 \downarrow}\right)^2~.
\end{eqnarray}

\subsection{Total renormalisation}
Adding the contributions from both \(V^*\) and \(J^*\),  the net renormalisation is the generation of a local correlation term \(-U^\prime\left(\hat n_{ 0\uparrow} - \hat n_{0 \downarrow}\right)^2\) on the zeroth site, where \(U^\prime\) is given by
\begin{eqnarray}
	U^\prime = \frac{-{J^*}^2/16}{\bar\omega + \frac{U^*}{2} + U_b + \frac{J^*}{4}} - \frac{1}{2}{V^*}^2\left(\frac{1}{\bar\omega + U^* + \frac{J^*}{4}} + \frac{1}{\bar\omega} - \frac{1}{\bar\omega + U^* + 2U_b + \frac{J^*}{2}} - \frac{1}{\bar\omega + 2U_b + \frac{J^*}{4}}\right)~.
\end{eqnarray}
The effective Hamiltonian for the zeroth site is therefore
\begin{eqnarray}
	H_\text{0+rest} = \underbrace{-\left(U^\prime + U_b\right)\left(\hat n_{0\uparrow} - \hat n_{0\downarrow}\right)^2}_\text{new correlated impurity} \underbrace{- t\sum_{j\in \text{n.n. of 0},\atop{\sigma}}\left(c^\dagger_{0\sigma}c_{j\sigma} + \text{h.c.}\right)}_\text{hopping between new impurity \& new bath} \underbrace{-t \sum_{\left<i,j \right>}\left(c^\dagger_{i\sigma}c_{j\sigma} + \text{h.c.}\right)}_\text{K.E. of new bath}~.
\end{eqnarray}

\section{Effective Hamiltonian for the excitations within the Hubbard sidebands}

Beyond \(r_{c1}\), the Hubbard sidebands can be seen to separate from the central Kondo resonance in the impurity spectral function. These sidebands are broad, indicating that the impurity site is hybridising with the bath from within the sidebands, at high energies \(\omega \sim U/2\). These processes lead to the presence of gapless excitations within the rest of the bath, outside the Mott gap. To find the effective Hamiltonian for the rest of the bath (sites 1 and onwards), we will take the eigenstates of the two-site Hamiltonian \(H\)that reside within the sideband as our ground-states, and treat the hopping from the zeroth site to the rest of the bath as a perturbation. The two-site Hamiltonian is
\begin{eqnarray}
	H_\text{two site} = V\sum_\sigma\left(c^\dagger_{d\sigma} c_{0\sigma} + \text{h.c}\right) + J \vec{S}_d\cdot\vec{S}_0 - \frac{U}{2} \left( \hat n_{d \uparrow} - \hat n_{d\downarrow} \right)^2 + \frac{U}{2}~.
\end{eqnarray}
In order to align the doubly-occupied impurity level at the more conventional value of \(U/2\), we have shifted the Hamiltonian by the same constant value. The effect of \(U_b\) has been ignored because it is small compared to \(U\) at the frequencies pertinent to the physics of the sidebands. 
The two-site states that reside within the Hubbard sidebands are
\begin{eqnarray}\label{gs-sideband}
	\ket{\Psi_{cc}} \equiv \ket{c}_d\ket{c}_0; ~ ~ c \in \left\{0,2\right\}; ~ ~\ket{\Psi_\text{CS}} \equiv \frac{1}{\sqrt 2}\left(\ket{2}_d\ket{0}_0 - \ket{0}_d\ket{2}_0\right);\nonumber \\
	\ket{\Psi_{\sigma,c}} = \alpha\ket{\sigma}_d\ket{c}_0 + \sqrt{1 - \alpha^2}\ket{c}_d\ket{\sigma}_0; ~ \sigma \in \left\{\uparrow,\downarrow\right\}; \nonumber\\
	\ket{\Psi_\text{CT}} = \beta\frac{1}{\sqrt 2}\left(\ket{\uparrow}_d\ket{\downarrow}_0 - \ket{\downarrow}_d\ket{\uparrow}_0\right) + \sqrt{1 - \beta^2}\frac{1}{\sqrt 2}\left(\ket{2}_d\ket{0}_0 + \ket{0}_d\ket{2}_0\right),
\end{eqnarray}
where the coefficients of the eigenstates are given by
\begin{eqnarray}
	\alpha \simeq \frac{V}{\sqrt{V^2 + \left(E_\text{X1} + \frac{U}{2}\right)^2 }},  ~ \beta = \frac{2V}{\sqrt{4V^2 + \left(E_\text{CS} + \frac{U}{2} + \frac{3J}{4}\right)^2}}~.
\end{eqnarray}
The energies of the states are
\begin{eqnarray}
	E_\text{gs} \equiv E_{00} = E_{22} = E_{CS} = \frac{U}{2}, E_\text{X1} \equiv E_{\sigma,c} = \frac{U}{4} + \sqrt{V^2 + \frac{U^2}{16}}~, \nonumber\\
	E_\text{X2} \equiv E_\text{CS} = \frac{U}{4} - \frac{3J}{8} + \sqrt{4V^2 + \left(\frac{U}{4} + \frac{3J}{8}\right)^2}~.
\end{eqnarray}
The ground-state (comprising the states shown in eq.~\eqref{gs-sideband}) is triply degenerate. In order to study the dynamics of the bath, we now introduce the perturbation \(H_I = -t\sum_\sigma \left( c^\dagger_{0\sigma}c_{1\sigma} + \text{h.c.}\right) \). We also take an expanded set of eigenstates \(\left\{\ket{\Psi_n}\otimes\ket{a}\right\}\), where \(\left\{\ket{\Psi_n}\right\}\) is the set of eigenstates of \(H_\text{two site}\), and \(\ket{a} \in \left\{ \ket{0}, \ket{\uparrow}, \ket{\downarrow}, \ket{2} \right\} \) represents the configuration of the first site. The eigenstates of \(H_\text{two site}\) are also eigenstates the total number operator \(\hat n_d + \hat n_0 + \hat n_1\), and since \(H_I\) does not conserve the total number, all odd order shifts are zero: \(\braket{\Psi_n | H_I^b | \Psi_n}, b=1,3,\ldots\), is zero for all the eigenstates. 

For the second order shift, we need to calculate the overlaps \(\braket{\Psi_m | H_I | \Psi_n}, m\neq n\) for \(n\) in the ground-state. Since all three ground-states involve the impurity and zeroth sites in charge configuration (0 or 2), \(H_I\ket{\Psi}_n\) will result in states of the form \(\ket{0(2)}_d\ket{\sigma}_0\ket{a}_1\). This resulting state has non-zero overlap only with the part \(\sqrt{1 - \alpha^2}\ket{c}_d\ket{\sigma}_0\ket{a}\) of the expanded state \(\ket{\Psi_{\sigma,c}}\ket{a}\). The square of each such overlap is, therefore, \(t^2\left( 1 - \alpha^2 \right) \). 
\begin{itemize}
	\item Excitations from the ground-states \(\ket{\Psi_{00}}\ket{0}\) and \(\ket{\Psi_{22}}\ket{2}\) have zero overlap with any other state, such that the second order shifts in these states is zero.
	\item Excitations from the ground-states \(\ket{\Psi_{00}}\ket{\sigma}\) and \(\ket{\Psi_{22}}\ket{\sigma}\) have non zero overlap with one excited state each, such that the second order shift in these states are \(\gamma^2 \equiv t^2\left( 1 - \alpha^2 \right) / \left(E_\text{X1} - E_\text{gs}\right) \).
	\item Excitations from the remaining ground-states have non-zero overlaps with two excited states each, such that the second order shift in these states are \(2\gamma^2\).
\end{itemize}
Upon dropping the zeroth order part (because it has no dynamics for the rest of the bath), the effective Hamiltonian at second order in \(t\) takes the form
\begin{eqnarray}
	H_\text{eff}^{(2)} = \gamma^2 &\left[\left(\ket{\Psi_{00}}\bra{\Psi_{00}} + \ket{\Psi_{22}}\bra{\Psi_{22}} + 2\ket{\Psi_\text{CS}}\bra{\Psi_\text{CS}}\right)\left(\sum_\sigma \ket{\sigma}_1\bra{\sigma}_1\right) \right.\nonumber \\
				      &\left.  + 2\left(\ket{\Psi_{22}}\bra{\Psi_{22}} + \ket{\Psi_\text{CS}}\bra{\Psi_\text{CS}}\right) \ket{0}_1\bra{0}_1+2\left(\ket{\Psi_{00}}\bra{\Psi_{00}} + \ket{\Psi_\text{CS}}\bra{\Psi_\text{CS}}\right)\ket{2}_1\bra{2}_1\right]~.
\end{eqnarray}
By using the completeness relation \(1 = \sum_\sigma \ket{\sigma}_1\bra{\sigma}_1 + \ket{0}_1\bra{0}_1 + \ket{2}_1\bra{2}_1\) of the Hilbert space of the first site, this evaluates to
\begin{eqnarray}
	H_\text{eff}^{(2)} = \text{constant} + \gamma^2 \left(\ket{\Psi_{22}}\bra{\Psi_{22}} - \ket{\Psi_{00}}\bra{\Psi_{00}} \right) \left(\ket{2}_1\bra{2}_1 - \ket{0}_1\bra{0}_1\right) ~,
\end{eqnarray}
where the "constant" term refers to operators that do not involve depend on the dynamics of the rest of the bath. By defining the charge isospin operator \(C^z\ket{2(0)} = +(-)\frac{1}{2}\ket{2(0)}\), the effective Hamiltonian can be written as (dropping the non-dynamical part)
\begin{eqnarray}
	H_\text{eff}^{(2)} = 4\gamma^2C_\text{tot}^z  C_1^z~,
\end{eqnarray}
where \(C_i^z\) is the charge isospin for site \(i\) and \(C_\text{tot}^z \equiv C_d^z + C_0^z\) is the z-component of the total charge isospin for the impurity and the zeroth sites.

The fourth order shift in the energy of the eigenstate \(\ket{\Psi_n}\) will be given by
\begin{eqnarray}
	E_{n}^{(4)} = \sum_{j,k,l} \frac{\left(H_I\right)_{nl}\left(H_I\right)_{lk}\left(H_I\right)_{kj}\left(H_I\right)_{jn}}{\left( E_l - E_n \right)\left( E_k - E_n \right)\left( E_j - E_n \right) } - E_n^{(2)} \left[\frac{\left(H_I\right)_{kn}}{E_k - E_n}\right]^2~,
\end{eqnarray}
where the matrix elements \(\left( H_I \right)_{in} \) are defined as \(\braket{i | H_I | n}\). This fourth order shift vanishes for almost all the ground-states. For the remaining states, it becomes important to treat their  degeneracy. The "appropriate" combinations of states that provide non-zero fourth order contribution are
\begin{eqnarray}
	\ket{+} = \sqrt{\frac{2}{3}}\ket{\Psi_{22}}\ket{0}_1 + \sqrt{\frac{1}{3}}\ket{\Psi_\text{CS}}\ket{2}_1, ~ ~\ket{-} = -\sqrt{\frac{2}{3}}\ket{\Psi_{00}}\ket{2}_1 + \sqrt{\frac{1}{3}}\ket{\Psi_\text{CS}}\ket{0}_1~.
\end{eqnarray}
Corresponding to these ground-states, the fourth-order shift in the eigenvalue is
\begin{eqnarray}
	E^{(4)} = \frac{3\gamma^4 \alpha^2 \beta^2}{\left( 1 - \alpha^2 \right) \left( E_\text{X2} - E_\text{gs} \right) }~,
\end{eqnarray}
such that the fourth order contribution to the effective Hamiltonian is
\begin{eqnarray}
	H_\text{eff}^{(4)} &= E^{(4)}\left(\ket{+}\bra{+} + \ket{-}\bra{-}\right) \nonumber\\
			   &= \frac{\gamma^4 \alpha^2 \beta^2}{\left( 1 - \alpha^2 \right) \left( E_\text{X2} - E_\text{gs} \right) } \left[- C_\text{tot}^z C_\text{tot}^2 C_1^z +\sqrt{2}\mathcal{P}_\text{tot}^{4}\left(C_0^+ - C_d^+\right)C_1^- + \text{h.c.}\right] ~.
\end{eqnarray}
The operator \(\mathcal{P}_\text{tot}^{4}\) acts on states belonging to the combined \(d+0\) Hilbert space and projects on to the \(\hat n_d + \hat n_0 = 4\) subspace.

\section{Behaviour of the Kondo scale $T_K$ close to the transition}
\label{app-Tk}
Towards obtaining the emergent Kondo temperature scale $T_K$ close to the transition, we follow the approach used by Moeller et al. 1995~\cite{moeller_1995} and Held et al. 2013~\cite{held_2013}. Near the transition, they obtained a Kondo model from the full AIM by applying a Schrieffer-Wolff transition that removes the charge fluctuations of the impurity site and retains the physics of only the s-d coupling \(J\). This amounts to removing the side-peaks from the impurity spectral function and focusing on the low-energy central peak. Held et al. then integrated the RG equation for this Kondo model by using a Lorentzian DOS in the bath. This is motivated by the fact that the central peak of the impurity spectral function is a Lorentzian, and the bath becomes equivalent to the impurity site under self-consistency. We implement the same approach but on our extended impurity model, such that the transformation then leads to a \(J-U_b\) model. The relevant RG equations are mentioned in Section IV of the main manuscript:
\begin{eqnarray}
	\Delta J = -\frac{n_j J\left(J + 4U_b\right)}{\omega - D/2 + U_b/2 + J/4},~~ \Delta U_b = 0~,
\end{eqnarray}
where \(n_j = \rho(D)|\Delta D|\) is proportional to the bath density of states. In Section VII of the main manuscript, we show that the spectral function of the zeroth site is tracked by that on the impurity site through the coherent spin-flip fluctuations, and this shows that the appropriate density of states for the bath in this regime is of the Lorentzian kind (same as the shape of the central peak of the impurity spectral function). We therefore set
\begin{eqnarray}
	\rho(D) = \frac{\rho_0 \Gamma^2}{D^2 + \Gamma^2}~,
\end{eqnarray}
where \(\Gamma\) represents the half-width of the Lorentzian. A similar line of arguments was presented by Held et. al.~\cite{held_2013}. Close to the transition, the RG flow of the Kondo coupling \(J\) is stunted by the competing pairing term, and the RG equation can be simplified into the Poor man's scaling \(1-\)loop form:
\begin{eqnarray}
	\frac{d J}{d D} \simeq -2\rho_0 \Gamma^2\frac{J\left(J + 4U_b\right)}{\left( \Gamma^2 + D^2 \right) D}~.
\end{eqnarray}
We now integrate this RG equation over the range \(D \in \left[D_0, D^*\right] \), where \(D_0\) is the bare bandwidth, and \(D^*\) is the fixed-point bandwidth. \(D^*\) represents the energy scale for the conduction electrons in the IR.
\begin{eqnarray}
	\int_{J_0}^{J^*} \frac{d J}{J\left(J + 4U_b\right)} = -\int_{D_0}^{D^*}\frac{2\rho_0 \Gamma^2 d D}{\left( \Gamma^2 + D^2 \right) D} \nonumber\\
	\implies \frac{1}{4 U_b \rho_0} \left(\ln \frac{J^*}{J_0} - \ln \frac{J^* + 4U_b}{J_0 + 4 U_b}\right) = \ln \frac{D^* + \Gamma^2}{D_0 + \Gamma^2} - \ln \frac{D^*}{D_0}~.
\end{eqnarray}
We will now take the limit of \(J_0 + 4U_b \to 0^-\) and {\it equate the singular terms on both sides} of the equation:
\begin{eqnarray}
	\ln \frac{D^*}{D_0} = -\frac{1}{4 U_b \rho_0} \ln \frac{J_0 + 4 U_b}{J_0} = -\frac{1}{4 U_b \rho_0} \ln \left(r_{c2} - r\right) + \text{non-singular part}~.
\end{eqnarray}
We have replaced the couplings \(J\) and \(U_b\) with the dimensionless parameters \(r = -U_b/J_0\) and \(r_{c2} = 1/4\). Dropping the non-singular part and solving for \(D^*\) then gives the IR energy scale:
\begin{eqnarray}
	D^* = D_0 \exp\left[-\frac{\ln\left( r_{c2} - r \right)}{4 U_b \rho_0}\right]~.
\end{eqnarray}
The dynamically generated Kondo temperature scale is then obtained as
\begin{eqnarray}
	T_K = \frac{D^*}{k_B} = \frac{D_0}{k_B} \exp\left[-\frac{\ln\left( r_{c2} - r \right)}{4 U_b \rho_0}\right]~.
\end{eqnarray}
Note that the prefactor of the logarithm is positive because \(U_b\) is negative: \(-4U_b\rho_0 = |4U_b \rho_0|\). As we approach the transition, the parameter \(r\) takes the limit \(r \to r_{c2}^-\), and the Kondo temperature scale vanishes:
\begin{eqnarray}
	\lim_{r \to r_{c2}^-} T_K = \frac{D_0}{k_B} \lim_{r \to r_{c2}^-}\left(r_{c2} - r\right)^{4|U_b\rho_0|} \to 0~.
\end{eqnarray}

\section{Effective Hamiltonian for the low-lying excitations at the critical point}
\label{app-nfl}

In order to obtain an effective Hamiltonian for the low-lying excitations in the bath, we will introduce a hopping term \(V\) between the zeroth site and the first site into the two-site Hamiltonian \(H_0\), and treat that term perturbatively. 
\begin{eqnarray}
	H = \underbrace{J \vec{S}_d\cdot\vec{S}_0 - U_b \left( \hat n_{0 \uparrow} - \hat n_{0 \downarrow} \right)^2 }_{H_0} \underbrace{- t\sum_\sigma\left( c^\dagger_{0\sigma}c_{1\sigma} + \text{h.c.}\right)}_V~.
\end{eqnarray}
As mentioned before, \(H_0\) is our zeroth Hamiltonian, while \(V\) is the perturbation. At the critical point, the zeroth Hamiltonian has a \(20-\)fold degenerate ground-state manifold, and we will show later that it is this degeneracy which leads to NFL excitations. The ground-states are
\begin{eqnarray}
	\left\{\ket{S=S^z=0}, \ket{\uparrow, 0}, \ket{\uparrow, 2}, \ket{\downarrow, 0}, \ket{\downarrow, 2}\right\}\otimes\left\{\ket{0},\ket{\uparrow},\ket{\downarrow},\ket{\uparrow \downarrow}\right\}~.
\end{eqnarray}
The ket to the left of \(\otimes\) describes the configuration of the impurity and zeroth sites; \(S=S^z=0\) represents the singlet, while the remaining four states \(\ket{\sigma, 0},\ket{\sigma,2}\) are local moment states. The ket to the right of \(\otimes\) represents the configuration of the first site. Eight of these states have zero matrix elements within the ground-state subspace, even in the presence of the perturbation, so we drop these states. Since the full Hamiltonian conserves the total number of particles \(N_\text{tot} = \sum_\sigma\left(\hat n_{d\sigma} + \hat n_{0\sigma} + \hat n_{1\sigma}\right)\) and the total spin \(S_\text{tot}^z = S_d^z + S_0^z + S_1^z\), the remaining 12 states can be split into decoupled blocks based on the values of these two operators:
\begin{eqnarray}
	N_\text{tot} = c+2, S_\text{tot}^z = 0&: \ket{S=S^z=0}\ket{c}, \ket{\uparrow, c, \downarrow}, \ket{\downarrow, c, \uparrow}; ~c=0,2, \nonumber\\
	N_\text{tot} = 3, S_\text{tot}^z = \pm\frac{1}{2}&: \ket{S=S^z=0}\ket{\sigma}, \ket{\sigma, 0, 2}, \ket{\sigma, 2, 0};\sigma=\uparrow,\downarrow~.
\end{eqnarray}
The perturbation \(V\) will lift the degeneracy in each block; to calculate the modified spectrum, we diagonalise each block in the presence of \(V\). We will first write down the action of the perturbation on the singlet ground-states: this will make it easier to identify the true eigenstates.
\begin{eqnarray}
	N_\text{tot} = c+2, S_\text{tot}^z = 0 &: V\ket{\text{SS}_c} = \frac{t}{\sqrt 2}\left( \ket{\downarrow,c,\uparrow} - \ket{\uparrow,c,\downarrow} \right),~ c=0,2,\nonumber\\
	N_\text{tot} = 3, S_\text{tot}^z = \frac{\sigma}{2} &: V\ket{\text{SS}_\sigma} = \frac{t}{\sqrt 2}\ket{\sigma}\otimes\left(\ket{0,2} + \ket{2,0}\right),~\sigma=\uparrow(1),\downarrow(-1)~,
\end{eqnarray}
where we have introduced the notation \(\ket{\text{SS}_{\alpha}} = \ket{S=S^z=0}\otimes\ket{\alpha},\alpha=0,\uparrow,\downarrow,2\). This shows that the action of the perturbation on the singlet state is to create linear combinations of the local moment states in the respective blocks. We therefore define the two possible linear combinations \(\ket{\text{LM}_\pm^{N_\text{tot},S_\text{tot}^z}}\) within each block:
\begin{eqnarray}
	N_\text{tot} = c+2, S_\text{tot}^z = 0 &: \ket{\text{LM}_\pm^{c+2,0}} = \frac{1}{\sqrt 2}\left(\ket{\uparrow, c, \downarrow} \pm \ket{\downarrow, c, \uparrow}\right),~c=0,2,\nonumber\\
	N_\text{tot} = 3, S_\text{tot}^z = \frac{\sigma}{2} &: \ket{\text{LM}_{\sigma}^{3,\frac{1}{2}}} = \frac{1}{\sqrt 2}\left(\ket{\sigma, 0, 2} \pm \ket{\sigma, 2, 0}\right),~\sigma=\uparrow(1), \downarrow(-1)~.
\end{eqnarray}
In terms of these linear combinations, we then get
\begin{eqnarray}
	V\ket{\text{SS}_0} = t\ket{\text{LM}_-^{2,0}}, ~ ~ V\ket{\text{LM}_-^{2,0}} = t\ket{\text{SS}_0}, ~ ~V\ket{\text{LM}_+^{2,0}} = 0; \quad V\ket{\text{SS}_\uparrow} = t\ket{\text{LM}_+^{3, \uparrow}},\nonumber \\
	V\ket{\text{LM}_+^{3, \uparrow}} = t\ket{\text{SS}_\uparrow}, ~ ~V\ket{\text{LM}_-^{3, \uparrow}} = 0, ~ ~ V\ket{\text{SS}_\downarrow} = t\ket{\text{LM}_+^{3, \downarrow}}, ~ ~ V\ket{\text{LM}_+^{3, \downarrow}} = t\ket{\text{SS}_\downarrow},\nonumber \\
	V\ket{\text{LM}_-^{3, \downarrow}} = 0; \quad V\ket{\text{SS}_2} = t\ket{\text{LM}_-^{4,0}}, ~ ~ V\ket{\text{LM}_-^{4,0}} = t\ket{\text{SS}_2}, ~ ~V\ket{\text{LM}_+^{4,0}} = 0~.
\end{eqnarray}
The zero matrix elements at the far right of each line should be understood as the fact that \(V\) takes those states out of the ground-state subspace. For e.g., the state \(V\ket{\text{LM}_-^{2,0}}\) is not really zero, but the overlap of \(V\ket{\text{LM}_-^{2,0}}\) with any other state in the ground-state subspace spanned by the 20 degenerate states is zero. We now see that each of the \(3\times3\) blocks decouples into a \(2\times2\) block \(\begin{pmatrix} 0 & t \\ t & 0 \end{pmatrix} \)and a \(1\times1\) block \(\begin{pmatrix} 0 \end{pmatrix} \) when written in terms of the linear combinations \(\ket{\text{LM}_\pm}\). The state in the final block is therefore already an eigenstate, but with zero energy, so that will drop out from the effective Hamiltonian. The remaining \(2\times 2\) block has eigenstates \(\ket{\pm^{N_\text{tot},S_\text{tot}^z}}\) that are linear combinations of the singlet states and the surviving local moment combination, with eigenvalues \(\pm t\):
\begin{eqnarray}
	\label{new g-states}
	N_\text{tot} = 2, S_\text{tot}^z = 0: \ket{\pm^{2,0}} = \frac{1}{\sqrt 2}\left(\ket{\text{SS}_0} \pm \ket{\text{LM}_-^{2,0}}\right),\nonumber \\
	N_\text{tot} = 3, S_\text{tot}^z = \frac{1}{2}: \ket{\pm^{3,\frac{1}{2}}} = \frac{1}{\sqrt 2}\left(\ket{\text{SS}_\uparrow} \pm \ket{\text{LM}_+^{3,\frac{1}{2}}}\right),\nonumber \\
	N_\text{tot} = 3, S_\text{tot}^z = -\frac{1}{2}: \ket{\pm^{3,-\frac{1}{2}}} = \frac{1}{\sqrt 2}\left(\ket{\text{SS}_\downarrow} \pm \ket{\text{LM}_+^{3,-\frac{1}{2}}}\right), \nonumber\\
	N_\text{tot} = 4, S_\text{tot}^z = 0: \ket{\pm^{4,0}} = \frac{1}{\sqrt 2}\left(\ket{\text{SS}_2} \pm \ket{\text{LM}_-^{4,0}}\right)~.
\end{eqnarray}
At this point, it is worth noting that the polarised ground-states \(\ket{S_\text{tot}^z=\sigma\frac{1}{2}},\sigma=\pm 1\) can be written as an equal superposition of the singlet states \(\ket{\text{SS}}_{d0}\otimes\ket{\sigma}_1\) and the local moment states \(\frac{1}{\sqrt 2}\left(\ket{\sigma, 0, 2} - \ket{\sigma, 2, 0}\right)\). The polarised symmetry-broken states therefore act as a bridge between the ground-states of the two phases, and are important in displaying the breakdown of the Kondo cloud at the QCP and the consequences arising from it (like inexact screening of the impurity and fractional magnetisation and entanglement entropy, which will be discussed in the next section).

By combining the two eigenstates \(\ket{\pm}\) for each block, one can then reconstruct the effective Hamiltonian \(\mathcal{H}^{N_\text{tot}, S_\text{tot}^z}_\text{eff}\) that describes the dynamics of the first site from within the ground-state subspace of the zeroth Hamiltonian. Formally, this is written as
\begin{eqnarray}
	\label{eff-ham-all}
	\mathcal{H}^{N_\text{tot}, S_\text{tot}^z}_\text{eff} &=& t\ket{+^{N_\text{tot}, S_\text{tot}^z}}\bra{+^{N_\text{tot}, S_\text{tot}^z}} - t\ket{-^{N_\text{tot}, S_\text{tot}^z}}\bra{-^{N_\text{tot}, S_\text{tot}^z}}~, \nonumber\\
							      &=& \begin{cases} t \left[\ket{\text{SS}_\alpha}\bra{\text{LM}_-^{N_\text{tot}, S_\text{tot}^z}} + \ket{\text{LM}_-^{N_\text{tot}, S_\text{tot}^z}}\bra{\text{SS}_\alpha}\right], ~ ~ \alpha=0,2~,\\
	t \left[\ket{\text{SS}_\alpha}\bra{\text{LM}_+^{N_\text{tot}, S_\text{tot}^z}} + \ket{\text{LM}_+^{N_\text{tot}, S_\text{tot}^z}}\bra{\text{SS}_\alpha}\right], ~ ~ \alpha=\uparrow, \downarrow~.\\
        \end{cases}
\end{eqnarray}
We first write down the effective Hamiltonian for \(\alpha=0\) which corresponds to \(N_\text{tot}=2, S_\text{tot}^z = 0\):
\begin{eqnarray}
	\mathcal{H}^{2, 0}_\text{eff} = t \left[\ket{\text{SS}_0}\bra{\text{LM}_-^{2,0}} + \ket{\text{LM}_-^{2,0}}\bra{\text{SS}_0}\right] ~.
\end{eqnarray}
In order to convert this into a second-quantised form, we will use the following identities:
\begin{eqnarray}
	\ket{\text{SS}_0} &= \frac{1}{\sqrt 2}\left(P_d^{ \uparrow} c^\dagger_{0 \downarrow} - P_d^{ \downarrow} c^\dagger_{0 \uparrow} \right) P_0^{(0)}P_1^{(0)}\ket{0}; \quad \ket{\text{LM}_-^{2,0}} &= \frac{1}{\sqrt 2}\left(P_d^{ \uparrow} c^\dagger_{1 \downarrow} - P_d^{ \downarrow} c^\dagger_{1 \uparrow} \right) P_0^{(0)}P_1^{(0)}\ket{0}~,
\end{eqnarray}
where \(P_i^{(m)} (m=0,1,2)\) projects onto the \(\hat n_i = m\) subspace on the i\(^\text{th}\) site, and \(P_d^\sigma\) projects the impurity spin into \(\ket{\sigma};\sigma=\uparrow,\downarrow\) configuration.
We then have
\begin{eqnarray}
	\ket{\text{SS}_0} \bra{\text{LM}_-^{2,0}} &= \frac{1}{2}\left(P_d^{\uparrow} c^\dagger_{0 \downarrow} - P_{d}^{\downarrow} c^\dagger_{0 \uparrow} \right)P_0^{(0)}P_1^{(0)}\left(P_{d}^{\uparrow} c_{1 \downarrow} - P_{d}^{\downarrow} c_{1 \uparrow} \right)\nonumber \\
						  &= P_1^{(0)}\frac{1}{2}\left(P_d^\uparrow c^\dagger_{0 \downarrow}c_{1 \downarrow} + P_d^\downarrow c^\dagger_{0 \uparrow}c_{1 \uparrow} - S_d^+ c^\dagger_{0 \downarrow} c_{1 \uparrow} - S_d^- c^\dagger_{ 0 \uparrow} c_{1 \downarrow}\right)P_0^{(0)}~,
\end{eqnarray}
where we have used \(P^2 = P\) if \(P\) is a projector, and that \(P_d^\uparrow P_d^ \downarrow = S_d^+\). Substituting these into the effective Hamiltonian and recognising that \(P_d^\uparrow = \frac{1}{2} + S_d^z, P_d^\downarrow = \frac{1}{2} - S_d^z\) gives
\begin{eqnarray}
	\mathcal{H}^{2, 0}_\text{eff} = \frac{t}{2}&\left[\left(\frac{1}{2} + S_d^z\right) c^\dagger_{0 \downarrow} P_0^{(0)} P_1^{(0)} c_{1 \downarrow} -S_d^+ c^\dagger_{0 \downarrow} P_0^{(0)} P_1^{(0)} c_{1 \uparrow} - S_d^- c^\dagger_{0 \uparrow} P_0^{(0)} P_1^{(0)} c_{1 \downarrow} \right. \nonumber\\
						   &\left.+ \left(\frac{1}{2} - S_d^z\right) c^\dagger_{0 \uparrow} P_0^{(0)} P_1^{(0)} c_{1 \uparrow} \right] + \text{h.c.}~.
\end{eqnarray}
In order to make this more illuminating, we define the holon excitation operators \(h^\dagger_{i\sigma} = c^\dagger_{i\sigma}P_i^{(0)}\). In terms of these operators, the effective Hamiltonian becomes
\begin{eqnarray}
	\mathcal{H}^{2, 0}_\text{eff} &=& \frac{t}{2}\left[\left(\frac{1}{2} + S_d^z\right) \left(h^\dagger_{0 \downarrow} h_{1 \downarrow} + h^\dagger_{1 \downarrow} h_{0 \downarrow}\right) - S_d^+ \left(h^\dagger_{0 \downarrow} h_{1 \uparrow} + h^\dagger_{1 \downarrow} h_{0 \uparrow}\right) - S_d^- \left(h^\dagger_{0 \uparrow} h_{1 \downarrow} + h^\dagger_{1 \uparrow} h_{0 \downarrow}\right) + \right. \nonumber\\
				      & & \left. \left(\frac{1}{2} - S_d^z\right) \left(h^\dagger_{0 \uparrow} h_{1 \uparrow}+ h^\dagger_{1 \uparrow} h_{0 \uparrow}\right) \right]\nonumber\\
				      &=& -t\left[S_d^z\frac{1}{2}\left(h^\dagger_{0 \uparrow} h_{1 \uparrow}+ h^\dagger_{1 \uparrow} h_{0 \uparrow} - h^\dagger_{0 \downarrow} h_{1 \downarrow} - h^\dagger_{1 \downarrow} h_{0 \downarrow}\right) + \frac{1}{2}S_d^+ \left(h^\dagger_{0 \downarrow} h_{1 \uparrow} + h^\dagger_{1 \downarrow} h_{0 \uparrow}\right) \right. \nonumber\\
				      &+& \left.\frac{1}{2}S_d^- \left(h^\dagger_{0 \uparrow} h_{1 \downarrow} + h^\dagger_{1 \uparrow} h_{0 \downarrow}\right) \right] + \frac{t}{4}\left(h^\dagger_{0 \uparrow} h_{1 \uparrow}+ h^\dagger_{1 \uparrow} h_{0 \uparrow} + h^\dagger_{0 \downarrow} h_{1 \downarrow} + h^\dagger_{1 \downarrow} h_{0 \downarrow}\right) \label{eff-ham-nfl}~.
\end{eqnarray}
The first part of the Hamiltonian (with a \(-t\) factor in front) is reminiscent of a two-spin Heisenberg Hamiltonian. To investigate this, we write down the potential second spin:
\begin{eqnarray}
	\mathcal{S}^z_1 \equiv \frac{1}{2}\left(h^\dagger_{0 \uparrow} h_{1 \uparrow}+ h^\dagger_{1 \uparrow} h_{0 \uparrow} - h^\dagger_{0 \downarrow} h_{1 \downarrow} - h^\dagger_{1 \downarrow} h_{0 \downarrow}\right)~.
\end{eqnarray}
Note that each term that comprises the full operator \(\mathcal{S}^z_1\) has a combined projector \(P_{01}^{(0)} = P_0^{(0)}P_1^{(0)} = P_1^{(0)}P_0^{(0)}\) in the middle with a single annihilation operator to the right. This means that \(\mathcal{S}^z\) projects on to the smaller set of only those states that satisfy \(\hat n_0 + \hat n_1 = 1\). That is also the meaning of the subscript \(1\) in \(\mathcal{S}^z_1\). There are four such states: \(\ket{\uparrow, 0}, \ket{\downarrow,0}, \ket{0,\uparrow}, \ket{0, \downarrow}\). In fact, it is easy to see that the eigenstates of \(\mathcal{S}^z\) are
\begin{eqnarray}
	\label{sigmaz-def}
	\ket{\Uparrow}_{1,\pm} &= \frac{1}{\sqrt 2}\left(\ket{\uparrow,0} \pm \ket{0, \uparrow}\right), \quad \mathcal{S}^z_1\ket{\Uparrow}_{1,\pm} = \pm\frac{1}{2}\ket{\Uparrow}_{1,\pm}, \\
	\ket{\Downarrow}_{1,\pm} &= \frac{1}{\sqrt 2}\left( \ket{\downarrow, 0} \pm \ket{0, \downarrow}\right),\quad \mathcal{S}^z_1\ket{\Downarrow}_{1,\pm} = \mp\frac{1}{2}\ket{\Downarrow}_{1,\pm}~.
\end{eqnarray}
The label \(\pm\) on top of the eigenstates indicates that these are also eigenstates of the parity operator \(\mathcal{P}_{01}\) that switches the site labels \(0\) and \(1\):
\begin{eqnarray}
	\label{parity-define}
	\mathcal{P}_{01}\ket{\alpha,\alpha^\prime} = \ket{\alpha^\prime,\alpha};~~ \alpha,\alpha^\prime \in \left\{0,\uparrow,\downarrow,2\right\} ~.
\end{eqnarray}
In fact, the eigenstates \(\ket{\Uparrow}_{1,+1}\) and \(\ket{\Downarrow}_{1,-1}\) both have \(\mathcal{S}^z = \frac{1}{2}\), but are orthogonal because they have different parities. Analogous to the \(\mathcal{S}_1^z\) operator, we can also read off the potential \(\mathcal{S}_1^\pm\) ladder operators from the effective Hamiltonian:
\begin{eqnarray}
	\label{sigmapm-def}
	\mathcal{S}_1^+ = h^\dagger_{0 \uparrow} h_{1 \downarrow} + h^\dagger_{1 \uparrow} h_{0 \downarrow}, ~ ~ ~\mathcal{S}_1^- = \left(\mathcal{S}_1^+\right)^\dagger~.
\end{eqnarray}
Applying these ladder operators on the eigenstates of \(\mathcal{S}_1^z\) gives
\begin{eqnarray}
	\mathcal{S}_1^+ \ket{\Downarrow}_{1,\pm} = \pm \ket{\Uparrow}_{1,\pm}; \quad\mathcal{S}_1^- \ket{\Uparrow}_{1,\pm} = \pm \ket{\Downarrow}_{1,\pm}; \quad\mathcal{S}_1^+ \ket{\Uparrow}_{1,\pm} = \mathcal{S}_1^- \ket{\Downarrow}_{1,\pm} = 0~.
\end{eqnarray}
This makes it clear that the pair of states \(\ket{\Uparrow}^{+}_1\) and \(\ket{\Downarrow}^{+}_1\) form a spin-half doublet, as does the other pair \(\ket{\Uparrow}^{-}_1\) and \(\ket{\Downarrow}^{-}_1\). We can define new spin operators that act only on a single parity subspace:
\begin{eqnarray}
	\mathcal{S}^a_{1,\pm} = \pm \mathcal{S}^a ~\frac{1}{2}\left(1 \pm \mathcal{P}_{01}\right);~ ~ a=x,y,z~.
\end{eqnarray}
The operator \(\frac{1}{2}\left(1 \pm \mathcal{P}_{01}\right)\) projects on to the subspace with \(\pm 1\) parity. In the notation \(\mathcal{S}^a_{1,\pm}\), the label \(\pm\) indicates the parity sector to which the operator applies, while 1 indicates that we are operating in the subspace with \(\hat n_0 + \hat n_1 = 1\). Two operators corresponding to different parity subspaces will commute because they act on disjoint Hilbert spaces: \(\left[\mathcal{S}^a_{+1}, \mathcal{S}^b_{-1}\right] = 0\). In terms of these projected spin operators, we recover the usual spin-half SU(2) algebra:
\begin{eqnarray}
	\mathcal{S}^z_{1,\pm} \ket{\Uparrow}_{1,\pm} = \frac{1}{2}\ket{\Uparrow}_{1,\pm}; \quad\mathcal{S}^z_{1,\pm}\ket{\Downarrow}_{1,\pm} = -\frac{1}{2}\ket{\Downarrow}_{1,\pm};\nonumber\\
	\mathcal{S}^+_{1,\pm} \ket{\Downarrow}_{1,\pm} = \ket{\Uparrow}_{1,\pm}; \quad\mathcal{S}^-_{1,\pm} \ket{\Uparrow}_{1,\pm} = \ket{\Downarrow}_{1,\pm}.\qquad
\end{eqnarray}
The part of the effective Hamiltonian that involves the impurity can now be written in terms of these new spin operators:
\begin{eqnarray}
	t\left(\vec{S}_d\cdot\vec{\mathcal{S}}_{-,1} - \vec{S}_d\cdot\vec{\mathcal{S}}_{+,1}\right); ~ ~ \vec{\mathcal{S}}_{\pm} = \left(\mathcal{S}_{\pm}^x ~ ~ \mathcal{S}_{\pm}^y ~ ~\mathcal{S}_{\pm}^z\right)~.
\end{eqnarray}
The last term in eq.~\eqref{eff-ham-nfl} (with a prefactor of \(\frac{t}{4}\)) can now be identified as the parity operator \(\mathcal{P}_{01}\) acting within the subspace \(\hat n_0 + \hat n_1 = 1\), because it has eigenstates \(\ket{\uparrow,0}\pm\ket{0,\uparrow}\) and \(\ket{\downarrow,0}\pm\ket{0,\downarrow}\) with eigenvalues \(\pm 1\). The full effective Hamiltonian in the \(N_\text{tot}=2\) subspace then takes the compact form
\begin{eqnarray}
	\mathcal{H}^{2, 0}_\text{eff} = t\left(\vec{S}_d\cdot\vec{\mathcal{S}}_{1,-1} - \vec{S}_d\cdot\vec{\mathcal{S}}_{1,+}\right) + \frac{t}{4}\mathcal{P}_{01} P_{01}^{(1)}~,
\end{eqnarray}
where \(P_{01}^{(1)} = P_{0}^{(1)}P_{1}^{(0)} + P_{0}^{(0)}P_{1}^{(1)}\) projects on to the \(\hat n_0 + \hat n_1 = 1\) subspace. In light of these new spin operators, the original eigenstates \(\ket{\pm^{2,0}}\) can be written in terms of these emergent spins \(\ket{\Uparrow}_{1,\pm}\) and \(\ket{\Downarrow}_{1,\pm}\):
\begin{eqnarray}
	\ket{\pm^{2,0}} &=& \frac{1}{2}\left(\ket{\uparrow,\downarrow,0} - \ket{\downarrow,\uparrow,0} \pm \ket{\uparrow,0,\downarrow} \mp \ket{\downarrow,0,\uparrow}\right) = \frac{1}{2}\left[\ket{\uparrow}\left(\ket{\downarrow,0} \pm \ket{0, \downarrow}\right) - \ket{\downarrow}\left(\ket{\uparrow,0} \pm \ket{0,\uparrow}\right)\right] \nonumber\\
			    &=& \frac{1}{\sqrt 2}\left( \ket{\uparrow}\ket{\Downarrow}_{1,\pm} - \ket{\downarrow}\ket{\Uparrow}_{1,\pm} \right)~.
\end{eqnarray}
Both the ground and excited states are therefore spin-singlet states, but they belong to the negative and positive parity sectors respectively.

We now proceed to the \(N_\text{tot}=4\) sector, which is equivalent to \(\alpha=2\) and \(\hat n_0 + \hat n_1 = 3\). From eq.~\eqref{eff-ham-all}, we note that the effective Hamiltonian of the \(\alpha=2\) sector is obtained from that of the \(\alpha=0\) sector by replacing the holons with doublons, \(\ket{0} \to \ket{2}\). This allows us to use the results of the \(N_\text{tot}=2\) sector simply by making the above-mentioned holon-doublon transformation. Accordingly, the composite up and down spins of the \(0+1\)-site system become
\begin{eqnarray}
	\ket{\Uparrow}_{3,\pm} = \frac{1}{\sqrt 2}\left(\ket{\uparrow,2} \pm \ket{2, \uparrow}\right), ~ ~\ket{\Downarrow}_{3,\pm} = \frac{1}{\sqrt 2}\left( \ket{\downarrow, 2} \pm \ket{2, \downarrow}\right)~,
\end{eqnarray}
which then allow us to define spin operators for the positive and negative parity sectors:
\begin{eqnarray}
	\mathcal{S}^z_\pm = \frac{1}{2}\left(\ket{\Uparrow}_{3,\pm} - \ket{\Downarrow}_{3,\pm}\right), ~~\mathcal{S}^+_\pm = \ket{\Uparrow}_{3,\pm}\bra{\Downarrow}_{3,\pm}~,
\end{eqnarray}
analogous to those of the \(N_\text{tot}=2\) sector in eqs.~\eqref{sigmaz-def} and \eqref{sigmapm-def}. These definitions then allow us to rewrite the effective Hamiltonian in terms of the eigenstates of \(\mathcal{S}^z_\pm\), and hence in terms of the spin operators \(\vec{\mathcal{S}}_{3,\pm}\). The eigenstates \(\ket{\pm^{4,0}}\) and the effective Hamiltonian \(H_\text{eff}^{4,0}\) for this sector can be written as
\begin{eqnarray}
	\ket{\pm^{4,0}} &= \frac{1}{\sqrt 2}\left( \ket{\uparrow}_d\ket{\Downarrow}_{3,\pm} -  \ket{\downarrow}_d\ket{\Uparrow}_{3,\pm}\right)~.
\end{eqnarray}
The effective Hamiltonian can then be written in terms of these emergent spins
\begin{eqnarray}
	H_\text{eff}^{4,0} = \frac{t}{2}&\left[\left( \ket{\uparrow}_d\ket{\Downarrow}_{3,+} -  \ket{\downarrow}_d\ket{\Uparrow}_{3,+}\right)\left( \bra{\uparrow}_d\bra{\Downarrow}_{3,+} -  \bra{\downarrow}_d\bra{\Uparrow}_{3,+}\right)\right.\nonumber \\
					&\left.- \left(\ket{\uparrow}_d\ket{\Downarrow}_{3,-} -  \ket{\downarrow}_d\ket{\Uparrow}_{3,-}\right)\left( \bra{\uparrow}_d\bra{\Downarrow}_{3,-} -  \bra{\downarrow}_d\bra{\Uparrow}_{3,-}\right)\right] ~.
\end{eqnarray}
By replacing the emergent spins with the spin operators \(\vec{\mathcal{S}}_{3,\pm}\), we get the final form of the Hamiltonian
\begin{eqnarray}
	H_\text{eff}^{4,0} = t\left( \vec{S}_d\cdot\vec{\mathcal{S}}_{3,-} - \vec{S}_d\cdot\vec{\mathcal{S}}_{3,+}\right) + \frac{t}{4}\mathcal{P}_{01}P_{01}^{(3)}~,
\end{eqnarray}
where \(\mathcal{P}_{01}\) is the parity transformation operator defined in eq.~\eqref{parity-define} and \(P_{01}^{(3)}\) projects on to the \(\hat n_0 + \hat n_1 = 3\) subspace. This completes the effective Hamiltonian for the \(S^z_\text{tot}=0\) sector.

We now come to the \(S^z_\text{tot}=+ 1/2\) sector. From eq.~\eqref{eff-ham-all}, the effective Hamiltonian for this sector is
\begin{eqnarray}
H_\text{eff}^{3,\frac{1}{2}} &= t \left[\ket{\text{SS}_\uparrow}\bra{\text{LM}_-^{3,\frac{1}{2}}} + \ket{\text{LM}_-^{3,\frac{1}{2}}}\bra{\text{SS}_\uparrow}\right]~,
\end{eqnarray}
where
\begin{eqnarray}
	\ket{\text{SS}_\uparrow} = \frac{1}{\sqrt 2}\left(P_d^{ \uparrow} c^\dagger_{0 \downarrow} - P_d^{ \downarrow} c^\dagger_{0 \uparrow} \right) c^\dagger_{1 \uparrow}\ket{0}; \quad\ket{\text{LM}_-^{3,\frac{1}{2}}} = \frac{1}{\sqrt 2} P_d^\uparrow \left(c^\dagger_{0 \uparrow} c^\dagger_{0 \downarrow} + c^\dagger_{1 \uparrow} c^\dagger_{1 \downarrow}\right) \ket{0}~,
\end{eqnarray}
such that
\begin{eqnarray}
	\ket{\text{SS}_\uparrow}\bra{\text{LAM}_-^{3,\frac{1}{2}}} &=& \frac{1}{2}\left(P_d^{ \uparrow} c^\dagger_{0 \downarrow} - P_d^{ \downarrow} c^\dagger_{0 \uparrow} \right) c^\dagger_{1 \uparrow}  \left(c_{0 \downarrow} c_{0 \uparrow} + c_{1 \downarrow} c_{1 \uparrow}\right)P_d^\uparrow \\
								   &=& -\frac{1}{2}\left[P_d^{ \uparrow} \left(c^\dagger_{1 \uparrow} c_{0 \uparrow}\hat n_{0 \downarrow} + c^\dagger_{0 \downarrow} c_{1 \downarrow} \hat n_{1 \uparrow}\right) + S_d^- \left(c^\dagger_{1 \uparrow}c_{0 \downarrow} \hat n_{0 \uparrow} - c^\dagger_{0 \uparrow} c_{1 \downarrow} \hat n_{1 \uparrow}\right)\right]~.
\end{eqnarray}
The full effective Hamiltonian then becomes
\begin{eqnarray}
	H_\text{eff}^{3,\frac{1}{2}} &=& -\frac{t}{2}\left[P_d^{ \uparrow} \left\{\left(c^\dagger_{1 \uparrow} c_{0 \uparrow} + c^\dagger_{0 \downarrow} c_{1 \downarrow}\right)\left( P_0^{(2)} P_1^{(0)} + P_0^{(0)} P_1^{(2)}\right) + \left(c^\dagger_{0 \uparrow} c_{1 \uparrow} + c^\dagger_{1 \downarrow} c_{0 \downarrow}\right) P_0^{(1)} P_1^{(1)}\right\} \right.\nonumber \\
				     &&\left.+ S_d^- \left(c^\dagger_{1 \uparrow}c_{0 \downarrow} - c^\dagger_{0 \uparrow} c_{1 \downarrow} \right)\left( P_0^{(2)} P_1^{(0)} + P_0^{(0)} P_1^{(2)}\right) + S_d^+ \left(c^\dagger_{0 \downarrow}c_{1 \uparrow} - c^\dagger_{1 \downarrow} c_{0 \uparrow} \right)P_0^{(1)} P_1^{(1)}\right] \nonumber\\
				      &=& t\left[P_d^{ \uparrow} \sqrt 2\mathcal{A}^z_\frac{1}{2} - \frac{1}{2}S_d^+ \mathcal{B}^-_\frac{1}{2} - \frac{1}{2}S_d^- \mathcal{B}^+_\frac{1}{2}\right] ~,
\end{eqnarray}
where we have defined
\begin{eqnarray}
	\mathcal{A}^z_\frac{1}{2} = -\frac{1}{2\sqrt 2}\left[\left(c^\dagger_{1 \uparrow} c_{0 \uparrow} + c^\dagger_{0 \downarrow} c_{1 \downarrow}\right)\left( P_0^{(2)} P_1^{(0)} + P_0^{(0)} P_1^{(2)}\right) + \left(c^\dagger_{0 \uparrow} c_{1 \uparrow} + c^\dagger_{1 \downarrow} c_{0 \downarrow}\right) P_0^{(1)} P_1^{(1)}\right]~,
\end{eqnarray}
and
\begin{eqnarray}
	\mathcal{B}^+_\frac{1}{2} &= \left(c^\dagger_{1 \uparrow}c_{0 \downarrow} - c^\dagger_{0 \uparrow} c_{1 \downarrow} \right)\left( P_0^{(2)} P_1^{(0)} + P_0^{(0)} P_1^{(2)}\right); \quad\mathcal{B}^-_\frac{1}{2} &= \left(c^\dagger_{0 \downarrow}c_{1 \uparrow} - c^\dagger_{1 \downarrow} c_{0 \uparrow} \right) P_0^{(1)} P_1^{(1)}~.
\end{eqnarray}

The effective Hamiltonian for the \(S_\text{tot}^z = -1/2\) sector is obtained by applying the transformations \(P_d^\sigma \to P_d^{\bar\sigma}, c_{i\sigma} \to c_{i\bar\sigma}\) on the Hamiltonian. This is motivated by the fact that applying the same transformations on the \(S_\text{tot}^z = 1/2\) eigenstates give the \(S_\text{tot}^z = -1/2\) eigenstates.
\begin{eqnarray}
	H_\text{eff}^{3,-\frac{1}{2}} = t\left[P_d^{ \uparrow} \sqrt 2\mathcal{A}^z_{-\frac{1}{2}} - \frac{1}{2}S_d^+ \mathcal{B}^-_{-\frac{1}{2}} - \frac{1}{2}S_d^- \mathcal{B}^+_{-\frac{1}{2}}\right] ~.
\end{eqnarray}
The operators \(\mathcal{A}^z_{-\frac{1}{2}}\) and \(\mathcal{B}^\pm_{-\frac{1}{2}}\) are obtained by applying the above mentioned transformation on their \(S^z_\text{tot} = +1/2\) counterparts.

\section{Non-Fermi liquid exponents in self-energy and two-particle correlations}
The one-particle self-energies relating to the impurity and zeroth sites, as well as some relevant two-particle correlation functions, can be shown to follow power-law behaviour close to the Brinkman-Rice transition at \(r_{c2}\), by mapping the impurity model to a classical Coulomb gas, similar to the work of Anderson, Yuval and Hamman for the Kondo model~\cite{anderson1969exact} and that of Si and Kotliar for a periodic Anderson model~\cite{si_kotliar_1993}. Indeed, the latter work uses an auxiliary that is similar to our extended Anderson impurity model. In the following, we adapt their results to our model. These non-universal power-laws are signatures of the non-Fermi liquid excitations that emerge at \(r_{c2}\).

Near \(r_{c2}\), extended SIAM can be reduced to a \(J-U_b\) model (shown in subsection V.A of the main manuscript). The conduction electrons scattering off the impurity suffer phase shifts because of the presence of the potentials \(J\) and \(U_b\). These phase shifts are given by \(\delta_\text{sp} = \arctan\left(\pi \rho_0 J/2\right)\) and \(\delta_\text{ch} = \arctan\left(\pi \rho_0 U_b/2\right)\)~\cite{si_kotliar_1993,giamarchi2004}, where \(\rho_0\) is the electronic density of states in the conduction bath.
The low-energy behaviour of the local impurity and bath self-energies can be expressed in terms of these phase shifts~\cite{si_kotliar_1993}:
\begin{eqnarray}
	\Sigma_{dd}(\omega) \sim \omega^{\gamma_{dd}}, ~ ~ \Sigma_{d0}(\omega) \sim \omega^{\gamma_{d0}}, ~ ~ \Sigma_{00}(\omega) \sim \omega^{\gamma_{00}}~,\nonumber \\
\gamma_{dd} = \frac{5}{4} - \left(\frac{\delta^*_\text{tot}}{\pi}\right)^2 - \left(\frac{\delta^*_\text{ch}}{\pi}\right)^2, ~ ~ \gamma_{d0} = 2\left[\left( \frac{\delta^*_\text{ch}}{\pi} \right)^2 + \left(\frac{\delta^*_\text{tot}}{\pi}\right)^2 - \frac{1}{4}\right], \nonumber\\
\gamma_{00} = \left(\frac{\delta^*_\text{tot}}{\pi} - \frac{1}{2}\right)^2 + \left(\frac{\delta^*_\text{ch}}{\pi} \right)^2 - 2~,
\end{eqnarray}
where we have defined a total phase shift parameter \(\frac{\delta_\text{tot}}{\pi} \equiv \frac{\delta_\text{sp} + \delta_\text{ch}}{\pi} - \frac{1}{2}\).
This can be extended to certain correlation functions as well. The impurity spin-flip correlation, the impurity-bath excitonic correlation and the zeroth-site pairing correlation~\cite{si_kotliar_1993}:
\begin{eqnarray}
	\left<S_d^+(\tau) S_d^-(0) \right>\sim \tau^{-\alpha_1} &\implies \left<S_d^+\right>(\omega) \sim \omega^{(\alpha_1 - 1)/2},  \nonumber\\
	\left<(c^\dagger_{d \uparrow} c_{0 \uparrow})(\tau) (c^\dagger_{0 \uparrow} c_{d \uparrow})(0)\right>\sim \tau^{-\alpha_2} &\implies \left<c^\dagger_{d \uparrow} c_{0 \uparrow} c^\dagger_{0 \uparrow} c_{d \uparrow}\right>(\omega) \sim \omega^{(\alpha_2 - 1)/2},  \nonumber \\
	\left<(c^\dagger_{0 \uparrow} c^\dagger_{0 \downarrow})(\tau) (c_{0 \downarrow} c_{0 \uparrow})(0)\right> \sim \tau^{-\alpha_3} &\implies \left<c^\dagger_{0 \uparrow} c^\dagger_{ 0\downarrow}\right>(\omega) \sim \omega^{(\alpha_3 - 1)/2}~,
\end{eqnarray}
where the exponents are given by
\begin{eqnarray}
	\alpha_1 = 2\left(1 - \frac{\delta_\text{sp}^*}{\pi}\right)^2, \alpha_2 &= \left(\frac{\delta^*_\text{tot}}{\pi} - \frac{1}{2}\right)^2 + \left(\frac{\delta^*_\text{ch}}{\pi} \right)^2,  \alpha_3 = \left(\frac{\delta^*_\text{tot}}{\pi} + \frac{1}{2}\right)^2 + \left(1 + \frac{\delta^*_\text{ch}}{\pi} \right)^2~.
\end{eqnarray}

\section{Entanglement entropy, magnetisation and phase shift, at the QCP}
As mentioned in the previous section, the polarised ground-states \(\ket{-}^{3,\pm\frac{1}{2}}\) of eq.~\eqref{new g-states} are special because they involve the strong-coupling ground-state as well as the local moment ground-state. The overall ground-states, therefore, contain both entangled and non-entangled impurity contributions, leading to a departure from maximal entanglement. We will now calculate the impurity entanglement entropy \(S_\text{EE}(d)\) in this polarised ground-state(s) that emerges exactly at the QCP, and show that \(S_\text{EE}(d)\) has a fractional value (in units of \(\log 2\)). This turns out to be linked to the fact that the impurity is screened partially in these states, leading to an impurity magnetisation that is non-vanishing but only half of that in the local moment phase.

The fractional entropy arises because of singular scattering of the gapless excitations at the Fermi surface leading to the destruction of the local Fermi liquid. In order to access these processes, we will work with a {\it semi-continuous} conduction bath beyond the zeroth site (total number of lattice sites \(N \to \infty\)): \(H_\text{bath} = \sum_{k,\sigma} \epsilon(k) \psi^\dagger_\sigma(k)\psi^\dagger_\sigma(k)\). This conduction bath that is composed of the \(k-\)states formed from the Hilbert space of the lattice sites beyond the zeroth site will be referred to as the reduced conduction bath. This is in addition to the Hamiltonian for the impurity and zeroth sites: \(H_{d0} = J \vec{S}_d\cdot\vec{S}_0 - \frac{U_b}{2}\left(c_{0\uparrow}^\dagger c_{0\uparrow} - c_{0\downarrow}^\dagger c_{0\downarrow}\right)^2\), where \(\vec S_d\) is the spin operator for the impurity site, and \(c_{0\sigma}^\dagger\) creates an electron of spin \(\sigma\) at the zeroth site. The full Hamiltonian also involves a single-particle hopping term that couples these two Hamiltonians: \(H = H_{d0} + H_\text{bath} + H_\text{int}\), where
\begin{eqnarray}
	H_\text{int} = -t \sum_\sigma\left( c^\dagger_{0 \sigma} c_{1\sigma} + \text{h.c.} \right) = -\frac{t}{N}\sum_{k,\sigma} \left[c^\dagger_{0\sigma} \psi_\sigma(k) + \text{h.c.}\right].
\end{eqnarray}
\(c^\dagger_{1\sigma} = \frac{1}{N}\sum_k \psi^\dagger_\sigma(k)\) is the creation operator for the first site of the conduction bath of \(N\) sites (but effectively the zeroth site for the reduced conduction bath defined by \(H_\text{bath}\)). We will treat \(H_\text{int}\) perturbatively in \(t/J\), in order to obtain the set of renormalised \(g-\)fold degenerate ground-states \(\left\{ \ket{\psi^{(n)}_\text{gs}}; n=1,\ldots,g\right\}\).

As mentioned before, the intention here is to calculate magnetisation and entanglement entropy in the non-Fermi liquid ground-states. In the case of a degenerate subspace with \(S_\text{tot}^z = \pm 1/2\) where \(S_\text{tot}^z = S_d^z + S_0^z + S_1^z\) is the total spin operator, the magnetisation \(m_d^z\) must be calculated in any {\it one} of the degenerate states for our problem. This is accomplished by solving the problem in the presence of an infinitesimal magnetic field term \(-h S_\text{tot}^z\) that lifts the degeneracy and chooses one of the symmetry-broken ground-states. This is similar to the "two self-energy description" introduced by David Logan for computing quantities in the local moment phase/Mott insulating phase in terms of different self-energies obtained by inserting positive and negative magnetic fields~\cite{Logan_2000,logan_2014,Logan_2015}. Due to the symmetry of the Hamiltonian under \(S_\text{tot}^z \to -S_\text{tot}^z\), both the limits \(h = 0^+\) and \(h=0^-\) give the same results for the density matrix and the entanglement entropy. We will work with \(h = 0^+\). Since the total Hamiltonian \(H\) and the perturbation \(H_\text{int}\) preserve the total spin \(S_\text{tot}^z = S_d^z + S_0^z + \int dk S_k^z\), it is sufficient to solve only for the sector with the most positive value of \(S_\text{tot}^z\), since those are the states that will be selected by the positive magnetic field \(h = 0^+\).

In the absence of \(H_\text{int}\), the set of ground-states with a positive \(S_\text{tot}^z\) will contain, to begin with, the local moment states \(\ket{\uparrow}_d\ket{0}_0\ket{\phi}\) and \(\ket{\uparrow}_d\ket{2}_0\ket{\phi}\), where \(\ket{\phi}\) represents the filled Fermi sea of the reduced conduction bath, acting as its spinless ground-state. These states have \(S_\text{tot}^z = 1/2\). We can label these states as \(\ket{n_\text{tot}, \mu_d^z}\), using the total number of particles \(n_\text{tot} = n_d + n_0 + n_\text{bath}\) (which is also conserved, and where we set \(n_\text{bath} = 1\) for the set \(\ket{\phi}\)), and the average magnetisation \(\mu_d^z\) of the impurity state: \(\ket{2,\frac{1}{2}} \equiv \ket{\uparrow}_d\ket{0}_0\ket{\phi}, \ket{4,\frac{1}{2}} \equiv \ket{\uparrow}_d\ket{2}_0\ket{\phi}\). Since the singlet state \(\ket{SS}_{d0}\) (on the \(d\) and zeroth sites) is degenerate with the local moment states \(\ket{\uparrow}_d\ket{0(2)}_0\), one can construct other states with \(S_\text{tot}^z = 1/2\) by creating magnetic excitations at the Fermi surface. The fact that we are working with a thermodynamically large bath (\(N \to \infty\)) means that there will be gapless excitations vanishingly close to the Fermi surface. Such states are of the form \(\ket{\text{SS}}_{d0}\ket{e_\uparrow}\) (labeled as \(\ket{4,0}\) in the \(\ket{n_\text{tot}, \mu_d^z}\) notation) and \(\ket{\text{SS}}_{d0}\ket{h_\downarrow}\) (labeled as \(\ket{2,0}\)), where \(\ket{e_\sigma}\) and \(\ket{h_\downarrow}\) are electron and hole states respectively:
\begin{eqnarray}
	\ket{e_\sigma} = \sum_k^\text{FS} \psi^\dagger_\sigma(k)\ket{\phi}, ~ ~\ket{h_\sigma} = \sum_k^\text{FS} \psi_\sigma(k)\ket{\phi}~.
\end{eqnarray}
The sum is over \(k-\)states on the Fermi surface. The singlet state was set to have zero spin on the impurity site in the state notation, because \(\left<S_d^z \right>\) vanishes in the singlet state. The full set of relevant degenerate ground-states, classified by the total number of particles, is
\begin{eqnarray}
	n_\text{tot} = 4: \ket{4, 0}, \ket{4, \frac{1}{2}}; ~ ~ ~ ~ n_\text{tot} = 2: \ket{2,0}, \ket{2, \frac{1}{2}}\label{imp_states}~.
\end{eqnarray}
Within each \(2\times 2\) block, the interaction matrix \(H_\text{int}\) takes the form ( in the basis of the states in eq.~\eqref{imp_states})
\begin{eqnarray}
	 \left(H_\text{int}\right)_\text{gs}^{n_\text{tot}=2} = \frac{t}{N} \times \begin{pmatrix}0 && 1/\sqrt 2 \\ \\ 1/\sqrt 2 && 0\end{pmatrix} = -\left(H_\text{int}\right)_\text{gs}^{n_\text{tot}=4}~,
\end{eqnarray}
and the \(S_\text{tot}^z = + \frac{1}{2}\) ground-states are then easily obtained to be of the form \(\ket{\psi_\text{gs}}^{(n_\text{tot})} = \frac{1}{\sqrt 2}\left(\ket{n_\text{tot}, 0} - \ket{n_\text{tot}, \frac{1}{2}}\right), n_\text{tot} = 2 ~\&~ 4\), independent of \(N\). Both ground-states involve an equal superposition of a \(\mu_d^z=1/2\) state and a \(\mu_d^z=0\) state, such that the net impurity magnetisation in either of the renormalised polarised states is
\begin{eqnarray}
	m_d^z = \frac{1}{2}\left( 0 + \frac{1}{2} \right) = \frac{1}{4}~.
\end{eqnarray}
The reduced density matrix \(\rho_\text{imp} = \text{Tr}^\prime\left(\ket{\psi_\text{gs}}^{(n_\text{tot})}\bra{\psi_\text{gs}}^{(n_\text{tot})}\right)\) (where \(\text{Tr}^\prime\) indicates partial trace over all states apart from those of the impurity) for the impurity spin in the polarised states can now be immediately obtained. It is independent of \(n_\text{tot}\), and actually turns out to be halfway between the maximally mixed density matrix \(\rho_\text{MM} = \frac{1}{2} I\) and the non-entangled density matrix \(\rho_\text{NE} = \begin{pmatrix} 1 & 0 \\ 0 & 0 \end{pmatrix} \):
\begin{eqnarray}
	\rho_\text{imp} = \begin{pmatrix} 3/4 && 0 \\ \\ 0 && 1/4 \end{pmatrix} = \frac{1}{2}\left(\rho_\text{MM} + \rho_\text{NE}\right), ~ ~ ~ S_\text{EE}(d) = -\text{Tr}\left(\rho_\text{imp} \ln \rho_\text{imp}\right) \simeq 0.81 \log 2 ~.
\end{eqnarray}
This fractional value of the entropy in units of \(\log 2\) is another signature of the non-Fermi liquid. We can express this in terms of an effective degeneracy \(\tilde g\) of the impurity degree of freedom. We define this effective degeneracy as \(\tilde g = 1 + 2|m_d^z|\), in order to recover the unique ground-state in the local Fermi liquid phase (that has \(m_d^z=0,\tilde g = 1\)) and the magnetic doublet in the local moment phase (that has \(m_d^z = 1/2, \tilde g =2\)). The proposed form also results in a non-integer impurity degeneracy for the NFL at the QCP, \(\tilde g^* = 3/2\). Using the effective degeneracy \(\tilde g\), one can also write down a general ground-state
\begin{eqnarray}\label{general-state}
	\ket{\psi_\text{gs}(g)} = \frac{1}{\sqrt{(2-\tilde g)^2 + (\tilde g-1)^2}}\left[(2-\tilde g)\ket{n_\text{tot},0} - (\tilde g - 1)\ket{n_\text{tot},1/2}\right]~,
\end{eqnarray}
that interpolates between the strong-coupling ground-state for \(r < r_{c2}\) (therefore \(\tilde g = 1\)) and the weak coupling ground-state \(r > r_{c2}\) (therefore \(\tilde g = 2\)), also describing the QCP in between (\(\tilde g = 3/2\)). The entanglement entropy, when written in terms of \(\tilde g\), takes the form
\begin{eqnarray}
	\rho_\text{imp} = \frac{1}{2}\begin{pmatrix} 2-\tilde g && 0 \\ \\ 0 && \tilde g \end{pmatrix} = \frac{1}{2} + \begin{pmatrix} m_d^z && 0 \\ \\ 0 && -m_d^z \end{pmatrix},\\
	S_\text{EE}(d) = -\left( 1 - \frac{\tilde g}{2} \right) \ln \left( 1 - \frac{\tilde g}{2} \right) - \frac{\tilde g}{2}\ln \frac{\tilde g}{2} ~.
\end{eqnarray}
The relation between the reduced density matrix \(\rho_\text{imp}\) and the impurity magnetisation \(m_d^z\) is consistent with that obtained by Kopp et al.,~\cite{kopp_chakravarty_2007}.

The general form \(\ket{\psi_\text{gs}(g)}\) of polarised interpolating ground-state also allows us to calculate the excess charge (\(n_\text{exc}\)) added, by the impurity, to the conduction bath Fermi surface (through gapless excitations). This then allows the scattering phase shift of the conduction electrons (obtained via the Friedel sum rule,~\cite{friedel_1956,langer1961friedel,langreth1966}) to be linked with the impurity magnetisation \(m_d^z\). From the definitions of the states \(\ket{n_\text{tot},0}\) and \(\ket{n_\text{tot}, 1.2}\), we note that the former state involves an impurity-bath singlet while the latter involves a decoupled local moment. The singlet state contributes an excess charge of unity to the conduction bath: \(n_\text{exc}^\text{SS} = 1\), because of the impurity-bath entanglement within it. The local state moment does not contribute any excess charge, because the impurity spin is no longer hybridising with the bath in that state: \(n_\text{exc}^\text{LM}=0\). Putting these together, the net excess charge contributed by the state in eq.~\eqref{general-state} is
\begin{eqnarray}
	n_\text{exc} = \frac{1}{(2-\tilde g)^2 + (\tilde g-1)^2}\left[(2-\tilde g)^2 n_\text{exc}^\text{SS} + (\tilde g - 1)^2n_\text{exc}^\text{LM}\right] = \left[1 + \left(\frac{2-\tilde g}{\tilde g-1}\right)^2\right]^{-1}~.
\end{eqnarray}
At the QCP, the excess charge acquires a fractional value of \(1/2\), indicating once more that only half of the impurity remains coupled with the bath at the transition. The Friedel sum rule can be used to equate this excess charge with the phase shift \(\delta^*\) suffered by conduction electrons as they scatter off the impurity, at the fixed point:
\begin{eqnarray}\label{phase-mag}
	\delta^* = \pi n_\text{exc} = \left[1 + \left(\frac{2-\tilde g}{\tilde g-1}\right)^2\right]^{-1} = \frac{4{m_d^z}^2}{1 - 4|m_d^z| + 8{m_d^z}^2}~.
\end{eqnarray}
Equation \eqref{phase-mag} unifies the effect of several important quantities that lead to the frustration of the impurity and the destruction of the Kondo  effect. The presence of \(U_b\) introduces local moment states into the spectrum and leads to states with non-vanishing impurity magnetisation \(m_d^z\) at the QCP; this non-zero \(m_d^z\) can be interpreted as an effective impurity degeneracy \(\tilde g\) that is only partially screened, resulting in only half the excess charge \(n_\text{exc}\) being contributed to the conduction bath. This reduction in the excess charge acts as a shift in the boundary conditions felt by the conduction electrons and manifests as a phase shift that is less than the unitarity limit of \(\pi\).

\section{Additional correlations near the ESQPT and the QCP}
\begin{figure}[htpb]
	\centering
	\includegraphics[width=0.48\textwidth]{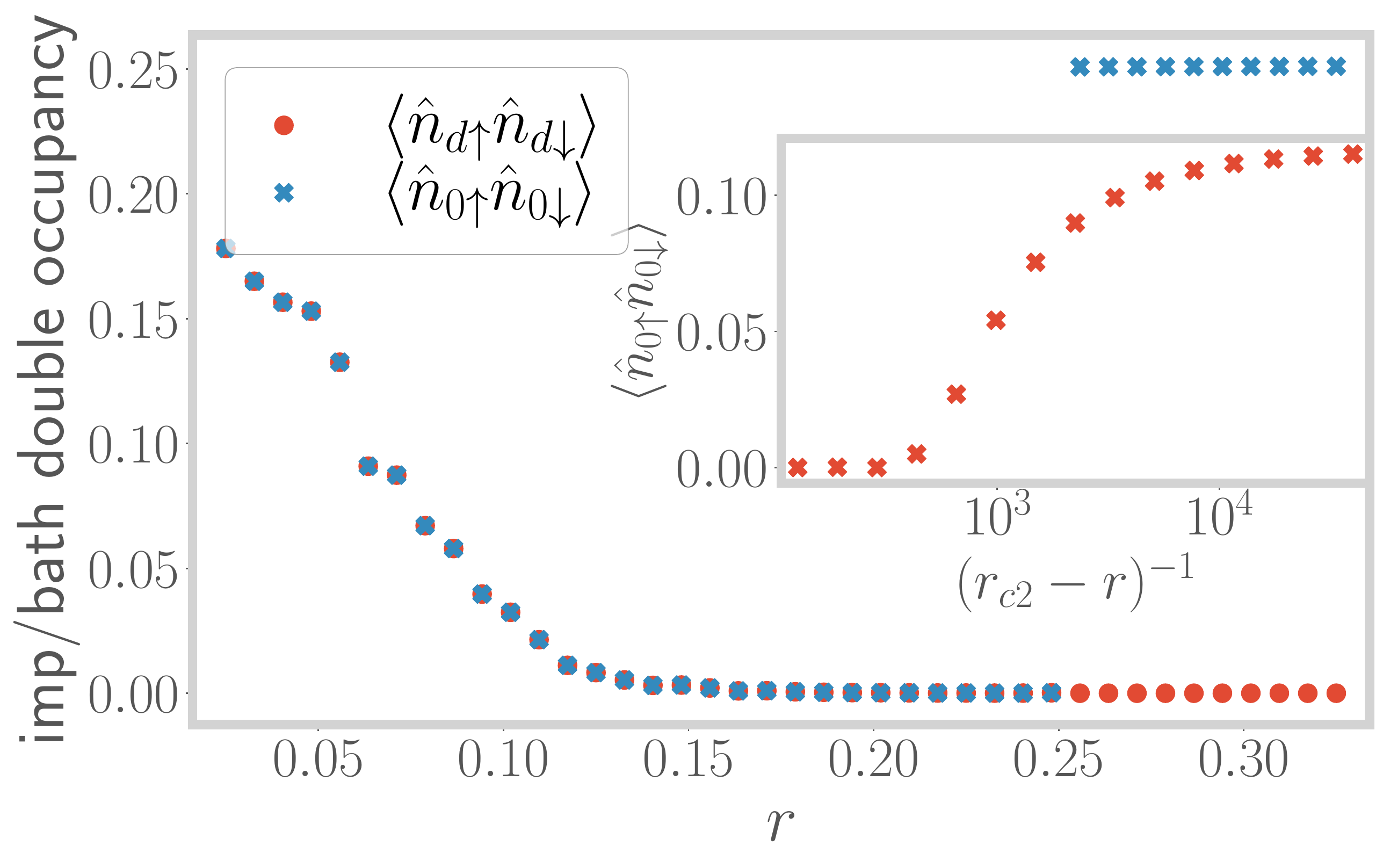}
	\includegraphics[width=0.48\textwidth]{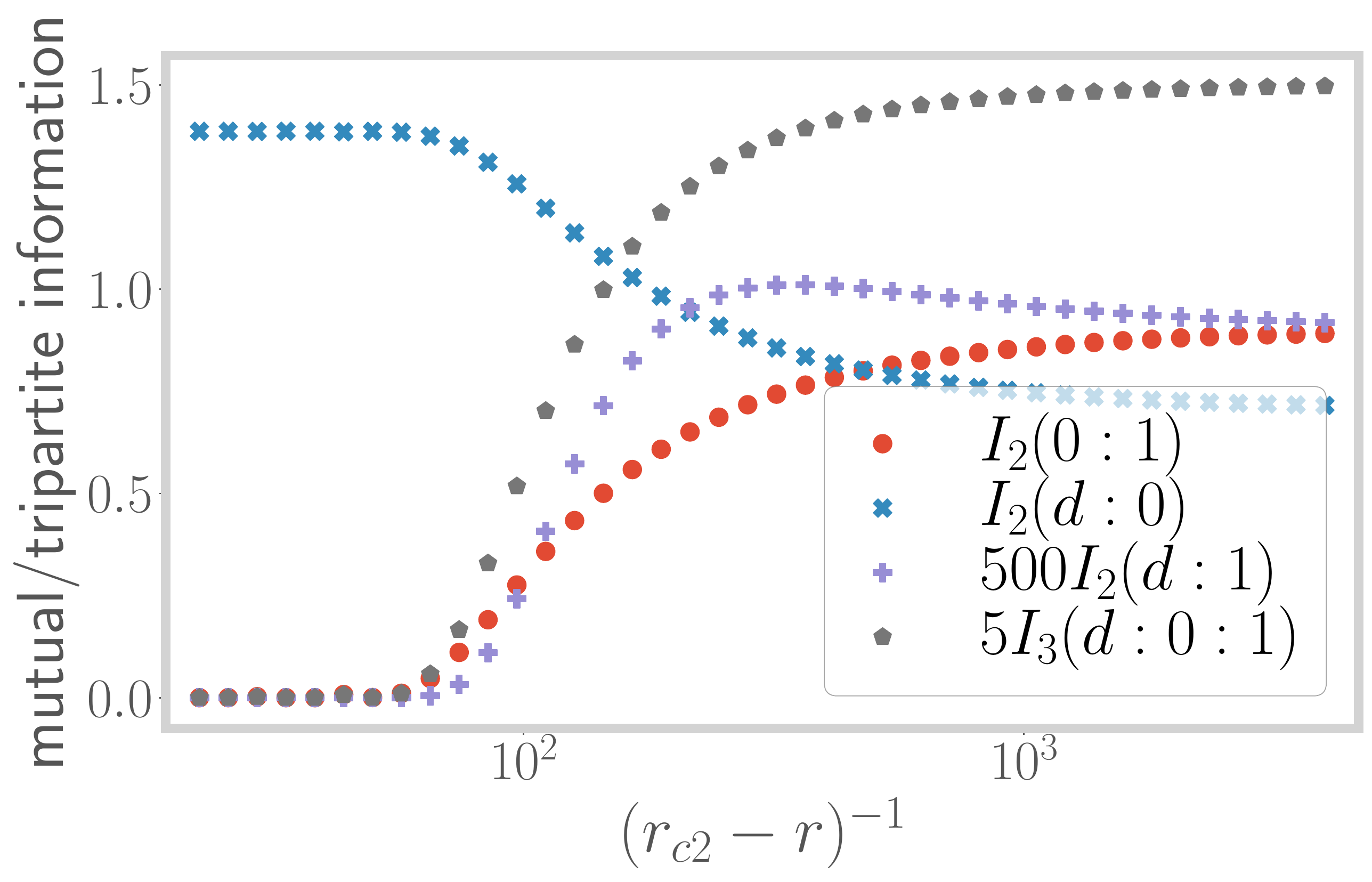}
	\caption{Left: Evolution of the double occupancy of impurity (red) and zeroth sites (blue) ground-state double occupancy, with \(r\). Right: Evolution of various quantum information-theoretic measures, very close to the transition. }
	\label{fig1}
\end{figure}

\begin{figure}[htpb]
	\centering
	\includegraphics[width=0.48\textwidth]{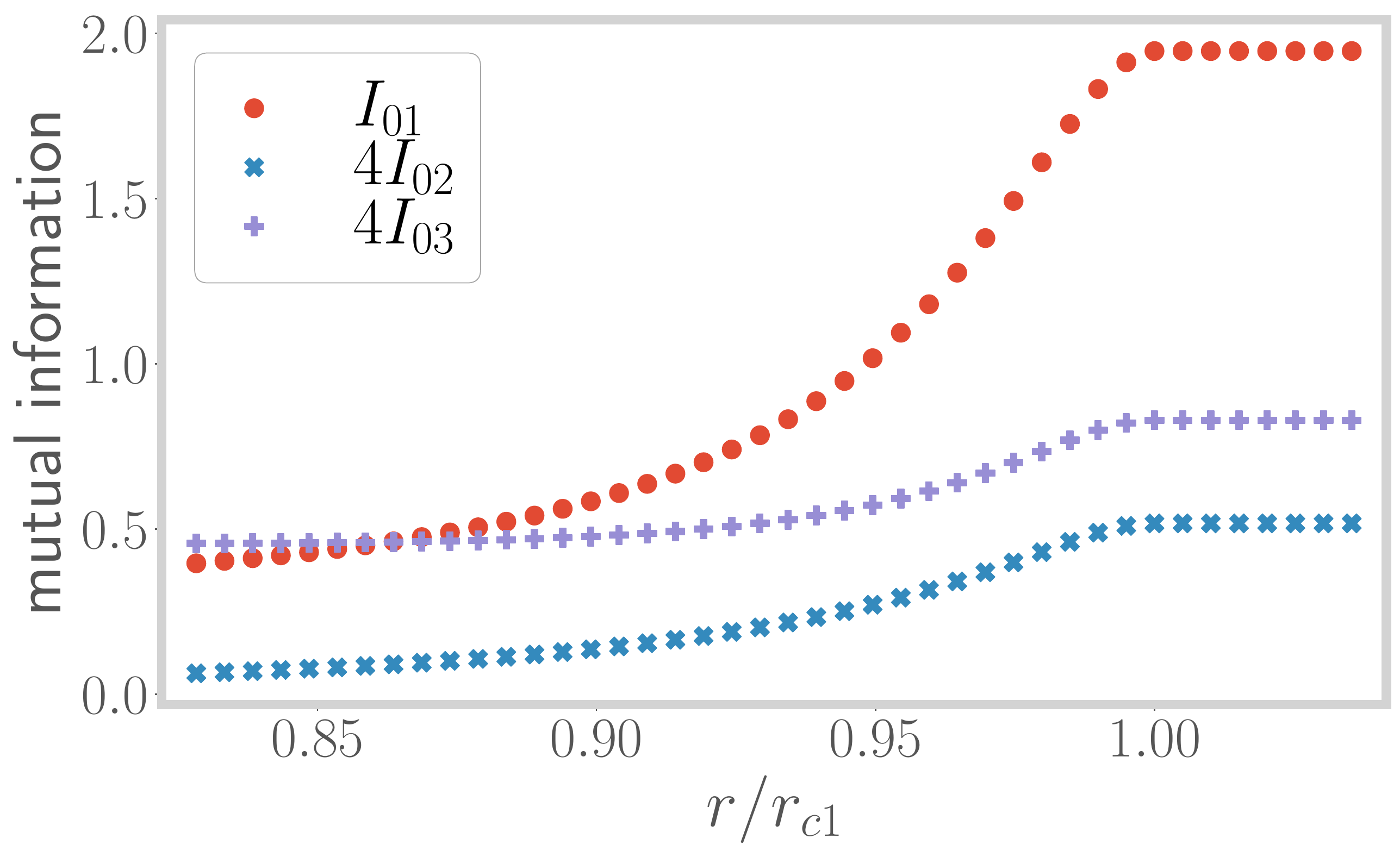}
	\includegraphics[width=0.48\textwidth]{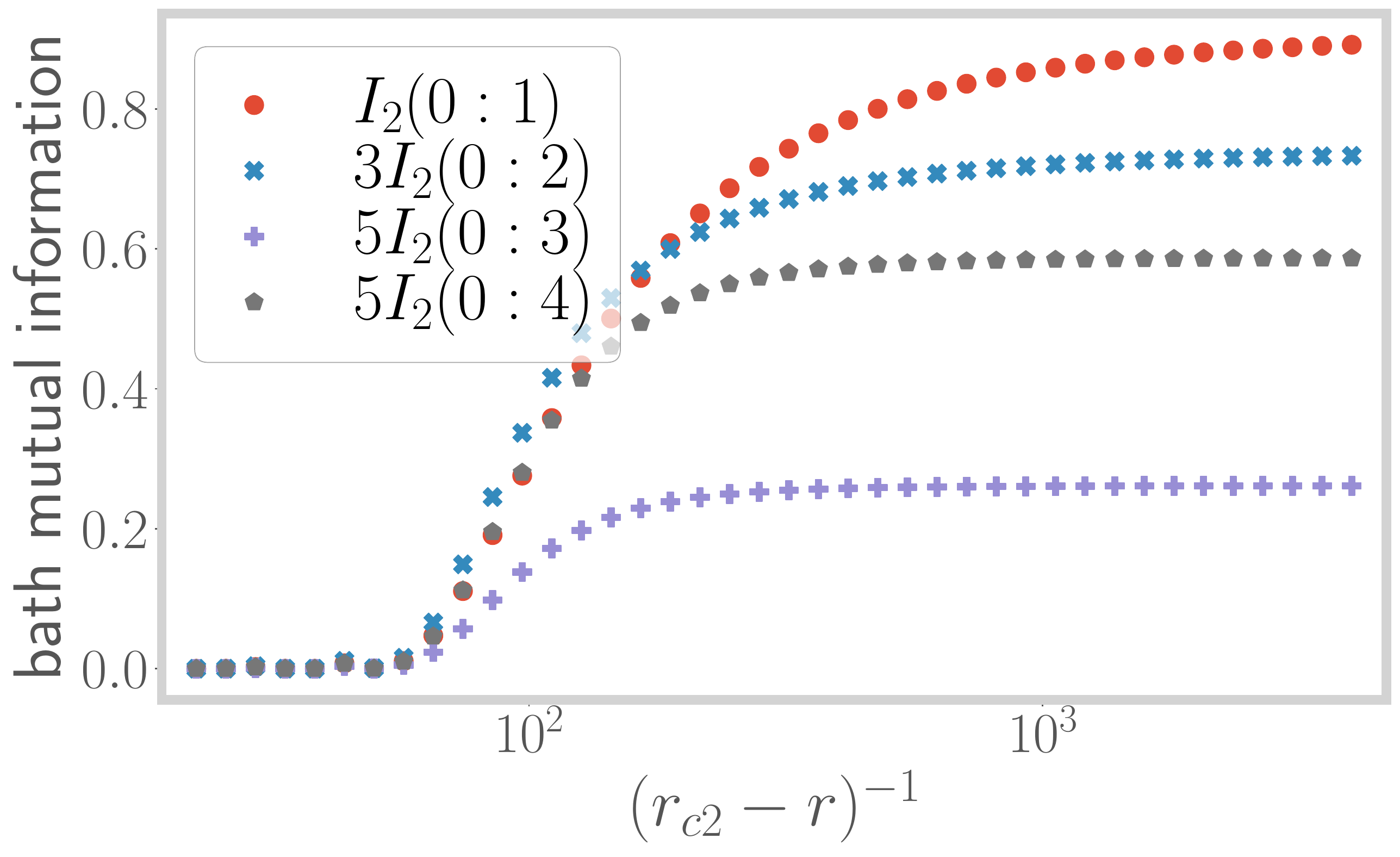}
	\caption{Variation of intra-bath mutual information between zeroth site and sites 1, 2 and 3, near \(r_{c1}\) (left) and \(r_{c2}\) (right).}
	\label{fig2}
\end{figure}

\begin{figure}[htpb]
	\includegraphics[width=0.48\textwidth]{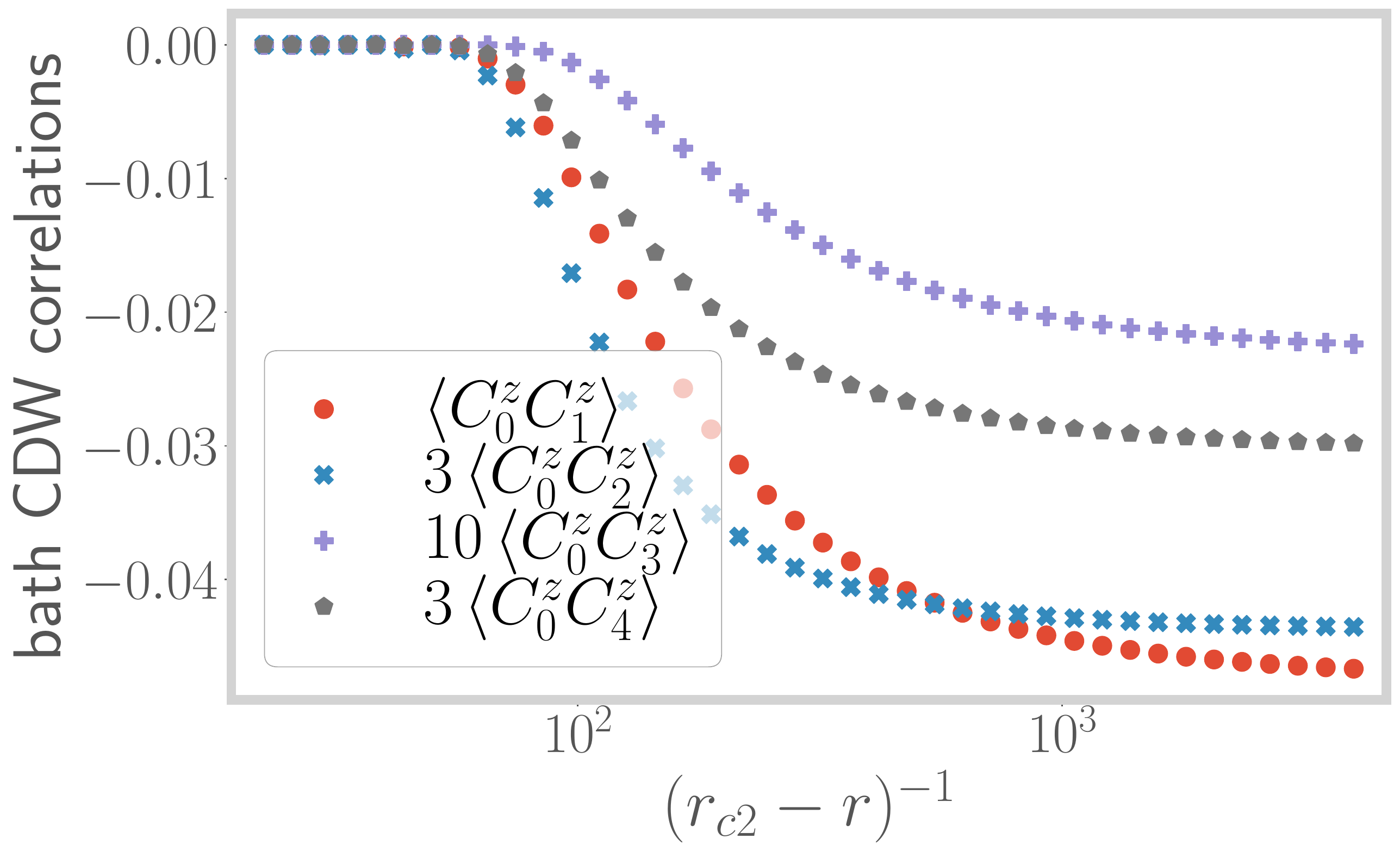}
	\includegraphics[width=0.48\textwidth]{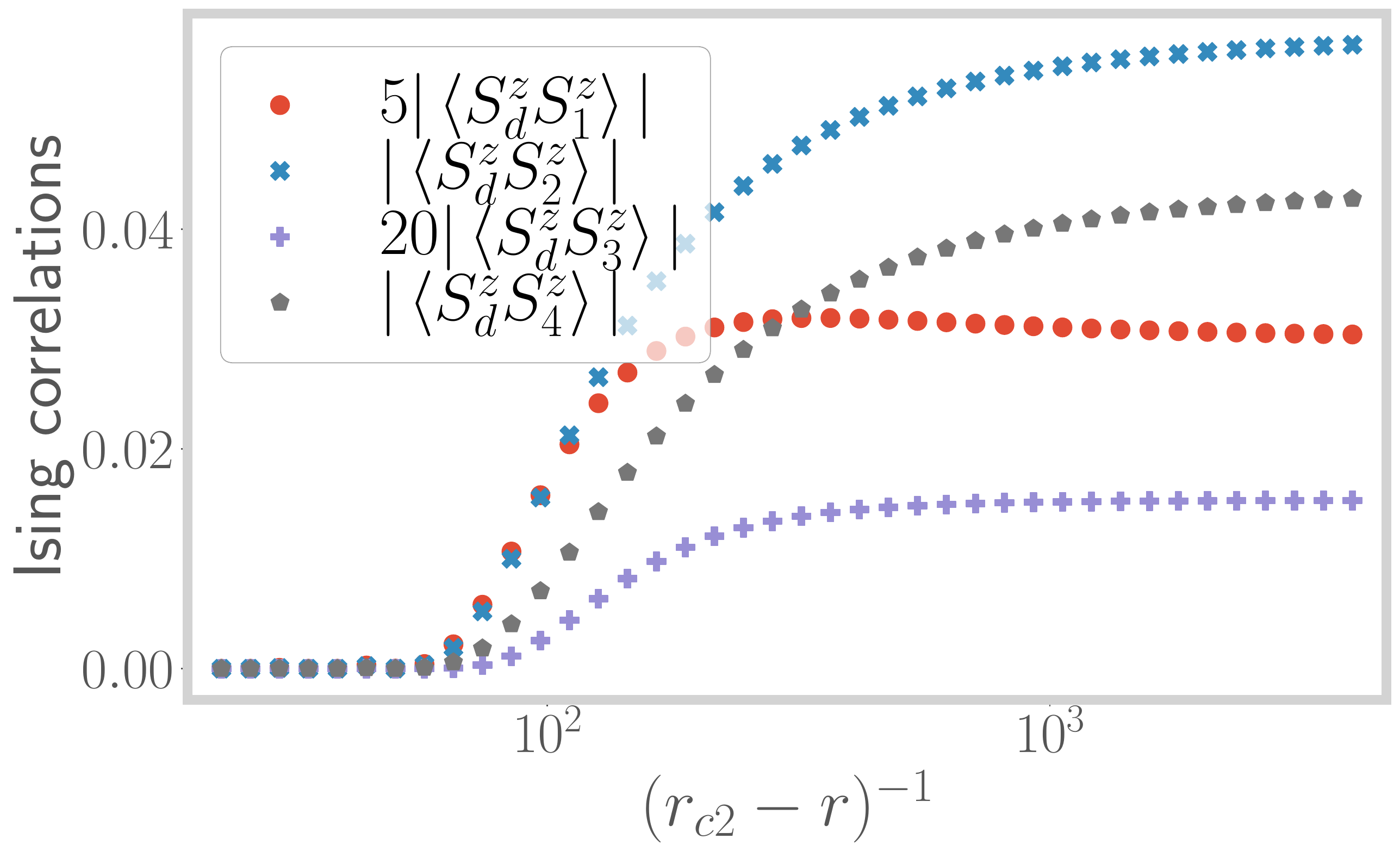}
	\caption{Left: Variation of charge isospin Ising correlations between the zeroth site and bath sites further down the chain. Right: Ising correlations between the impurity site and bath sites beyond the zeroth site. Both sets of correlations show an overall increase as \(r \to r_{c2}\).}
	\label{fig3}
\end{figure}

We have included some additional correlations for the extended SIAM here to highlight the features that are mentioned in the main manuscript. Left panel of Fig.~\eqref{fig1} shows the impurity and zeroth site double occupancy. Both vanish at \(r_{c1}\), indicating the emergence of the Kondo model. Very close to \(r_{c2}\), the zeroth site doublon occupancy picks up (see inset), showing the destruction of the Kondo cloud. Right panel of Fig.~\eqref{fig1} shows various measures of entanglement close to \(r_{c2}\). The impurity-bath mutual information (blue) decreases as the transition is approached, showing the breakdown of impurity screening. The impurity and first site MI (violet), the zeroth site and first site MI (red) and the impurity, zeroth site and first site tripartite information (gray) increase, showing the emergence of long-ranged correlations near the transition. Fig.~\eqref{fig2} shows the emergence of long-ranged intra-bath mutual information near \(r_{c1}\) and \(r_{c2}\), signaling critical behaviour near the ESQPT and QCP respectively. Intra-bath CDW correlations and impurity-bath Ising correlations in Fig.~\eqref{fig3} also show the previously-mentioned emergence of long-ranged correlations. 

\section*{References}

\bibliography{esiam-manuscript}